\documentclass[a4paper,12pt]{mythesis}

\setboolean{masters}{false}             
\setboolean{hasfigures}{false}
\setboolean{hastables}{false}

\usepackage{booktabs} 
\usepackage{array} 
\usepackage{paralist} 
\usepackage{verbatim} 
\usepackage[parfill]{parskip} 

\usepackage[footnotesize , center, labelfont=bf]{caption}
\usepackage{graphicx}
\usepackage{amssymb}
\usepackage{amsmath}
\usepackage{amsthm}
\usepackage{mathrsfs}
\usepackage{varioref}
\usepackage{dsfont}
\usepackage{color}
\usepackage{subfig}
\usepackage[sectionbib, authoryear]{natbib}
\usepackage{chapterbib}
\usepackage{afterpage}
\usepackage[section]{placeins}
\usepackage[a4paper,left=2cm,right=2cm,top=3cm,bottom=3cm]{geometry}
\usepackage[colorlinks = false, citecolor=green, linkcolor=red]{hyperref}
\input{Qcircuit}
\DeclareMathAlphabet{\mathpzc}{OT1}{pzc}{m}{it}

\usepackage{fancyhdr} 
\usepackage{sectsty}
\allsectionsfont{\sffamily\mdseries\upshape} 

\usepackage[nottoc,notlof,notlot]{tocbibind} 
\usepackage[titles,subfigure]{tocloft} 

\bibpunct {[} {]} {;} {a} {,} {,}

\newcommand{\<}{\langle} 
\renewcommand{\>}{\rangle}
\newcommand{\h}{{\mathcal{H}}}
\newcommand{\lo}{{\mathcal{L}}}
\newcommand{\s}{{\mathcal{S}}}
\newcommand{\e}{{\mathcal{E}}}
\newcommand{\oo}{{\mathcal{O}}}
\newcommand{\uu}{{\mathcal{U}}}
\newcommand{\su}{{\mathcal{SU}}}
\newcommand{\ma}{{\mathcal{A}}}
\newcommand{\mb}{{\mathcal{B}}}
\newcommand{\p}{{\mathcal{P}}}
\newcommand{\m}{{\mathcal{M}}}
\newcommand{\co}{\mathds{C}}

\newcommand{\one}{\mathds{1}}
\newcommand{\imag}{\mathfrak{i}}

\theoremstyle{plain}

\title{Nuclear-Electronic Spin systems, Magnetic resonance, and Quantum information processing}
\author{Mohammad Hamed Mohammady}
\department{The Department of Physics and Astronomy}   

\date{}                   

\spacing{1}
\declarationfile{declaration.tex}
\spacing{1}
\acknowledgementsfile{acknowledgement.tex}
\spacing{1}
\abstractfile{abstract.tex}

\begin{document}

\makefrontmatter

%
%

\spacing{1}
\chapter{Introduction}
\section{The dream of quantum computation}
Digital computers have been the defining technology of the second half of the $20^{\mathrm{th}}$ century, whose commercial viability became possible due to the advent of the transistor and the integrated circuit. Due to the decrease in manufacturing costs of integrated circuits, coupled with improvements in miniaturisation of the components,  the power of computers -- characterised by the time taken to solve a particular problem -- has been growing incessantly up to the present day, following closely the famous Moore's law \citep{Mooreslaw}. Some problems, however, require an exponential increase in computational time with respect to a linear increase in the size of the problem. Richard Feynman  made the observation that tracing  the evolution of a quantum state with a computer grows exponentially hard with the size of the Hilbert space; we can solve Schr\"odinger's equation for a two-level atomic system quite efficiently, but simulating a complex virus with quantum degrees of freedom numbering in the millions would take eons. So why not use a quantum system to \emph{simulate} another one? Such a quantum system can be called a \emph{quantum computer} \citep{FeynmanQIP}, in contrast with the current digital computers that operate under the laws of classical physics and are thus named classical computers. Quantum computation is also referred to as  \emph{quantum information processing} (QIP). This observation by Feynman was made more concrete by David Deutsch who asked whether the laws of physics, which to the best of our knowledge are quantum mechanical, could be used to derive the laws of computation \citep{Deutsch85}. He asserted that a computational device built using the laws of quantum physics will be able to simulate arbitrary physical systems efficiently, whereas classical computers can only do so with classical systems. As classical physics is a subset of quantum physics, then, a quantum computer is a generalisation of a classical computer. In the following years there was a surge of interest in developing specific quantum algorithms  that would offer an advantage to the corresponding algorithm running on a classical computer. This culminated in the discovery by Peter Shor   \citep{shors-algorithm} of a quantum algorithm for discovering the prime factors of  numbers, for which no efficient counterpart in classical computing is known to exist.     

Although theoretical research in quantum computing did not stop here, and continues to be a vibrant field of research to this day, more and more people started to contemplate  building a quantum computer in the laboratory capable of performing algorithms such as that developed by Peter Shor. To this end, physical systems were sought that offered access to quantum degrees of freedom. A system with a degree of freedom that can take one of two discrete values, in analogy with the bits of classical computation, is called a \emph{qubit}. By bringing together many such systems and effecting interactions between them, we can have a many-qubit quantum computer. The physical systems considered can be categorised  with respect to the quantum degree of freedom used. The most common  fall into three categories: (a) photon (b) charge and (c) spin.
Two implementations using photons are  
\begin{enumerate}[(i)]
\item
\emph{Linear optics}

This scheme  uses single photons as qubits. Single qubit operations are carried out by beam splitters and phase shifters, and the qubits can be measured destructively with photo detectors. Interaction between the photons can be implemented  deterministically, provided the availability of materials with strong enough  Kerr non-linearities, or stochastically using ancillary photons and measurements \citep{KLM,LOQC-review}.

\item \emph{Cavity QED}

Again, this scheme will use beam splitters and phase shifters on the individual photons, but interactions are effected using an optical cavity containing atoms that couple to the photons \citep{CavityQED-QIP}.  

\end{enumerate}

Charge based quantum computers can be built using

\begin{enumerate}[(i)]
\item \emph{Quantum dots}

The quanta of charge, electrons, are localised in three-dimensional space via electrostatic potentials, such that the number of electrons forms a quantum degree of freedom so as to provide a qubit. Electrostatic gates are used to perform operations on a single qubit, and  the  Coulomb interaction is used to establish  coupling between multiple qubits. The charge is detected via transistors \citep{quantum-dots}.  

\item \emph{Superconductors}

In  superconducting materials at certain temperatures, two electrons may bind to form a Cooper pair which, as with electrons, may be confined in an electrostatic potential. The  individual charge superconducting qubits are controlled by electrostatic gates, and interaction between them is brought about by use of Josephson junctions. As with quantum dots, the qubits are measured by detection of the charge using transistors. It should be noted, however, that due to the limitations posed by charge superconducting qubits, recent efforts in superconducting QIP have focused on  the   phase and flux qubit implementations instead  \citep{superconductingqubit}.

\end{enumerate}

Spin based quantum computing largely falls into the two camps of 

\begin{enumerate}[(i)]
\item \emph{Trapped ions}    

The ions are trapped with lasers, and offer their   internal energy states as the relevant quantum degree of freedom. The ions are individually manipulated by laser pulses, and  the  interactions between them are induced by means of phonons. Measurement is performed by detecting the fluorescence seen when a probe laser pulse is resonant with the hyperfine levels    \citep{IonTrapMBQC}. 

\item \emph{Magnetic resonance}

In this case, the spin is associated with either the electron or nuclear spin (or both) of a system. Individual spins are manipulated by means of magnetic resonance, which is nuclear magnetic resonance (NMR) for nuclear spins, electron spin resonance (ESR) for electron spins, and electron nuclear double resonance (ENDOR) for both. Interactions are induced by exchange, or dipolar, coupling between the spins. Measurements are performed weakly by the free induction induced by the spin's Larmor precession or projectively with, for example, magnetic resonance forced microscopy. A good review article for magnetic resonance QIP  is \citep{NMR-QIP}. 
\end{enumerate}

All of these schemes have their own pros and cons, and it is beyond the scope of this work to compare them all. Here, we shall focus on the magnetic resonance implementation.

\section{Interactions in open systems: the inherent contradiction}
No matter what the physical implementation is, there are two necessary criteria  for robust quantum computation which cannot both be satisfied \emph{simultaneously}. To be able to manipulate the quantum system of interest, it must be an \emph{open quantum system} where it interacts with its environment, such as the measuring apparatus. However, for the computation to be quantum mechanical, the coherence established must not be destroyed and, resultantly, the system must be a \emph{closed quantum system} where it does not interact with its environment.  Here lies one of the key problems facing quantum computation, the resolution of which rests upon the ability to control the interaction of the quantum system of interest with its environment. In other words to alternate it, at will, between an open and closed quantum system.

Spin-based QIP encapsulates this issue quite neatly. We may choose as our qubits either the electron spin or the nuclear spin where the former interacts much more strongly with its environment than the latter. Consequently, while the electron spin can be measured and manipulated easily and at a rapid rate, it also decoheres  very quickly. In contrast, while the nuclear spin has longer coherence times, it is manipulated and measured much more slowly, and  with much greater difficulty. One possible solution that immediately presents itself is to be able to switch between the two systems at will: to transfer the quantum information to the electron spin for manipulation and measurement, and to then transfer it again to the nuclear spin for storage.   
\section{Magnetic resonance QIP and Kane's proposal}
 Nuclear magnetic resonance  QIP, using the nuclear spins of an ensemble of molecules in solution at room temperature to provide  qubits,  was one of the first to be demonstrated experimentally. This implementation was rife with problems, however, which include among them: 
\begin{enumerate}[(i)]
\item Because the energy difference between the  spin states is very small, they are generally in a highly mixed thermal equilibrium state. Cooling techniques cannot be used to alleviate this problem as the spins are in solution. 
\item Since measurements  allowed on such ensemble states are statistical averages of traceless observables,  increasing the number of qubits in the system decreases the detected signal. This renders such systems as inherently unscalable \citep{NMR-QIP-bad}. 
\item  It was shown that as the number of physical qubits is increased,  the entanglement of the system vanishes. As entanglement is thought to be one of the key factors that distinguishes QIP from classical computation, NMR implementations in liquid solution with more than a few physical qubits are entirely classical \citep{NMR-QIP-separable}. \end{enumerate}

Solid-state NMR seemed to offer many advantages to solution NMR \citep{solid-state-NMR-QIP}. These advantages include: (a)  longer coherence times, (b) higher susceptibility to polarisation, (c)  stronger coupling between spins, thus enabling faster multi-qubit gates, and (d) ability to dynamically reset the qubits so as to enable error correction protocols. Also, solid-state spin architectures allow for the possibility of performing strong, projective measurements on single spins \citep{MRFM-single-spin}, as opposed to the weak ensemble measurements of traditional magnetic resonance.  The best known proposal for scalable QIP in solid-state was put forward by Bruce Kane \citep{kane1998}, who suggested the use of phosphorus-doped silicon (Si:P).  Such a system at low temperatures offers a localised donor nuclear spin -- which for phosphorus is spin one-half -- coupled to a localised donor electron spin, also of spin one-half, by the hyperfine interaction. Each localised nuclear-electronic spin system can be considered as two coupled qubits. The nuclear and electronic degrees of freedom can be manipulated by global ENDOR pulses. To allow for selectivity, $A$-gates at each site control the hyperfine coupling strength between the electron and nuclear spins, and hence the  transition frequencies. Nearest neighbour interactions are mediated via the electron spins, which have an exchange interaction due to the overlap of their wavefunctions. The strength of such interactions can be controlled by $J$-gates that alter the degree of electronic wavefunction overlap. At the end of the computation the electron spins at each site are measured. In such a scheme the nuclear spin, having much longer coherence times, houses the quantum information, and the electron spin is used as an ancillary system to enable ENDOR pulse selectivity, and nearest neighbour  interactions. 

\begin{figure}[!htb]
\centering
\includegraphics[width=3.5 in]{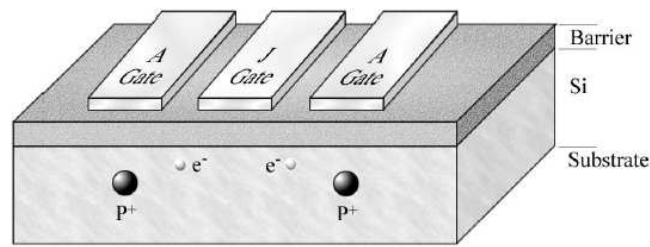}
\caption{The Kane quantum computer, using the donated nuclear and electron spins of phosphorus impurities in  silicon (Si:P). $A$-gates control the hyperfine coupling between the electron and nuclear spins of a single donor, and $J$-gates control the interaction between nearest neighbour electron spins. Image taken from \url{http://www.ccms.uq.edu.au/research_materials.htm}.}
\end{figure}

Although present technology does not allow for the manufacture of such an intricate device, research in Si:P with QIP in mind has been flourishing over the last decade.  The research has included:
precision placement of dopants in silicon \citep{schofield2003}, Ensemble measurement of spins by electrical detection \citep{mccamey-2006,Steger-2006,Morley-2008-1,morishita-2009}, detection of single or small number of spins  \citep{Fu-2004,Morello-2009,single-shot-spin-measurement-Si}, decoherence of  the spins \citep{Tyryshkin-2003,Morley-2008-1,McCamey-2012}, initialisation of spins in a highly polarised, or hyperpolarised state \citep{mccamey-hyperpolarisation,Steger2011}, two-qubit interactions \citep{Greenland-2010}, entanglement \citep{Simmons-2011}, and the transfer of quantum information between the nuclear and electronic spins  \citep{Morton2008}.

One of the obstacles to QIP is ``decoherence''. In silicon impurity architectures, the decoherence mechanisms mainly fall under two categories; the temperature dependent spin-boson, and temperature independent spin-spin mechanisms. The spin-boson mechanism involves an interaction between the donor spin and the quanta of vibration in the silicon crystal, or \emph{phonons}, which is dependent on the temperature of the system and can be lowered (but not entirely removed) by cryogenic cooling. At temperatures above 7 K, the interaction is dominated by the Orbach process \citep{Orbach}, whereas at temperatures of $\sim 5$ K, the dominant  mechanisms are due to single-phonon, or ``direct'',  and two-phonon processes. Further lowering of the  temperature results in the single phonon mechanism  dominating \citep{FeherRelaxation}.   Such  processes lead to depolarisation of the spins, which follows an exponential decay of the form $e^{-\tau/T_1}$. The parameter $T_1$ is referred to as the \emph{longitudinal}, or \emph{spin-lattice} relaxation time. At temperatures where only the single phonon mechanism prevails,  for electron spins of silicon impurities $T_1 \propto 1/ (T B_0^4)$ where $T$ is the temperature and $B_0$ the applied magnetic field. Therefore, the smaller the magnetic field, the longer the $T_1$ time for the electron becomes \citep{Rothrelaxation,hasegawa}. Furthermore $T_1$ poses an upper bound on the  \emph{transverse}, or \emph{spin-spin}, relaxation time $T_2$, itself also characterised  by the exponential decay  $e^{-\tau/T_2}$,   as $T_2\leqslant 2T_1$. 
Finally,  the spin-boson mechanism   affects the electron spin more strongly than it does the nuclear spin;  for Si:P at temperatures of 6-12 K the $T_1$ times for the two spin types obey the relationship $T_{1n}\approx 10^2 T_{1e}$. 

The spin-spin mechanism for decoherence comes from undesirable interactions between the system spin and its surrounding spins, which in natural silicon is due to both the $^{29}$Si isotopes that have spin one-half, and the undesirable interaction between the donors themselves. To a very good approximation, it is only the electron spin of the donor which is involved in this interaction as the nuclear donor spin is far more localised.  Such a process generally only leads to dephasing. The mechanism by which  dephasing is brought about by the surrounding \emph{spin bath} of $^{29}$Si  is called \emph{spectral diffusion}, delineated by \citep{Witzel-spectral-diffusion,Yang-spectral-diffusion}. Colloquially, spectral diffusion is brought about when the members of a spin bath which the qubit interacts with \emph{flip-flop}. The region of the crystal in which the flip-flop process has an effect is called the \emph{active region}, in which the dipolar coupling among the members of the spin bath are comparable to their interaction strength with the donor spin, termed the \emph{super hyperfine coupling}.

 For QIP it is not so much the coherence time itself that is important, but the ratio of the coherence time with respect to computation time; if the error probability of a quantum operation, determined by this ratio, is sufficiently small, then we may perform our quantum computation fault-tolerantly. We perform our quantum gates on single donors by using magnetic resonance pulses whose speed, given a certain pulse strength, depends on both the gyromagnetic ratio of the spin species considered and the degree to which the desired transition frequency differs from those of unwanted transitions: the larger the difference between these frequencies, the faster the pulse may be such that we maintain \emph{selective} control of the desired transition. The gyromagnetic ratio for the electron spin, in Si:P, is  $\sim 10^3$ times larger than that of the nuclear spin. Also, at sufficiently large magnetic fields the ESR transition frequencies differ to a greater degree from the other transitions than do the NMR transition frequencies. These factors, together,  indicate that in principle ESR pulses can be many times faster than NMR ones, which has lead many researchers to consider using the electron spin for quantum computation.  However, as shown by \citep{Morton2008}, it is possible to use the best of both worlds, and use the nuclear spins for quantum memory and electron spins for processing.

There is a limit to how low the concentration of donors in silicon can be to allow for traditional, weak ensemble ENDOR detection, with the threshold for natural silicon being at $\sim10^{13}$ cm$^{-3}$  phosphorus nuclei. NMR requires even higher concentrations. This poses a fundamental limitation on how much the undesirable donor-donor induced dephasing can be reduced by.
Recently \citep{Tyryshkin-2011} studied the coherence times of highly enriched Si, with the concentration of $^{29}$Si lowered to less than 50 ppm, and saw the electron spin coherence times $T_2$ of Si:P raised to $\sim 1$ s at a temperature of $\sim 5$ K. Even more recently \citep{Steger-Si-180s} demonstrated that such  highly purified silicon allows for rapid hyperpolarisation of the nuclear spins of phosphorus. This allows for an improvement in detection of the donor spins, allowing the phosphorus concentrations to be lowered to $\sim 10^{12}$ cm$^{-3}$,  that leads to an even further increase in $T_2$ times, which for the nuclear spins  were measured to be $\sim 180$ s.

\section{Enter bismuth}

The difficulty of manufacturing the Kane quantum computer, which requires precise  controllable spin-spin interactions by electrodes, has spawned an interest in other donor species in silicon. It has been shown by \citep{Lloyd-1993,Benjamin-global-qubit-qubit} that, for QIP, it suffices to control the spin-spin interactions of a many-body spin system collectively, and not individually, so long as more than one spin species is used.  \citep{Stoneham-2003} proposed the use of different donor species in silicon to allow for such a global control, as each species will have different resonance frequencies. The two species are placed such that in their ground states none will interact. By optically exciting one species to their excited Rydberg state, we can effect an intermediated interaction between members of the other species, in which we store our quantum information. This scheme allows us to dispense with the requirement of  $J$-gate electrodes.  In the same year, \citep{Sougato-Benjamin-2003} demonstrated that it is possible to use  multi-species spin systems to perform quantum computation with the spin-spin interactions always being on, provided we are able to collectively tune the energies of each individual spin species. 

In more recent times, one specific donor species with favourable properties has been identified.   Bismuth belongs to the same group in the periodic table as phosphorus, group V, and has many extremal properties. Bismuth doped silicon (Si:Bi) offers a nuclear-electronic spin system very similar to Si:P and indeed other group V impurities in silicon, but with some crucial differences.
\begin{enumerate}[(i)]
\item It is the heaviest group V donor with the highest  binding energy of $\sim 71$ meV, compared with that of Si:P which is $\sim 2$ meV.
\item  The effective Hamiltonian for Si:Bi at low temperatures gives an isotropic hyperfine interaction between the nuclear and electron spins with the highest strength of $1.4754$ GHz, compared to that of Si:P which is $117.5$ MHz.
\item Both Si:Bi and Si:P donate electrons of spin one-half. However, while the latter has a nuclear spin which is also one-half, the former has the largest nuclear spin of $9/2$.
\end{enumerate}

 Decoherence is still an issue to be dealt with; if bismuth impurities have much shorter coherence times, their inclusion with phosphorus will not be advantageous. It has not been until  recently that the relaxation processes of Si:Bi have become the subject of intensive study   \citep{Belli-bismuth-eseem}.    \citep{Morley-2010} showed that Si:Bi has  electron spin coherence times at least as long as Si:P at comparable temperatures and $^{29}$Si concentrations.  Indeed,  owing to the dominant thermalisation mechanism at temperatures above 7 K being the Orbach process, whose effect is mitigated by the binding energy of the donor, Si:Bi has longer electron $T_1$ times than Si:P in this temperature regime. Furthermore, as shown by  \citep{George-2010}, at low temperatures where the electron $T_2$ time is limited by spectral diffusion, Si:Bi has electron $T_2$ times approximately $30 \% $ longer than Si:P in natural silicon, owing again to the  greater binding energy which reduces the donor Bohr radius, thereby shrinking the active region. \citep{sekiguchi-2010} demonstrated a high capacity for hyperpolarisation of Si:Bi, $\sim90 \ \%$, which  is likely to be improved upon in the near future. Also, as was shown by my collaborators and me both theoretically  \citep{Mohammady-2010,Mohammady-2012} and experimentally \citep{Morley-hybrid},  the large nuclear spin and hyperfine coupling of Si:Bi  results in some interesting spectroscopic properties. The large hyperfine interaction strength brings about entanglement of the nuclear and electronic spin degrees of freedom  in the magnetic field region of $B_0 \lesssim\ 0.6$ T.  In this regime, transitions that at high magnetic fields are classified as  NMR can be achieved with speeds of the same order of magnitude that is characteristic  of ESR.  Provided that the relaxation times for the transitions that, at high fields, are labeled NMR do not significantly decrease in this field regime, Si:Bi  has the potential for offering a  QIP platform that is more robust than Si:P.  The large nuclear spin offered by Si:Bi offers a further possible advantage over Si:P. This allows for certain optimal working points (OWPs) wherein pure dephasing processes such as spectral diffusion are reduced to a negligible amount, and under certain limiting conditions can be removed entirely. Theoretical work has been done in this regard by my colleagues and me both in \citep{Mohammady-2012} and \citep{Balian-diffusion}.

\section{Thesis outline}

This thesis aims to explore the prospect of using Si:Bi, or indeed any nuclear-electronic spin system obeying a similar Hamiltonian, as a platform for QIP. Proof of principle arguments composed of analytic, numerical, and experimental studies will be considered. It is my aim to keep this work as self contained as is feasible, and progress my arguments from the general and abstract down to  concrete examples. Furthermore, as this study is quantum mechanical in nature, I will attempt to describe the relevant phenomena in the elegant formalism of operational quantum mechanics.  Consequently, this thesis will be divided into three parts. 

{\bf{PART I}}: Here,  I will lay down the mathematical formalism for quantum theory of finite dimensional systems, measurement and control with magnetic resonance, and the fundamentals of quantum computing. 

{\bf{PART II}}: Here, you will find an exposition of  my published research. This will be composed of four chapters. In the first two chapters I will keep my arguments as general as possible, and provide a purely analytical study pertaining to nuclear-electronic spin systems. In the first chapter I shall provide a study of the Hamiltonian and coherent dynamics alone, and complete the study in the second chapter by considering the nuclear-electronic spin system as an open system subject to decoherence. In the third chapter, I shall then move to a more concrete setting and use the established theory to provide numerical predictions for the magnetic resonance properties of Si:Bi and Si:P. In the fourth chapter, I shall conclude by providing experimental data pertaining to Si:Bi and compare with our theoretical predictions.

 {\bf{PART III}}: This final section will consist of arguments as to the application of Si:Bi for quantum information processing, and concluding remarks.

As a further note, throughout this thesis, I set $\hbar=1$. Indeed, as all relevant calculations yielding numerical values are in units of frequency (spectroscopy), $\hbar$ need not be considered at all.

\spacing{1} 
 \bibliographystyle{plainnat}                                 
 \bibliography{references}

\part{Theoretical Background}
\spacing{1}
\chapter{Quantum theory}\label{quantum theory}
\section{Introduction}

\begin{figure}[!htb]
\centering
\includegraphics[width=5in]{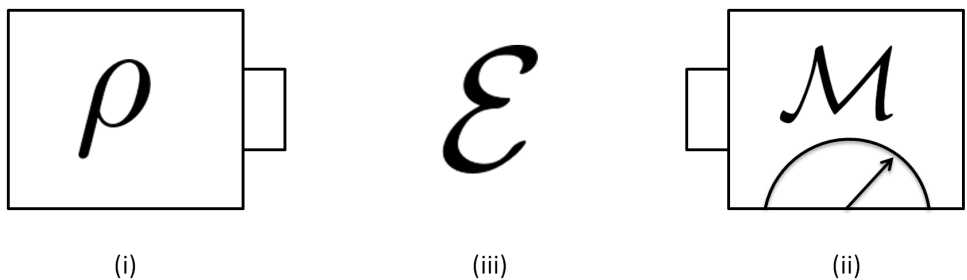} \caption{ A general schematic of an experiment in the paradigm of operational quantum mechanics: (i) preparation, (ii) measurement, and (iii) transformation.   } \label{operationalquantumfig}
\end{figure}

When asked to describe what quantum mechanics is, most scientifically literate individuals will answer with a statement equivalent to ``\emph{the study of the physical behaviour of microscopic things}''. Such an answer is not unexpected, given that the development of the theory was largely\footnote{The pioneer of quantum theory Max Planck, immortalised by  the  constant named after him, $h$,  postulated that the energy of electro-magnetic waves was \emph{quantised} proportional to their frequency. This finding was used to solve the ultraviolate catastrophe of black body radiation, which is clearly a domain of macroscopic objects as large as stars. Albert Einstein, however, sharpened this quantisation in his study of the photoelectric effect to introduce the idea of the \emph{photon} which is a \emph{tiny} corpuscular object.} motivated by the experimental study of \emph{microscopic objects} such as photons and electrons in the early half of the 20$^{th}$ Century. Upon further reflection, however, such a description proves to be unsatisfactory as  quantum theory is used to study certain aspects of the behaviour of macroscopic objects such as semiconductors and metals.   This motivates another description of what quantum theory is at its core. The minimal understanding of quantum theory, also known as operational quantum theory,  is exemplified by  Asher Peres in the sentence:

\begin{centering}

``\emph{Quantum theory gives   probabilities for measurement outcomes    following a specified preparation of a quantum system}.''

\end{centering}

Let us ponder the meaning of the key elements within this statement.   A  \emph{preparation} is an equivalence class of  well-defined and repeatable tasks such as turning levers on a machine, whilst a \emph{measurement} entails probing the system in some systematic  way so as to extract information from it. These
\emph{measurement outcomes}, characterised by the permanent record they leave either in the apparatus itself or the lab note of the physicist, are described by entirely classical means. However,  for this classical measurement outcome to be used to make a claim about the  object of study, a theoretical model must be used. This model establishes a connection between this classical measurement outcome to some underlying degrees of freedom of the object of study. In quantum theory   we may call this the \emph{quantum system}.  
 Finally, the meaning of \emph{probability} here takes the usual sense as the relative frequency of a particular measurement outcome as a fraction of the total number of measurements, in the limit of the number of measurements tending to infinity.

An illustrative concrete example   is the Stern-Gerlach experiment \citep{Stern-Gerlach}. Here, the physicist performs some well-defined set of tasks to prepare the  particles in a particular way, and chooses his measurement procedure by passing these particles through a magnetic field gradient of a specified orientation. He then completes the measurement by observing the position of the particles after they leave the magnetic field, which is an entirely classical quantity. The probability of the measurement outcome is obtained by repeating this process many times, while the way this experiment is interpreted is by invoking the concept of a degree of freedom called the spin. Therefore, the quantum system on which the experiment is conducted is the spin.

 In this chapter, I aim to provide an overview of operational quantum mechanics. An experiment  described by operational quantum mechanics can be separated into three parts, as shown in Fig.\ref{operationalquantumfig}, where the components are:
\begin{enumerate}[(i)]
\item \emph{Preparation}

 An equivalence class of preparations of the system are described by a quantum state $\rho$, which is an operator on a Hilbert space.

\item \emph{Measurement}

Measurements are carried out on quantum states, and give a measurement value, or event, coupled with a probability.

\item \emph{Transformation}

Between the preparation and measurement the quantum state can be altered by a linear quantum operation.
\end{enumerate}
It should be noted, however,  that \emph{how} such a separation is made is somewhat arbitrary; the processes of  transformation and measurement themselves constitute a preparation, while the evolution process can also be absorbed into the measurement process. Regardless, we may always conceptually separate a quantum mechanical experiment in this way.
In most introductory courses on quantum theory quantum states are always assumed to be pure, measurements are always assumed to be projective, and operations are always assumed to be unitary. Here, I will give descriptions for what general states, measurements and operations are. In each case it turns out that if we are allowed to consider our system as a small subspace of a larger Hilbert space, then all general states, measurements and operations on this subspace can indeed be seen as pure states, projective measurements and unitary transformations on the larger Hilbert space.

The literature available on quantum theory is indeed vast, and I cannot hope to provide an exhaustive list here. However, I shall name a small selection of books, each with different aims, that should provide a good overview of the topic. Good modern textbooks for computation of numerical values are \citep{sakurai}  and \citep{audretsch}, while \citep{von-Neumann} and \citep{Heinosaari} concerns the  mathematical structure of quantum theory. The theory of measurement is covered extensively in \citep{Busch-operational} and \citep{Busch-measurement}.  The books \citep{Wheeler-Zurek}, \citep{Perez} and \citep{Bub} deal  with foundational issues.

\section{Basic concepts}

\subsection{The Hilbert space}
A Hilbert space, $\h$, is a complete  inner product vector space. To explain this, let us make the following observations regarding  inner product vector spaces $\mathds{V}$:  

\begin{enumerate}[(i)]
\item 

The inner product  for  $\mathds{V}$  is a function defined as 
$\langle \cdot |\cdot\rangle : \mathds{V}\times\mathds{V} \to \mathds{C}
$ which maps any pair of vectors \footnote{In the Dirac notation, a vector $\phi$ is denoted as $|\phi\>$. } $\psi  ,  \varphi \in \mathds{V}$ to an element $c$  of the complex numbers $\mathds{C}$.
\item
 The inner product can be used to define a norm $\|\cdot \|$  of a vector $\psi$  as $  \| \psi \| := \sqrt{|\langle \psi|\psi\rangle|}$, which can itself be used to define a distance measure between two vectors as $d(\psi,\varphi):=\|\psi-\varphi\|$. Therefore, an inner product space is also a normed space, as well as a metric space.

\end{enumerate}
 
   Since an inner product space has metric properties, we can talk about convergent sequences\footnote{A sequence of vectors $\{\phi_n\}_{n=1}^\infty$ converges to a vector $\psi$ if for all $\epsilon > 0$ there exists an integer $N$ such that for every $n>N$, $d(\phi_n,\psi)< \epsilon$.} of vectors $\{\phi_i \}_{i=1}^\infty$ in this space. This inner product space  is said to be complete, and hence a Hilbert space, if  and only if every  absolutely convergent sequence,  meaning that it satisfies $\sum_{i=1}^\infty \| \phi_i  \|^2  < \infty$, is convergent. The dimension, $d$, of a Hilbert space  is the cardinality of the largest set of orthonormal vectors in that space $\{\phi_i \in \h : \<\phi_i|\phi_j\>=\delta_{ij} \}$, where $\delta_{ij}$ is a Kronecker delta function. Such a set is referred to as the orthonormal basis\footnote{The inner product can be used to expand any vector $\psi$    with respect to an orthonormal basis $\{\phi_i\}$ as $\psi=\sum_i\<\phi_i|\psi\>\phi_i$. } that spans the Hilbert space.
All finite dimensional $\mathds{V}$ are complete and therefore also Hilbert spaces. In fact, it turns out that all finite dimensional Hilbert spaces are isomorphic to complex inner product spaces $\mathds{C}^d$. Not all infinite dimensional inner product spaces are Hilbert spaces, however.   Hilbert spaces of infinite dimension that are of interest are \emph{separable}, being spanned by the delta functions $\{\delta_i: i \in \mathds{N}\}$ where $\mathds{N}$ are the natural numbers, which means  that their dimension is countably\footnote{A set is countably infinite if there exists a one-to-one correspondence between this set and the natural numbers $\mathds{N}$.} infinite. These are most commonly encountered as the Hilbert spaces of square-integrable functions on $\mathds{R^N}$,  denoted as  $\ell^2(\mathds{R^N})$,  where $\mathds{R}$ represents the real numbers.

\subsection{Operators on Hilbert space}

Now that we have established the Hilbert space, we may talk of linear mappings, or \emph{operators}, on this space. Consider  an operator on a Hilbert space defined as $L:\h \to \h$.   Such operators are said to be bounded if there exists a  $0 \leqslant t< \infty$ such that $\| L\psi\|\leqslant t \| \psi \| \ \forall \ \psi \in \h$.  The space of bounded operators itself forms a vector space, and all operators on the finite dimensional space $\co^d$ are bounded. The norm of a bounded operator can therefore be given as $\|L \|=\sup_{\|\psi\|=1}\|L \psi \| < \infty$.  The trace of a bounded operator, $\mathrm{tr}[\cdot]$, is a function defined as 
\begin{equation}\mathrm{tr}[L]:=\sum_{i}\<\phi_i|L\phi_i\> 
\end{equation}
 which is independent of the basis $\{\phi_i\}$ used.   This defines a subclass of the space of bounded operators, called the trace-class operators $\lo(\h)$, such that $\mathrm{tr}[L] < \infty \ \forall \ L \in \lo(\h)$. As before, all finite dimensional operators  are automatically also  trace-class operators.  The trace operation gives us a method of calculating  the Hilbert-Schmidt inner product between any  $L,Y\ \in\lo(\h)$ as $\<L|Y\>_{\mathrm{HS}}:=\mathrm{tr}[L^\dagger Y]$, and the Hilbert-Schmidt norm as $\|L\|_{\mathrm{HS}}:=\sqrt{\mathrm{tr}[L^\dagger L]}$. Here, the operator $L^\dagger$ is the \emph{adjoint}, or \emph{Hermitian conjugate} of $L$, characterised by the identity $\<\varphi|L\psi\>=\<L^\dagger \varphi|\psi\> \ \forall \ \psi ,\varphi \in \h$. Just as the inner product can be used to expand a vector in $\h$ with respect to an orthonormal basis, so too can the Hilbert-Schmidt inner product be used to expand an operator in $\lo(\h)$. Given $\h$, with basis $\{\phi_i\}$, we may obtain an orthonormal basis in $\lo(\h)$  as $\{\mathbb{E}_{ij}\}$ \footnote{In the Dirac notation  $\mathbb{E}_{ij}$ is represented as $|\phi_i\>\<\phi_j|$ such that $|\phi_i\>\<\phi_j| \psi=\<\phi_j|\psi\>\phi_i \ \forall \ \psi \in \h$.} which, for finite dimensional Hilbert spaces $\co^d$, has a dimension $d^2$. An operator $L$ can thus be expanded as
\begin{equation}
L=\sum_{i,j} \<\mathbb{E}_{ij}|L\>_{\mathrm{HS}} \mathbb{E}_{ij} .
\end{equation}
 Two important operators are the identity operator $\mathds{1}$ and the null operator $\mathds{O}$, characterised respectively by $\mathds{1}\psi=\psi \ \forall \ \psi \in \h$, and $\mathds{O}\psi= \phi_\mathrm{null} \ \forall \ \psi \in \h$. Here, $\phi_\mathrm{null}$ represents the null vector which has the property $\|\phi_\mathrm{null}\|=0$.  

Let us denote the subset of $\lo(\h)$ in which all operators are self-adjoint   as $\mathcal{L}_s(\h)$. An operator $L$ is self-adjoint\footnote{   An  operator $L:\mathcal{D}(L)  \to \h$, where $\mathcal{D}(L)\subseteq \mathcal{D}(L^\dagger)\subseteq\h$ is a dense domain of $L$, is self-adjoint if $\<\varphi|L\psi\>=\<L\varphi|\psi\> \ \forall \ \psi,\phi \in\mathcal{D}(L)$ and $\mathcal{D}(L)=\mathcal{D}(L^\dagger)$. This operator is also Hermitian if it is bounded, which is true when $\mathcal{D}(L)=\mathcal{D}(L^\dagger)=\h$. } if $\<\varphi |L\psi\>=\<L \varphi|\psi\> \ \forall  \ \psi,\varphi \in \h$, equivalently stated as $L=L^\dagger$.  The positive operators where $\< \psi|L\psi\> \geqslant 0 \ \forall \ \psi \in \h$ are a subset of the self-adjoint operators. The positivity condition can be used to determine the ordering relation between two positive operators $L$ and $Y$; we may say that   $L\geqslant Y$ if $\<\psi|(L-Y)\psi\>\geqslant 0 \ \forall \ \psi \in \h$. Furthermore, any positive operator $Y$ can be composed  as $Y=L^{\dagger}L$ for some linear operator $L \in\lo(\h)$ that need not itself be positive. This is easy to prove as $\<\psi|L^\dagger L\psi\>=\<L\psi|L\psi\>\geqslant0  \ \forall \ \psi \in \h, L \in \lo(\h)$.   
\subsection{The state space}

If the trace of a positive self-adjoint operator is one, it is called a \emph{density operator} and represents a quantum state. The state space of density operators is the convex set 
\begin{equation}
\mathcal{S}(\h):=\{\rho \in \lo_s(\h) : \rho \geqslant \mathds{O},  \mathrm{tr}[\rho]= 1 \}
\end{equation} which means that, for any $\rho_1,\rho_2 \in \mathcal{S}(\h)$,  any convex combination thereof also exists in that space; $\lambda\rho_1+(1-\lambda)\rho_2 \in \mathcal{S}(\h) \ \forall \ \lambda\in [0,1]$. An important property of $\rho$ is the purity$:=\mathrm{tr}[\rho^2]$. For \emph{pure states}, purity $=1$, whereas for \emph{mixed states}, purity $<1$. Pure states are the extremal points of the convex set $\mathcal{S}(\h)$, and all mixed states in $\mathcal{S}(\h)$ can be formed by a convex combination of pure states. This pure state decomposition of a density operator is not unique. Indeed, there are infinite such decompositions. One useful decomposition of $\rho \in\mathcal{S}(\h)$ is given by the \emph{canonical decomposition} of    orthogonal rank-1 projector operators  $\Pi(\phi_i)\equiv |\phi_i\>\<\phi_i|$, satisfying $\Pi(\phi_i)\Pi(\phi_j)=\delta_{ij}\Pi(\phi_i)$, which is given by   
\begin{equation}
\rho = \sum_{i}P(i)\Pi(\phi_i) \ \ \ \text{such that} \sum_iP(i)=1.
\label{canonical decomposition of mixed state}
\end{equation}  
This representation is unique if all the $P(i)$ differ from one another, and corresponds to a pure state if there is only one non vanishing $P(i)$. The $\{P(i)\}$ are thus interpreted as probabilities that the pure states $\{\Pi(\phi_i)\}$ are prepared.  Since  $\Pi(\phi_i)$ projects onto the  equivalence class of vectors $\{c\phi_i : c \in \mathds{C},|c|=1 \}$, called a \emph{ray}, pure states may be referred to as such. Indeed, many physics texts simply refer to a pure state as the vector $\psi$ itself, and I will often use this short-hand description throughout this thesis. 

 A useful representation of density operators $\rho\in\mathcal{S}(\co^d)$, in terms of an orthonormal basis in $\lo_s(\co^d)$,  is given by 
\begin{equation}
\rho = \frac{1}{d}(\mathds{1}+\vec{n}.\vec{F})\label{orthogonal basis representation of quantum states}
\end{equation}     
where $\vec{F}$ is a vector of $d^2-1$  traceless,  self-adjoint, and \emph{unitary}\footnote{A unitary operator $U$ has the  property that $\|U\psi\|=\|\psi\| \ \forall \ \psi \in  \h$. Equivalently, we may say that $UU^\dagger = U^\dagger U=\mathds{1}$.} operators $F_i$, such that $\mathrm{tr}[F_iF_j]=d\delta_{ij}$. Because these operators are  both unitary as well as self-adjoint,  their eigenvalues are one of $\{+1,-1\}$. Furthermore, $\vec{n}$ is a vector in $\mathds{R}^{d^2-1}$, where $\mathds{R}$ denotes the real numbers,  such that $\|\vec{n}\|\leqslant1$.  Hence, the state space $\mathcal{S}(\co^d)$ can be seen as a convex space in the real vector space $\mathds{R}^{d^2-1}$. This representation of the density operator facilitates the understanding of purity. Pure states are those for which $\|\vec{n}\|=1$, and as $\|\vec{n}\|$ decreases, so too does the purity. The    \emph{maximally mixed state} is $\frac{1}{d}\mathds{1}$, obtained when $\|\vec n\|=0$,  for which the  purity takes the minimal value of $1/d$. Such a state is trivially unique. 

There are two methods that are usually used to determine how close two quantum states are. The first is given by the \emph{trace distance} 
\begin{equation}
D[\rho_1,\rho_2]:=\frac{1}{2}\|\rho_1-\rho_2\|_{\mathrm{tr}}
\end{equation}
which uses the \emph{trace norm} $\|\cdot\|_{\mathrm{tr}}:L \mapsto \mathrm{tr}|L|$. Here, $|L| = \sqrt{L^\dagger L}$ is the \emph{absolute value} of $L$ which is positive for all linear operators $L$. This uses the square root lemma which states that,    for every positive operator $Y$, there exists a unique positive operator $\sqrt{Y}$ such that $\sqrt{Y}\sqrt{Y}=Y$.  

The other  measure  is called the \emph{fidelity}, which is given by  
\begin{equation}
\mathrm{Fid}[\rho_1,\rho_2] =\mathrm{tr}\left[\sqrt{\sqrt{\rho_1}\rho_2\sqrt{\rho_1}}\right]\equiv\mathrm{tr}\left[\sqrt{\sqrt{\rho_2}\rho_1\sqrt{\rho_2}}\right]
\end{equation}
 and, for pure states $\Pi(\psi)$ and $\Pi(\varphi)$, is equivalent to the quantity $|\<\varphi|\psi \>|$. \footnote{Uhlmann's theorem states that $\mathrm{Fid}[\rho_1,\rho_2]=\sup_{\{\psi,\phi\}}|\<\psi|\phi\>|$ where $\psi$ and $\phi$ are \emph{purifications} of $\rho_1$ and $\rho_2$ respectively. This is trivial if $\rho_1$ and $\rho_2$ are themselves pure states.}
 The two measures have the properties:
 
 \begin{center}
 \begin{tabular}{|c||c|}
 \hline
Trace distance & Fidelity \\\hline
$0\leqslant D[\rho_1,\rho_2] \leqslant 1$ & $0\leqslant \mathrm{Fid}[\rho_1,\rho_2] \leqslant 1$ \\\hline
$D[\rho_1,\rho_2]=0 \Leftrightarrow \rho_1=\rho_2$  & $\mathrm{Fid}[\rho_1,\rho_2]=1 \Leftrightarrow \rho_1=\rho_2$ \\\hline
\end{tabular}
\end{center}

 The trace distance and fidelity can be seen as being complements of each other; while the trace distance gives a value of 0 for two identical states, the fidelity gives a value of 1. Conversely, while the trace distance gives a value of 1 for two orthogonal states, the fidelity gives a value of 0.

\subsection{Composite systems} \label{composite systems}

We may combine the Hilbert spaces $\h^A$ and $\h^B$ to form a new composite Hilbert space $\h^A\otimes \h^B$ via the  \emph{tensor product}. The  basis vectors of $\h^A\otimes \h^B$ can be constructed as $\{\phi_i\otimes\varphi_j\}$, where $\{\phi_i\}$ is an orthonormal basis in $\h^A$ and likewise $\{\varphi_j\}$ is an orthonormal basis in $\h^B$, and the inner product on $\h^A\otimes \h^B$ is defined as
$
\<\phi_i\otimes\varphi_k|\phi_j\otimes\varphi_l\>=\<\phi_i|\phi_j\>\<\varphi_k|\varphi_l\>
$.
   For finite dimensional cases, if $\mathrm{dim}(\h^A)=d_A$ and $\mathrm{dim}(\h^B)=d_B$, then $\mathrm{dim}(\h^A\otimes \h^B)=d_Ad_B$. 

Any operator  $L \in\lo(\h^A\otimes\h^B)$ can be expanded with respect to the Hilbert-Schmidt inner product as\begin{equation}
L=\sum_{i,j}c_{ij} \ma_i\otimes \mb_j
\end{equation}
where $\{\ma_i\}$ and $\{\mb_j\}$ are respectively orthonormal bases in $\lo(\h^A)$ and $\lo(\h^B)$ with respect to the Hilbert-Schmidt inner product, and $c_{ij}=\<\ma_i\otimes\mb_j|L\>_{\mathrm{HS}} \in \mathds{C}$. Clearly, this can be done with respect to any basis.  What is known as the \emph{operator-Schmidt} decomposition -- in analogy with the Schmidt decomposition of a vector in Hilbert space mentioned in Appendix \ref{von Neumann entropy} -- is the decomposition of $L$ into $k$ orthonormal \emph{product operators} $\ma_i\otimes\mb_i$ as
\begin{equation}
L=\sum_{i=1}^k q_i \ma_i\otimes\mb_i
\end{equation}   
using a specific orthonormal basis,  where $q_i > 0$ and $k$ represents the Schmidt-rank. For a composite system $\co^d\otimes\co^{d'}$, where $d\leqslant d'$, the maximum value the Schmidt-rank can take is $d^2$.  If the Schmidt-rank is one, then $L$ is itself a product operator.

  The \emph{partial trace}  over $\h^B$ is a linear mapping $\mathrm{tr}_B : \lo(\h^A\otimes\h^B) \to \lo(\h^A)$ such that \begin{equation}
\mathrm{tr}[Y\mathrm{tr}_B[L]]=\mathrm{tr}[(Y \otimes \mathds{1})L] \ \forall \ L \in \lo(\h^A\otimes \h^B), Y \in \lo(\h^A)  
\end{equation}
and the partial trace over $\h^A$ is similarly defined. It follows from this definition that the partial trace is a positivity preserving operation. It is also a trace preserving operation, as 
\begin{equation}
\mathrm{tr}[\mathrm{tr}_B(L)]=\mathrm{tr}[\mathrm{tr}_A(L)]=\mathrm{tr}[L] \ \forall \ L \in \lo(\h^{A}\otimes\h^B).
\end{equation} 
Consequently, for any $\rho \in \s(\h^A\otimes \h^B)$, the partial trace provides a uniquely defined \emph{reduced density operator} $\rho^A:=\mathrm{tr}_B[\rho]$ and $\rho^B:=\mathrm{tr}_A[\rho]$.

\subsection{Measurement}\label{observables}

\begin{figure}[!htb]
\centering
\includegraphics[width=5in]{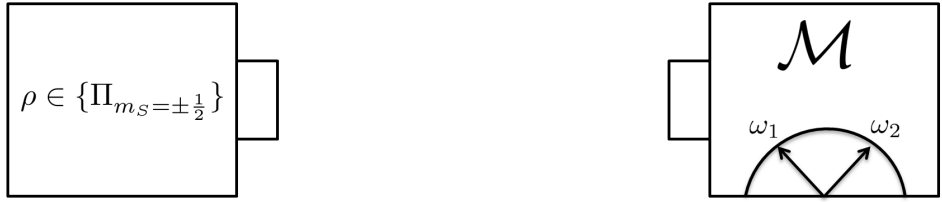} \caption{ A simplified depiction of the Stern Gerlach experiment. A preparation device can switch between preparing a spin one-half particle with $m_S=+\frac{1}{2}$ and $m_S=-\frac{1}{2}$. If the measurement is sharp, then an event  $\omega_{1}$ will determine that the particle was of spin $m_S=+\frac{1}{2}$ and conversely for the other event.   } \label{SternGerlach}
\end{figure}

As mentioned at the start of this chapter, the central aspect of  quantum theory is the prediction of probabilities for measurement outcomes. In the previous section I covered what constitutes a quantum state, which is an equivalence class of experimental preparation procedures that uniquely determines the probability distribution of all possible measurements. In this section, then, I will give an outline of what constitutes a measurement.   

In measure theory a measurable space is defined as $\<\Omega,\mathcal{F}\>$ where $\Omega$ is a sample space and $\mathcal{F}$ is a collection of subsets of $\Omega$ where every $\omega \in \mathcal{F}$ is identified with an \emph{event},  or a measurement outcome. Furthermore, $\mathcal{F}$ satisfies the conditions
\begin{enumerate}[(i)]
\item  $\Omega$ and the empty set $\{\emptyset\}$ are in $\mathcal{F}$. 
\item for every $\omega \in \mathcal{F}$, its complement $\Omega \setminus\omega:=\{x \in \Omega: x \notin \omega\}$ is also in $\mathcal{F}$.
\item For a sequence of pairwise disjoint  subsets $\{\omega_i: \omega_i\in \mathcal{F}, \omega_i\cap\omega_j=\emptyset\}$, where  $ i \in \mathds{N}$, then  $\bigcup_{i} \omega_i \in \mathcal{F}$. 
\end{enumerate}
 A measure $\mu$ on $\<\Omega,\mathcal{F}\>$ is a probability measure if
\begin{enumerate}[(i)]
\item $\mu(\omega) \in [0,1] \ \forall \ \omega \in \mathcal{F}$.
\item $\mu(\bigcup_{i} \omega_i )=\sum_{i} \mu(\omega_i)$ for any sequence of pairwise disjoint subsets $\{\omega_i: \omega_i\in \mathcal{F}, \omega_i\cap\omega_j=\emptyset\}$ where $i\in \mathds{N}$. This is known as the  countable additivity  condition.
\item  $\mu(\Omega)$=1.   
\end{enumerate}
In quantum theory a \emph{positive operator valued measure}, abbreviated as POVM, is characterised as a mapping $\m: \omega \mapsto M_\omega$ where  $M_\omega$, which  is   an \emph{effect} operator associated with the event $\omega$,  is an affine\footnote{An affine mapping $M[\cdot]$ is defined such that $M[\lambda\rho_1+(1-\lambda)\rho_2]=\lambda M[\rho_1]+(1-\lambda)M[\rho_2]$.} mapping from the state space $\mathcal{S}(\h)$ to the interval $[0,1]$, and is determined by the  \emph{Born rule} as
\begin{equation}
M_\omega:\rho \mapsto \mathrm{tr}[M_\omega\rho]\in [0,1] \ \forall \ \rho \in \mathcal{S}(\h) .
\end{equation}
Here $\mathrm{tr}[M_\omega \rho]$ takes the meaning of a conditional probability of detecting event $\omega$ given a measurement of $\m$ on a state $\rho$, denoted $P(\omega|\rho,\m)$. A POVM, also known as an \emph{observable}, has the following features:
\begin{enumerate}[(i)]
\item $\mathds{O}\leqslant M_\omega \leqslant \mathds{1}$ for all $\omega \in \mathcal{F}$.
\item $M_{\bigcup_{i} \omega_i}=\sum_{i}M_{\omega_i}$ for any pairwise disjoint sequence $\{\omega_i: \omega_i\in \mathcal{F}, \omega_i\cap\omega_j=\emptyset\}$ where $i\in \mathds{N}$.
\item $M_\Omega=\mathds{1}$.
\end{enumerate}

 The situations that are of interest for the remainder of this thesis are those where  the number of possible events are    finite. In such a situation we may make the substitution $M_{\omega_i}\equiv M_i$, and   loosely identify the POVM with the collection of  the effects \footnote{Hence why, in the physics literature, an effect is  often referred to as a POVM element.}  as 
\begin{equation}
\mathcal{M}=\{M_i\}_{i=1}^N  \ \text{with the property that} \ \sum_{i=1}^NM_i=\mathds{1}. \end{equation}

\subsubsection{Sharp observables}
A subclass of POVMs are projective valued measures (PVM), which are also known as \emph{sharp observables}. Here, the effects are described by  orthogonal  projector operators, and the events can be associated with the support of the projectors on $\h$. Furthermore, these projectors need not be rank-1. Indeed, while a PVM with   $d$ projector effects requires a Hilbert space with a minimum dimension of $d$, a $d$-dimensional Hilbert space  allows for a PVM with fewer projector effects.  We denote a PVM on finite dimensional Hilbert spaces $\co^d$, constituted of $d$ rank-1 projector effects associated with the orthonormal basis $\{\phi_i\}_{i=1}^d$, as $\mathcal{P}=\{\Pi(\phi_i)\}_{i=1}^d$.     Naimark's dilation theorem \citep{Naimark} relates POVMs and PVMs by stating that  any  POVM  acting on a $d$-dimensional Hilbert space can be realised by a  PVM acting on an \emph{extended Hilbert space} with dimension $d'\geqslant d$. 

In many experiments we only wish to determine the \emph{expectation value} of a given observable, given by the average event value. For the case of sharp observables, we may facilitate this by identifying  $\mathcal{P}=\{\Pi(\phi_i)\}_{i=1}^d$ with a self-adjoint operator $O\in \lo_s(\co^d)$, which has the spectral decomposition
\begin{equation}
O=\sum_{i=1}^d\omega_i\Pi(\phi_i)
\end{equation}
with the real eigenvalues $\{\omega_i\}_{i=1}^d$ associated with the different events.  The expectation value of measuring $O$ on $\rho$ is then given by \begin{equation}
\<O\>:=\sum_{i=1}^d\omega_iP(\omega_i|\rho)\equiv\mathrm{tr}[O\rho] 
\end{equation}
  which for pure states is $\mathrm{tr}[O\Pi(\psi)]\equiv \<\psi|O \psi\>$.

\subsubsection{Instruments}

A  measurement  of observables, whereby the probability distribution over the events can be used to distinguish between different quantum states $\rho_1$ and $\rho_2$, and hence extract \emph{information} from the system,    disturbs  the quantum system in question \citep{Busch-measurement-disturbance}. It is useful, therefore, to consider measurements in relation to how they transform quantum states or, equivalently, how they can be used as preparation devices. To this end, we may introduce the concept of an \emph{instrument} $\mathcal{I}^\m$ defined as 
\begin{align}
\mathcal{I}^{\m}_{\omega_i}: \rho \mapsto \mathcal{I}^\m_{\omega_i}[\rho] \end{align}
with the properties
\begin{enumerate}[(i)]
 \item $P(\omega_i|\rho,\m)= \mathrm{tr}[\mathcal{I}^\m_{\omega_i}(\rho)].$
 
 \item $\mathcal{I}^\m_{\bigcup_i\omega_i}[\rho]= \sum_i \mathcal{I}^\m_{\omega_i}[\rho]$ for any pairwise disjoint sequence  $\{\omega_i: \omega_i\in \mathcal{F}, \omega_i\cap\omega_j=\emptyset\}$ where $i \in \mathds{N}$.

\item
$ \mathrm{tr}[\mathcal{I}^\m_{\Omega}(\rho)]=1.$
\item The post-measurement state is $\rho_{\omega_i}^\m=\mathcal{I}^\m_{\omega_i}[\rho]/\mathrm{tr}[\mathcal{I}^\m_{\omega_i}(\rho)].$
\end{enumerate}

    Most generally, the  action of the instrument can be  written as
\begin{equation}
\mathcal{I}^\m_{\omega_i}[\rho]= \sum_{j=1}^NP(j)K_{i,j}\rho K_{i,j}^\dagger \label{measurement state transformation}
\end{equation}
such that 
\begin{equation}
M_i = \sum_{j=1}^N K_{i,j}^\dagger K_{i,j}.
\end{equation}
The case where $N=1$ is referred to as an \emph{efficient measurement}, and that where $N >1$ is an \emph{inefficient \footnote{ Inefficient measurements include statistical uncertainties, whereby several different state transformations are registered as the same measurement outcome.  } measurement}.
 In the case of efficient measurements,  we may use the \emph{polar decomposition} \footnote{The polar decomposition states that for every $L \in \lo(\co^d)$, there exists a unitary operator $U$ such that $L = U |L|$. $U$ is uniquely determined only if $L$ is invertible. } to represent the \emph{Kraus operator} $K_i$ \citep{Kraus} as 
\begin{equation}
K_i = U \sqrt{K_i^\dagger K_i} \equiv U\sqrt{M_i}
\end{equation}
where $U$ can be one of many unitary operators. Hence, we may conceive of the measurement transformation as occurring in two parts; the state is first transformed under the action of $\sqrt{M_i}$, followed by some unitary transformation. It should be clear that, generally, an instrument uniquely determines an observable, but that an observable may be implemented by many instruments. 

What is known as a \emph{minimal measurement} is an instrument  where   $U = \mathds{1}$.  Minimal measurements of sharp observables are achieved by \emph{L\"uders} instruments
\begin{equation}
\mathcal{I}^\p_{\omega_i}[\rho]=\Pi(\phi_i)\rho \Pi(\phi_i).
\end{equation}
 Here, after a measurement outcome $\omega_i$, the post-measurement state is $\rho^\p_{\omega_i}=\Pi(\phi_i)$. It is often said that the state $\rho$ \emph{collapses} to $\Pi(\phi_i)$.  Resultantly, any proceeding measurement of the system by the \emph{same} sharp observable $\p$  will give the result of $\omega_i$ with a probability
\begin{equation}
P(\omega_i|\rho^\p_{\omega_i},\p)=\mathrm{tr}[\Pi(\phi_i)]=1.
\end{equation}
  One can  say that a sharp observable due to a L\"uders instrument constitutes a \emph{repeatable measurement}.  
 
\subsubsection{Measurement models}\label{Measurement model}
In the previous section we saw that an observable $\m=\{M_{\omega_i}\}$ on a Hilbert space $\h$ is uniquely determined by an instrument $\mathcal{I}^\m$. These are in turn uniquely determined by a \emph{measurement model} $\mathbf{M}^\m$, where the observable $\m$ on the system of interest, often referred to as the \emph{object} in such a context, is measured indirectly by observing a \emph{probe}, or measurement apparatus, after it has interacted with the object. A measurement model may generally be described by the 5-tuple  
\begin{equation}
\mathbf{M}^\m=\<\mathcal{K},\varrho,U,Z,f\>
\end{equation} where $\mathcal{K}$ is the probe Hilbert space, $\varrho$ is the initial state of the probe, $U$ is a unitary operator acting on $\h\otimes\mathcal{K}$, and $Z$ the self-adjoint operator associated with the sharp observable on the  probe. The ``\emph{pointer function}'' $f$ is an invertible mapping between the measurable space of the probe $\<\Omega',\mathcal{F}'\>$ and that of the object $\<\Omega,\mathcal{F}\>$ such that $f:\omega_i' \mapsto \omega_i$. For this model to determine the observable $\m$, it must satisfy the ``\emph{probability reproducibility condition}''
\begin{equation}
\mathrm{tr}\left[(\mathds{1}\otimes M_{f^{-1}(\omega_i) })U \rho\otimes\varrho U^\dagger\right]=\mathrm{tr}\left[M_{\omega_i}\rho \right] \ \forall  \  \rho \in \s(\h),  \label{probability reproducibility condition}
\end{equation} 
 and for it to also be repeatable, it must further satisfy
\begin{equation}
\mathrm{tr}\left[(M_{\omega_i}\otimes M_{f^{-1}(\omega_i) })U \rho\otimes\varrho U^\dagger\right]=\mathrm{tr}\left[M_{\omega_i}\rho \right] \ \forall  \  \rho \in \s(\h).\label{repeatability condition}
\end{equation}
 The most general measurement model for a  sharp observable associated with the  self-adjoint operator $O$, which is diagonal with respect to the basis $\{\phi_i\}$, is the von Neumann-L\"uders measurement model \citep{von-Neumann} where the initial probe state $\varrho\equiv \Pi(\varphi)$ is pure,  the action of the joint unitary operator $U$ is
\begin{equation}
U: \phi_i\otimes\varphi \mapsto \phi_i\otimes\varphi_i,
\end{equation}   
and the probe observable $Z$ is diagonal with respect to the basis $\{\varphi_i\}$.

\subsubsection{Ensemble measurements of sharp observables }\label{ensemble state section}
\begin{figure}[!htb]
\centering
\includegraphics[width=3in]{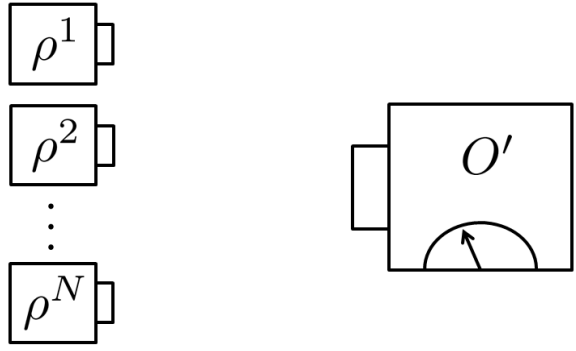} \caption{A measurement device performing an ensemble measurement of a sharp observable $O$, denoted  $O'$, on a composite Hilbert space $\h^{\otimes N}$.  } \label{ensemble}
\end{figure}

It is possible to perform local measurements on a composite system $\h^A \otimes \h^B$, with observables $\m^A$ on subspace $\h^A$ and $\m^B$ on subspace $\h^B$. The events of such an experiment are given by the cartesian product of the events of the individual observables, denoted $\<\omega_i^A,\omega_j^B\>$, with the corresponding effect operators $M_i^A\otimes M_j^B$. The joint probability of the these two events is  
\begin{equation}
P(\omega_i^A,\omega_j^B|\m^A,\m^B,\rho)=\mathrm{tr}[M_i^A\otimes M_j^B\rho].
\end{equation}

However, in many experimental situations we do not have access to the events in the individual subspaces, but only  their average. The composite system in such situations is referred to as an \emph{ensemble} or \emph{assembly}. If  each $d$-dimensional subsystem of a composite Hilbert space $\bigotimes_{n=1}^N\h^n\equiv \h^{\otimes N}$ can be measured by the same sharp observable 
\begin{equation}
O=\sum_{i=1}^d \omega_i\Pi(\phi_i)
\end{equation}
   we may identify the following self-adjoint operator
 \begin{align}
O'&=\frac{1}{N}\sum_{n=1}^N O^n\otimes\mathds{1}^{\neg n} 
\end{align}    
with the \emph{ensemble} measurement of $O$,
where each $O^n$ acts on subspace $\h^n$ and   $\mathds{1}^{\neg n}$ signifies an identity operator on all subspaces other than $\h^n$.  The    eigenvectors of this observable are given by $\bigotimes_{n=1}^N \phi_n $
with the generally degenerate eigenvalues $\frac{1}{N}\sum_{n=1}^N\omega_n$.

If we wish to only determine the expectation value of the ensemble observable, we notice that
 \begin{equation}
\mathrm{tr}[O'\rho]=\mathrm{tr}\left[\frac{1}{N}\sum_{n=1}^NO^n \otimes \mathds{1}^{\neg n}\rho\right] =\mathrm{tr}\left[O\left(\frac{1}{N}\sum_{n=1}^N \rho^n\right)\right]
\end{equation}
where $\rho^n=\mathrm{tr}_{\neg n}[\rho]$. As such, we may say that the  composite state $\rho$ on a $d^N$-dimensional Hilbert space has associated with it an \emph{effective ensemble state} on a $d$-dimensional Hilbert space given by 
\begin{equation}
\bar\rho= \frac{1}{N}\sum_{n=1}^N \rho^n 
\end{equation}
which has a clear interpretation as the \emph{averaged} reduced density operator. 

Let us now consider a von Neumann-L\"uders measurement of this ensemble observable, where the  object is initially in a pure separable state $\Psi=\bigotimes_{i=1}^N\psi_n$. The measurement Hamiltonian is
\begin{equation}
H_I=f(t)P \otimes \frac{g}{N}\sum_{n=1}^N O^n\otimes \mathds{1}^{\neg n}
\end{equation}
where $P$ is a self-adjoint  operator acting on the probe  from the conjugate pair $\{P,Q\}$, $f(t)$ is a function that is non vanishing only during the measurement process and is normalised such that $\int dt f(t)=1$,  and $g$ is the strength of the measurement   (See Appendix \ref{von Neumann measurement Gaussian pointer}) . The probe is initially in a Gaussian state which, in the $Q$ representation, is  given by  $\psi(q)$. After the measurement interaction, the object and probe are  generally entangled, with the reduced state of the probe being in the statistical mixture $\sum_aP(a)\Pi(\psi(q-ga))$, where $\{a\}$  are the eigenvalues of the ensemble observable $O'$. The shift in the expectation value of $Q$ on the probe, $\Delta\<Q\>$, will correspond with the measured eigenvalue.

An interesting question to ask is what  the measurement statistics of the ensemble observable will be in the limit of $N \to \infty$.   As shown by \citep{weak-measurements-Aharanov-1990} we may always write \begin{equation}
O\psi_n=\alpha\psi_n + \beta \psi_n^\perp
\end{equation}
 where $\<\psi_n^\perp|\psi_n\>=0$. From this we can determine that $\alpha=\<\psi_n|O\psi_n\> \equiv \<O_n\>$ and  $\beta=\sqrt{\<O\psi_n| O \psi_n\>-\<\psi_n|O \psi_n\>^2} \equiv \Delta_{ O_n}$.  It follows that the projection of $\Psi$ by $O'$  is  
\begin{align}
O'\Psi&=\left(\frac{1}{N}\sum_{n=1}^N O^n\otimes \mathds{1}^{\neg n}\right) \bigotimes_{n=1}^N\psi_n=\frac{1}{N}\sum_{n=1}^N\left( \ \<O_n\>\psi_n + \Delta_{ O_n}\psi_n^\perp\right)\bigotimes_{m \ne n} \psi_m, \nonumber \\
&=\left(\frac{1}{N}\sum_{n=1}^N  \<O_n\> \right) \bigotimes_{n=1}^N\psi_n + \frac{1}{N}\sum_{n=1}^N \Delta_{ O_n} \psi_n^\perp \bigotimes_{m\ne n}\psi_m,\nonumber \\ &=\<\Psi|O'\Psi\>\Psi+\Psi^\perp\equiv\tilde\Phi,
\end{align}

 which can be normalised as $\Phi=\tilde\Phi/\|\tilde \Phi\|$. In the limit of $N \to \infty$, we explicitly calculate $\|\tilde \Phi \|$ to be
\begin{align}
\lim_{N \to \infty}\| \tilde\Phi\|    &=\lim_{N \to \infty}\left\| \<\Psi|O'\Psi\> \bigotimes_{n=1}^N\psi_n+  \frac{1}{N}\sum_{n=1}^N \Delta_{ O_n} \psi_n^\perp \bigotimes_{m\ne n}\psi_m \right\|, \nonumber \\&= \lim_{N \to \infty} \sqrt{ \<\Psi|O'\Psi\>^2 +\frac{1}{N^2}\sum_{n=1}^N (\Delta_{ O_n})^2}, \nonumber \\
&= \<\Psi|O'\Psi\>, \label{ensemble identical state limit large N}
\end{align}  where we have relied on the fact that every vector in the summation  $\sum_n\psi_n^\perp \bigotimes_{m\ne n}\psi_m$ is mutually orthogonal. Using this result,  the trace    distance between $\Psi$ and $\Phi$ is calculated as
\begin{align}
\lim_{N \to \infty}d(\Psi,\Phi)=\lim_{N \to \infty}\left|2\left(1-\frac{\<\Psi|O'\Psi\>}{\|\tilde \Phi\|} \right)\right|^{1/2}=0
\end{align}

  and, thus, we arrive at the striking conclusion that
\begin{equation}
\lim_{N \to \infty}O'\Psi = \<\Psi|O'\Psi\>\Psi \ \forall \ \Psi=\bigotimes_{n=1}^N\psi_n \in \h^{\otimes N}.
\end{equation}

If the object is initially in the state $\rho=\sum_i P(i)\Pi(\Psi_i)$,  after the measurement interaction the state of the system and apparatus will be   
\begin{equation}
\sum_i P(i)\Pi(\Psi_i) \otimes\Pi(\psi(q-g\<\Psi_i|O'\Psi_i\>)) 
\end{equation} with the probe's reduced state given as  $\sum_i P(i)\Pi(\psi(q-g\<\Psi_i|O'\Psi_i\>))$. In the limit of $\psi(q)$ being a delta function, a single measurement of the probe will reveal $\<\Psi_i|O'\Psi_i\>$ with a probability of $P(i)$.  Additionally, if all the probe states $\psi(q-g\<\Psi_i|O'\Psi_i\>)$ are  (effectively) orthogonal, then the state of the object will not be (significantly) altered  by the measurement.
 
It should  be stressed that the proof above has rested upon the separability of the object state. If the initial pure state $\Psi$ is entangled, then even in the $N \to \infty$ limit it will not be an eigenstate of $O'$.  

\subsection{Entanglement}\label{entanglement}

In a composite Hilbert space, there are states that cannot be written as a convex combination of product states. In this section I shall provide a historical overview for why such states are of interest with respect to the correlations they possess.

Consider two sharp observables $O_x$ and $O_p$  which do not commute, i.e. where  $[O_x,O_p ]_-:=O_xO_p - O_pO_x \neq \mathds{O}$.   In their seminal paper \citep{EPR-1935} Einstein, Podolsky and Rosen (EPR) argued that, as a quantum state  cannot instantaneously predict the measurement outcome of two non commuting observables $O_x$ and $O_p$ with certainty, that there are two possibilities:
\begin{enumerate}[(i)]
\item
The values of these non commuting observables cannot have simultaneous reality. \item
Quantum mechanics  does not have a one-to-one correspondence between the elements of the theory and the elements of reality, and is thus \emph{incomplete}.  
\end{enumerate}
 
A quantity described by a physical theory is said to correspond with an  \emph{element of reality}  if its value can be predicted with certainty without disturbing the system. EPR proposed a thought experiment, where two observers at the space-like separated\footnote{Space-like separated events are those which lie outside one another's light cones and, by the principle of relativistic causality, cannot be causally related.} positions  $A$ and $B$ -- who we call Alice and Bob respectively -- share a pure quantum state of two particles emitted simultaneously from a source, and moving in opposite directions such that their relative momentum is zero, and their relative distance is $L$  (the non commuting observables here are position and momentum). Therefore, the measurement statistics is highly correlated such that if a measurement outcome for the position of the particle at $A$ is $x$, then the position at $B$ is $-x$ with certainty, and similarly for the momentum.  Although this example uses an infinite dimensional Hilbert space, we may consider the argument in a finite dimensional case which is conceptually more simple. Imagine if the vector in $\h=\mathds{C}^2\otimes\mathds{C}^2$  corresponding with  a bipartite pure state is given by
\begin{equation}
\Psi^{A+B}=\frac{1}{\sqrt{2}}(\phi_0^A\otimes\phi_0^B +\phi_1^A\otimes\phi_1^B) \label{EPR state vector}
\end{equation}
where $\{\phi_0, \phi_1\}$ forms an orthonormal basis for both $\h^A$ and $\h^B$.  Suppose also that both Alice and Bob can measure the non commuting observables $O_x=\{\Pi(\phi_0),\Pi(\phi_1)\}$ with events $\{x_0,x_1\}$ and $O_p=\{\Pi(\phi_+),\Pi(\phi_-)\}$ with events $\{p_+,p_-\}$. We designate $\phi_\pm=\frac{1}{\sqrt{2}}(\phi_0\pm \phi_1)$. If both Alice and Bob  measure the same observable, then their measurement statistics are fully correlated. For example, since 
\begin{align}
&P(x_0^A,x_0^B)=\mathrm{tr}[(\Pi(\phi_0)^A\otimes\Pi(\phi_0)^B)\Pi(\Psi)]=\frac{1}{2} \end{align}
and
\begin{align} & P(x_0^A)=\mathrm{tr}[(\Pi^A_{\phi_0}\otimes\mathds{1}^B) \Pi(\Psi)]=\frac{1}{2}
\end{align}
we may use Bayes' rule to infer that $P(x_0^B|x_0^A)=1$.
Hence, if the measurement of $O_x$ by Alice yields the event $x_0$ then  Alice knows with certainty that the measurement of $O_x$ by Bob will yield result $x_0$ also, without Bob ever needing to perform that measurement. Similar arguments hold for measurement of $O_p$. EPR argue that as    Bob's system can be measured indirectly without disturbing it (because the two observers are space-like separated)  the value of $x$ and $p$ at $B$ are simultaneously elements of reality. \footnote{It is true that only one of $O_x$ and $O_p$ can be measured in any given experiment. However, in the EPR framework, because the two observers are space-like separated, Alice has freedom to choose which measurement she performs without disturbing Bob's system, and hence the state of Bob must contain the information for both $x$ and $p$ simultaneously.}    Therefore  EPR conclude that Quantum mechanics is incomplete.

Following this result, many people have approached the apparent incompleteness of quantum theory by trying to construct \emph{hidden variable} theories which stipulate that  the perceived randomness of quantum theory is due to some hidden variables that we cannot control in experiments, and hence sample a random distribution thereof. In the 1960s  John Bell presented a no-go theorem on the nature of such hidden variable theories \citep{bellspeakable-1987}. If we consider the EPR paradox, it relies on two propositions; realism and locality. These have the following descriptions           
\begin{enumerate}[(i)]
\item \emph{Realism}: As in classical physics, the property of the object that we wish to measure, such as the momentum, \emph{exists} in that object independently of our measurement. As such, the measurement process is an act of discovering what is already out there in the universe. Such a view takes the probabilistic nature of quantum theory in the same light as that of statistical mechanics; the probabilities are a product of our ignorance of the true state of the system in question.
\item
\emph{Locality}: An observation of a physical quantity at position $A$ depends only on the parameters at position $A$ and not any other. For an event at $A$ to influence an event at $B$, there must be a physical intermediary \emph{message} which transports this influence between the two positions. Such a message cannot travel faster than the speed of light according to the principle of relativistic causality. As such, for the event at $B$ to be caused by the event at $A$, the former must exist in the future light cone of the latter.   
\end{enumerate}

\begin{figure}[!htb]
\centering
\includegraphics[width=4in]{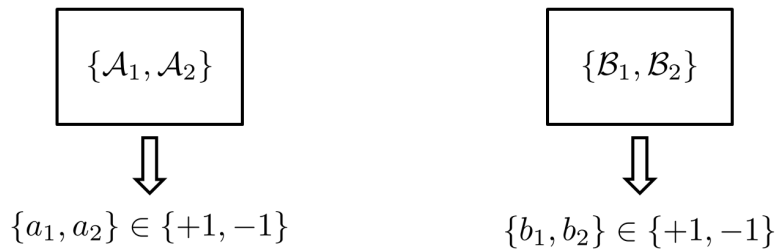} \caption{The setup for a CHSH Bell inequality.  The measurements $\{\ma_1,\ma_2 \}$ and $\{\mb_1,\mb_2 \}$ are carried out at two space-like separated locations $A$ and $B$  so as to ensure their statistics are independent from one another under the locality assumption. Each measurement will give an event value $\{+1,-1\}$.} \label{Bell setup}
\end{figure}

John Bell constructed a general measurement scheme of two space-like separated systems $A$ and $B$, and \emph{assuming} local realism, determined an upper bound on the correlation of measurements conducted in these two positions. It is important to note that  quantum theory has not been assumed here, but only a general scheme of measurements which can have a binary outcome of $\pm 1$ according to some underlying real parameter. A more general version of the Bell inequality is the CHSH inequality \citep{CHSH-1969}, where the pair of local observables $\{\mathcal{A}_1,\mathcal{A}_2\}$ and $\{\mathcal{B}_1,\mathcal{B}_2\}$ can be measured on system $A$ and $B$ respectively.  These observables have event values $\{a_1,a_2\} \in \pm1$ and $\{b_1,b_2\}\in\pm 1$ respectively, and are dependent on some hidden variable. The correlation of a measurement, say, $\ma_1$ on $A$ and $\mb_1$ on $B$, is given by the expectation value  $\<a_1b_1\>$. This \emph{correlation function} can give values in the range of $[-1,1]$. If it is $+1$ then $\ma_1$ and $\mb_1$ are fully correlated, and if it is $-1$ they are fully anti-correlated. If, on the other hand, it is $0$, then the two are uncorrelated. The local realistic assumptions imply that the following always holds true
\begin{equation}
(a_1+a_2)b_1 +(a_1 - a_2)b_2 = \pm 2
\end{equation}
 because if $a_1+a_2=0$ then $a_1-a_2=\pm1$ and vice versa. Note that this equation has counterfactual measurements; in any one experiment, only one of the joint event values 
\begin{equation}
\{a_1b_1, a_1b_2,a_2b_1,a_2b_2\} \in \pm 1
\end{equation} can be determined. In this argument, we are essentially assuming that if, say, $a_1=+1$ in the joint event $a_1b_1$ owing to some hidden variable, then its value \emph{would also be} $+1$ in the joint event $a_1b_2$. In other words, we are assuming that $a_1$ has an objective, predefined value which is independent of the measurement performed at $B$. This is where the local-realistic assumption manifests itself.  By taking expectation values for such measurements and taking the modulus, we obtain the CHSH inequality 
\begin{equation}
|\< a_1b_1\> +\<a_1b_2 \> + \<a_2b_1 \> - \<a_2b_2 \>|\leqslant 2. \label{CHSH}
\end{equation}

However, quantum theory can, depending on the state and choice of observables used, violate the CHSH upper bound of $2$. \footnote{It should be noted, however, that to date no experiment has been able to prove that \emph{reality} violates the CHSH inequality, because experimental imperfections and impracticalities introduce loopholes, which would increase the upper bound of 2. Although a few experiments  have closed some of these loopholes, to prove that reality violates Eq.\eqref{CHSH}, one must close \emph{all} of these loopholes simultaneously. } This result means that quantum theory is not a local-realistic theory, and consequently any hidden variable theory which aims to replicate the predictions of quantum theory must abandon either realism, locality, or both.  Another important inequality is Cirel'son's inequality \citep{cirelson-bound} 
\begin{align}
C&=\ma_1\otimes\mb_1+\ma_1\otimes\mb_2+\ma_2\otimes\mb_1 -\ma_2\otimes\mb_2, \nonumber \\ \| C \| &\leqslant 2\sqrt{2},\label{Cirel'son}
\end{align}

stating that the maximum amount by which the CHSH inequality can be violated given quantum theory is $2\sqrt{2}$.     

So what states $\rho^{A+B} \in \s(\h^A\otimes\h^B)$ are the CHSH inequality violated for? It turns out that a violation of the CHSH inequality is a  sufficient reason for  a bipartite state to be \emph{inseparable}. \footnote{Violation of the CHSH inequality is not necessary for inseparability, however, as is evident by the inseparable Werner states \citep{Werner-1989-Werner-states} that do not violate the CHSH inequality. This shows that inseparability is a more general phenomenon than Bell inequalities.}    
Separability of a bipartite state $\rho^{A+B}$ is assured whenever it can be written as a convex combination of product states 
\begin{equation}
\rho^{A+B} = \sum_{i}P(i)\rho^A_i\otimes\rho^B_i. \label{seperability criterion}
\end{equation}
When a bipartite state cannot be written in such a form, as is the case for Eq.\eqref{EPR state vector}, it is inseparable or \emph{entangled}.  Two good review articles for entanglement are given by \citep{plenio-2007-7} and \citep{Horodecki2007}. I give a brief description of three  measures which quantify entanglement in Appendix \ref{entanglement measures}.

\subsection{Quantum dynamics}\label{quantum dynamics}

\begin{figure}[!htb]
\centering
\includegraphics[width=5in]{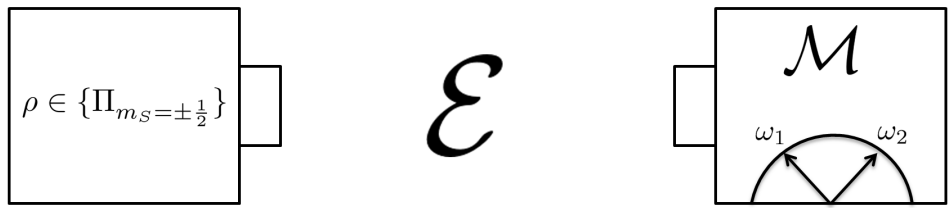} \caption{The simplified Stern Gerlach experiment with the inclusion of quantum channels. After the preparation of a specific quantum system $\rho$, and before the state reaches the detectors, it may evolve according to a quantum channel $\e[\rho]$. This is known as the Schr\"odinger picture. Alternatively, instead of the state, the detectors may be seen to evolve by the dual channel $\e^\dagger$. This is the Heisenberg picture.} \label{quantumchannels}
\end{figure}

So far our description of a quantum mechanical experiment has involved two conceptual parts; the preparation of a quantum state, and the measurement of said state. We may, however, include another element in our description to allow for the possibility of both measuring a different state than that which was prepared initially, or to change the measurement from that which was initially chosen. To this end, we introduce the concept of \emph{quantum operations}\footnote{Quantum operations are also called super-operators; just as operators map between vectors in a Hilbert space, super-operators map between operators in the operator vector space. }, defined as $\e[\cdot]:\s(\h)\to\tilde\s(\h')$, which map states from the state space  $\s(\h)$ to the \emph{subnormalised} state space $\tilde\s(\h')=\{\tilde\rho \in \lo_s(\h') :  0\leqslant \mathrm{tr}[\tilde\rho]\leqslant 1 \}$. The dimension of $\h'$ need not be the same as that of $\h$;  $\mathrm{dim}(\h')>\mathrm{dim}(\h)$ implies the addition of an ancillary Hilbert space, and $\mathrm{dim}(\h')< \mathrm{dim}(\h)$ implies the discarding of a subspace. A quantum operation must satisfy the following conditions:

\begin{enumerate}[(i)]
\item
\emph{Linearity}: This results from the requirement that we may stochastically switch between different experimental parameters.    Hence for $\lambda \in [0,1]$ we have   $\lambda \e_1[\rho]+(1-\lambda)\e_2[\rho]=\e_3[\rho]$, and   $\e[\lambda \rho_1 +(1-\lambda)\rho_2]=\lambda \e[\rho_1]+(1-\lambda) \e[\rho_2]$.

\item
\emph{Trace non-increasing}: All quantum operations must satisfy $\mathrm{tr}[\e(\rho)]\leqslant 1$. A quantum operation for which $\mathrm{tr}[\e(\rho)]=1$ is a deterministic quantum operation, also known as a \emph{quantum channel}. Conversely, a quantum operation for which $\mathrm{tr}[\e(\rho)]<1$ is a \emph{stochastic} or \emph{probabilistic} quantum operation.

\item
\emph{Complete positivity}:
A quantum state is positive, and hence $\tilde\rho'=\e[\rho]$ must also be positive for all $\rho \in \s(\h)$. However this is not sufficient. Given a composite system $\h^A\otimes \h^B$, a quantum operation $\e^A:\s(\h^A)\to\tilde\s(\h'^A)$ is said to be \emph{completely positive} if the mapping $\e^A\otimes\mathds{1}^B$ on $\s(\h^A\otimes \h^B)$ is also positive for all finite dimensional extensions  $\h^B$. This is a necessary requirement  for  physical processes,  as operating locally on a subspace of an entangled state $\rho^{A+B}$ must provide a valid physical state. As  shown in Appendix \ref{negativity}, an example of a positive map which is not completely positive is the partial transposition.   
\end{enumerate} 

 All quantum operations can be written down in the \emph{operator sum form}, also known as the \emph{Kraus decomposition} \citep{Kraus}  
\begin{equation}
\e[\cdot]:\rho \mapsto \sum_{i}^{}K_i\rho K_i^\dagger \; \; \; \; , \; \; \; \; \; \sum_{i}^{}K_i^\dagger K_i \leqslant \mathds{1}
\end{equation}
with the Kraus operators $\{K_i\}$. In the case of finite dimensional Hilbert spaces $\co^d$, the number of Kraus operators cannot be greater than $d^2$. It should be immediately apparent that measuring a POVM  using the instrument $\mathcal{I}^\m_{\Omega}$  results in a quantum channel. \footnote{Or a  stochastic quantum operation for the instrument $\mathcal{I}^\m_{\bigcup_i\omega}$ where $\bigcup_i\omega_i \subset \Omega$.} 

So far we have discussed quantum operations as mappings on quantum states. However, since quantum  theory only concerns itself with the probability of measurement outcomes, we can consider the dual of a quantum operation, $\e^\dagger$, as a mapping on effects
\begin{align}
P(\omega_i|\e[\rho])&=\mathrm{tr}[M_i\e[\rho]]=\mathrm{tr}\left[\sum_jM_iK_j\rho K_j^\dagger\right]\nonumber \\ &=\mathrm{tr}\left[\sum_j K_j^\dagger M_iK_j\rho\right] =\mathrm{tr}[\e^\dagger[M_i]\rho]=P(\omega_i'|\rho)
\end{align}       
where $\omega_i'$ denotes the event associated with  the effect $\e^\dagger[M_i]$. The framework in which states are transformed is the \emph{Schr\"odinger picture}, and the equivalent framework in which the effects change is the \emph{Heisenberg picture}.

The simplest class of a quantum channel is the unitary channel, where $\e_U[\cdot]:\rho \mapsto U\rho U^\dagger$. This channel has the unique property that it is \emph{reversible}, meaning that it has an inverse channel -- which for a unitary channel is given by its dual -- such that concatenating this channel with its inverse gives $\e_U^{-1}\circ\e_U [ \rho]=U^\dagger U \rho  U^\dagger U = \rho$.        

An aspect of quantum channels which will be of relevance to the study of quantum noise is their \emph{contractivity}  
\begin{align}
&\mathrm{Fid}[\e(\rho_1),\e(\rho_2)]\geqslant \mathrm{Fid}[\rho_1,\rho_2], \  \nonumber \\
&D[\e(\rho_1),\e(\rho_2)]\leqslant D[\rho_1,\rho_2], \ 
\end{align}
 which means that deterministic quantum channels can never improve our ability to distinguish between two quantum states.   \footnote{It should be noted that this does not need to hold for stochastic quantum operations, which can probabilistically decrease the fidelity between quantum states. An example is the stochastic perfect state discrimination \citep{Chefles-QSD} .  } The equality here holds for the unitary channels or when $\rho_1$ and $\rho_2$ are both stationary states of the  channel $\e[\cdot]$. \footnote{The stationary states of a channel are defined as the set of states $\{\rho_i\}$ such that $\e[\rho_i]=\rho_i$. } Some channels  that increase the fidelity between two states, and hence lead to quantum noise, are the depolarising channel, the dephasing channel, and  the random unitary channel
\begin{equation}
\e_{\mathrm{depolarising}}[\cdot]: \rho \mapsto \lambda \frac{1}{d}\mathds{1}+(1-\lambda)\rho. \label{depolarising channel} \end{equation}
\begin{equation}
\e_{\mathrm{dephasing}}[\cdot]:\rho \mapsto \sum_i \<\phi_i|\rho \phi_i\> \Pi(\phi_i). \label{dephasing channel}
\end{equation}
\begin{equation}
\e_{\mathrm{random\ unitary}}[\cdot]:\rho \mapsto \sum_iP(i)U_i\rho U_i^\dagger . \label{random unitary channel}
\end{equation}

The depolarising channel is a convex combination of the identity channel and the contraction of $\rho$ to the maximally mixed state $\frac{1}{d}\mathds{1}$. The  dephasing channel reduces a state $\rho$ to its diagonal components, given a specific basis $\{\phi_i\}$, where all elements of the matrix $\<\phi_i|\rho\phi_j\>\Pi(\phi_i)$ such that $i\ne j$ are eliminated. The random unitary channel has a clear meaning, which is a convex combination of different unitary channels, such that the full channel itself is no longer unitary and hence irreversible.   

\subsubsection{Separable operations}
The concept of separability, introduced in Sec.\ref{entanglement}, can also be used for quantum operations. A quantum operation $\e^{A+B}[\cdot]$ is said to be \emph{separable} with respect to the $A:B$ partition if it can be written as
\begin{equation}
\mathcal{E}^{A+B}[\rho]=\sum_i( \mathcal{A}_{i}\otimes \mathcal{B}_{i} )\rho^{A+B}( \mathcal{A}_{i}^\dagger \otimes \mathcal{B}_{i}^\dagger ).
\end{equation}
Given a separable input state $\rho^{A+B}$, such separable operations cannot generate an entangled state. 

A separable unitary operator is given by $U^{A+B}=U^A\otimes U^B$. However, as shown in \citep{Measurement-entanglement}, A SWAP unitary on $\h^A\otimes\h^B$, possible only when $\h^A$ and $\h^B$ have the same dimension, also does not generate any entanglement given an initial product state. Such a unitary operation is characterised as 
\begin{equation}
U\phi^A\otimes\varphi^B=V_{BA}\varphi^A\otimes W_{AB}\phi^B
\end{equation}
 with the isometries $V_{BA}:\h^B \to \h^A$ and $W_{AB}:\h^A \to \h^B$. Notwithstanding such a unitary map is strictly not separable, but separability preserving. This is a weaker condition \citep{HarrowNielsen2003}, stipulating that although a SWAP map cannot generate entanglement given a separable input state, one may generate entanglement between $\h^A$ and $\h^B$ utilising ancillary systems $\h^{A'}$ and $\h^{B'}$. Additionally, if the SWAP map is the result of some continuous unitary operator $U_\tau$ at $\tau = t$, then at some other value of $\tau$ it must be entangling. \footnote{Consider, for example, the entangling $\sqrt{\mathrm{SWAP}}$ gate.}

\subsubsection{Stinespring's dilation theorem}\label{Stinespring's dilation theorem}
 Two examples of quantum operations, where they correspond respectively  to an increase and decrease in Hilbert space dimension,  and are also trace preserving and hence quantum channels, are
\begin{enumerate}[(i)]
\item \emph{Independent addition of an ancillary system}:  The addition of an ancillary system, or a \emph{probe}, $\varrho \in \s(\h^B)$ to the system space $\h^A$, independently of the state in the system space, is characterised by the map $\e_{\varrho}:\rho \mapsto \rho\otimes \varrho \ \forall \ \rho \in \s(\h^A) $. The complete positivity and linearity of this map are evident. This is  also a quantum channel  as it is trace preserving, since $\mathrm{tr}[\rho\otimes\varrho]=\mathrm{tr}[\rho]\mathrm{tr}[\varrho]=1$. 
\item \emph{Partial trace}:   In the Heisenberg picture, it is the effects that are acted upon by the dual of the partial trace,    defined as $\mathrm{tr}_B^\dagger:\lo(\h^A)\to\lo(\h^A\otimes\h^B)$, which acts on an effect $M_i^A$ as $\mathrm{tr}_B^\dagger[M_i^A]=M_i^A\otimes\mathds{1}^B$. This  is  completely positive,  linear, as well as  trace preserving. It is therefore a  quantum  channel.   
\end{enumerate}

We may therefore concatenate these, together with a unitary channel $\e_U$ with $U \in \lo(\h^A\otimes\h^B)$, to obtain a quantum channel $\e[\cdot]$ acting on a system $A$ as

\begin{equation}
\e^A[\cdot]=\mathrm{tr}_B\circ \e_{U}^{} \circ \e_{\varrho}.
\end{equation} 

Stinespring's dilation theorem \citep{Stinespring}, analogous to Neimark's dilation theorem, states that \emph{any} quantum channel can be decomposed in such a way. Equivalently, any unitary evolution acting on  a system and  a probe, initially in a product state, leads to a deterministic quantum channel on the system. If the unitary channel forms a dynamical semigroup, where given a positive  parameter $\tau$ we have  $U_{\tau_1}U_{\tau_2}=U_{\tau_1+\tau_2}$ , then we may equivalently say that the doublet $\<\e_{U_\tau}, \e_{\varrho} \>$ characterises a family of quantum channels $\{\e_\tau: \tau \in \mathds{R}^+\}$, where the  channels therein differ only due to the positive  parameter $\tau$ of the unitary channel. 

Let us consider a concrete example.  We start with the initial addition of a probe by the map $\e_\varrho:\rho^A\mapsto\rho^A\otimes\varrho^B$. Using the canonical pure state decomposition of $\varrho^B$, this can be expressed as 
\begin{equation}
\sum_iP(i)\rho^A\otimes\Pi(\phi_i)^B\equiv \sum_iP(i)|\phi_i\>\rho^A\<\phi_i|.
\end{equation}
  The composite system subsequently evolves due to the unitary operator $U_\tau$, followed by  a partial trace over $\h^B$   which results in the following quantum channel 
\begin{align*}
\e^A_\tau\left[\rho^A\right]&=\sum_{i,j}P(i)\<\varphi_j|U_\tau|\phi_i\>\rho^A\<\phi_i|U^{\dagger}_\tau |\varphi_j\>, \\
&=\sum_{i,j} K_{i,j}(\tau) \rho^A K_{i,j}^\dagger(\tau).
\end{align*}

 Here, the Kraus operators \footnote{Note that the quantity $\<\varphi_j|U_\tau |\phi_i\>$ is not an inner product, but rather a shorthand for the operator
$\sum_lq_l\<\varphi_j|\mb_l \phi_i\> \ma_l$ acting on $\h^A$, where $\sum_l q_l\ma_l\otimes \mb_l$ is the operator-Schmidt decomposition of $U_\tau$.} are given by $K_{i,j}(\tau)=\sqrt{P(i)}\<\varphi_j|U_\tau|\phi_i\>$. This quantum channel can be thought of as a result of a measurement model, as discussed in Sec.\ref{Measurement model}, where the partial trace is in fact the result of carrying out a projective measurement on the probe with respect to the basis $\{\varphi_j\}$.  If the joint unitary channel  is separable, meaning that $U_\tau=U^A_\tau\otimes U^B_\tau$, then no information can be transferred from the object to the probe, and system $A$ simply undergoes a reversible unitary evolution.

\section{Closed quantum systems}\label{closed quantum system dynamics}
A closed quantum system is a mathematical idealisation reminiscent of the free point mass in classical mechanics. It assumes that the quantum degrees of freedom of the system are completely isolated from the rest of the universe, and any interaction that they can have with the outside world are through classical means, such that the external object is not affected at all by the quantum system. An isolated spin in a classical magnetic field is a clear example, where it  interacts not with other quantum objects, but rather with the external magnetic field which is a purely classical quantity, and not affected by the state of the spin.

   The differential equation governing the evolution of a quantum state in a closed system is  the Liouville-von Neumann equation 
\begin{equation}
\mathscr{L}[\rho(t)]\equiv \frac{d}{d t}\rho(t)=\mathfrak{i}[\rho(t),H(t)]_-
\end{equation}  
where $\mathscr{L}$ is the Liouville super-operator, and  $H(t)$ is the -- generally time-dependent -- Hamiltonian of the system, which is a self-adjoint operator. This is reminiscent of Hamilton's equations of motion expressed in terms of the Poisson brackets in classical mechanics. The final time state $\rho(t)$,  given an initial state $\rho(t_0)$,  is deterministically obtained as
\begin{align}
\rho(t)=\mathpzc{T}_{\leftarrow}e^{\int^t_{t_0}dt_1\mathscr{L}(t_1)}\rho(t_0)&=\rho(t_0)+\mathfrak{i}\int_{t_0}^tdt_1\left[\rho(t_0),H(t_1)\right]_- \nonumber \\ & \ \ \ -\int_{t_0}^tdt_1 \int_{t_0}^{t_1}dt_2\left[\left[\rho(t_0),H(t_1)\right]_-,H(t_2)\right]_-+...,\nonumber \\ &=U_{t,t_0}\rho(t_0)U^\dagger_{t,t_0}
\end{align}
where $\mathpzc{T}_{\leftarrow}$ is the time ordering operator, and $t_1>t_2>...>t_n$. In other words, the time-dynamics of closed quantum systems are governed by unitary channels that are generated by the Hamiltonian.  The time evolution unitary operator $U_{t,t_0}$ is the solution to Schr\"odinger's equation
\begin{equation}
\frac{d}{dt}U(t)=-\mathfrak{i}H(t)U(t)
\end{equation}
and is in general given by the Dyson series as
\begin{align}
U_{t,t_0}&:=\mathpzc{T}_{\leftarrow}e^{-\mathfrak{i}\int_{t_0}^tdt_1 H(t_1)},\nonumber \\&=\mathds{1}+\sum_{n=1}^\infty(-\imag)^n \int_{t_0}^tdt_1...\int_{t_0}^{t_{n-1}} dt_n \mathpzc{T}_{\leftarrow} H(t_1)...H(t_n) \nonumber \\
&= \mathds{1}+\sum_{n=1}^\infty\frac{(-\mathfrak{i})^n}{n!}\int_{t_0}^tdt_1...\int_{t_0}^t dt_n  H(t_1)...H(t_n). \label{Dyson series unitary}
\end{align}

\subsection{Unitary evolution given a time-independent Hamiltonian}
In the special case of a time-independent Hamiltonian, the unitary operator simplifies to $U_{t,t_0}=e^{-\mathfrak{i}(t-t_0) H}$. From now on, for time-independent Hamiltonians, we shall    use the simplified notation $U_\tau:=e^{-\imag \tau H}$ where $\tau=t-t_0$.   The stationary states of the unitary channel $\e_{U_\tau}$ are convex combinations of the pure states $\{\Pi(\phi_i)\}$, where $\{\phi_i\}$ form a basis in which $H$ is diagonal. This  is often referred to as the eigenbasis of $H$ and the individual  pure states are referred to as the eigenstates, or eigenvectors, of the Hamiltonian.

The unitary operators generated by time-independent Hamiltonians form dynamical semigroups, because $U_{\tau_1}U_{\tau_2}=U_{\tau_1+\tau_2}$. This is easily understood when considering the unitary operators in their diagonal form
\begin{equation}
U_\tau=\sum_ie^{-\imag \tau E_i}\Pi(\phi_i)
\end{equation}
given the Hamiltonian's eigenbasis $\{\phi_i\}$ and energies $\{E_i\}$. If the Hamiltonian is the same for both unitaries, then they are diagonal in the same basis, and their concatenation  results in the addition of their phases. Furthermore, given a finite dimensional Hilbert space, we may always find a $\mathpzc{T} < \infty$ such that 
\begin{equation}
\mathpzc{T}E_n=2\pi \ \forall \ E_n.
\end{equation}  
At times $\mathpzc{T}$, then,  $U_\tau$ becomes an identity operator. As $U_\tau$ is continuous and differentiable, we may say that it forms an \emph{orbit} in the state space $\s(\co^d)$. Any initial input state $\rho(t_0)$ will be taken along a smooth and differentiable loop and, at times $N\mathpzc{T}$, where $N$ is an integer, will come back to $\rho(t_0)$. The orbit time $\mathpzc{T}$ depends upon the dimension of the Hilbert space, growing longer as the Hilbert space gets larger. 
\subsection{Unitary evolution given a time-dependent Hamiltonian}   
A useful technique for dealing with time-dependent Hamiltonians  is to separate them into a time-independent and time-dependent part as $H(t)=H_0+H_I(t)$. The dynamics can then be taken into the \emph{interaction picture} by the transformation of both the Hamiltonian and the state by the unitary operator $U_{0,\tau}:=e^{-\mathfrak{i}\tau H_0}$ as 
\begin{align}
H(t) &\mapsto \tilde H(t)=U_{0,\tau}^\dagger H(t)U_{0,\tau},\nonumber \\ \rho(t)&\mapsto \tilde \rho(t)=U_{0,\tau}^\dagger\rho(t)U_{0,\tau}. \label{interaction picture substitution}
\end{align}
The Liouville-von Neumann equation in the interaction picture involves only the interaction Hamiltonian $\tilde H_I(t)$, and is given by
\begin{equation}
 \frac{d}{dt}\tilde\rho(t)= \mathfrak{i}[\tilde \rho(t),\tilde H_I(t)]_-
\end{equation} 
which gives $\tilde U_{t,t_0}= U_{0,\tau}^\dagger U_{t,t_0} U_{0,\tau}$. To see that this is equivalent to the Schr\"odinger picture, we insert the  substitutions in Eq.\eqref{interaction picture substitution} such that
\begin{align}
& \frac{d}{dt}\left(e^{\mathfrak{i}\tau H_0}\rho(t)e^{-\mathfrak{i}\tau H_0}\right)= \mathfrak{i}[e^{\mathfrak{i}\tau H_0} \rho(t)e^{-\mathfrak{i}\tau H_0}, e^{\mathfrak{i}\tau H_0}H_I(t)e^{-\mathfrak{i}\tau H_0}]_- \nonumber \\ &\Rightarrow e^{\mathfrak{i}\tau H_0}\frac{d}{dt}\rho(t)e^{-\mathfrak{i}\tau H_0}=\mathfrak{i}[e^{\mathfrak{i}\tau H_0}\rho(t)e^{-\mathfrak{i}\tau H_0},H_0]_-+\mathfrak{i}[e^{\mathfrak{i}\tau H_0} \rho(t)e^{-\mathfrak{i}\tau H_0}, e^{\mathfrak{i}\tau H_0}H_I(t)e^{-\mathfrak{i}\tau H_0}]_- \nonumber \\
&\Rightarrow  e^{\mathfrak{i}\tau H_0}\frac{d}{dt}\rho(t)e^{-\mathfrak{i}\tau H_0}=e^{\mathfrak{i}\tau H_0}\left(\mathfrak{i}[\rho(t),H_0+H_I(t)]_-\right)e^{-\mathfrak{i}\tau H_0}.
\end{align} 
\section{Open quantum system dynamics}

An open quantum system is the extension of a closed system to a more realistic setting. Here, the quantum system in question is embedded in a  larger system that is itself considered to be closed.\footnote{Perhaps only the universe can  reasonably be assumed to be closed. It is difficult to conceive of a system that is \emph{fully} isolated from the rest of the universe, as even a scattering event by a single photon, where the latter is unaccounted for, would render it an open system. }  As such, the dynamics of an open system is given by the reduced dynamics of the larger one
\begin{equation}
\mathscr{L}[\rho^{A}(t)]\equiv \frac{d}{d t}\rho^A(t)=\mathrm{tr}_B\left(\mathfrak{i}[\rho(t),H(t)]_- \right) \label{trace liouville von neumann}.
\end{equation}

The component of the full Hilbert space which excludes the system is one on which we are unable to perform any measurements, and therefore is designated the term \emph{environment} or  \emph{bath} as inspired by thermodynamics. It is possible to write the full system Hamiltonian as 
\begin{equation}
H=H^A\otimes\mathds{1}^B+H_I^{A+B}+\mathds{1}^A\otimes H^B \label{open system hamiltonian}
\end{equation}
where the interaction term is isolated from the Hamiltonians that govern the system $A$ and environment $B$ only. As shown in the discussion of Stinespring's dilation theorem in Sec.\ref{Stinespring's dilation theorem}, if the system and environment are initially in a product state, then the evolution undergone by the system's reduced density operator is described by a quantum channel. This is a non unitary channel if and only if the unitary evolution the composite system undergoes is inseparable, which results from the presence of the interaction Hamiltonian. As such, we may characterise non unitary evolution of open quantum systems by the entanglement established between them and their environment.  

 Solving Eq.\eqref{trace liouville von neumann} is generally very difficult, however, and grows more so exponentially as the size of the environment increases. Approximate techniques do exist to treat the reduced dynamics of specific systems, however, with the most common one being the  master equation technique. In what follows, we give a brief overview of this. 
   
\subsection{From dynamical semigroups to the Lindblad master equation }

As mentioned previously, the unitary operators generated by time-independent Hamiltonians form dynamical semigroups. This is a special subclass of dynamical semigroups in general, defined as  a family of quantum channels $\{\e_\tau : \tau \in \mathds{R}^+ \}$    satisfying $\e_{\tau_2}\circ \e_{\tau_1}=\e_{\tau_1+\tau_2}$ for all $\tau_1,\tau_2 \in \mathds{R}^+$, and trivially $\e_0=\mathds{1}$.   As outlined in \citep{breuer}  a family of quantum channels that form  a dynamical semigroup, under certain mathematical conditions which we shall not cover here, can be expressed in exponential form as 
\begin{equation}
\e_\tau=e^{\tau\mathscr{L}}
\end{equation}  
where the Liouville super-operator $\mathscr{L}$ is the generator of this map. Such a generator can be expressed most generally as the \emph{Lindblad master equation} which has the form
\begin{equation}
\mathscr{L} [\rho(t)]\equiv \frac{d}{dt}\rho(t)=\imag \left[\rho(t),H\right]_-+\mathscr{D}[\rho(t)]
\end{equation} 
where $H$ is the time-independent Hamiltonian governing the unitary part of the dynamics and $\mathscr{D}$ is the \emph{dissipator} given by
\begin{equation}
\mathscr{D}[\rho(t)]=\sum_i^{N \leqslant d^2-1}\gamma_i\left(L_i\rho(t)L_i^\dagger-\left[\rho(t),L_i^\dagger L_i \right]_+ \right).
\end{equation}

Here, $[\cdot,\cdot]_+$ is the anti-commutator defined as $[A,B]_+:=AB+BA$. The $\gamma_i$ are positive,  having the dimension of inverse time, and can be interpreted as decay rates. The associated Kraus operators $L_i$ are also called  Lindblad operators in such a context. $d$ has its usual meaning as the dimension of the Hilbert space. Such a differential equation preserves the positivity, trace, and self-adjoint properties of density operators.  

Due to the semigroup structure of this form of dynamics, the operators in the differential equation are time-independent, and the evolution of state $\rho(t)$ does not depend on its history but rather only on its  configuration at the infinitesimal time $t$.  This type of dynamics is called \emph{Markovian}. 

\subsection{A    microscopic derivation of the Lindblad master equation}\label{master equation derivation} 
In the previous section we discussed how quantum channels that form a dynamical semigroup  are generated by the Lindblad master equation. Here we wish to start off with a microscopic description of an open quantum system, and from this derive a Markovian master equation which, to ensure it preserves the properties of density operators, we must be able to give in Lindblad form. Provided we have knowledge of the total Hamiltonian of the system and its environment, we may use the method of the \emph{weak-coupling approximation} which, as the name implies, assumes the system and its environment are weakly coupled.  The arguments provided follow those from   \citep{breuer} and \citep{Masterequation-tutorial}.  

 Let the time-scale for the relaxation of the environment to the thermal equilibrium state $\varrho$ such that $[\varrho,H^B]_-=\mathds{O}$ be  $ \tau_\mathrm{env}$,  and the time-scale for the relaxation of the system due to its interaction with the environment be $ \tau_\mathrm{sys}$. If  the interaction strength between the system and environment,  $V:=\|H_I \|$, is \emph{weak} in  comparison with the spectral width given by the environment relaxation time, $V \tau_\mathrm{env} \ll 1$,  then the coarse-grained time-scale for the system evolution, $\Delta t$, satisfies the condition $ \tau_\mathrm{sys} \gg \Delta t \gg  \tau_\mathrm{env}$. We are therefore justified in expressing the  system-environment composite state for all time $t$  as the product state  $\rho_{\mathrm{tot}}(t)=\rho(t)\otimes\varrho$, which is known as the Born approximation.

We will develop our master equation by using the Liouville-von Neumann equation in the interaction picture. Firstly, we note that  the rate of change in the total system in our coarse-grained picture is given by \emph{averaging out} the small-time fluctuations of the state in the interval $\Delta t$; $\frac{\Delta\tilde\rho_{\mathrm{tot}}(t)}{\Delta t}=\frac{1}{\Delta t}\int_t^{t+\Delta t}ds\frac{d}{ds}\tilde \rho_{\mathrm{tot}}(s)$. To begin, we use the fact that the density operator can be written as $\tilde\rho_{\mathrm{tot}}(t)=\tilde\rho_{\mathrm{tot}}(t_{0})+\imag\int_{t_0}^{t} ds[\tilde\rho_{\mathrm{tot}}(s),\tilde H_{I}(s)]_-$ to write the Liouville-von Neumann equation for the reduced dynamics in   differentio-integral form.   
\begin{align}
\frac{\Delta\tilde\rho(t)}{\Delta t}&=\frac{1}{\Delta t}\left(\imag\int_{t}^{t+\Delta t}dt_1\mathrm{tr}_B\left[\tilde\rho(t)\otimes \varrho,\tilde H_I(t_1)\right]_- \nonumber \right) \\ &-\frac{1}{\Delta t}\left( \int_{t}^{t+\Delta t}dt_1\int_{t}^{t_1}dt_2\mathrm{tr}_B\left[\left[\tilde \rho(t)\otimes \varrho,\tilde H_I(t_2)\right]_-,\tilde H_I(t_1)\right]_-\right) \nonumber \\
&-\frac{1}{\Delta t}\left(\imag  \int_{t}^{t+\Delta t}dt_1\int_{t}^{t_1}dt_2\int_t^{t_2}dt_3 \mathrm{tr}_B\left[\left[\left[\tilde \rho(t_3)\otimes \varrho,\tilde H_I(t_3)\right]_-,\tilde H_I(t_2)\right]_-,\tilde H_I(t_1)\right]_-\right).
\end{align}  

Because we have assumed the interaction Hamiltonian is weak, we may neglect the third term of this equation, proportional to $V^3$, which contains the contribution of the \emph{history} of the state in the interval $[t,t+\Delta t]$. The resultant equation will only depend on the state of the system at the start of the coarse-graining -- $\tilde \rho(t)$ -- and hence we can make the approximation $\frac{\Delta\tilde\rho(t)}{\Delta t}=\frac{d}{dt}\tilde\rho(t)$ which treats the rate of change of the state as a differential operator dependent on the state of the system  at time $t$ only. This is known as the Markovian or coarse-graining approximation. 

As the commutators  still contain the full interaction Hamiltonian, this equation is not very useful.  We can proceed by writing the interaction Hamiltonian in its operator-Schmidt decomposition
\begin{align*}
\tilde H_I(t)&= \sum_\alpha\mathcal{\tilde A}_\alpha(t)\otimes \mathcal{\tilde B}_\alpha(t)\\
&=\sum_\alpha e^{\imag t H^A}\mathcal{A}_\alpha e^{-\imag t H^A}\otimes e^{\imag t H^B}\mathcal{B}_ \alpha e^{-\imag t H^B}
\end{align*}
where the individual $\ma_\alpha$ and $\mb_\alpha$ need not be themselves self-adjoint operators. \footnote{Here we have set $t_0=0$ so as to change  our usual notation of $U_\tau$ to $U_t$. This is done to avoid confusion with the use of a different symbol $\tau$ later on. } This makes our differential operator
\begin{align}
\frac{d}{dt}\tilde\rho(t)&=\imag\sum_\alpha \frac{1}{\Delta t}\int_{t}^{t+\Delta t}dt_1 \mathrm{tr}_B\left[\tilde\rho(t)\otimes\varrho, \mathcal{\tilde A}_\alpha(t_{1})\otimes \mathcal{\tilde B}_\alpha(t_{1}) \right]_- \nonumber \\&-\sum_{\alpha,\beta}\frac{1}{\Delta t}\int_{t}^{t+\Delta t}dt_1\int_{t}^{t_1} dt_{2}\mathrm{tr}_B\left[ \left[\tilde \rho(t)\otimes\varrho, \mathcal{\tilde A}_\beta(t_{2})\otimes \mathcal{\tilde B}_\beta(t_{2}) \right]_- ,\mathcal{\tilde A^\dagger}_\alpha(t_{1})\otimes \mathcal{\tilde B^\dagger}_\alpha(t_{1})  \right]_- \ 
\end{align}       
where we have used the fact that $H_I$ is a self-adjoint operator to make the substitution   $\sum_\alpha\mathcal{\tilde A}_\alpha(t)\otimes \mathcal{\tilde B}_\alpha(t)= \sum_\alpha\mathcal{\tilde A^\dagger}_\alpha(t)\otimes \mathcal{\tilde B^\dagger}_\alpha(t)$, which will prove to be a useful mathematical tool later on. Owing to the fact that the state of the environment is stationary, and that we can always shift the energy scales, we may make the non-restrictive assumption that $\<\mathcal{B}_\alpha\>:=\mathrm{tr}_B[\mathcal{B}_\alpha \varrho]=0 \ \forall \ \alpha$ \footnote{We may write $H_I'=\sum_\alpha \mathcal{A}_\alpha\otimes(\mathcal{B}_\alpha-\<\mb_\alpha\>)=\sum_\alpha \ma_\alpha\otimes\mb_\alpha - \sum_\alpha \<\mb_\alpha\> \ma_\alpha\otimes\mathds{1}$ for any arbitrary $\<\mb_\alpha\>$. Taking the expectation value of the interaction Hamiltonian with respect to subsystem B then gives $\<H_I'\>_B=\sum_\alpha \ma_\alpha(\<\mb_\alpha\>-\<\mb_\alpha\>)=0 $, and we may take the full Hamiltonian as $H'=(H^A+\sum_\alpha\<\mb_\alpha\>\ma_\alpha)\otimes\mathds{1}^B+\mathds{1}^A\otimes H^B+H_I'$. }.  This has the consequence that the first part of the differential equation can be ignored. 

Before continuing further, it is useful to make some mathematical modifications to this differential equation. Firstly, let us express the operators $\ma_\alpha$ in the eigenbasis of $H^A$, given as $\{\phi_i\}$. This is done by noting that
\begin{align}
\ma_\alpha&=\sum_\Omega\ma_\alpha(\Omega) \nonumber \\
\ma_\alpha(\Omega) &=\sum_{i,j}\delta(\Omega -\omega_{ij})\<\phi_j|\ma_\alpha \phi_i\>|\phi_j\>\<\phi_i|
\end{align}    

where $\omega_{ij}=\<\phi_i|H^A\phi_i\>-\<\phi_j|H^A \phi_j\>$ and $\delta(\cdot)$ is the Kronecker delta function. Consequently  the interaction picture operators are given by 
$\tilde \ma_{\alpha}(\Omega,t)=e^{-\imag t\Omega}\ma_\alpha(\Omega)$ and $\tilde \ma_{\alpha}^\dagger(\Omega,t)=e^{\imag t\Omega}\ma_\alpha(\Omega)$ . Furthermore, let us make the observation that for self-adjoint operators $\{A,B,C\}$ we have 
\begin{equation}\left[\left[A,B\right]_-,C\right]_-=ABC-CAB + \mathbb{H.C}
\end{equation} where $\mathbb{H.C}$ denotes the Hermitian conjugate of the terms appearing before it.  Finally, we note that as $t_1\geqslant t_2$, we can make  the substitution $t_1-t_2=\tau \geqslant 0$. Changing the integration  variables and limits accordingly leads to $\int_{t}^{t+\Delta t}dt_1\int_{t}^{t_1}dt_2=\int_{0}^{\Delta t}d\tau\int_{t+\tau}^{t+\Delta t}dt_1$ and our differential equation takes the form
\begin{align}
\frac{d}{dt}\tilde\rho(t)=\sum_{\alpha,\beta}\sum_{\Omega,\Omega'} &\frac{1}{\Delta t} \int_{0}^{\Delta t}d\tau e^{\imag\Omega \tau}\int_{t+\tau}^{t+\Delta t}dt_1 e^{\imag(\Omega'-\Omega)t_1}\<\tilde\mb_\alpha^\dagger(t_1)\tilde \mb_\beta(t_1-\tau)\> \nonumber \\ 
&\times \left(\ma_\beta(\Omega)\tilde \rho(t)\ma_\alpha^\dagger(\Omega')-\ma_\alpha^\dagger(\Omega')\ma_\beta(\Omega)\tilde\rho(t)\right)+\mathbb{H.C} \ \ . \end{align}

 Because of the cyclicity of the trace operator and the commutatitivity of $H^B$ with $\varrho$, we have the identity 
\begin{align}
\<\tilde\mb_\alpha^\dagger(t_1)\tilde \mb_\beta(t_1-\tau)\>&=\mathrm{tr}[\tilde\mb_\alpha^\dagger(t_1)\tilde \mb_\beta(t_1-\tau)\varrho] \nonumber \\
&=\mathrm{tr}\left[e^{\imag t_1H^B}\mb_\alpha^\dagger e^{-\imag t_1H^B}e^{\imag(t_1-\tau)H^B}\mb_\beta e^{-\imag(t_1-\tau)H^B}\varrho\right] \nonumber \\
&= \mathrm{tr}\left[e^{\imag\tau H^B}\mb_\alpha^\dagger e^{-\imag\tau H^B}\mb_\beta \varrho\right]\nonumber \\ &=\<\tilde\mb_\alpha^\dagger(\tau)\tilde \mb_\beta(0)\> \end{align}
which we may call the \emph{bath time correlation function}. This only depends on the time  $\tau$ during which the initially uncorrelated system and environment have been undergoing the joint evolution process. Because of our coarse-graining assumption that $\Delta t \gg \tau_{\mathrm{env}}$  this function vanishes \emph{sufficiently} fast such that we may take the  upper limit of the integral of $d\tau$ to infinity, and the lower limit of the integral of $dt_1$ to $t$.  \footnote{The coarse-graining $\Delta t$ is bounded by the bath correlation time $\tau_{\mathrm{env}}$, which is itself limited by the  interaction strength $V$ owing to the weak-coupling approximation  $V \tau_{\mathrm{env}}\ll 1$. Hence the period  in which the correlation function can be appreciable, such that the Markovian approximation is still valid, grows longer the weaker the interaction strength becomes.  }

  We now bring our differential equation back into the Schr\"odinger picture to obtain
\begin{align}
\frac{d}{dt}\rho(t)&=\imag\left[\rho(t),H^A\right]_- \nonumber \\&+\sum_{\alpha,\beta}\sum_{\Omega,\Omega'}   G_{\alpha,\beta}(\Omega)J_{\alpha,\beta}(\Omega,\Omega')\left(\ma_\beta(\Omega) \rho(t)\ma_\alpha^\dagger(\Omega')-\ma_\alpha^\dagger(\Omega')\ma_\beta(\Omega)\rho(t)\right) +\mathbb{H.C}.
\end{align}
where we have made the substitutions
\begin{align}
G_{\alpha,\beta}(\Omega)&=\int_{0}^{\infty}d\tau\<\tilde\mb_\alpha^\dagger(\tau)\tilde \mb_\beta(0)\>  e^{\imag\Omega \tau}, \nonumber \\
J(\Omega,\Omega')&=e^{\imag(\Omega-\Omega')t}\int_{t}^{t+\Delta t}dt_1 \frac{e^{\imag(\Omega'-\Omega)t_1}}{\Delta t}.
\end{align}
This Markovian differential equation is incomplete however, as it is not guaranteed that it forms the generator of dynamical semigroups. Indeed, there are cases where such a differential operator has been shown to fail the complete positivity criterion.  
An additional step required to get a Markovian master equation that generates a dynamical semigroup is the \emph{secular approximation}.
 We make the observation that
\begin{equation}
J(\Omega,\Omega')=e^{\imag(\Omega'-\Omega)\frac{\Delta t}{2}}\frac{\sin[(\Omega'-\Omega)\Delta t/2]}{(\Omega'-\Omega)\Delta t/2}
\end{equation}
 which is independent of the time variable $t$.  The absolute value of this function becomes infinitesimally narrow around the region $\Omega'-\Omega=0$ for all the frequency terms $\{\Omega\}$ if the time scale $\Delta t$ is \emph{sufficiently long} \footnote{Because $\Delta t \ll \tau_{\mathrm{sys}}$ we may relax this condition to $ \Delta t\gtrsim 1/|\Omega'-\Omega|  \ \forall \ \Omega\ne\Omega' $}  such that $\Delta t \gg 1/ |\Omega'-\Omega| \ \forall \ \Omega\ne\Omega' $, which may also be expressed as $V/|\Omega' - \Omega|\ll 1 \ \forall \ \Omega \ne \Omega' $.  If this condition is met, we may make the secular approximation with the replacement  $J(\Omega,\Omega')=\delta(\Omega-\Omega')$. This  can be qualitatively expressed as the energy states of the system fluctuating many times during the period of appreciable change caused by the environment, such that terms with different frequencies will be \emph{averaged out}. 

 Taking advantage of the fact that we are free to swap the indices $\{\alpha,\beta\}$, we may incorporate the Hermitian conjugate component into the Master equation to obtain  \begin{align}
\frac{d}{dt} \rho(t)=\imag\left[\rho(t),H^A\right]_- +\sum_{\alpha,\beta}&\sum_\Omega \left[G_{\alpha,\beta}(\Omega)+G_{\beta,\alpha}^*(\Omega)\right] \ma_\beta(\Omega) \rho(t)\ma_\alpha^\dagger(\Omega) \nonumber\\&  -\left(G_{\alpha,\beta}(\Omega) \ma_\alpha^\dagger(\Omega)\ma_\beta(\Omega)\rho(t)+G_{\beta,\alpha}^*(\Omega) \rho(t)\ma_\alpha^\dagger(\Omega)\ma_\beta(\Omega)\right).
\end{align}
  We now introduce some new notation.  $G_{\alpha,\beta}(\Omega)$ can be written as
\begin{equation}
G_{\alpha,\beta}(\Omega)=\frac{1}{2}\Gamma_{\alpha,\beta}(\Omega)+\imag\Lambda_{\alpha,\beta}(\Omega) \end{equation}
where we have used
\begin{align}
\Gamma_{\alpha,\beta}(\Omega)&= G_{\alpha,\beta}(\Omega)+G_{\beta,\alpha}^*(\Omega), \nonumber \\
\Lambda_{\alpha,\beta}(\Omega)&=\frac{1}{2\imag}\left( G_{\alpha,\beta}(\Omega)-G_{\beta,\alpha}^*(\Omega)\right).
\end{align}

The matrices $\boldsymbol\Gamma(\Omega)$ and $\boldsymbol\Lambda(\Omega)$ are Hermitian and positive semi-definite. Given that for an operator $L$ we have the identity $\mathrm{tr}[L]^*=\mathrm{tr}[L^\dagger]$, we note that \begin{align*}
G_{\beta,\alpha}^*(\Omega) &= \left(\int_0^\infty d\tau e^{\imag\Omega\tau} \mathrm{tr}_B[\tilde \mb_\beta^\dagger(\tau)\tilde \mb_\alpha(0)\varrho] \right)^*= \int_0^\infty d\tau e^{-\imag\Omega\tau} \mathrm{tr}_B[\tilde \mb_\alpha^\dagger(0)\tilde \mb_\beta(\tau)\varrho]\\
&= \int_0^\infty d\tau e^{-\imag\Omega\tau} \mathrm{tr}_B[e^{-\imag\Omega\tau} \mb_\alpha^\dagger e^{\imag\Omega\tau} \mb_\beta\varrho]=  \int_0^\infty d\tau e^{-\imag\Omega\tau} \mathrm{tr}_B[\tilde \mb_\alpha^\dagger(-\tau)\tilde \mb_\beta(0)\varrho] \\ &=  \int_{-\infty}^0 d\tau e^{\imag\Omega\tau} \mathrm{tr}_B[\tilde \mb_\alpha^\dagger(\tau)\tilde \mb_\beta(0)\varrho]
\end{align*}
where in the last line we make the substitution of variables $\tau \to -\tau$. This leads to 
\begin{equation}
\Gamma_{\alpha,\beta}(\Omega) = \int_{-\infty}^\infty d\tau e^{\imag\Omega\tau} \<\tilde\mb_\alpha^\dagger(\tau)\tilde \mb_\beta(0)\>
\end{equation}

which is the Fourier transform of the bath correlation function $\<\tilde\mb_\alpha^\dagger(\tau)\tilde \mb_\beta(0)\>$, and

\begin{equation}
\Lambda_{\alpha,\beta}(\Omega)=\frac{1}{2\imag}\left(\int_0^\infty d\tau e^{\imag\Omega \tau} \<\tilde\mb_\alpha^\dagger(\tau)\tilde \mb_\beta(0)\> - e^{-\imag\Omega \tau} \<\tilde\mb_\alpha^\dagger(-\tau)\tilde \mb_\beta(0)\>\right).
\end{equation}
Substituting these into our differential equation yields a  Markovian master equation in its first standard form, which is
\begin{equation}
\mathscr{L}[\rho(t)]\equiv\frac{d}{dt}\rho(t)=\imag\left[\rho(t),H^A+H_{LS} \right]_-+ \mathscr{D}[\rho(t)]
\end{equation}
where 
\begin{equation}
H_{LS}=\sum_{\alpha,\beta}\sum_\Omega \Lambda_{\alpha,\beta}(\Omega)\ma_\alpha^\dagger(\Omega) \ma_\beta(\Omega)
\end{equation}
is the \emph{Lamb shift}, whose action is the change in the energy levels of the system free Hamiltonian $H^A$, and the dissipator is given by 
\begin{equation}
\mathscr{D}[\rho(t)]=\sum_{\alpha,\beta}\sum_\Omega\Gamma_{\alpha,\beta}(\Omega)\left(\ma_{\beta}(\Omega)\rho(t)\ma_\alpha^\dagger (\Omega) -\frac{1}{2}\left[\ma_\alpha^\dagger(\Omega)\ma_\beta(\Omega), \rho(t) \right]_+  \right).
\end{equation}
This master equation preserves all properties of density operators, as it can always be written in Lindblad form. This can be done by choosing an appropriate unitary operator $U$ which diagonalises the positive decay rate matrix $\boldsymbol\Gamma(\Omega)=\sum_{\alpha,\beta}\Gamma_{\alpha,\beta}(\Omega)$ as $ U\boldsymbol \Gamma(\Omega) U^\dagger=\sum_\mu \Gamma'_\mu (\Omega)$. The resultant Lindblad operators in the new basis are given by   $L_\mu=\sum_\alpha^{}U_{\alpha,\mu}^\dagger \ma_\alpha$ , and so the dissipator for the Lindblad Master equation is given by 
\begin{equation}
\mathscr{D}[\rho(t)]=\sum_{\mu}\sum_\Omega\Gamma'_{\mu}(\Omega)\left(L_{\mu}(\Omega)\rho(t)L_\mu^\dagger (\Omega) -\frac{1}{2}\left[L_\mu^\dagger(\Omega)L_\mu(\Omega), \rho(t) \right]_+  \right).
\end{equation}

\subsection{Decoherence in open quantum systems}\label{decoherence in open systems section}

Let us make one final remark that relates our discussion thus far with the concept of decoherence \citep{introduction-decoherence}. Decoherence is the term used to describe the irreversible process of environment-induced effective superselection rules \citep{Zurek-einselection} in a Hilbert space, whereby superpositions can only be established within specified subspaces, and not between them. Although developed in its inception to explain quantitatively the lack of interference effects in the macroscopic world (Schr\"odinger's cat), it has come to be used in the domain of mesoscopic, and even microscopic systems. The exact microscopic model that best describes the decoherence process is reliant upon the system in question. Nonetheless, a paradigmatic model that aides in conceptualising the problem is that of an environmental measuring process, which was covered briefly in Sec.\ref{Stinespring's dilation theorem}. Although no person performs a measurement on the environment, so long as the system and environment become entangled such that  measurements on the latter would reveal information pertaining to the state of the prior, the system will undergo an irreversible quantum channel. It should be noted, however, that without the process of measurement on the environment, the loss of coherence will not be irreversible, as the unitary evolution operator will eventually \footnote{Except, of course, if the system is infinite-dimensional.} bring the system back to its initial configuration. In this formulation, then, the generation of entanglement between system and environment is necessary but not sufficient for decoherence.

 There are, generally speaking, two types of decoherence: Pure decoherence and dissipative decoherence.
\begin{enumerate}[(i)]
\item
 Pure decoherence, in the language of quantum channels,  is caused by the pure dephasing channel Eq.\eqref{dephasing channel}. Given some preferred basis the diagonal elements of the density operator  are left intact but all off-diagonal elements, or coherences, are destroyed. Usually, the preferred basis is the eigenbasis of the free Hamiltonian describing the system. In the language of open quantum systems, a  sufficient condition for pure decoherence is that the interaction Hamiltonian  between the system and environment commutes with the system Hamiltonian.\footnote{In the case that the interaction Hamiltonian $H_I$ does not commute with the system Hamiltonian $H^A$, decoherence will be \emph{perturbatively} pure if $\| H^A \| \gg \|H_I\|$.} 

\item Dissipative decoherence, in addition to destroying coherences, also changes the populations of the system. The depolarising channel Eq.\eqref{depolarising channel} is an example of such a mechanism.
 In the language of open quantum systems, this usually results in the exchange of energy between the system and environment. A necessary  condition for dissipative decoherence is that the interaction Hamiltonian between the system and environment does not commute with the system Hamiltonian. \end{enumerate}

 \spacing{1} 
 \bibliographystyle{plainnat}                                 
 \bibliography{references}

\spacing{1}
\chapter{Quantum measurement and control with magnetic resonance}
\section{Introduction}
Magnetic resonance  is a  paradigmatic experimental framework for investigating the interaction of the intrinsic spin of a material with electromagnetic radiation; it is a well established experimental field, with a history of over 50 years, explained by quantum theory. The  intrinsic spin of the system -- first demonstrated by the Stern-Gerlach experiment \citep{Stern-Gerlach} -- in  most magnetic resonance experiments  is  associated either with the electronic or nuclear degrees of freedom of the material. The intrinsic spin is the prototypical example of a quantum system described by a finite dimensional Hilbert space. Indeed, many texts will refer to a finite dimensional quantum system, that is not spin proper, as a \emph{pseudo-spin}. Regarding the radiation, as the name implies, it is the (oscillating) magnetic component thereof which is of interest. Given typical  high  intensities of the electromagnetic field, we can treat it classically, and for a large wavelength of the field in comparison with the size of the sample under study, we may stipulate that the strength of the magnetic field is independent of position. The effective Hamiltonian  governing the spin degree of freedom in such a paradigm can therefore be given as 
\begin{equation}
H(t)= \sum_{i\in \{x,y,z\}}\gamma_i\left[ B_{0i}+B_{1i}(t)\right]J_i
\end{equation}    
where $\{ B_{0x}, B_{0y}, B_{0z}\}$ and $\{B_{1x}(t),B_{1y}(t),B_{1z}(t)\}$ are, respectively, static and fluctuating  external magnetic fields.\footnote{Although typically $B_1 \ll B_0$, the photon occupancy of the electromagnetic field is high. This is why we are justified in treating this classically.} Furthermore,  $\{J_x,J_y,J_z\}$ are the spin observables, whose properties are covered in Appendix \ref{angular momentum appendix},  and $\{\gamma_x,\gamma_y, \gamma_z\}$ are the gyromagnetic ratios that determine how strongly the spin operators couple to the external field.  As the static magnetic field  defines the coordinate frame of the spin system, we may define the $z$ axis with respect to this, and relabel it $B_0$. The situations that we will be studying require the electromagnetic field to be propagating in a perpendicular direction to $B_0$, and consequently only $\{B_{1x}(t),B_{1y}(t)\}$ will remain. The resulting Hamiltonian can thus be rewritten as
\begin{align}
H(t)&= \gamma_zB_0J_z  + \gamma_xB_{1x}(t)J_x+\gamma_yB_{1y}(t)J_y,\nonumber \\
&=H_0+H_I(t).
\end{align} 
  Magnetic resonance, then, is the phenomenon where  an  $H_I(t)$ whose time dependence is exhibited by a sinusoidally varying magnetic field, with frequency $\omega$,  induces transitions between eigenstates of $H_0$; the pure stationary states of the unitary channel $\e_{U_{0,\tau}}$. For a given pair of eigenstates $\{\phi_i,\phi_j\}$, these transitions are possible if (a) $\<\phi_i|H_0\phi_i\>-\<\phi_j|H_0 \phi_j\>=\omega$, hence the resonance in the name, and (b)  $|\<\phi_i|J_x\phi_j\>|>0$ or, equivalently, that $\Delta m_J=\pm 1$ where $m_J$ is the spin quantum number. We say that the transition is allowed by the \emph{dynamical selection rule for $J_x$}.      

If the transition is that of nuclear spin states, it is called nuclear magnetic resonance (NMR), and if it is electron spins that have these transitions the effect is named electron spin resonance (ESR). \footnote{In the literature this is often also referred to as electron paramagnetic resonance (EPR). I opt not to use this term so as to avoid any confusion with the Einstein-Podolsky-Rosen (EPR) paradox. } As far as the mathematical model used to describe the phenomena is concerned, these two processes are identical, save for the different gyromagnetic ratios for the two spin types, where that of electron spins is around three orders of magnitude larger than the nuclear one. A consequence of this is that, for values of the static field that can be achieved in most experiments today,  the transition frequencies observed in NMR are in the radio frequency (RF) range, whereas those of ESR experiments are in the micro wave (m.w.) domain. \footnote{RF frequencies are $\sim$ MHz and m.w. frequencies are $\sim$ GHz.} 
NMR as it is understood today was first observed by two groups in America who both published their results in Physical Review in the year 1946 \citep{NMR-Bloch,NMR-Purcell}.  ESR was first observed in the USSR in 1945 \citep{Zavoisky}, and independently developed two years later by researchers in the United Kingdom \citep{Bagguley}. Systems with both nuclear and electronic degrees of freedom were shown to be manipulatable using electron nuclear double resonance (ENDOR) techniques, demonstrated by \citep{FeherEndor}. A good textbook covering techniques for ESR and ENDOR is \citep{schwieger}, while NMR  is covered by \citep{NMR-relaxation}.

\section{Controlled dynamics in magnetic resonance }

\subsection{Lie algebras, Lie groups, and controllability}\label{Lie algebras and groups}\label{Lie algebras}
In Sec.\ref{closed quantum system dynamics} it was shown that the dynamics of a closed quantum system is governed by a unitary operator that is the solution to Schr\"odinger's equation. If we want to achieve controllability, defined as the ability to perform  unitary operations capable of mapping between any two pure states in a given state space, we must be able to control the system Hamiltonian. Naively approaching the problem of controllability would seem to require an infinity of Hamiltonians an experimentalist needs to construct in the lab, one associated with each \emph{orbit}. In the case of finite dimensional Hilbert spaces, however, we may use the concept of Lie algebras and groups to understand how controllability can be achieved given a finite set of Hamiltonians an experimentalist can switch between at will \citep{Domenico-control-dynamics}. 

A Lie algebra $\mathfrak{L}$ is defined as a vector space with the addition of a binary operation $\mathfrak{L}\times\mathfrak{L} \to \mathfrak{L}$ defined as the Lie bracket, or commutator $[\cdot,\cdot]$, which for any ordered pair of elements $\{A,B\} \in \mathfrak{L}$ provides another element $[A,B]\in \mathfrak{L}$. If   the Lie algebra is one of complex matrices, which is called a \emph{linear Lie algebra}, then the commutator is defined as $[A , B]_-:=AB-BA$  and  the Lie algebra  has the following properties. \footnote{It is only condition (iii) that is particular to the Lie algebra of complex matrices. For general Lie algebras this is replaced by $[A,A]=\mathds{O} \ \forall \ A \in \mathfrak{L}$.}
\begin{enumerate}[(i)]
\item \emph{Bilinearity}: 
\begin{align}
[\alpha A+\beta B,C]_-=\alpha[A,C]_-+\beta[B,C]_-  \ \ &, \ \ [C,\alpha A+\beta B]_-=\alpha[C,A]_-+\beta[C,B]_-   \\
\forall A,B,C \in \mathfrak{L} \ \ , \ \ \alpha,\beta \in \mathds{C} \nonumber
\end{align}
\item \emph{Jacobi identity}: 
\begin{equation}
\left[A,\left[B,C\right]_{-}\right]_{-}+\left[B,\left[C,A\right]_{-}\right]_{-}+\left[C,\left[A,B\right]_{-}\right]_{-}=0
\end{equation}
\item \emph{Skew-symmetry condition}: 
\begin{equation}
[A,B]_-=-[B,A]_-
\end{equation}
\end{enumerate}

The linear Lie algebras of finite dimensional vector spaces can be generated by a finite set of elements. The generators of  $\mathfrak{L}$ are the smallest subset of its linearly independent elements $\{A_i\}$ such that  every element of the whole algebra can be obtained from this set (or any linear combination thereof) by the repeated use of the defined commutator. In the case of $d\times d$ complex matrices, the generators are a subset of  a given basis that spans $\lo(\mathds{C}^d)$.

A Lie group \footnote{In the case of  linear Lie algebras, the  group operation for $e^\mathfrak{L}$ is matrix multiplication.} $e^{\mathfrak{L}}$ is obtained by exponentiating the elements of a Lie algebra $\mathfrak{L}$, and is defined as
\begin{equation}
e^{\mathfrak{L}}=\{e^{B} : B \in \mathfrak{L}\}.
\end{equation}

Given the generators $\{ A_i\}_{i=1}^s$ of the linear Lie algebra $\mathfrak{L}$, we may find a finite number $r$ such that every  element of $e^{\mathfrak{L}}$ can be given as
\begin{equation}
e^{B}=\prod_{i=1}^r e^{t_iA_i}
\end{equation} 

where $\{t_i\}$ are positive scalar quantities. This follows from the fact that, as $[A_{i},A_{j}]_-\ne\mathds{O}$ for any pair of generators $\{A_i,A_j\}$, then the group multiplication $e^{tA_i} e^{tA_j}\ne e^{tA_i+tA_j}$. Instead, it is supplied by the Baker-Campbell-Hausdorff formula
\begin{equation}
e^{tA_i} e^{tA_j}=e^{tA_i+tA_j+\frac{t^2}{2}[A_i,A_j]_-+O(t^3)}.\label{BCH formula}
\end{equation}
 
Two linear Lie algebras,  of interest in quantum control theory, are the algebra of $d\times d$ skew-Hermitian\footnote{An operator $T$ is skew-Hermitian if $T^\dagger=-T$. Any $T$ can be constructed as $\pm\imag L$ given a Hermitian operator $L$. If $L$ is the Hamiltonian, then convention chooses $-\imag L$ as the element of $\mathfrak{su}(d)$ to give the generated unitary operator that is the solution to Schr\"odinger's equation.   } operators $\mathfrak{u}(d)$, and its \emph{subalgebra}\footnote{The subspace $\mathfrak{A}\subseteq \mathfrak{L}$ is a subalgebra of $\mathfrak{L}$ if it also forms an algebra given the commutator defined on $\mathfrak{L}$.} of skew-Hermitian operators of zero trace $\mathfrak{su}(d)$. The subalgebra  $\mathfrak{su}(d)$ is obtained by simply omitting the generator of $\mathfrak{u}(d)$ which is proportional to the identity operator and, as a consequence, has a dimension of $d^2-1$.   $\mathfrak{u}(d)$ generates the Lie group of unitary matrices $\uu(d)$, and  $\mathfrak{su}(d)$ generates the subgroup of unitary matrices with unit determinant $\mathcal{SU}(d)$, called the special unitary group. As the identity operator commutes with all the other generators of $\mathfrak{u}(d)$, any member of $\mathcal{SU}(d)$ can be turned to one of $\uu(d)$ by a multiplication of a phase factor, but this clearly has no effect on the evolution of quantum states. 

In summary,  if an experimentalist has access to a set of Hamiltonians $\{H_i\}$ that generate the Lie algebra $\mathfrak{su}(d)$, which gives the Lie group  $\mathcal{SU}(d)$, then he can achieve controllability in the state space  $\s(\mathds{C}^d)$.

\subsubsection{The Lie algebras $s\mathfrak{u}(2)$ and $s\mathfrak{o}(3)$ and the Bloch sphere} 

\begin{figure}[!htb]
\centering
\includegraphics[width=1.5in]{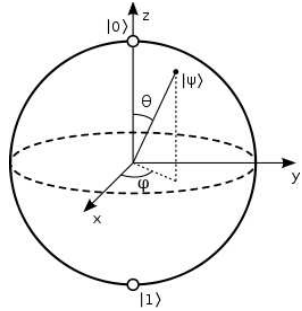} \caption{The Bloch sphere allows for an elegant geometrical interpretation of quantum states and dynamics in $\s(\co^2)$. Image taken from \citep{nielsenchuang}. } \label{blochsphere}
\end{figure}

The Pauli matrices which span $\lo(\mathds{C}^2)$ are given by $\{\sigma_i\}_{i=0}^3$, where each $\sigma_i$ is given as
\begin{equation}
\sigma_0\equiv \mathds{1}=\begin{pmatrix}1 & 0 \\
0 & 1 \\
\end{pmatrix} 
, \ \sigma_1\equiv \sigma_x=\begin{pmatrix}0 & 1 \\
1 & 0 \\
\end{pmatrix} 
, \  \sigma_2 \equiv  \sigma_y =\begin{pmatrix}0 & -\imag \\
\imag & 0 \\
\end{pmatrix},
\ \sigma_3\equiv \sigma_z=\begin{pmatrix}1 & 0 \\
0 & -1 \\
\end{pmatrix}
\end{equation} 
with respect to the basis $\{\phi_0,\phi_1\}$. $\sigma_x$ is diagonal with respect to the basis $\{\phi_+,\phi_-\}$ where $\phi_\pm=\frac{1}{\sqrt{2}}(\phi_0\pm\phi_1)$ and $\sigma_y$ is diagonal with respect to the basis $\{\varphi_{+},\varphi_{-}\}$ where $\varphi_\pm=\frac{1}{\sqrt{2}}(\phi_0 \pm \imag\phi_1)$. The Lie algebra $\mathfrak{su}(2)$ is spanned by the skew-Hermitian Pauli matrices\footnote{These are in fact, ignoring scalar multiplications,  all elements of the Pauli group $\mathcal{G}=\{\sigma_i\}_{i=0}^3\times \{\pm1,\pm i\}$.  } 
\begin{equation}
 \ \bar\sigma_1=-\frac{\imag}{2}\sigma_{1}, \  \bar\sigma_2  =-\frac{\imag}{2}\sigma_{2},
\ \bar\sigma_3=-\frac{\imag}{2}\sigma_3
\end{equation} 

 where $[\bar \sigma_i,\bar \sigma_j]_-=\epsilon_{ijk}\bar\sigma_k$.  Any two-element subset of these suffices to generate   $\mathfrak{su}(2)$ which in turn  generates the Lie group $\su(2)$. The Lie group of $3\times 3$ orthogonal matrices $S\mathcal{O}(3)$ that perform rotations in $\mathds{R}^3$ is generated by the Lie algebra $\mathfrak{so}(3)$ which is related to $\mathfrak{su}(2)$ by the isomorphism \footnote{A map $\vartheta:\mathfrak{su}(2)\to \mathfrak{so}(3)$ that preserves the commutator  of $\mathfrak{su}(2)$ such that $\vartheta([A,B]_-)=[\vartheta(A),\vartheta(B)]_-$ is a homomorphism. A homomorphism that is bijective (mapping is one-to-one and onto),  is an isomorphism.} $\vartheta$. The effect of this isomorphism  is the homomorphism  $\varTheta: \su(2) \to \mathcal{SO}(3)$, where $\varTheta[e^A]=e^{\vartheta[A]} \ \ \forall \ \ A \in \mathfrak{su}(2)$. The reason this is a homomorphism and not an isomorphism is that $\varTheta[e^A]=\varTheta[-e^A]$, and as such is a two-to-one and onto mapping. 

We can therefore treat  unitary evolutions acting on the state space $\s(\mathds{C}^2)$ as rotations in $\mathds{R}^3$, and hence use the Euler decomposition to write any unitary operator $U \in \su(2)$ as
\begin{equation}
U = e^{\imag\alpha}e^{\theta_3 \bar\sigma_z}e^{\theta_2 \bar\sigma_y}e^{\theta_1 \bar\sigma_z} \label{SU(2) Euler decomposition}
\end{equation}
where $\alpha$ is a phase factor.  An arbitrary $V \in \su(2)$ can also be represented by the matrix
\begin{equation}
V = \begin{pmatrix}\cos(\theta)e^{-\imag\xi} & -\sin(\theta)e^{\imag\zeta} \\
\sin(\theta)e^{-\imag\zeta} & \cos(\theta)e^{\imag\xi} \\
\end{pmatrix}\label{general 2by2 U}
\end{equation}
with $\theta \in [0,\pi]$ and $\xi,\zeta \in [0,2\pi)$. $U=V$ when the conditions $\theta=\frac{\theta_2}{2}, \xi=\frac{\theta_1+\theta_3}{2},\zeta=\frac{\theta_1-\theta_3}{2}$ are met.  In fact, any two orthogonal generators of $\mathfrak{su}(2)$ can be used to construct $U$, and the Euler angles can be determined as demonstrated above by performing a unitary transformation on these generators to obtain $U$ in the form of Eq.\eqref{SU(2) Euler decomposition}.

We may express any quantum state $\rho \in \s(\mathds{C}^2)$ in the Pauli basis as 
\begin{equation}
\rho = \frac{1}{2}(\mathds{1}+n_x \sigma_x+n_y \sigma_y+ n_z \sigma_z)
\end{equation}
and hence the state can be completely parameterised with respect to the vector $\vec n$ in $\mathds{R}^3$. The extremal states with $|\vec n|^2=1$ are the pure states, and the state with $|\vec n|=0$ is the maximally mixed state. Because any $U \in \su(2)$ will translate an extremal state to only other extremal states, and that  the homomorphism between $\mathcal{SO}(3)$ and $\su(2)$ describes this unitary evolution as a \emph{rotation} in $\mathds{R}^3$, which leaves the length of a vector with respect to the center of rotation invariant, then the state space of $\s(\mathds{C}^2)$ can be completely characterised as a unit sphere in $\mathds{R}^3$, known as the Bloch sphere, shown in Fig.\ref{blochsphere}. Any point on the surface of the Bloch sphere describes a pure state whose associated vector $\psi \in \mathds{C}^2$  can be represented as
\begin{equation}
\psi=\cos(\theta)\phi_0+e^{\imag\theta'}\sin(\theta)\phi_1
\end{equation}
where the polar angle has the range $\theta \in [0,\pi]$ and the azimuthal angle has the range $\theta' \in [0,2\pi)$. However, because the relationship between $\su(2)$ and $\mathcal{SO}(3)$ is a homomorphism we need only consider half of the Bloch sphere to describe states. This is because a vector $\psi'$ on the opposite side of the Bloch sphere is shown to obey  
\begin{equation}
\psi'=\cos(\pi-\theta)\phi_0+e^{\imag(\theta'+\pi)}\sin(\pi-\theta)\phi_1=-\left(\cos(\theta)\phi_0+e^{\imag\theta'}\sin(\theta)\phi_1\right)=-\psi
\end{equation}    
such that it varies only by a phase factor which is an unobservable quantity. Therefore, by convention the state vectors spanning the Hilbert space are described by imposing half polar angles $\theta\mapsto\theta/2$ to give
\begin{equation}
\psi=\cos\left(\frac{\theta}{2}\right)\phi_0+e^{\imag\theta'}\sin\left(\frac{\theta}{2}\right)\phi_1.
\end{equation} 
This geometrical picture offered by the Bloch sphere provides a  method of explaining states and dynamics in $\s(\co^2)$ in a language that is intuitive. 
\subsection{The magnetic resonance control scheme}\label{magnetic resonance control}
 The dynamics in magnetic resonance can occur only in a two-dimensional subspace   where the spin quantum numbers of the eigenstates differ  by one, assuming no degeneracy. Hence, for the sake of clarity, we may consider   a spin-half object in a static unidirectional magnetic field, whose axis we label $z$, with the Hamiltonian 
\begin{equation}
H_0^\pm=\pm\frac{\gamma B_0}{2} \sigma_z.
\end{equation}
Depending on the coupling of the spin and the external field, this will have the high energy (low energy) eigenstate $\phi_0$ and the low energy (high energy) eigenstate $\phi_1$ with transition frequency $\Omega=\gamma B_0$. 

In order to achieve controllability, we wish to obtain the generators of $\mathfrak{su}(2)$. To this end we  introduce a circularly polarised oscillating magnetic field, or \emph{driving field},  in a plane perpendicular to $z$ with its polarity dependent on the sign of $H_0^\pm$
\begin{equation}
H_I^\pm(t)=\frac{\omega_1}{2}f(t)\left( \cos(\omega t)\sigma_{x}\pm\sin(\omega t)\sigma_{y}\right).
\label{magnetic resonance driving field}\end{equation} 
Here, $\omega_1=\gamma B_1$ is the strength of the driving field, where $B_1$ denotes the strength of the magnetic field in this plane. $\omega$ is the frequency of the field, and the $+$ sign signifies a right handed (RH) circularly polarised field and the $-$ sign denotes a left handed (LH) one. $f(t)$ is the ``pulse'' function, which is non-vanishing only during the time that the driving field is switched on. We only consider a square pulse, satisfying
\begin{equation}
f(t)=\begin{cases}0 & \mathrm{when} \ t<t_0 \\
1 & \mathrm{when} \ t_0 \leqslant t \leqslant t_0+\tau \\
0 & \mathrm{when} \ t>t_0+\tau \\
\end{cases}.
\end{equation}
  We can now take $H_I^\pm(t)$ in the rotating frame of $H_0^\pm$ to give
\begin{align}
\tilde H_I^\pm(t) = \frac{\omega_1}{2}f(t)\left( \cos(\omega t)e^{\pm\imag t\frac{\Omega}{2} \sigma_z}\sigma_{x}e^{\mp \imag t\frac{\Omega}{2} \sigma_z}\pm\sin(\omega t)e^{\pm\imag t\frac{\Omega}{2} \sigma_z}\sigma_ye^{\mp \imag t\frac{\Omega}{2} \sigma_z}\right). \end{align} 

Noting that 
\begin{align}
 e^{\pm\imag t\frac{\Omega}{2} \sigma_z}\sigma_{x}e^{\mp\imag t\frac{\Omega}{2} \sigma_z}=\cos(\Omega t)\sigma_x \mp \sin(\Omega t)\sigma_y, \nonumber \\
 e^{\pm\imag t\frac{\Omega}{2} \sigma_z}\sigma_ye^{\mp\imag t\frac{\Omega}{2} \sigma_z}=\cos(\Omega t)\sigma_y \pm \sin(\Omega t)\sigma_x,
\end{align}

this becomes
\begin{equation}
\tilde H_I^\pm(t)= \frac{\omega_1}{2}f(t)\left( \cos[(\omega-\Omega) t]\sigma_x+\sin[(\omega-\Omega) t]\sigma_y \right)\label{rotating frame interaction Hamiltonian pauli}.
\end{equation}
Given the resonance condition $\omega=\Omega$,  a RH(LH) driving field will give $\tilde H_I^\pm(t) =\frac{\omega_1}{2}f(t) \sigma_x$, and  a phase shift $\omega t \mapsto \omega t + \pi/2$ will give $\tilde H_I^\pm(t) =\frac{\omega_1}{2} f(t)\sigma_y$. We therefore have the required generators for $\mathfrak{su}(2)$ in the rotating frame, and as such can generate the Lie group $\su(2)$ by the Euler decomposition shown in Eq.\eqref{SU(2) Euler decomposition}. Because the absorption of a RH(LH) photon of the driving field leads to the increase(decrease) of angular momentum by one, we can say that the  RH and LH fields have, respectively,  positive and negative angular momentum photons.

\subsubsection{The rotating wave approximation}\label{rotating wave approximation section}

\begin{figure}[!htb]
\centering
\includegraphics[width=3in]{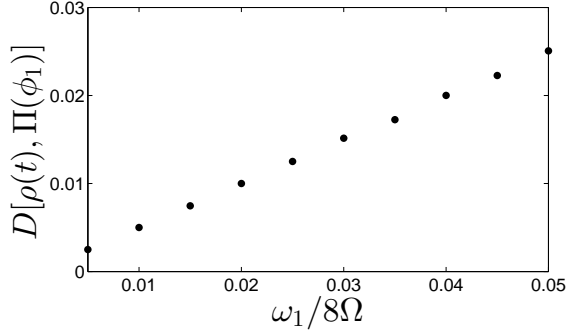} \caption{The trace distance between $\Pi(\phi_1)$ and the numerical solution to the Liouville-von Neumann equation using the interaction Hamiltonian in Eq.\eqref{RWA interaction Hamiltonian}, given an initial input state $\Pi(\phi_0)$ and evolution time $\tau=4\pi/\omega_1$. The trace distance converges to zero as $\omega_1/8\Omega \to 0$. } \label{RWA trace distance two level system}
\end{figure} 
In the example above, where circularly polarised fields are used, we may exactly    establish, in principle,  our desired generators in the rotating frame,  provided the resonance condition is met. No approximations are necessary here.  In many experimental situations, due to engineering limitations, it is not possible to establish circularly polarised magnetic fields. Rather, a linearly polarised field is used. In such a situation, however, even when the resonance condition is met, we do not have a rotating frame Hamiltonian that exactly gives the desired generator. Consequently, we need to make the rotating wave approximation (RWA) which is valid up to an arbitrary accuracy when the approximate solution converges with the exact one.
This is analogous to the secular approximation made in the microscopic derivation of a Lindblad master equation, where the oscillating terms of the Hamiltonian are ignored.

A linearly polarised driving field  can be considered as being  composed in equal parts of a RH and a LH field
\begin{equation}
H_I(t)=\frac{\omega_1}{2} f(t) \cos(\omega t) \sigma_x\equiv\frac{\omega_1}{4}f(t)\left([\cos(\omega t) \sigma_x + \sin(\omega t) \sigma_y] +[\cos(\omega t) \sigma_x - \sin(\omega t) \sigma_y]  \right).
\end{equation} 
For the system Hamiltonians $H_0^\pm$, the resulting rotating frame interaction Hamiltonian is given by  
\begin{equation}
\tilde H_I^\pm(t)= \frac{\omega_1}{4}f(t)\left\{( \cos[(\omega-\Omega) t]+ \cos[(\omega+\Omega) t])\sigma_x+(\sin[(\omega-\Omega) t]\mp\sin[(\omega+\Omega) t])\sigma_y \right\}.
\end{equation}
At resonance, the interaction Hamiltonian is given by
\begin{equation}
\tilde H_I^\pm(t)= \frac{\omega_1}{4}f(t)\left( \sigma_x + \cos[(2\Omega) t]\sigma_x\mp\sin[(2\Omega) t]\sigma_y \right).\label{RWA interaction Hamiltonian}
\end{equation}
This will be approximately close to the desired rotating frame Hamiltonian $\frac{\omega_1}{4}f(t) \sigma_x$  given the rotating wave approximation. The rotating wave approximation  is justified if the pulse has a small time variation,  $df(t)/dt \ll 1$,  and if it is sufficiently weak. The former criterion is met by using a square pulse. For the latter, we may use time-dependent perturbation theory. The unitary operator generated by the rotating frame Hamiltonian of Eq.\eqref{RWA interaction Hamiltonian} is determined by the Dyson series, shown in Eq.\eqref{Dyson series unitary}, as 
\begin{align}
U&=\mathds{1}-\imag \frac{\omega_1}{4}\left(\tau\sigma_x \mp\imag\frac{(e^{\pm \imag2 \Omega\tau}-1)}{2\Omega}|\phi_0\>\<\phi_1| \pm \imag \frac{(e^{\mp \imag2 \Omega\tau}-1)}{2\Omega}|\phi_1\>\<\phi_0|\right)+...,\nonumber \\ &=\mathds{1}+\frac{1}{l!}\sum_{l=1}^\infty \left(-\imag \frac{\tau\omega_1}{4}\sigma_x\right)^l+\frac{1}{l!}\sum_{l=1}^\infty \left(-\imag \frac{\omega_1}{8\Omega}\right)^l O_l, \nonumber \\
&=U_{\mathrm{RWA}}+\tilde O.
\end{align}
 The first term,  $U_{\mathrm{RWA}}$,  is just the unitary operator generated by the rotating wave approximation Hamiltonian, while the second term, $\tilde O$, is the correction  containing all the sinusoidal terms. 
 Let the system initially  be in the pure state  with the associated vector  $\psi$. We may calculate the   distance between the solution using the RWA unitary operator and that using the exact unitary operator as
\begin{align}
d(U\psi,U_{\mathrm{RWA}}\psi  )&=\sqrt{\left\<(U-U_{\mathrm{RWA}})\psi\big|(U-U_{\mathrm{RWA}})\psi\right\>} \nonumber \\
 &=\sqrt{\left\< \tilde O \psi\big|\tilde O\psi\right\>}\nonumber \\
&=\sqrt{\frac{1}{l!m!}\sum_{l,m=1}^\infty \left( \frac{\omega_1}{8\Omega}\right)^{l+m}  \left\< O_l \psi\big| O_m \psi \right \>}.
\end{align}

In the limit $\omega_1/8\Omega \to 0$, the distance  vanishes, and the rotating wave approximation becomes valid.    

In practice $\omega_1/8\Omega$ will have some finite value, and the rotating wave approximation will have some error $\epsilon$. For general mixed states, where we assign $\rho(t)$ as the exact solution, and $\varrho(t)$ as the solution using the rotating wave approximation,  this error  can be defined as the trace distance  
\begin{equation}
\epsilon=D[\rho(t),\varrho(t)] \in [0,1].
\end{equation} 
 Given an arbitrarily small error threshold  $\epsilon$ and a convergence threshold\footnote{If $\omega_1/8\Omega$ is too large the systems evolution will not be convergent with that of the desired Hamiltonian;  decreasing $\omega_1/8\Omega$ will not necessarily result in a decrease in the trace distance.  } $N$, there are values of $\omega_1/8\Omega< N$ such that  $D[\rho(t),\varrho(t)]< \epsilon$. This is illustrated in Fig.\ref{RWA trace distance two level system} which uses the Runge-–Kutta-–Fehlberg method to numerically integrate the Liouville-von Neumann equation and determine $\rho(t)$. The initial state is set to $\rho(t_0)=\Pi(\phi_0)$ and the final state, given the rotating wave approximation, is determined by a $\pi$ rotation about the $x$ axis of the Bloch sphere, which gives $\varrho(t)=\Pi(\phi_1)$.  As $\omega_1/8\Omega$  gets smaller, the rotating wave approximation becomes increasingly more accurate.

\subsection{The Hahn echo and dynamical decoupling}\label{Hahn echo sec}
  
Dynamical decoupling \citep{dynamical-decoupling} is the application of dynamics on a quantum system with the aim of decoupling its evolution from that of its environment, thereby reducing or altogether removing decoherence. The Hahn echo \citep{Hahn-echo} is one simple example of such a scheme used in magnetic resonance and is applicable to two-dimensional systems, or dynamical selection rule allowed two-dimensional subspaces of systems. There are two cases in which we may apply the Hahn echo, only one of which constitutes dynamical decoupling, but both of which counteract  phenomenological  dephasing:

\begin{enumerate}[(i)]
\item \emph{Dephasing of ensemble state}

In this case, there are no environmental degrees of freedom from which we wish to decouple our system's evolution and hence the Hahn echo does not constitute a form of dynamical decoupling. Instead, this phenomenon arises  in situations where  we only have access to  an ensemble state $\bar \rho$, which was  introduced in Sec.\ref{ensemble state section}. Consider the case where the ensemble state belongs to $\s(\mathds{C}^2)$ and is composed of individual spin half objects $\rho^n$, each of which is experiencing a  magnetic field $B_0(t)$ in \emph{the same direction}: they are all generated by $\bar \sigma_z$.  As a result, we may write the unitary operators in their diagonal form as
\begin{equation}
U_{t,t_0}^n=e^{\frac{\imag}{2}\int_{t_0}^t \omega^n(s) ds}\Pi(\phi_0)+e^{-\frac{\imag}{2}\int_{t_0}^t \omega^n(s) ds}\Pi(\phi_1)
\end{equation} 
where each $U^n_{t,t_0}$ differs only by the generally time-dependent frequency function $\omega^n(t)$. The state at times $t$ is given by the random unitary channel
\begin{equation}
\e_{\mathrm{random \ unitary}}:\bar\rho\mapsto\frac{1}{N}\sum_{n=1}^NU^n_{t,t_0}\bar\rho U_{t,t_0}^{n\dagger}
\end{equation}
which will bring about dephasing. The Hahn echo, acting on each ensemble member, is defined as the sequence 
\begin{align}
\sigma_xU_{t_1,t}^n\sigma_x U_{t,t_0}^n&=e^{\frac{\imag}{2}\left(\int_{t_0}^t \omega^n(s) ds-\int_{t}^{t_1} \omega^n(s) ds\right)}\Pi(\phi_0) \nonumber \\ &+e^{-\frac{\imag}{2}\left(\int_{t_0}^t \omega^n(s) ds-\int_{t}^{t_1} \omega^n(s) ds\right)}\Pi(\phi_1)\label{Hahn echo ensemble}   
\end{align}
where $t-t_0=t_1-t$. This gives the identity operator if $\omega(t)$ is time-independent. Here the $\sigma_x$ operation, which is  also called a $\pi$ pulse about the $x$ axis, has the effect of \emph{reversing} the phase evolution effected by each $U_{t,t_0}^n$, thereby canceling the dephasing undergone by the ensemble.  If $\omega(t)$ does have a time dependence, but changes slowly, we may expand the   Taylor series of the integrands in Eq.\eqref{Hahn echo ensemble} to first order so as to obtain 
\begin{align}
&\int_{t_0}^t \omega^n(s) ds-\int_{t}^{t_1} \omega^n(s) ds \nonumber \\&\simeq \int_{t_0}^t \left( \omega^n[t_0]+(s-t_0)\dot \omega^n[t_0] \right) ds-\int_{t}^{t_1} \left(\omega^n[t_0]+(s-t_0)\dot \omega^n[t_0]\right) ds\nonumber \\
&=(t-t_0) \omega^n[t_0].
\end{align}
As such,  we may suppress dephasing by the Hahn echo in the limit $t-t_0\to 0$:  in the limit of continuous  $\pi$ pulses.
\item \emph{Decoupling a  spin-spin  interaction of Schmidt-rank one }

Consider a composite system $A+B$ where system $A$  is a spin one-half particle and system $B$, designated as the environment, has arbitrary spin, with the  Hamiltonian written in the operator-Schmidt decomposition  as $H=H^A\otimes H^B$, where $H^A$ and $H^B$ are self-adjoint operators in $\lo_s(\co^2)$ and $\lo_s(\co^d)$ respectively.   Because the  Hamiltonian has a Schmidt-rank of one, it  does not result in  exchange of energy between the two subspaces. This Hamiltonian may be written as $H=(E_0\Pi(\phi_0)+E_1 \Pi(\phi_1))\otimes H^B$ where $\{\phi_0,\phi_1\}$ is the basis in which $H^A$ is diagonal, and $\{E_0,E_1\}$ are the corresponding  energies. The unitary operator that is the solution to Schr\"odinger's equation for this Hamiltonian can then be shown to be of the form
\begin{equation}
U_\tau=\Pi(\phi_0)\otimes U^{\phi_0}_\tau+\Pi(\phi_1)\otimes U^{\phi_1}_\tau
\end{equation}
where the conditional unitaries on $B$ are given by $U^{\phi_{i}}_\tau=e^{-\imag\tau E_{i}H^B}$, and are both generated by $H^B$ and hence commute. This is an entangling operation and causes pure decoherence in system A, with the preferred basis being the eigenbasis of $H^A$. By acting on $A$ with an operator $\sigma_x$, chosen with respect to the basis  $\{\phi_0,\phi_1\}$, we have
\begin{equation}
(\sigma_x\otimes\mathds{1})U_\tau (\sigma_x\otimes\mathds{1})U_\tau=\Pi(\phi_0)\otimes U_\tau^{\phi_1}U_\tau^{\phi_0}+\Pi(\phi_1)\otimes U_\tau^{\phi_0}U_\tau^{\phi_1}=\mathds{1}\otimes U_\tau^{\phi_0}U_\tau^{\phi_1}
\end{equation}
which is  decoupled with respect to the $A:B$ divide. 

Consider a specific example of such a   Hamiltonian as  $H=\frac{1}{4}\mathcal{J}\sigma_z\otimes \mb_z$. This is often referred to as an Ising interaction, and $\{\phi_0,\phi_1\}$ is the basis in which $\sigma_z$ is diagonal, and the conditional unitaries on the environment system $B$ are $U_\tau^{\phi_0}=U_\tau^-$ and $U_\tau^{\phi_1}=U_\tau^+$ and are given as 
 \begin{equation}
 U_\tau^\pm=e^{\pm \imag \frac{\mathcal{J}}{4} \tau \mb_z}.
 \end{equation}
 
 This is clearly an entangling operation, as can be seen by having the initial pure product  state represented by the vector $\psi=\frac{1}{2}\left(\phi_0+\phi_1\right)\otimes\left(\varphi_0+\varphi_1\right)$, where $\<\varphi_0|\varphi_1\>=0$. After the evolution for a time period $\tau = \frac{\pi}{\mathcal{J}}$ the state evolves, ignoring an overall phase factor, to
\begin{equation}
\psi'=\frac{1}{2}\left(\phi_0\otimes[\varphi_0+\imag \varphi_1]+\phi_1\otimes[\varphi_1+\imag \varphi_0]\right)
\end{equation} 
 which is a maximally entangled state. The reduced density operator for system A is consequently given by a maximally mixed state $\frac{1}{2}\mathds{1}$ due to the decoherence that has taken place. However, by using the Hahn echo sequence on subsystem $A$  we can reverse this entanglement generation as
\begin{equation}
(\sigma_x\otimes\mathds{1})U_\tau (\sigma_x\otimes\mathds{1})U_\tau=\mathds{1}.
\end{equation}
\end{enumerate}  

It should be noted that the Hahn echo sequence can only remove dephasing in case (i) and decoherence in case (ii) if there are no other noise processes occurring.  The presence of noise may be  modelled by the replacement of the unitary operator with an  irreversible quantum channel $\e$, in which case  
\begin{equation}
(\e_{\sigma_x}\otimes\mathds{1}) \circ \e_{\tau_2} \circ (\e_{\sigma_x}\otimes\mathds{1}) \circ \e_{\tau_1} \ne \mathds{1}
\end{equation} 
where $\e_{\sigma_x}[\rho]=\sigma_x\rho\sigma_x$.

\section{Magnetic resonance experiments with weak measurements}\label{magnetic resonance weak measurement}
There are two main types of experiment that can be conducted in magnetic resonance. The first is  continuous wave (c.w.) spectroscopy, which was the exclusive method in the early days of magnetic resonance. Here, the system is exposed to a weak driving field -- of a set frequency -- continuously for a long period of time. As the static magnetic field is altered, the absorption  (emission) of radiation from (to) the driving field is detected, providing a spectrum. More recently, the method of pulsed spectroscopy has been developed, where the system is driven by short, powerful pulses of radiation. The magnetic field is no longer swept, and the measurement over time will provide a generally sinusoidal time-varying signal called the free induction. The Fourier transform of the free induction will give a signal peak in the frequency domain with similar properties to the associated peak given by the c.w. method. In both cases, measurement is presently performed on ensembles of spin systems, although  much effort is being invested in realising measurements of single systems, which will hopefully prove fruitful in the near future.    

In Chap.\ref{quantum theory} I described the notion of conceptually separating a quantum mechanical experiment into the three components of preparation, transformation, and measurement. We may use such a conceptual compartmentalisation to study these two types of experiments in magnetic resonance.  We need not worry too much about the preparation for now, and take for granted that spin systems may be produced by the press of a ``button''. The transformation part of the experiment, on the other hand, is clearly described by    magnetic resonance. What remains, then, is an account of measurement.  The measurement that we will consider is described by weak ensemble measurements, discussed in  Sec.\ref{ensemble state section}. The von Neumann-L\"uders measurement scheme  here uses a single probe state coupled weakly with each ensemble member, and the only information available about the system is the expectation value of the ensemble observable.  
     
\subsection{Continuous wave spectroscopy}\label{c.w. spectroscopy}
When a spin system interacts coherently with electromagnetic radiation of frequency $\omega$, any increase in energy of the spin system is coupled with a decrease of the same energy in the radiation, and vice versa. We say that the system absorbs quanta of energy from the radiation field, or emits quanta of energy to it.   Continuous wave spectroscopy  on a system, governed by the  time-independent Hamiltonian $H_0(B_0)$, gives three pieces of information about such interactions.
\begin{enumerate}[(i)]
 \item 
 The first is the so called \emph{transition rate} proportional to  $|\<\phi_k|J_x\phi_j\>_{B_0}|^2$ where $\{\phi_j({B_0}),\phi_k({B_0})\}$ are two eigenstates of $H_0(B_0)$.   
\item 
The second is the magnetic field values $B_0$ for which such states have an energy difference (or frequency) $\Omega=\omega$. \item 
 Finally,  c.w. spectroscopy tells us whether a transition is an absorption or an emission process.
\end{enumerate}
 The quantum mechanical treatment of the interaction involved in c.w. spectroscopy is calculated perturbatively using what is known as Fermi's golden rule, which was largely developed by Paul Dirac \citep{Dirac-Fermi-golden-rule}. Here, the system is evolved by a \emph{weak} driving field which  ensures that if the system is driven for a long time -- hence the  \emph{continuous} in the name -- so that transitions occur only at resonance, the final state has a high fidelity with its original configuration.

 Let us consider a two-dimensional subspace $\{\phi_j({B_0}),\phi_k({B_0})\}$ in which transitions are permitted by the dynamical selection rule for $J_x$. This has the transition frequency given by   $\Omega_{kj}^{B_0}=|E_k(B_0)-E_j(B_0)|$, where   $E_j(B_0)=\<\phi_j|H_0(B_0)\phi_j\>_{B_0}$. Initially, we prepare the system to be in the  pure state $\Pi(\phi_j[B_0])$ associated with the vector $\phi_j({B_0})$. Subsequently we turn on the time-dependent Hamiltonian, or driving field,  to get the total Hamiltonian
\begin{equation}
H^\pm(t)=H_0(B_0)+\lambda f(t)\left(\cos[\omega t]J_x\pm\sin[\omega t]J_y\right)
\end{equation}
  where $\lambda=\gamma B_1$ is the strength of this field, $\omega$ is the frequency, and the $\pm$ term designates a RH and LH field. We wish to determine the probability of finding the system in state $\Pi(\phi_k(B_0))$ after some time $\tau$. To this end, we note that we may expand the state vector of the spin system at any time with respect to the eigenbasis of $H_0$, and take note that the effect of the  driving field is to merely change the coefficients of these basis vectors. Therefore, the interaction picture state vector is given by 
\begin{equation}
\tilde\psi(t)({B_0}) = \sum_n \alpha_n(t)\phi_n({B_0})
\end{equation}
where we note that $\alpha_j(t_0)=1$ so as to satisfy $\tilde \psi(t_0)({B_0})=\phi_j({B_0})$. We may write the Dyson series, shown in Eq.\eqref{Dyson series unitary}, for the interaction picture unitary operator as 
\begin{equation}
\tilde U_{t,t_0}=\mathds{1}+\sum_{l=1}^\infty\frac{(-\imag\lambda)^l}{l!}\int_{t_0}^{t}dt_1...\int_{t_0}^{t} dt_l \left(\cos[\omega t_l]\tilde J_x(t_l)\pm\sin[\omega t_l]\tilde J_y(t_l)\right).
\end{equation}

 It is then possible to write the solutions of $\alpha_n(t)$ determined by $\<\phi_n|\tilde U_{t,t_0} \psi(t_0)\>_{B_0}$, including only up to the $l^\mathrm{th}$ term of the Dyson series, as the infinite sequence $\{\alpha_n^l(t)\}_{l=0}^\infty$ which converges to $\alpha_n(t)$.     The distance between $\alpha_n^1(t)$ and $\alpha_n(t)$, denoted $\epsilon$, can be made arbitrarily small by  reducing the size of  $\lambda/\Omega_{kj}^{B_0}$.  Provided  $\lambda/\Omega_{kj}^{B_0} \ll 1$ we can, with a small error $\epsilon$, approximate the dynamics by the first order perturbation theory.  Explicitly we calculate
\begin{align}
 \<\phi_k| \psi(t)\>_{B_0}&=\<\phi_k|\tilde U_{t,t_0}\phi_j\>_{B_0},\nonumber \\ &\simeq 
-\imag\lambda\int_{t_0}^{t_0+\tau}dt_1\left(\cos[\omega t_1]\<\phi_k|J_x\phi_j\>_{B_0}\pm\sin[\omega t_1]\<\phi_k|J_y\phi_j\>_{B_0}\right)e^{\imag t_1\Omega_{kj}^{B_0}}, \nonumber \\
&= -\imag\lambda\<\phi_k|J_x\phi_j\>_{B_0} \int_{t_0}^{t_0+\tau}dt_1\left(\cos[\omega t_1]- \imag\sin[\omega t_1]\right)e^{\imag t_1\Omega_{kj}^{B_0}}, \nonumber \\ &=-\lambda\<\phi_k|J_x\phi_j\>_{B_0}\left(\frac{1- e^{-\imag\tau(\omega - \Omega_{kj}^{B_0})}}{\omega - \Omega_{kj}^{B_0}}  \right),\nonumber \\
&=-2\imag e^{\imag\frac{\tau}{2}(\omega- \Omega_{kj}^{B_0})}\lambda  \<\phi_k|J_x\phi_j\>_{B_0} \frac{\sin\left[\frac{\tau}{2}(\omega-\Omega_{kj}^{B_0})\right]}{\omega-\Omega_{kj}^{B_0}}.
 \end{align}
  Such an approximation always yields $\<\phi_j|\psi(t)\>_{B_0}=1$. Hence, after renormalisation,   the transition probability given a magnetic field value of $B_0$ is  given by
\begin{equation}
\frac{|\<\phi_k|\tilde\psi(t)\>_{B_0}|^2}{1-|\<\phi_k|\tilde\psi(t)\>_{B_0}|^2}\simeq\left|\tau\lambda  \<\phi_k|J_x\phi_j\>_{B_0}\mathrm{sinc}\left[\frac{\tau}{2}\Delta \omega^{B_0}\right]\right|^2
\end{equation}
if $\tau\lambda \ll 1$. Here,  we have made the substitution $ (\omega - \Omega_{kj}^{B_0})=\Delta\omega^{B_0}$.    Let us note the  reciprocal Fourier transform relationship between the time of the driving field's action and the range of frequencies on which it acts \footnote{This is also referred to as the time-frequency uncertainty relation in the literature, even though different in nature to that between position and momentum, because time is a parameter and not an observable in quantum mechanics.} given as 
\begin{equation}
\tau \Delta \omega^{B_0} \sim 1.
\end{equation} 
 Because we have made the weak driving strength and long driving time assumptions, we may take the limit $\tau\to \infty$ by choosing $\tau$ to be  arbitrarily long (limited by the strength of $\lambda$) so as to make the width of the Sinc function arbitrarily narrow, thus enabling us to develop  a coarse-grained picture where we may treat \emph{segments} of the continuous magnetic field variable as discrete values, providing the discrete set $\{B_0^l\}_{l=1}^L$. Therefore, we finally arrive at the following expression for the transition probability after driving the system for time $\tau$   
\begin{equation}
|\<\phi_k|\psi(t)\>_{B_0^l}|^2 = |\lambda\tau\<\phi_k|J_x\phi_j\>_{B_0^l}|^2 \label{cw transition probability}
\end{equation}  
which is non vanishing only when  $\Delta \omega^{B_0^l}=0$ : when the driving field is in resonance with the frequency between the two eigenstates. We will omit the $l$ superscript from now on. 
Because Eq.\eqref{cw transition probability} is quadratic in time, this still does not give a transition rate.    
   We may, however, integrate this function  over the magnetic field variable to get \begin{equation}
\lim_{\tau \to \infty}\int_{B_0} dB_0' \ \ \left|\tau\lambda  \<\phi_k|J_x\phi_j\>_{B_0'}\mathrm{sinc}\left[\frac{\tau}{2}\Delta \omega^{B_0'}\right]\right|^2=\tau 2\pi |\lambda  \<\phi_k|J_x\phi_j\>_{B_0}|^2.
\end{equation}
which is the total probability of transition over the continuum of the magnetic field. This is a quantity that is linear in time, and  is used to define the transition rate $2\pi|\lambda  \<\phi_j|J_x\phi_k\>_{B_0}|^2$.

Because this experiment  is performed on an ensemble, however, we must take into account the possibility that not all members
of the ensemble are in the same initial state. Assuming that the detection of an absorption event is marked by  $+$, and likewise that of an emission is marked by  $-$, and that $E_k(B_0)>E_j(B_0)$, the measured transition rate will be modified to
\begin{equation}
2\pi|\lambda  \<\phi_j|J_x\phi_k\>_{B_0}|^2\left(P(j)-P(k) \right)
\end{equation}
where $P(j)$ and $P(k)$ are respectively the probabilities that the system is initially in the state $\Pi(\phi_j(B_0))$ and $\Pi(\phi_k(B_0))$.

\subsection{Pulsed spectroscopy}
A precessing magnet generates a magnetic field which, if surrounded by a wire,  induces a current therein. This process is called the free induction. An ensemble of spins in a magnetic field, when prepared in a superposition of the free Hamiltonian's eigenvectors, also generate a free induction. So long as this superposition exists in a two-dimensional subspace, the effective observable can be given by the self-adjoint operator  $\sigma_{\underline m}=\Pi_{+\underline m}-\Pi_{-\underline m}$ with the  projector effects 
\begin{equation}
\Pi_{\pm \underline m}=\frac{1}{2}\left(\mathds{1}\pm \underline m.\underline\sigma \right)
\end{equation}
where $\underline m=\{m_x,m_y,0_z\}$ is a vector of unit length in the $x-y$ plane of the Bloch sphere. \footnote{In an alternative formulation, we may write the vectors corresponding to the projector effects as $\phi_{\pm\theta'}=\frac{1}{\sqrt{2}}(\phi_0\pm e^{\imag\theta'}\phi_1)$ for $\theta' \in [0,\pi)$.} We may write the Pauli operators with respect to the energy eigenvector basis of the two-dimensional subspace of our system, $\{\phi_0,\phi_1\}$, such that $\phi_0$ denotes the excited state and $\phi_1$ the ground state.  As the vector $\underline m$ can be chosen freely,  for simplicity we may  consider the two cases of $ \sigma_x$ and $ \sigma_y$. Since these sharp observables are measured weakly on an ensemble, we only have access to the expectation value of said observables on the effective ensemble state. 

Unlike  c.w. spectroscopy, pulsed spectroscopy utilises strong ``pulses'' of electromagnetic radiation to induce any desired unitary from the group $\su(2)$, within any two-dimensional subspace that obeys the relevant dynamical selection rule. Therefore, with the pulses at our disposal we may at first prepare every member of the ensemble in the  state $\rho=\frac{1}{2}(\mathds{1} + \underline n.\underline \sigma)$.  If the state  is  allowed to evolve according to the  sub-Hamiltonian $H_{eg}=\frac{\Omega}{2}\sigma_z$, the expectation value of $\sigma_x$ calculated at time $t=t_0+\tau$ will be given by
\begin{equation}
\mathrm{tr}[\sigma_x \rho(t)]:=n_x(t)=n_x(t_0) \cos(\Omega \tau)-n_y(t_0)\sin(\Omega \tau).
\end{equation}  
Similarly, the measurement of $\sigma_y$ will yield
\begin{equation}
\mathrm{tr}[\sigma_y \rho(t)]:=n_y(t)=n_y(t_0) \cos(\Omega\tau)+n_x(t_0)\sin(\Omega \tau).
\end{equation}
 If the phase of the measurement  is varied so as to measure $\sigma_x$, or $\sigma_y$, in the rotating frame of the two-level subspace, then this sinusoidal dependence will be omitted and the measurement will only reveal $n_x(t_0)$, or $n_y(t_0)$.
Alternatively, if we merely want to measure the length of the Bloch vector component in the $x-y$ plane,  we may weakly measure the two sharp observables $\sigma_x$ and $\sigma_y$ in the lab frame, such that we may simultaneously\footnote{Even though the two observables do not commute, we may still measure both simultaneously by using weak measurements which, by definition, do not disturb the quantum state in question.} determine the sinusoidally varying Bloch vector components $n_x$ and $n_y$. This is the quadrature detection technique used in pulsed spectroscopy. The time-independent Bloch vector component  parallel to  the $x-y$ plane can then be determined  by $\sqrt{n_x^2(t)+n_y^2(t)}=\sqrt{n_x^2(t_0)+n_y^2(t_0)}$.

 Furthermore, we may use the weak measurement of $\sigma_x$ to, in principle, determine the expectation value of any PVM on our two-dimensional subspace. Recall the equivalence between the Heisenberg and Schr\"odinger pictures where   $\mathrm{tr}[U^\dagger \sigma_xU \rho]=\mathrm{tr}[\sigma_xU \rho U^\dagger]$, such that the unitary transformation $U^\dagger \sigma_xU=\sigma_{\underline n}$ gives a two element PVM along any axis $\underline n$ of the Bloch sphere. Therefore, we can effect this change in measurement basis by simply performing the unitary transformation $U\rho U^\dagger$ prior to weakly measuring $\sigma_x$.

\section{Pulsed Fourier transform  spectroscopy and signal broadening}\label{signal broadening}

Fourier transform spectroscopy, as the name implies, calculates the Fourier transform of the measured time varying free induction to give a frequency domain signal.  In the ideal case discussed above, the frequency domain signal will be a  delta function centred around the frequency $\Omega$ of the two-level subspace.   In realistic experimental situations, however, these are \emph{broadened} in the frequency  domain owing to the free induction decay (FID), which is caused by dephasing. Here, we consider two cases of signal broadening.

\subsection{Homogeneous broadening}

\begin{figure}[!htb]
\centering
\subfloat[FID]{\label{homogeneousFID}\includegraphics[width=0.4\textwidth]{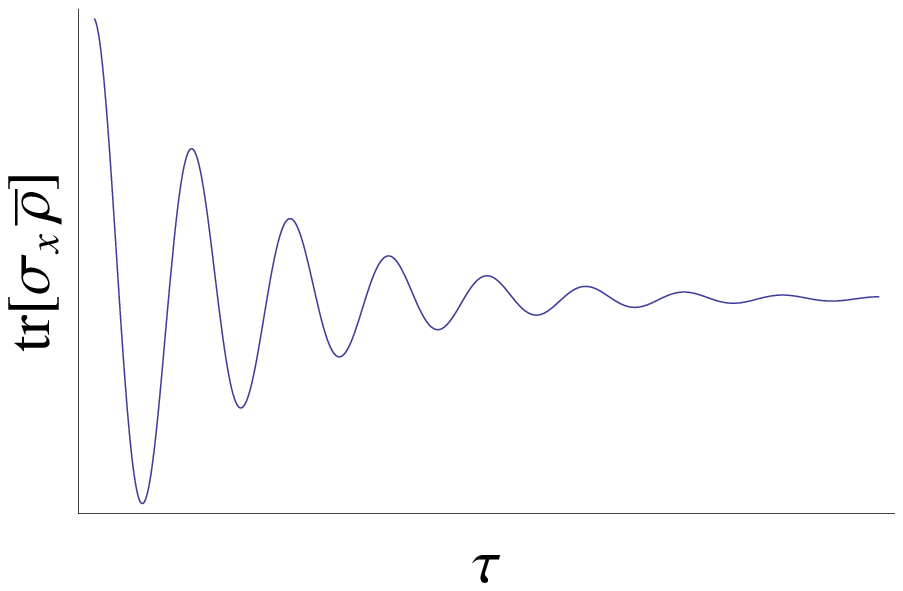}}
\ \ \ \ \ \ \ \  \subfloat[Fourier transform signal]{\label{homogeneousFourier}\includegraphics[width=0.4\textwidth]{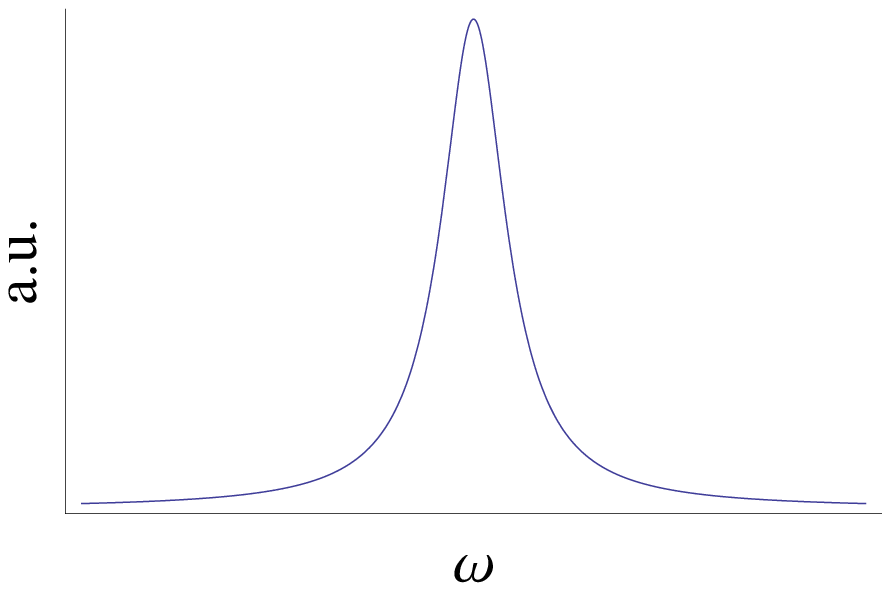}}
\caption{Exponential FID will lead to a Fourier transform signal broadening of Lorentzian form.}
\end{figure}

Assuming dephasing processes that are identical for every member of the ensemble, the resultant broadening is called homogeneous. Consider the case where the effective ensemble state undergoes the pure dephasing channel
\begin{equation}
\e: \bar \rho \mapsto (1-\lambda[\tau])\bar \rho+\lambda[\tau]\sigma_z\bar \rho \sigma_z
\end{equation}

where $\lambda(\tau)=(1-e^{-\alpha \tau})/2$.  Hence the FID signal of an  input state $\bar \rho(t_0)=\Pi(\phi_+)$ with associated vector $\phi_+=\frac{1}{\sqrt{2}}(\phi_0+\phi_1)$ is given by 
\begin{equation}
\mathrm{tr}[\sigma_x \bar \rho(t)]=\cos(\Omega \tau)e^{-\alpha \tau}.
\end{equation}
   The Fourier transform of this signal gives a Lorentzian function
\begin{equation}
\frac{1}{\pi}\frac{\alpha}{\alpha^2+(  \omega - \Omega)^2}
\end{equation}      

centred around $\Omega$, and whose width is proportional to the decay rate $\alpha$.

\subsection{Inhomogeneous broadening}
\begin{figure}[!htb]
\centering
\subfloat[FID]{\label{inhomogeneousFID}\includegraphics[width=0.4\textwidth]{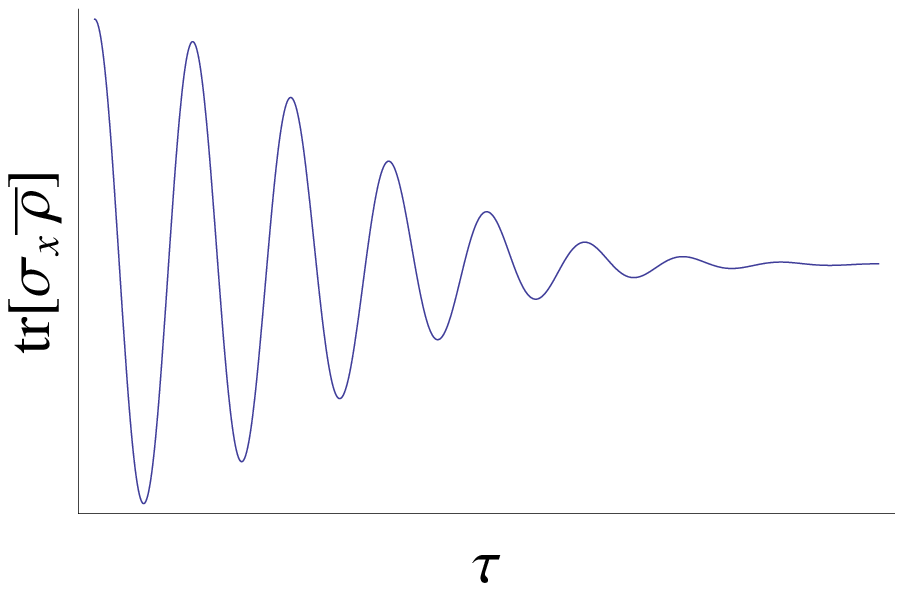}}
\ \ \ \ \ \ \ \  \subfloat[Fourier transform signal]{\label{inhomogeneousFourier}\includegraphics[width=0.4\textwidth]{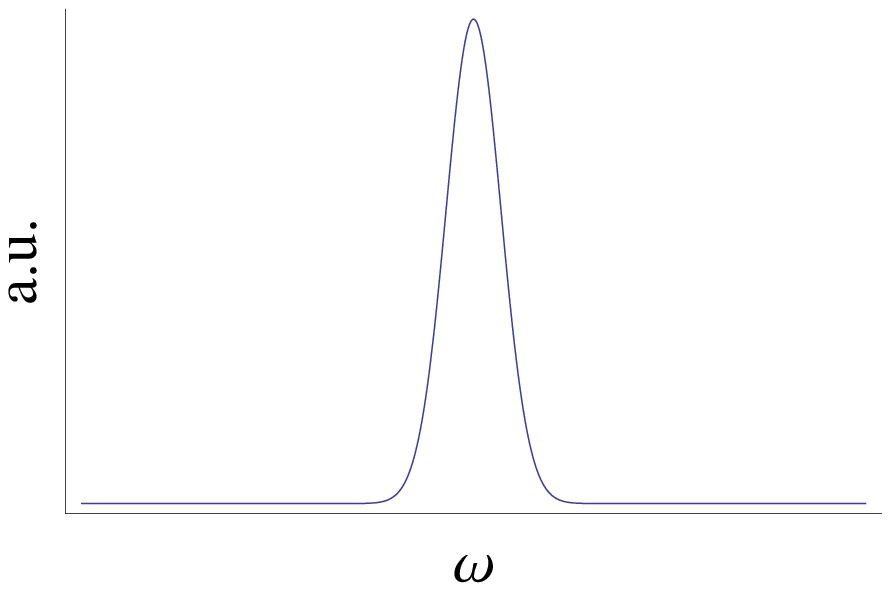}}
\caption{Gaussian FID will lead to a Fourier transform signal broadening of Gaussian form.}
\end{figure}
In this scenario, it is assumed that the ensemble on which the experiment is conducted has effective magnetic field inhomogeneities. \footnote{This inhomogeneity is time invariant and only dependent on the position of the ensemble member.} As a result there will be a distribution of free induction oscillation frequencies $\{\omega_i\}$ which in most situations  will have a Gaussian distribution with mean $\Omega$ and variance $\alpha^2$.  In the thermodynamic limit of infinite ensemble members, the effective ensemble state undergoes a random unitary channel 
\begin{equation}
\e: \bar \rho \mapsto  \frac{1}{\alpha\sqrt{2 \pi}}\int_{-\infty}^\infty d \omega e^{-\frac{1}{2}\frac{\Delta\omega^2}{\alpha^2}}e^{- \tau \omega \bar \sigma_z}\bar \rho e^{ \tau \omega \bar \sigma_z}
\end{equation}
where $\Delta \omega = \omega - \Omega$. Given  an  input state $\bar \rho(t_0)=\Pi(\phi_+)$ with associated vector $\phi_+=\frac{1}{\sqrt{2}}(\phi_0+\phi_1)$, the free induction varies as
\begin{equation}
\mathrm{tr}[\sigma_x \bar \rho(t)]=\frac{1}{\alpha\sqrt{2 \pi}}\int_{-\infty}^\infty d \omega e^{-\frac{1}{2}\frac{\Delta\omega^2}{\alpha^2}}\cos(\omega \tau)=\cos(\Omega \tau)e^{-\frac{1}{2}(\tau\alpha)^2}.
\end{equation}

The Fourier transform of this signal is given by a Gaussian function 
\begin{equation}
\frac{1}{\alpha\sqrt{2\pi}}e^{-\frac{(\omega-\Omega)^2}{2\alpha^2}}.
\end{equation}

\section{ Pulsed magnetic resonance experiments for studying dynamics}\label{pulsed dynamics measurement scheme}
Here we will cover the main pulsed spectroscopy experiments aimed at studying the dynamics of a spin system within a two-dimensional dynamical selection rule allowed subspace.   As before, these procedures are split up into the three stages of preparation, transformation, and measurement.
\subsection{Dephasing }\label{dephasing measurement scheme}
The dephasing in a dynamical selection rule allowed two-dimensional subspace of a spin system  can be easily ascertained by measuring the decay of the Bloch vector component $\sqrt{n_x(t)^2+n_y(t)^2}$ in the lab frame. If it is the dephasing due to irreversible processes that is to be investigated, as opposed to reversible ones such as the ensemble-caused random unitary channel, or Ising interactions with a static environment, then the Hahn echo can be used  to remove those contributions. The dephasing measurement protocol can be conducted as follows

\begin{enumerate}[(i)]
\item  Prepare the initial state $\rho=\frac{1}{2}(\mathds{1}-\sigma_x)$. 
\item Allow system to evolve for a time $\tau$, and then carry out the Hahn echo sequence.
\item At the end of the Hahn echo sequence, which is after a period $2\tau$ has passed, weakly measure the expectation value of $\sigma_x$ and $\sigma_y$.
\item Repeat (i)-(iii) $N$ times, and on each occasion increase $\tau$ by a constant value $\Delta \tau$ 
\end{enumerate}
The equivalent rotating frame  sequence of operations and measurements can be denoted as follows
\begin{equation}
 \ \frac{1}{2}(\mathds{1}-\sigma_x) \ \underrightarrow{e^{-\imag\tau H}\circ e^{\pi \bar \sigma_x}\circ e^{-\imag\tau H}} \ \frac{1}{2}(\mathds{1}+f(2\tau)\sigma_x) \ \underrightarrow{\ \langle\sigma_x\rangle\ } \ f(2\tau)
\end{equation}

By plotting $f(2\tau)\equiv\sqrt{n_x(2\tau)^2+n_y(2\tau)^2}$ as given by the free induction decay, the dephasing properties can be investigated. If the dephasing is exponential, a fitting function of the form $e^{-\tau/T_2}$ is used, where $T_2$ parameterises the exponential dephasing time. This is also referred to as the \emph{transverse relaxation time} or the \emph{spin-spin relaxation time} in the literature. In the absence of any other noise process, the  corresponding quantum channel is
\begin{equation}
\e_\tau[\tilde\rho(t)] =[1-\lambda(\tau)]\tilde\rho(t)+\lambda(\tau)\sigma_z \tilde\rho(t) \sigma_z\label{dephasing channel sigmaz}
\end{equation}
where $\lambda(\tau)=(1-e^{-\tau/T_2})/2$ is the probability of performing a $\sigma_z$ operation on the state under conjugation.  This is known as the dephasing channel and  forms a dynamical semigroup. Consequently, its Lindblad master equation is of the form
\begin{equation}
\frac{d}{dt}\rho(t)= \imag\left[\rho(t),H_0\right]_-+ \frac{1}{2 T_2}\left(\sigma_z\rho(t) \sigma_z - \rho(t)\right).
\end{equation} 
The state fully dephases when $\tau \to \infty$ where $\lambda(\tau)=1/2$. In some circumstances the dephasing process is non-Markovian. This is captured by the use of a  fitting function $e^{-\tau/T_2 -(\tau/T_S)^n}$ where the added parameter $T_S$ is an indication of non-exponential dephasing which is a result of non-semigroup dynamics.

\subsection{Amplitude damping  }\label{Amplitude damping measurement scheme}

In the magnetic resonance literature, amplitude damping is known as \emph{spin-lattice relaxation}, or \emph{longitudinal relaxation}.  The ensemble system in thermal equilibrium will relax to  the thermal state $\rho_{\mathrm{th}}$ given as
\begin{align}
\rho_{\mathrm{th}}=\frac{e^{-\frac{ H_0}{k_B T}}}{\mathrm{tr}[e^{-\frac{ H_0}{k_B T}}]} &\equiv \frac{1}{\mathrm{tr}[e^{-\frac{ H_0}{k_B T}}]} \sum_{i=1}^d e^{-\frac{E_i}{k_BT}}\Pi(\phi_i)   \label{thermal state}
\end{align}
where $k_B$ is Boltzmann's constant and $T$ is the temperature.  In any two-dimensional subspace with an allowed dynamical selection rule, the renormalised state will be 
\begin{align}
\rho&=\frac{e^{-\frac{1}{k_B T} (\Delta E)}}{Z}\left(\Pi(\phi_0)+\Pi(\phi_1)\right) +\frac{1-e^{-\frac{1}{k_B T} (\Delta E)}}{Z}\Pi(\phi_1) \nonumber \\
&= \frac{2e^{-\frac{1}{k_B T} (\Delta E)}}{Z}\frac{1}{2}\mathds{1}+\frac{1-e^{-\frac{1}{k_B T} (\Delta E)}}{Z}\frac{1}{2}(\mathds{1}-\sigma_z)
\end{align}  
where $\Pi(\phi_0)$ is the excited state and $\Pi(\phi_1)$ the ground state, with $\Delta E=E_{0}-E_1>0$ the energy difference. $Z=1+e^{-\frac{1}{k_B T} (\Delta E)}$ is the renormalised partition function. The second line shows that the term on the left is proportional to a maximally mixed state and the term on the right is proportional to the pure ground state configuration. This latter component is referred to as a \emph{pseudo pure state}. 
The component which is proportional to the maximally mixed state is invariant under any quantum operation, and does not contribute to the free induction. We may therefore consider only the  pseudo pure state. As such, the free induction signal will be improved at low temperature environments or at higher magnetic fields $B_0$.

To determine the amplitude damping process, we must measure the decay of $n_z$. In such a case, we may take the pseudo pure state at thermal equilibrium, rotate it to its orthogonal state in the Bloch sphere, and observe how fast it decays back to its original state.  As discussed previously, to measure the $n_z$ component of the Bloch sphere we need to first rotate it onto the $x-y$ plane, which is done by the $\pi/2$ pulse $e^{\frac{\pi}{2}\bar \sigma_y}$. To ascertain the decay rate we simply need to increment the time we allow the system to evolve before performing this pulse. As before, other reversible dephasing mechanisms can be removed with the Hahn echo.  The steps for the amplitude damping measurement can therefore be decomposed in the following way

\begin{enumerate}[(i)]
\item After allowing the system to  relax to its thermal equilibrium, carry out a $\pi$ pulse about the $y$ axis to take the pseudo pure state to its orthogonal state on the Bloch sphere.
\item Wait for a time $\tau$ to allow the system to evolve, and then Perform a $\pi/2$ pulse to take the state onto the $x-y$ plane. Following this, carry out  the Hahn echo sequence using a constant time period $\tau'$ to remove reversible dephasing noise mechanisms. 
\item  Weakly measure the expectation values of  $\sigma_x$ and $\sigma_y$.
 
\item Repeat (i)-(iii) $N$ times, and on each occasion increase $\tau$ by a constant value $\Delta \tau$ .
\end{enumerate}

The equivalent rotating frame  sequence of operations can be denoted as follows
\begin{align}
&\frac{1}{2}(\mathds{1}-\sigma_z) \ \underrightarrow{e^{\pi\bar \sigma_y}} \ \frac{1}{2}(\mathds{1}+\sigma_z) \ \underrightarrow{\tau} \ \frac{1}{2}(\mathds{1}+f(\tau)\sigma_z)  \nonumber \\ &\underrightarrow{e^{-\imag\tau' H}\circ e^{\pi \bar \sigma_x}\circ e^{-\imag\tau' H}\circ e^{\frac{\pi}{2}\bar \sigma_y}} \ \frac{1}{2}(\mathds{1}-f(\tau)\sigma_x) \ \underrightarrow{\ \<\sigma_x\> \ } \ -f(\tau)
\end{align}

This process is usually exponential such that in the absence of any other relaxation processes $f(\tau)=2e^{-\frac{\tau}{T_1}}-1$. The relaxation time scale here, $T_1$, is usually referred to as the longitudinal relaxation time or the spin-lattice relaxation time. Mathematically, it can be described by the amplitude damping channel
\begin{equation}
\e_\tau[\tilde\rho(t)]=K_{0,\tau}\tilde\rho(t) K_{0,\tau}^\dagger+K_{1,\tau} \tilde\rho(t) K_{1,\tau}^\dagger
\end{equation}
with the Krauss operators
\begin{align}
&K_{0,\tau}=\begin{pmatrix}1 & 0 \\
0 & e^{-\frac{\tau}{2T_1}} \\
\end{pmatrix} & K_{1,\tau}= \begin{pmatrix}0 & \sqrt{1-e^{-\frac{\tau}{T_1}}} \\
0 & 0 \\
\end{pmatrix}
\end{align}
which forms a dynamical semigroup, and as such has the following Lindblad master equation
\begin{equation}
\frac{d}{dt}\rho(t)=\imag\left[\rho(t),H_0\right]_-+\frac{1}{T_1}\left(|\phi_1\>\<\phi_0|\rho(t)|\phi_0\>\<\phi_1|-\frac{1}{2}[|\phi_0\>\<\phi_0|,\rho(t)]_+ \right).
\end{equation}
This relaxation process also leads to an exponential decay of the Bloch vector component in the $x-y$ plane which could be detected by the dephasing measurement protocol. In such a case, the $T_2$ time is twice the $T_1$
\begin{equation}
T_2=2T_1.
\end{equation}

\subsubsection{Depolarisation}

In the case of depolarisation, where a state is taken to the maximally mixed state, there is no pseudo pure state. Instead of the initial $\pi$ pulse from the amplitude damping measurement, we must prepare our system in a pure state and observe how $n_z$ vanishes. The depolarisation channel is given as
\begin{equation}
\e_\tau[\rho]=(1-\lambda(\tau))\rho+\frac{\lambda(\tau)}{3}\left(\sigma_x \rho \sigma_x +\sigma_y \rho \sigma_y + \sigma_z \rho \sigma_z \right)
\end{equation}

which forms a dynamical semigroup and has the Lindblad master equation

\begin{equation}
\frac{d}{dt}\rho(t)=\imag[\rho(t),H_0]_-+\sum_{i=1}^3\gamma\left(\sigma_i\rho(t)\sigma_i- \rho(t) \right).
\end{equation}

where $\lambda(\tau)=\frac{3}{4}(1-e^{-4\gamma \tau})$. The system fully depolarises as $\tau \to \infty$ so that $\lambda(\tau)= 3/4$. 

\subsection{Nutation}\label{nutation measurement scheme}

If we wish to observe the Rabi oscillations, or \emph{nutation},   caused by our driving field, we may use the following protocol.

\begin{enumerate}[(i)]
\item Prepare the system in the state $\rho=\frac{1}{2}(\mathds{1}+\sigma_z)$.

\item Perform the operation $e^{\omega \tau \bar \sigma_y}$ to effect a $\omega \tau$ rotation about the $y$ axis, followed by a Hahn echo sequence.

\item Weakly measure the expectation value of $\sigma_x$ and $\sigma_y$.

\item Repeat (i)-(iii) $N$ times, on each occasion increasing $\tau$ by $\Delta \tau$.
\end{enumerate}

The equivalent rotating frame  sequence of operations can be denoted as follows
\begin{align}
&\frac{1}{2}(\mathds{1}+\sigma_z) \ \underrightarrow{e^{\omega \tau\bar \sigma_y}} \ \frac{1}{2}(\mathds{1}+\sin(\omega\tau)\sigma_x+\cos(\omega \tau)\sigma_z) \   \nonumber \\ & \underrightarrow{e^{-\imag\tau' H}\circ e^{\pi \bar \sigma_x}\circ e^{-\imag\tau' H}} \ \frac{1}{2}(\mathds{1}+\sin(\omega\tau)\sigma_x-\cos(\omega \tau)\sigma_z) \  \underrightarrow{\ \<\sigma_x\> \ } \ \sin(\omega\tau)
\end{align}

The Fourier transform of this FID signal will give information pertaining to the Rabi frequency $\omega$.

\section{The equipment}\label{ESR equipment}
Although complex and varied in practice, the ESR(NMR) spectrometer can be conceptually simplified to its essential components. First, the components common to both c.w. and pulsed spectrometers are

 \begin{enumerate}[(i)]
\item 
\emph{External static magnetic field}
\

As the only tunable term in the free, time-independent Hamiltonian, it allows us to set the  transition frequencies to the desired value.

\item \emph{m.w.(RF) source}
\

This generates a sinusoidally oscillating electromagnetic field which drives the electronic (nuclear) transition at resonance.
The strength of the source is constant, but the strength of the m.w.(RF) reaching the sample is controlled by an attenuator. The m.w.(RF) may be transported to the sample by a rectangular pipe called a waveguide.

\item\emph{ m.w.(RF) resonator}
\

For ESR, utilising m.w. fields which have short wavelengths, the resonator may be a small metallic box which houses the sample, called a cavity. The iris placed between the cavity and the m.w. waveguide  controls the intensity of waves that  enter and are reflected away from the cavity, by affecting the impendence of the cavity to the radiation. The size of the iris, and hence the coupling between the m.w. and cavity, is often controlled by a screw. The cavity is critically coupled if the size of the iris is such that all of the incident  radiation enters the cavity. If the m.w. frequency is in resonance with the cavity (dependent on its size) then the cavity absorbs all of the radiation.  As a result, an ESR machine which is built to detect transitions of a particular frequency must have a cavity of the appropriate dimensions. The Q-factor determines how efficiently the cavity stores the radiation energy as opposed to dissipating it away, with an increase in Q-factor leading to an increase in signal-to-noise ratio. Furthermore, the radiation inside the cavity produces a standing wave at resonance, which has its electric and magnetic fields out of phase. As the electric field causes off-resonance absorption by most samples and the heat dissipation thereof leads to a reduction in the Q-factor, we can position the sample at the electric field minimum so as to ensure only the magnetic field component of the m.w. leads to resonant absorption, and hence improve the signal quality.

NMR spectrometers, needing access to RF fields of long wavelength, use a different type of resonator, known as LCR (or RLC) resonators. These incorporate a resistor, inductor, and capacitor to expose the sample to the required long wavelength RF fields.

\item \emph{Cryostat}
\

To cool the sample, we place the  resonator inside a cryostat, through which we pump liquid helium. Both the cryostat and the tubing through which the liquid helium is pumped are isolated from the environment by a vacuum. The temperature of the cryostat is measured, and if temperatures above those of liquid helium are desired, a heater within the cryostat is used to elevate the temperature accordingly. 

\

\item \emph{Detector}
\

  As the sample  resonantly absorbs the radiation, the impendence of the resonator changes and hence  the degree by which radiation is reflected is altered. This can be detected by a diode.  

\end{enumerate}

The components which differ between c.w. and pulsed spectrometers are in the detection component. Firstly, for c.w. spectrometers these are
\begin{enumerate}[(i)]
\item \emph{Field modulator}
\

   To improve the sensitivity of the signal, what is actually detected is the first derivative of the absorption spectrum, by use of field modulation. This works by sending a sinusoidally oscillating magnetic field -- with a particular modulation amplitude (MA) and frequency -- in the same direction as the static magnetic field. This produces an amplitude-modulated signal which will have a sine wave shape. The larger the modulation amplitude, the better the signal-to-noise ratio will be, but modulation amplitudes  larger than  the line width of the absorption spectrum will produce signals whose line widths are larger, and whose shapes are distorted. The modulation frequency  must also be chosen with careful consideration, as the Fourier transform relationship means that for ESR(NMR) signals that are close, a modulation frequency that is too large will reduce the resolution.  

\item
\emph{Reference arm}
\

To ensure that the detector diode is operating in the linear regime (where the m.w.(RF) power is proportional to the square of the diode current), a reference arm is used which, when operating, supplies the detector with auxiliary m.w.(RF) power.  For an ESR spectrometer used in ETH Zurich, for example,  the diode operates in the linear regime for high currents, and so we set the power of the reference arm to increase the zero-value of the diode current to 200 $\mu A$. The phase of the reference arm is also important, with larger currents detected when the reflected radiation from the cavity and the radiation from the reference arm are in phase.   

\end{enumerate}

Pulsed spectrometers differ in their detector technology in the following way

\begin{enumerate}[(i)]

\item \emph{Amplifier}

As pulsed spectroscopy requires powerful pulses, an amplifier is needed to increase the power of the microwave source.

\item \emph{Shielding}

The FID signals are weak compared to the background noise and need to be amplified, using a preamplifier,  in order to be detected. However, because the m.w.(RF) radiation in pulsed spectrometers  is strong, the preamplifier can easily be destroyed. Therefore, the preamplifier must be shielded from the pulses until they dissipate. Consequently, there is a \emph{dead time} between the driving of the system and the detection of the free induction, the duration of which is dependent on the resonator's frequency and Q-factor.   

\item \emph{Pulse programmer}

An electronic interface is required so as to program the spectrometer to conduct its sequence of pulses and measurements, which occur at such a fast pace that they cannot be controlled in situ by a human being.

\end{enumerate}

\spacing{1}                                  
 \bibliographystyle{plainnat}
 \bibliography{references}

\spacing{1}
\chapter{Quantum Information Processing}

\section{Introduction}
The digital computer is a device that can perform a function $f:\{0,1\}^N\to\{0,1\}^M$. Here, $\{0,1\}^N$ means a string of characters of length $N$, where  each character  can take either a value of $0$ or $1$. These characters are known as \emph{bits}, while the function $f$ is known as an \emph{algorithm}. A \emph{universal Turing machine} is a mathematical construction, named after the mathematician Alan Turing, which can simulate \emph{any} digital computer \citep{Turing-machine} and hence perform any algorithm $f$. This Turing machine consists of an infinite memory tape which can contain symbols, and  a device that can read and alter these symbols. The action of altering these symbols is known as an \emph{elementary operation}. In the language of complexity theory, we may define the efficiency with which a Turing machine can simulate a given algorithm. An algorithm is said to be computed \emph{efficiently} if it is soluble in polynomial time, and it is computed \emph{inefficiently} if it is soluble in super-polynomial (often exponential) time. For an algorithm $f$ solving a problem of size $n$, it is computed in polynomial time if the number of elementary operations -- each of which take an equal time to perform -- grow\footnote{If the polynomial function is $\alpha n^k + \beta n^l$ such that $l<k$, then as $n\to \infty$ the second term makes a negligible contribution and, ignoring the constant multiplicative factor $\alpha$, this polynomial function grows in the order of $n^k$, denoted as $O(n^k)$.} as $O(n^k)$ ,where $k$ is some integer, in the asymptotic limit of $n \to \infty$ . Conversely, if the size of the problem grows as $O(\alpha^n)$ for any real value $\alpha$, then it does so in exponential time and is computed inefficiently. 

The complexity class $BPP$ is one where every algorithm belonging to it can be simulated efficiently (with a bounded probability of error) using a Turing machine augmented by a random number generator. The discovery of the prime factors of  numbers is one which is not known to exist in $BPP$ at this time, which is an open problem in complexity theory. As demonstrated by \citep{Shors-algorithm}, however, one can  efficiently factorise an integer $N$ in $O((\mathrm{log} N)^3)$   by using the laws of quantum mechanics to devise a \emph{quantum computer}. The complexity class of algorithms that can be simulated efficiently on a \emph{universal quantum computer} is $BQP$, and as classical physics is a subset of quantum physics, then we know that $BPP \subseteq\ BQP$. Although this has not been proven,   examples such as Shor's algorithm suggest that the equality here does not hold, and $BQP$ is in fact larger than $BPP$. This intuition has spurred an interest in quantum computer science, with the discovery of various other algorithms that are in $BQP$, but not yet known to exist in \emph{BPP}. 

A parallel rise in  interest in quantum information theory was also seen at this time. One of the pioneers in this field was Benjamin Schumacher, who in \citep{quantum-coding}  used the von Neumann entropy covered in   appendix \ref{von Neumann entropy}  to give quantum information an operational meaning as the amount by which a composite quantum system may be faithfully encoded into a smaller Hilbert space. In light of the relationship between information theory and computer science, quantum computation is also referred to as quantum information processing (QIP). 

A good introductory text book for quantum information and quantum computer science is \citep{nielsenchuang}.  A more recent text focusing on quantum information processing is \citep{quantum-computing}. Good online resources are the lecture notes by John Preskill \citep{preskillnotes} and John Watrous \citep{watrousnotes}.

\section{Universal quantum computation}

\begin{figure}[!htb]
\begin{equation}
\Qcircuit  @C=1em @R=0.6em{\lstick{\phi_{0}^1} & \multigate{4}{\ \ \mathcal{E} \ \ } & \qw   &&&& \lstick{\phi_0^1} & \multigate{6}{\ \ U \ \ } & \qw  \\ \lstick{\phi_0^2}  &  \ghost{\ \ \mathcal{E} \ \ } &  \qw &&&& \lstick{\phi_0^2} & \ghost{\ \ \mathcal{U} \ \ } & \qw   \\
\lstick{.} & \ghost{\ \ \mathcal{E} \ \ } & \qw & \push{\rule{.3em}{0em} = \rule{.3em}{0em}} &&&\lstick{.}& \ghost{\ \ \mathcal{U} \ \ } & \qw \\
\lstick{.} & \ghost{\ \ \mathcal{E} \ \ } & \qw &&&&\lstick{.}& \ghost{\ \ \mathcal{U} \ \ } & \qw\\
\lstick{\phi_0^N} & \ghost{\ \ \mathcal{E} \ \ } & \qw  &&&& \lstick{\phi_0^N}  & \ghost{\ \ \mathcal{U} \ \ } & \qw \\ &&&&&& \lstick{.} & \ghost{\ \ \mathcal{U} \ \ } & \meter \\ &&&&&& \lstick{\phi_0^{N+M}} & \ghost{\ \ \mathcal{U} \ \ } & \meter }
\nonumber\end{equation} 
\caption{ The circuit model of quantum computation. The most general quantum computer performs a quantum operation (generally stochastic) on an input of $N$ qubits each set to the pure state with the associated vector $\phi_0$ in the computational basis. This can also be achieved by increasing our system by $M$ ancillary qubits and performing a unitary map on the whole system. As per Stinespring's dilation theorem, the desired quantum operation on our desired $N$-qubit state will be completed when a projective measurement is carried out on the $M$ ancillary qubits. In the case of stochastic quantum operations, we would post select the transformed state for only a subset of the possible measurement outcomes on the ancillary systems. At the end of the computation, we \emph{read out} the result by measuring each qubit in the computational basis. }\label{QCircuit}
\end{figure}
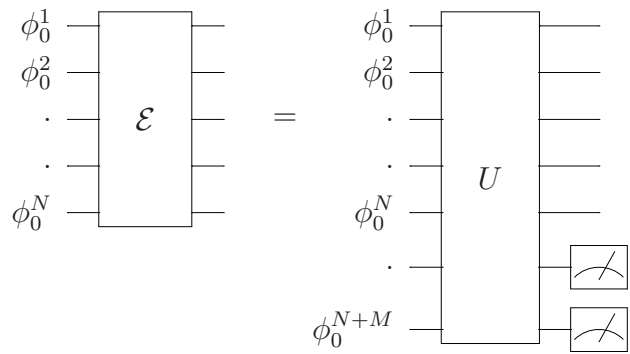

Analogous to the fundamental unit of classical computation -- the bit $x \in\{0,1\}$ -- in quantum information we have the quantum bit, or qubit, which is a state $\rho \in \s(\co^2)$ with the computational basis states  $\{\phi_0,\phi_1\} $.
A universal quantum computer (or a quantum information processing device) is a device which can perform any quantum operation $\mathcal{E} : \rho \mapsto \mathcal{E}[\rho] $ on an $N$-qubit input state $\rho\in\s(\co^{2^N})$. As this state can be prepared by a quantum channel acting on the pure, separable initial input state $\Pi(\phi_0)^{\otimes N}$ we may, without loss of generality, absorb the preparation map into $\mathcal{E}$, and redefine our quantum computer as a device which performs a quantum operation on the input state $\Pi(\phi_0)^{\otimes N} $. We may further restrict our quantum computer to unitary operators as any quantum operation $\mathcal{E}$ acting on $N$ qubits may be obtained by a unitary map acting on our system coupled with  $M$ ancillary qubits, followed by projective measurements on said qubits, as per Stinespring's dilation theorem. At the end of the computation, the result is obtained by measuring each qubit in the computational basis. This is referred to as the computational \emph{read out}. This quantum computer can be drawn as a circuit diagram shown in Fig.\ref{QCircuit}, where each \emph{wire} corresponds to $\co ^2$ and time flows from left to right. Hence the box corresponding to the unitary map acts on the input state coming in from the left, and outputs another state which exits on the right.  Consequently, this is referred to as the circuit model of quantum computation.

A universal quantum computer must be able to perform any transformation in the state space $\s(\co^{2^{N}})$. In analogy with classical digital computers which can be formed by a set of universal logic gates, we want to identify a finite set of \emph{universal quantum gates}, which we can  concatenate in order to achieve controllability within a state space $\s(\co^{2^{N}})$ of arbitrary $N$. As we discussed in Sec.\ref{Lie algebras}, we can have controllability in $\s(\co^d)$ if we have access to the generators of the Lie algebra $\mathfrak{su}(d)$ from which we can obtain the Lie group of unitary matrices $\mathcal{SU}(d)$.   In the case of one-qubit unitaries, which correspond to rotations on the Bloch sphere, we can use two orthogonal Pauli matrices to obtain a unitary operator $R_{\underline{n}.\underline{\sigma}}(\theta):=e^{-\imag\frac{\theta}{2}\underline{n}.\underline{\sigma}}$. 

\begin{equation}
\Qcircuit @C=1em @R=1em{\lstick{\phi_0}& \gate{R_{\underline{n}.\underline{\sigma}}(\theta)} & \qw &  \psi} \nonumber
\end{equation}     

 As has been shown by \citep{divincenzo-1995} we can generate the group $\mathcal{SU}(2^{N})$ given the group $\mathcal{SU}(2)$ for each qubit subspace, and any entangling two-qubit gate which we can perform between the qubits. Two such gates are the controlled NOT (CNOT) and controlled phase (CZ) gates
\begin{align}
&\text{CZ}= \Pi(\phi_0)\otimes \one+\Pi(\phi_1)\otimes \sigma_z, &\text{CNOT}= \Pi(\phi_0)\otimes \one+\Pi(\phi_1)\otimes \sigma_x.
\end{align}  
Here, the qubit subspace which the projectors act on is called the \emph{control qubit}, and the subspace where either the identity or the Pauli matrix acts on is the \emph{target qubit}.
These two are equivalent under the action of the Hadamard gate in conjugation on the target qubit subspace
\begin{align}
\mathrm{CZ}=(\one\otimes \mathrm{H})&\mathrm{CNOT}(\one\otimes \mathrm{H}), \nonumber \\
\mathrm{H}=\frac{1}{\sqrt{2}}&(\sigma_x + \sigma_z).
\end{align}
The Hadamard gate transforms Pauli matrices to Pauli matrices under conjugation, such that
\begin{align}
&\mathrm{H}\sigma_x\mathrm{H}=\sigma_z, \nonumber \\
&\mathrm{H}\sigma_y\mathrm{H}=-\sigma_y, \nonumber \\
&\mathrm{H}\sigma_z\mathrm{H}=\sigma_x.
\end{align}
We may restrict the number of gates needed to be built experimentally by utilising the  \emph{standard universal set} given by $\{\mathrm{H},P_{\pi/8},\mathrm{CNOT}\}$ where the $\pi/8$ phase gate is given by
\begin{equation}
P_{\pi/8}=\Pi(\phi_0)+e^{\imag\frac{\pi}{4}}\Pi(\phi_1)=e^{\imag\frac{\pi}{8}}e^{-\imag\frac{\pi}{8} \sigma_z}.
\end{equation}
 Here, we note that the $\mathrm{H}$ and $P_{\pi/8}$ gates can generate any unitary operation in the state space of a single qubit. 

In what sense do we say that the CNOT (or CZ) gate is an entangling one? We may observe that these unitaries  cannot be factorised into $U^A\otimes U^B$. The CNOT gate can produce an entangling gate $U_E$ given an initial product state in the computational basis,  if the control qubit is first taken to the $\sigma_x$ basis. This can be done by preceding a CNOT by an $\mathrm{H}^A\otimes\one^B$ to get 

\begin{equation}
U_E =   \mathrm{CNOT}^{A+B}(\mathrm{H}^A\otimes\mathds{1}^B)  
\end{equation}

which performs the maps 
\begin{align}
U_E:\begin{cases} \phi_0\otimes\phi_0 \mapsto \frac{1}{\sqrt{2}}\left(\phi_0\otimes\phi_0+\phi_1\otimes\phi_1 \right) \equiv \Phi^+  \\
\phi_0\otimes \phi_1 \mapsto\frac{1}{\sqrt{2}}\left(\phi_0\otimes\phi_1+\phi_1\otimes\phi_0 \right) \equiv \Psi^+  \\
\phi_1\otimes\phi_0 \mapsto \frac{1}{\sqrt{2}}\left(\phi_0\otimes\phi_0-\phi_1\otimes\phi_1 \right) \equiv \Phi^-  \\
\phi_1\otimes\phi_1 \mapsto \frac{1}{\sqrt{2}}\left(\phi_0\otimes\phi_1-\phi_1\otimes\phi_0 \right) \equiv \Psi^-  \\
\end{cases}  .
\end{align}

\subsection{Alternative models of quantum computation}  
Other models of quantum computation exist which are equivalent to the circuit model, but may have advantages in terms of implementation. One such model is the measurement based quantum computer, which takes a highly entangled state such as the cluster state as a resource, and the quantum computation is then  carried out by adaptive single-qubit measurements \citep{MBQC}. To generate the cluster state we may take an array of qubits, all prepared in the state $\phi_+:=\frac{1}{\sqrt{2}}(\phi_0+\phi_1)$, and then perform a CZ gate between all the neighbours. In general terms, the rows of the cluster state may be seen as representing a logical qubit, and hence for the cluster state to allow for universal quantum computation that is advantageous to a classical computer, we require multiple rows of qubits that are interconnected; a two-dimensional cluster state is needed.

Another model is that of adiabatic quantum computation \citep{adiabatic-quantum}, where the quantum information is stored in the ground state of a certain Hamiltonian. The  algorithm is then solved by adiabatically  changing this Hamiltonian, thereby performing a unitary map that transforms the initial ground state to a new one. This model has garnered interest because it has been shown to have some inherent protection from decoherence \citep{adiabatic-robust}.

\section{Fault tolerance}
So far our description of a quantum computer has assumed the lack of noise, which is of course an unreasonable expectation for any realistic device. When noise is introduced into our system, which could be due to the interaction of a qubit with its environment, or the imprecise realisation of a unitary gate, our quantum computer fails with some probability.  We may model the noise process as acting after our computation. In the case of depolarising noise, for example, the output of our quantum computation, $\rho$, is taken to $(1-P)\rho + P\frac{1}{2}\mathds{1}$.   The computation is then said to fail with a probability $P$. What we require is a method of performing our computation \emph{fault tolerantly}, such that by investing resources, we may lower the probability of failure for a computation to an arbitrarily small amount $\epsilon$. This requires a method of performing error correction. In classical digital computers, the \emph{logical bits} are encoded into \emph{physical bits} to allow for error correction. The simplest example is 
\begin{align}
&0_L=0000...   &1_L=1111...
\end{align} 
If the noise process is \emph{independent} for the physical bits, such as a  bit flip occurring with an independent probability for each of the physical bits, we may use this introduced redundancy to correct the bit flip errors by taking the logical qubit to be that of the most common of the physical bits. For example, if we have the physical bit $1000...$, then we may make the assumption that because most of the physical bits are 0, then this must be our original logical bit $0_L$. In quantum computing, there is a  more sophisticated method of encoding logical qubits into physical ones, and it is called the \emph{stabiliser code}. I will give a \emph{very} brief overview of this, but interested readers may learn more by referring to the texts I introduced at the beginning of this chapter.

\subsection{Stabiliser formalism}
\subsubsection{Stabiliser space  and the Clifford group}
  Consider the  $N$-qubit Pauli group    $\mathcal{G}_N$ and its subset $G_N$ which  are observables \footnote{The $N$-qubit Pauli group is defined as $\mathcal{G}_N:=(\{\sigma_i\}_{i=0}^3\times\{\pm1,\pm i\})^{\otimes N}$.  The subset of this which are self-adjoint are n-qubit Pauli observables, $G_n$, whose elements are given as  $\bigotimes_{n=1}^N \sigma_n$ .  These are all both unitary and self-adjoint, so  have eigenvalues from the set $\{+1,-1\}$. } on $\co^{2^N}$. The subset of $G_N$, denoted $S_t$, which is generated by the set $\{g_i\}_{i=1}^r$, defines the subspace $\co^{2^K}$ that it stabilises as\begin{equation}
\co^{2^K}:=\{\psi\in \co^{2^N} : g\psi=\psi \ \forall \ g \in S_t\}
\end{equation}
where $K=N-r$. Every physical state $\rho \in \s(\co^{2^N})$ which is spanned by the stabilised subspace is an encoding of our quantum information, and the corresponding logical state is given as $\varrho \in \s(\co^{2^K})$. Let us assume that the operators $\{E_i\}$, which can cause  errors on the physical states $\rho \in \s(\co^{2^N})$ by the map $\rho \mapsto \rho'=(1-P)\rho + P E_i\rho E_i$,  are a subset of $G_N$. It can be shown that every $E_i$ either commutes or anticommutes with each $g\in S_t$, and    as $\rho$ and $\rho'$ are both  an eigenstate of the Stabiliser observables, we may measure these on the system, without disturbing it, to reveal the eigenvalues $\pm 1$. For each  $g$ that gives an eigenvalue $-1$, or an \emph{error syndrome}, we are informed of the set of possible  errors  $\{E_i:[E_i,g]_+=\mathds{O}  \}$ that could have occurred.  Acting on $\rho$ by \emph{any} $E_j$ such that $[E_j E_i, g]_-=\mathds{O} $   can then correct  for the error. 

What kind of unitary operations are allowed in this stabilizer formalism? We require that our unitaries transform a stabilized vector to another stabilised vector, and hence
\begin{equation}
U\psi=Ug\psi\iff U\psi=UgU^\dagger U\psi.
\end{equation} Therefore, $U\psi$ must be stabilised by $Ug U^\dagger$ for all $g\in S_t$. The set of such unitary operations, that transform members of the Pauli group to other members of the Pauli group under conjugation, are known as the \emph{Clifford group}, and can be generated by the set $\{\mathrm{H},P_{\pi/4},\mathrm{CNOT}\}$.  This differs from the standard universal set in that $P_{\pi/8}$ has been replaced by $P_{\pi/4}$, given by
\begin{equation}
P_{\pi/4}=\Pi(\phi_0)+\imag\Pi(\phi_1)=e^{\imag\frac{\pi}{4}}e^{-\imag\frac{\pi}{4} \sigma_z}.
\end{equation} 

\subsubsection{Local noise, concatenated codes, and the threshold theorem}

Let us consider the case where we  encode one logical qubit into $N$ physical ones: $[[N,1]]$.   The simplest noise model is local, where there are no correlations in the errors that may occur on the qubits. In such a case, given the probability of failure for each physical qubit being $P$, this encoding reduces the probability of failure for the encoded qubit  to be  $cP^2$ for a constant $c$ which is dependent on the code used. We can further reduce the probability of failure by concatenating the code, whereby we develop a hierarchy of codes for a time block \footnote{A time block here refers to a unit of time in which at most one operation is carried out on any given physical qubit.} of computation ad infinitum. For $i$ levels of concatenation, then, the error probability for the concatenated code is $P(c,i)=c^{2^i-1}P^{2^i}$. 

Consider now an ideal (noiseless) circuit containing $p(n)$ gates -- where $p(\cdot)$ is a polynomial function and $n$ specifies the size of the problem -- with the output  $\rho$, and the noisy encoding  $[[N,1]]$ which prepares the state $\rho'$. We define the error with respect to the trace distance of these states as  $D[\rho , \rho']= \epsilon$. The \emph{threshold theorem} states that there is an error threshold $P_T\equiv 1/c$ for the physical qubits, where provided that $P < P_T$, there exists an $i$ such that for any $\epsilon>0$ we can satisfy $P(c,i)\leqslant \epsilon / p(n)$.  The encoded circuit will contain $O(\mathrm{poly}[\mathrm{log}_2p(n)/\epsilon]p(n))$ gates, where $\mathrm{poly}[\cdot]$ is a polynomial function independent of $n$ \footnote{Here, $P(c,i)=P_T\left(\frac{P}{P_T}\right)^{2^i}$. This is a doubly exponential function in $i$.}. As such, a doubly exponential reduction in the error of the encoded circuit requires only a polylogarithmic increase in the physical size of the computation.

 The threshold theorem provides a symbiotic relationship between design of fault tolerant codes and experimental realisation of quantum computation. The  larger $P_T$ is for a given code, the better the code is. Conversely the smaller $P$ is for a physical realisation, the better the physical realisation is. Early codes gave  $P_T$ in the range of $10^{-5}-10^{-6}$, but more recent codes have pushed this higher to $\sim10^{-3}$. For a particular physical realisation, with usually an exponential coherence time of $T_2$  and the longest quantum gate taking $\Delta t$ to perform, we may estimate the upper bound for $P$ for a single gate to be $P\sim( 1-e^{-\Delta t/T_2})/2$. This can be lowered to reach the threshold requirement by either decreasing $\Delta t$, increasing $T_2$, or both.   

A simple example of a stabiliser code  is the $[[5,1]]$ code, which encodes one logical qubit into five physical ones \citep{Gottesman-thesis}. The stabilizers are $\{g_i\}_{i=1}^{16}$, with the sixteen linearly independent elements generated  from the set
\begin{align}
g_1=\sigma_x\otimes\sigma_z\otimes\sigma_z\otimes \sigma_x\otimes \mathds{1}, \nonumber \\
g_2=\sigma_z\otimes\sigma_z\otimes\sigma_x\otimes\mathds{1}\otimes \sigma_x, \nonumber \\
g_3=\sigma_z\otimes\sigma_x \otimes \mathds{1}\otimes\sigma_x\otimes\sigma_z, \nonumber \\
g_4=\sigma_x\otimes\mathds{1}\otimes\sigma_x\otimes\sigma_z\otimes\sigma_z.
\end{align}
The basis of the logical qubit is defined as $\phi_0^L$ being  an equal superposition of physical qubit basis states with an even number of $\phi_1$ (such as $\phi_1\otimes\phi_1\otimes\phi_0\otimes\phi_0\otimes\phi_0$), and $\phi_1^L$ being an equal superposition of physical qubit basis states with an odd number of $\phi_1$ (such as $\phi_0\otimes\phi_0\otimes\phi_1\otimes\phi_1\otimes\phi_1$). The logical qubit operators are given as\begin{align}
&Z=\sigma_z\otimes\sigma_z\otimes\sigma_z\otimes\sigma_z\otimes \sigma_z & X=\sigma_x\otimes\sigma_x\otimes\sigma_x\otimes\sigma_x\otimes \sigma_x
\end{align}
   The logical CNOT gate is implemented by performing a CNOT between the corresponding physical qubits of each logical qubit. As can be verified, such a code does not propagate any local error as the computation proceeds.  
\begin{figure}[!htb]
\begin{align}
&\Qcircuit @C=1em @R=.7 em { & \ctrl{1}& \qw \\ & \targ & \qw }  \ \ \ \  = \ \ \ \ 
\Qcircuit @C=1em @R=.7 em {&\ctrl{5}& \qw &\qw& \qw & \qw & \qw \\
 & \qw &\ctrl{5} & \qw  & \qw  & \qw & \qw\\
  & \qw & \qw &  \ctrl{5} & \qw & \qw & \qw \\
   & \qw & \qw & \qw  &\ctrl{5} & \qw & \qw \\
    & \qw & \qw & \qw & \qw  &\ctrl{5} & \qw \\
     & \targ & \qw & \qw & \qw & \qw & \qw \\
&\qw & \targ & \qw & \qw & \qw & \qw\\ 
 & \qw & \qw     & \targ & \qw & \qw & \qw \\ 
 & \qw & \qw & \qw     & \targ & \qw & \qw \\
 & \qw & \qw & \qw & \qw & \targ & \qw } \nonumber
  \end{align}
  \caption{The logical CNOT operation in the $[[5,1]]$ code is performed by performing a physical CNOT gate  on each corresponding physical qubit of the two logical qubits.}
\end{figure}
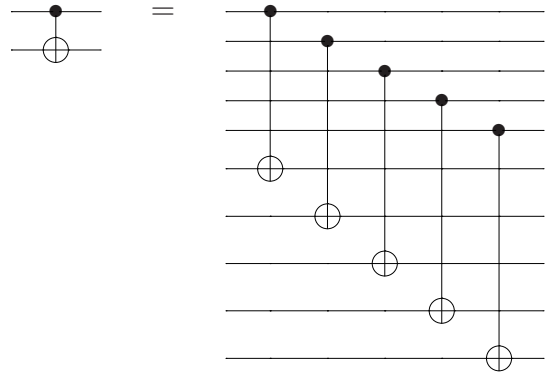

\subsection{Gottesman-Knill theorem}

\begin{figure}[!htb]
\centering
\includegraphics[width=0.3\textwidth]{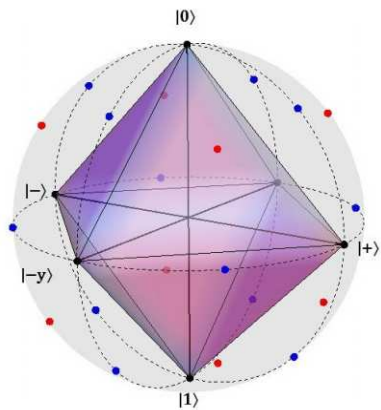}
\caption{The Stabiliser state space $\mathcal{ST}(\co^2)$ represented as an octahedron inside the Bloch sphere. The twelve blue pure states are the $H$-type magic states and the eight red pure states are the $T$-type magic states. Image taken from \citep{Hussain-magic-states}.}
\end{figure}

Let us consider the state space of a logical qubit in the stabilizer code.  The state space which can be obtained by the one-qubit Clifford group $\mathcal{C}_1$ -- generated by $\{P_{\pi/4},\mathrm{H}\}$ -- acting on  an  initial pure  state chosen from one of the eigenstates of the logical Pauli observables $\{X,Y,Z\}$,  is the convex hull whose extremal states are said eigenstates. This convex space is an octahedron in the Bloch sphere, with its vertices given by the vectors $\{\phi_0,\phi_1,\phi_\pm,\varphi_\pm \}$ where $\phi_\pm=\frac{1}{\sqrt{2}}(\phi_0\pm \phi_1)$ and $\varphi_\pm =\frac{1}{\sqrt{2}}(\phi_0\pm \imag \phi_1)$. The state space of $N$ qubits is similarly given by the convex  hull whose vertices are given by the eigenstates of $\{X,Y,Z\}^{\otimes N}$. It should be clear that the stabilizer state space $\mathcal{ST}(\co^{2^N})$ is a subset of the full state space $\s(\co^{2^N})$, and that our description of a quantum computer in the stabiliser formalism does not equate with our previous definition of a universal quantum computer. Indeed, we lack the ability to generate the $P_{\pi/8}$ gate, found in the standard universal set, with the one qubit Clifford group $\mathcal{C}_1$. The question remains whether or not a computation in the stabiliser state space lies outside the complexity class $BPP$, and thus provides an advantage to a classical computer. This can be rephrased thusly: can such a computation be \emph{simulated efficiently} on a classical digital computer?  A quantum computer and a  classical computer produce probability distributions over measurement outcomes $p(x)$ and $\tilde p(x)$ respectively. In order for the classical computer to simulate the quantum computer,  the $L_1$ distance between the probability distributions must be  able to be brought below an arbitrarily small positive amount $\epsilon$.
\begin{equation}
L_1[p(x),\tilde p(x)]:=\sum_x\frac{ |p(x)- \tilde p(x)|}{2} \leqslant \epsilon
\end{equation}

   A quantum computer is said to be efficiently classically simulatable if  the number of computational tasks required to bring the $L_1$ distance below $\epsilon$  grows polynomially with an increase in the size of the problem. The Gottesman-Knill theorem \citep{Gottesman-1998} states that, any quantum computer which consists of preparation in the computational basis, unitary gates from the Clifford group, and measurement of observables in the Pauli group, is efficiently classically simulatable. As such, a quantum computer which generates maps within the stabilizer space $\mathcal{ST}(\co^{2^N})$ offers no advantage to a classical digital computer with access to a random number generator. However, as shown by \citep{Kitaev-magic-states}, universal quantum computation can be obtained in the stabiliser formalism by distilling so called \emph{magic states} from many copies of a mixed non-stabiliser state $\rho'$.  Consider the pure states $\Pi(\psi_H)$ and $\Pi(\psi_T)$ 
\begin{align}
&\Pi(\psi_H)=\frac{1}{2}\left(\mathds{1}+\frac{1}{\sqrt{2}}(\sigma_x+\sigma_y)\right) &\Pi(\psi_T)=\frac{1}{2}\left(\mathds{1}+\frac{1}{\sqrt{3}}(\sigma_x+\sigma_y+\sigma_z)\right).
\end{align}
 There are twelve $H$-type magic states given by $\{U\psi_H:U \in \mathcal{C}_1\}$ and eight $T$-type magic states given by $\{U\psi_T:U \in \mathcal{C}_1\}$.  The $H$-type magic states are those that are eigenstates of  Hadamard-like gates which correspond with $180^\circ$ rotations about the edges of the stabiliser octahedron. The $T$-type magic states are eigenstates of $T$-like gates (the $P_{\pi/8}$ gate is also known as a $T$ gate) which correspond with $120^\circ$ rotations about the faces of the stabiliser octahedron.

\section{DiVincenzo's criteria and scalable QIP}\label{DiVincenzo's criteria}

In the early days of quantum computer science, \citep{divincenzo-1998} identified five criteria that any physical implementation of a QIP device -- within the circuit model -- must satisfy to enable universal, robust and fault tolerant quantum computation. These are famously known as DiVincenzo's criteria, and are
\begin{enumerate}[(i)]
\item \emph{Access to a  scalable Hilbert space}

Any system which we hope to use for quantum information processing must have degrees of freedom which are described by the mathematics of Hilbert spaces. To perform arbitrarily large calculations we will need arbitrarily large Hilbert spaces, so the system must be \emph{scalable}. This means that  we can bring several copies of the system and perform interactions between them so as to gain access to a larger Hilbert space. For this to be efficiently scalable, the operations needed for computation must grow logarithmically with the size of the Hilbert space; increasing the number of energy levels within a single system is not efficiently scalable.

\item \emph{Initialisation}

The system must have the capacity of being initialised in a fiducial state, such as the pure product state with vector $\phi_0^{\otimes n}$. For most physical systems this equates with cooling the system to its ground state.  

\item \emph{Universal set of quantum gates}

We must have access to a universal set of quantum gates to achieve controllability in the entire state space. In fault tolerant schemes, magic states are required to turn the Clifford group gates universal.

\item \emph{Long coherence times}

These need to be sufficiently long  in comparison with gate times  to allow for   fault tolerance.

\item \emph{Measurements}

We must be able to perform strong, projective measurements on our qubits. 

\end{enumerate}

\spacing{1}                                  
 \bibliographystyle{plainnat}
 \bibliography{references}

\part{Quantum dynamics of nuclear-electronic spin systems}
\spacing{1}

\chapter{Closed system dynamics }
\section{Introduction}
Nuclear-electronic spin systems consist of coupled electronic and nuclear spin degrees of freedom. The experimentally accessible states of such a system are the eigenstates of the free Hamiltonian, and dynamics can be established between these states using magnetic resonance. In this chapter I aim to delineate the properties of the eigenstates, and the transitions that can be established between them using magnetic resonance. This largely follows from the work done by my colleagues and me in \citep{Mohammady-2010,Mohammady-2012}.  The system shall be treated as a closed quantum system, wherein all dynamics are described by unitary transformations.

\section{The  Hamiltonian and state space}

\begin{figure}[!htb]
\centering
\includegraphics[width=5in]{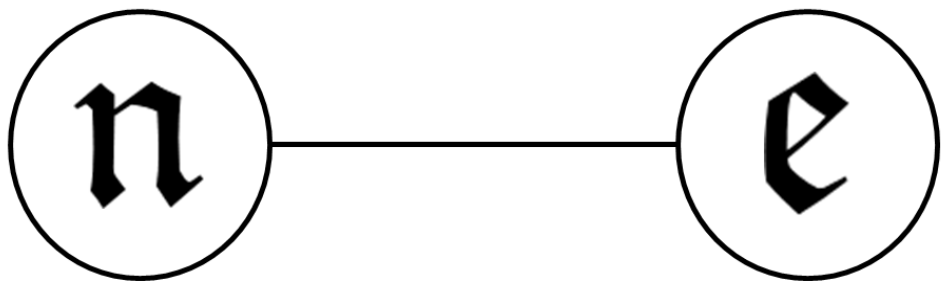} \caption{ The coupled nucleus and electron spin constitute a nuclear-electronic spin system}
\end{figure}

We define a nuclear-electronic spin system as a bipartite composite system consisting of an electronic  degree of freedom and a nuclear degree of freedom, where the total system has zero orbital angular momentum, and is thus described only by the intrinsic spin.  The electron spin operators are given as $\{S_x,S_y,S_z\}$ which act on the Hilbert space $\h^\mathfrak{e}:=\co^{2S+1}$ where $S$ is the total electron spin, and the nuclear spin  operators are $\{I_x,I_y,I_z\}$ and act on $\h^\mathfrak{n}:=\co^{2I+1}$ where $I$ is the total nuclear spin. Consequently, the dimension of the full  Hilbert space $\h^{\mathfrak{e+n}}$ is $d=(2S+1)(2I+1)$. The closed system dynamics is governed by the Breit-Rabi Hamiltonian \citep{Breit-Rabi}
\begin{equation} 
H_0=B_0\left( \gamma_eS_z-\gamma_n I_z\right) + A_{\mathrm{iso}} \sum_{i\in\{x,y,z\}}S_i\otimes I_i
\label{nuclear-electronic spin system hamiltonian} 
\end{equation}    
 where $B_0$ is the static magnetic field in the $z$ direction, $\gamma_e$ and $\gamma_n$ are the electron and nuclear gyromagnetic ratios respectively, and $A_{\mathrm{iso}}$ is the isotropic hyperfine interaction strength.  Such a Hamiltonian can describe  atomic systems \citep{atomic-Breit-Rabi}, endohedral fullerenes \citep{fullerene}, as well as silicon impurities \citep{kane1998} at low enough temperatures that the donor electrons are sufficiently localised and not in the conduction band. We shall restrict our discussion  by setting the electron spin  to $S=1/2$, whilst allowing for the nuclear spin $I$ to  take an arbitrary  integer or half-integer value. While this is sufficient for describing the three major silicon impurity architectures where the dopants used are phosphorus, antimony, and bismuth, it will no longer be able to describe endohedral fullerenes, where $S=3/2$ and $I=1$. Notwithstanding, our approach can be extended to the more general setting, albeit with more cumbersome analytical expressions.

Before continuing with our analysis, let us introduce some  notation. We define the electron Zeeman frequency as $\omega_0=\gamma_e B_0$, and the rescaled Zeeman frequency as $\tilde \omega_0=\omega_0/A_{\mathrm{iso}}$. The ratio of the nuclear to electronic gyromagnetic ratios is defined as $\delta_\gamma=\gamma_n/\gamma_e$ which is very small, usually being in the range $10^{-4}-10^{-3}$. Finally, we denote the eigenvectors of $S_z$ using the Dirac notation as $\{|m_S\>: S_z|m_S\>=m_S|m_S\>\}$ and similarly the eigenvectors of $I_z$ as $\{|m_I\>:I_z|m_I\>=m_I|m_I\>\}$. Because of our restriction on $S$, $m_S$ can take one of only two values $ \{+1/2,-1/2\}$. 

We first observe that the Hamiltonian obeys the following commutation relation
\begin{equation}
[H_0,S_z+I_z]_-=\mathds{O}.
\end{equation} 
Hence, the eigenvectors of the Hamiltonian must be  superpositions of the product eigenvectors of the individual spin operators,  $|m_S\>\otimes|m_I\>$,  such that the total magnetisation quantum number $m=m_S+m_I$ is conserved. The Hamiltonian therefore decomposes into a direct sum of sub-Hamiltonians $H_m$
\begin{equation}
H_0=\bigoplus_m H_m.
\end{equation}
By inspection, we can see that there is only one unique product spin eigenvector $|m_S\>\otimes|m_I\>$ with $m=\pm(I+1/2)$, which is given by $|\pm1/2\>\otimes|\pm I\>$. Therefore, the corresponding sub-Hamiltonian is represented by a $1\times1$ matrix denoted as $H_{m=\pm(I+1/2)}$. For any  $|m| <(I+1/2)$, the eigenvectors are given as a superposition of the product vectors
$|1/2\>\otimes| m-1/2\>$ and $|-1/2\>\otimes| m+1/2\>$
and, consequently, the corresponding sub-Hamiltonians are represented by $2\times 2$ matrices, which we denote as $H_{m,2}$. These can be evaluated easily using the basic rules of angular momentum operators covered in Appendix \ref{angular momentum appendix}. Furthermore, as any matrix in $\lo(\co^2)$ can be spanned by the Pauli matrices $\{\sigma_i\}_{i=0}^3$, we do so for $H_{m,2}$. Here, the Pauli matrices are defined with respect to the basis $\{|1/2\>\otimes|m-1/2\>,|-1/2\>\otimes| m+1/2\>\}$. The sub-Hamiltonians are thus given as 
\begin{align} 
H_{m=\pm(I+1/2)}&= \frac{A_{\mathrm{iso}}}{2}(\pm\mathcal{W}_m-\epsilon_m), \nonumber \\
H_{m,2} &= \frac{A_{\mathrm{iso}}}{2}\left(\mathcal{W}_m{ \sigma}_z  + \mathcal{O}_m{ \sigma}_x   -\epsilon_m\mathds{1}\right), 
\end{align}
where
\begin{align}
 \mathcal{W}_m &= m+{\tilde \omega_0}(1+ \delta_\gamma),\nonumber \\
\mathcal{O}_m &= \sqrt{I(I+1)+\frac{1}{4}-m^2},\nonumber \\
\epsilon_m &= \frac{1}{2}(1+4{\tilde \omega_0}m\delta_\gamma).
\label{sub-Hamiltonians}
\end{align}
 To keep the expressions compact, we define a positive parameter 
$R_m =\sqrt{ \mathcal{W}_m^2 +\mathcal{O}_m^2}$, with the ad hoc specification that  $R_{m=\pm(I+1/2)}$ is given as $\mathcal{W}_{m=\pm(I+1/2)}$, and not the absolute value thereof. Furthermore, the \emph{angle} $\theta_m$ is  defined  such that
\begin{align}
 & \cos (\theta_m):=\frac{\mathcal{W}_m}{R_m} 
 &\sin( \theta_m):=\frac{\mathcal{O}_m}{R_m}.
 \end{align} 
 As $\oo_m$ is a constant for any given $m$ subspace, any variation in these angular quantities is due to the magnetic field $B_0$.  The magnetic field constitutes the reference coordinate frame for our spin system, by the direction of which  the quantities $\{m,m_S,m_I\}$ have meaning. Hence, we may restrict it to take only positive values: $B_0 \geqslant 0$. Due to this consideration, for subspaces where $|m| < (I+1/2)$, the range of values that $\theta_m$ can take are given by
\begin{eqnarray}
\theta_m\in\begin{cases}[0,\arctan\left(\frac{\oo_m}{|m|}\right)] & \text{when} \ m > 0, \\ [0,\frac{\pi}{2}] & \text{when} \ m = 0,\\
[0,\frac{\pi}{2}+\arctan\left(\frac{\oo_m}{|m|}\right)] & \text{when} \ m < 0, \\
\end{cases}
\end{eqnarray}

\noindent where the minimal value occurs as $B_0\to \infty$ and the maximal value is actualised when $B_0=0$. For finite $I$, then, it follows that   $\theta_m < \pi \ \forall \ B_0$. For subspaces where $|m| = (I+1/2)$, on the other hand, since $\oo_{m=\pm(I+1/2)}=0$   it follows that $\theta_{m=\pm(I+1/2)}$ is always zero. 

Going back to the angular representation of the sub-Hamiltonians, as $\cos(\theta_{m=\pm(I+1/2)})=1$ for all magnetic fields, the form of $H_{m=\pm(I+1/2)}$ will be unaltered. This representation, however, will allow us to rewrite
 $H_{m,2}$   as   
\begin{eqnarray} 
H_{m,2}= \frac{A_{\mathrm{iso}}}{2}\left(R_m\cos[ \theta_m] {\sigma}_z +R_{m}\sin [\theta_m]{ \sigma}_x-\epsilon_m\mathds{1}\right.)\label{Hm1} 
\end{eqnarray}
with the eigenvectors
\begin{equation} 
 \phi^\pm_m = a_m\left|\pm1/2\right\>\otimes\left| m\mp1/2\right\> \pm  b_m  \left|\mp1/2\right\>\otimes\left| m\pm1/2\right\>
\label{nuc-elec entangled eigenstate} 
\end{equation} 
\noindent where 
\begin{equation}
a_m = \cos\left(\frac{\theta_m}{2}\right)\equiv \frac{\mathcal{W}_m+R_m}{\sqrt{\mathcal{O}_m^2+(\mathcal{W}_m+R_m)^2}} \ \ , \ \ b_m=\sin\left(\frac{\theta_m}{2}\right)\equiv \frac{\mathcal{O}_m}{\sqrt{\mathcal{O}_m^2+(\mathcal{W}_m+R_m)^2}}.\label{coeffs}
\end{equation}

From this, we see that $\cos(\theta_m)$ and $\sin(\theta_m)$ can be expressed equivalently as 
\begin{align}
&\cos(\theta_m)=a_m^2-b_m^2 &\sin(\theta_m)=2a_mb_m ,
\end{align}
and make the further identification that
\begin{equation}
\<\phi^\pm_m|S_z\phi^\pm_m\>=\pm \frac{1}{2}\cos(\theta_m)
\label{costheta expectationvalue}\end{equation}
and 
\begin{equation}
\<\phi^\mp_m|S_z\phi^\pm_m\>=- \frac{1}{2}\sin(\theta_m).
\label{sintheta expectationvalue}\end{equation}
 This representation of the eigenstates facilitates an understanding of the different magnetic field regimes of the system.
\begin{enumerate}[(i)]
\item
We define the \emph{high-field regime} when the condition  $B_0(\gamma_e+\gamma_n)/A_{\mathrm{iso}}\gg 1 $ is satisfied. 
In this regime $\theta_m\to0$, and hence $a_m\to 1$ and $b_m\to 0$ for all $m$;  the electron and nuclear spins are said to \emph{decouple}. 
\item
 We define the \emph{low-field regime} when the condition   $B_0(\gamma_e+\gamma_n)/A_{\mathrm{iso}} \lesssim 1$  is satisfied. Here, an appreciable superposition of the product spin  eigenvectors is established; the electron and nuclear spins are entangled. 
\end{enumerate}
The eigenstate representation of Eq.\eqref{nuc-elec entangled eigenstate}, henceforth referred to as the \emph{adiabatic basis}, may also be used to describe the eigenvectors of $H_{m=\pm(I+1/2)}$. This is done by the identification $|\pm1/2\>\otimes|\pm I\>\equiv\phi^\pm_{ \pm (I+1/2)}$, noting that  $b_{m=\pm(I+1/2)}=0$ for all magnetic fields, and that the vectors $\phi^\pm_{\mp(I+1/2)}$ do not represent eigenstates of the system. Furthermore, in addition to $\{\phi^\pm_m\}$, we may also represent the eigenbasis of the Hamiltonian as $\{\varphi_i: i \in \mathds{N}, 1\leqslant i \leqslant 2(2I+1)\}$ such that $\varphi_1$ is the ground state, and $\varphi_{2(2I+1)}$ the maximally excited state.

As for the energies, we determine those belonging to  the eigenstates of $H_{m,2}$ as
\begin{equation} 
E^{\pm}_m = \frac{A_\mathrm{iso}}{2}\left[-\frac{1}{2}(1+4{\tilde \omega_0}m\delta_\gamma) \pm R_m \right] 
\label{H2 energies} 
\end{equation}

while the energies of the eigenvectors $\phi^\pm_{\pm(I+1/2)}$ can be given  more simply as 
\begin{equation}
E_{m=\pm(I +1/2)} =\frac{A_\mathrm{iso}}{2}\left[-\frac{1}{2}(1\pm4{\tilde \omega_0}m\delta_\gamma) \pm \mathcal{W}_m \right] \equiv\pm\frac{\omega_0}{2}(1 - 2\delta_\gamma I)+\frac{A_{\mathrm{iso}}I}{2}.\label{H1 energies}
\end{equation}

\subsection{Energy ordering phases} 
\begin{figure}[!htb]
\centering
\includegraphics[width=3in]{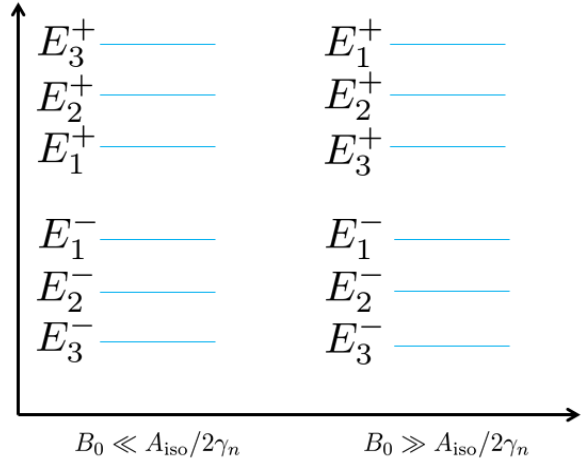} \caption{ Pictorial representation of the energy ordering phases. At all fields, we have $E^+_m > E^-_{m'}$ and $E^-_m < E^-_{m-1}$. However, the ordering between $E^+_m$ and $E^+_{m-1}$ reverses at $B_0 \simeq A_{\mathrm{iso}}/2\gamma_n$.  } \label{energyordering}
\end{figure}

We define the term \emph{energy ordering phase} as a regime of the system defined with respect to the ordering of energies $E_m^\pm$. At all non-zero magnetic fields, the energies obey the ordering
\begin{align}
E_m^+ &> E_{m'}^- \ \ \forall \ \ \{m,m'\},\nonumber \\
E_m^- &<E_{m-1}^- \ \ \forall \ \ m.
\end{align}
However, there are two phases in which the  ordering of the energies $E_m^+$ and $E_{m-1}^+$ is reversed. At the magnetic fields $B_0 \lesssim A_{\mathrm{iso}}/2\gamma_n$, these energies obey the ordering
\begin{align}
E_m^+ &>E_{m-1}^+ \ \ \forall \ \ m
\end{align}
which is  opposite to the ordering of energies $E_m^-$ and $E_{m-1}^-$. At magnetic fields $B_0\gtrsim A_{\mathrm{iso}}/2\gamma_n$, however, the ordering reverses to
\begin{align}
E_m^+ &<E_{m-1}^+ \ \ \forall \ \ m
\end{align}
which is the same as the ordering of the energies $E_m^-$ and $E_{m-1}^-$. Consequently, we may define the magnetic field region $B_0\simeq A_{\mathrm{iso}}/2\gamma_n$ as a transition point between the two phases of the energy ordering. Note that the system will be in the high-field regime at magnetic fields much smaller than this energy ordering phase transition. 
\subsection{Cancellation resonances}
We introduce the term \emph{cancellation resonance} to describe the  magnetic field regimes that  simplify the sub-Hamiltonians. There are two types of cancellation resonance:

\begin{description}
\item[Type I] : Takes place when
 $\mathcal{W}_m=0$
 
This can only occur for subspaces where $m \leqslant 0$. At these field values, and for subspaces with $-I+1/2 \leqslant m\leqslant 0$, the  term in $H_{m,2}$ dependent on $\sigma_z $  vanishes entirely, such that $\theta_m=\pi/2$.
For the subspace $m=-(I+1/2)$, on the other hand, the sub-Hamiltonian simplifies to the magnetic field-independent value  
\begin{equation}
-A_{\mathrm{iso}}\left(\frac{1}{4}+\frac{\delta_\gamma}{1+\delta_\gamma}(I^2+I+\frac{1}{4})\right)  
\end{equation}
\item[Type II]: Takes place when $\mathcal{W}_m=\mathcal{O}_m$

This only affects the sub-Hamiltonians $H_{m,2}$, but is no longer restricted to negative $m$ subspaces. Here, the sub-Hamiltonian becomes proportional to $\sigma_x+\sigma_z$, meaning that $\theta_m=\pi/4$.

 \end{description}

\subsection{Avoided crossings}
The stationary points of the energy are defined as the magnetic field values where   $d E/d B_0=0$.  For states $\phi^\pm_{\pm(I+1/2)}$ the derivative of  the energy with respect to $B_0$ is given by 
\begin{equation}
\frac{dE_{m=\pm(I+\frac{1}{2})}}{dB_0}=\pm\frac{\gamma_e}{2}(1 - 2\delta_\gamma I)
\end{equation}
which is constant for all magnetic fields and never vanishes. For all other states, this is calculated to be
\begin{equation}
\frac{dE^\pm_m}{dB_0}=-m\gamma_e\delta_\gamma \pm\frac{\gamma_e}{2}\cos(\theta_m)(1+\delta_\gamma)
\end{equation} 
and disappears when 
\begin{equation}
\cos(\theta_m)=\pm\frac{2m\delta_\gamma}{1+\delta_\gamma}.\label{stationary energy}
\end{equation}
 Since $\delta_\gamma \ll1$,  it follows that only the energies of subspaces with $-I+1/2\leqslant m \leqslant 0$ can be made stationary, as $\cos(\theta_m)$ for all other subspaces can never be made small, or even worse, negative. Other than the subspace $m=0$, the two energies $E_m^+$ and $E_m^-$ are not made stationary at the same magnetic field, as is the case  for traditional Landau-Zener avoided crossings \citep{Landau-1932, Zener-1932}. Rather, these occur at different  field values that are equally close to the  type I cancelation resonance. We may equivalently express this as the system being an equal angular distance from $\pi/2$; $E^+_m$ is made stationary when $\theta_m=\pi/2 +\zeta_m$ and $E_m^-$ is made stationary when $\theta_m = \pi/2 - \zeta_m$, where
 \begin{equation}\zeta_m= \arccos\left(\frac{2|m| \delta_\gamma}{1+\delta_\gamma}\right).
 \end{equation} 
\subsection{Entanglement}\label{nuc-elec entanglement}

  All  eigenvectors of $H_0$, other than  $\phi^\pm_{\pm(I+1/2)}$, are entangled with respect to the nuclear and electronic spin Hilbert spaces, except in the asymptotic limit of $(\gamma_e+\gamma_n)B_0/A_{\mathrm{iso}}\to \infty$. The reduced density operators of any eigenstate $\Pi(\phi^\pm_m)\equiv |\phi^\pm_m\>\<\phi^\pm_m|$ where $|m| <(I+1/2)$ are
\begin{align}
\mathrm{tr}_\mathfrak{n}[\Pi(\phi^\pm_m)]&=a_m^2\Pi_{{\pm1/2}}+b_m^2 \Pi_{{\mp 1/2}}, \nonumber \\
\mathrm{tr}_\mathfrak{e}[\Pi(\phi^\pm_m)]&=a_m^2\Pi_{{m\mp1/2}}+b_m^2 \Pi_{{m\pm 1/2}},\label{reduced states nuc-elec}
\end{align} 
with $\Pi_{n}\equiv |n\>\<n|$. From this we can calculate the entropy of entanglement, discussed in Appendix \ref{von Neumann entropy}, as
\begin{equation}
E[\Pi(\phi^\pm_m)]=-a_m^2\log_2(a_m^2)-b_m^2\log_2(b_m^2).\label{nuc-elec entropy of entanglement}
\end{equation}
The eigenvectors $\phi^\pm_m$ are the maximally entangled Bell states 
\begin{equation}
\Psi^\pm:=\frac{1}{\sqrt{2}}\left(\phi_0\otimes\phi_1\pm\phi_1\otimes\phi_0\right)
\end{equation}  when $a_m^2=b_m^2=1/2$, which occurs when $\theta_m=\pi/2$. This is satisfied at the type I cancelation resonance, meaning that only subspaces with $-I+1/2 \leqslant m\leqslant0$ have maximally entangled eigenstates.

For a statistical mixture, the entropy of entanglement no longer satisfies the requirements of an entanglement measure. If $\h^\mathfrak{n}=\co^2$ then we may use the concurrence, described in Appendix \ref{concurrence}. For $\h^\mathfrak{n}$ of larger dimensions we can use the negativity, described in Appendix \ref{negativity}.  In this case care must be taken in the interpretation of results; for $\h^\mathfrak{n}=\co^{d>3}$ a negativity greater than zero shows the presence of entanglement but a zero negativity does not show its absence.  

In addition to the mathematical criteria for entanglement to be present, care must be taken in the interpretation of this ``entanglement'' in operational terms. We are only justified in speaking of entanglement between the components of a composite system if we can speak of the system as being composite in the first place; a Hilbert space can be factorised as $\h^A\otimes\h^B$ if we can perform  local measurements on  $\h^A$ and/or $\h^B$. There are, consequently, two cases in which we may speak of entanglement:
\begin{enumerate}[(i)]
\item Entanglement in the composite system $\h^A\otimes\h^B$ may be defined with respect to local operations and classical communication  (LOCC) if we are able to perform local  measurements on each subspace, supplemented with classical communication. 
\item If we are able to perform measurements on just a single subspace, say $\h^A$, then we may treat it as an open quantum system. If we can assume that the composite system is pure, and the measurement statistics of system $\h^A$ reveals a statistical mixture, then we may infer that the composite system is entangled. This is the justification behind the traditional explanation of decoherence, as well as of the von Neumann-L\"uders measurement model.
\end{enumerate}

 In the current context, then, the eigenstates that are mathematically shown to be entangled with respect to the electron and nuclear spin Hilbert spaces are \emph{operationally entangled} only if we can locally measure  either the electron or nuclear spin, or both. If we can measure both of the systems locally, then we may use the entanglement as a resource to, say, teleport quantum information. 

In the prime examples of nuclear-electronic spin systems -- silicon impurities --    only local measurements on the electronic degree of freedom has been reported to date. Therefore the entanglement cannot, for now at least, be operationally defined with respect to LOCC.  This notwithstanding,  the entanglement of the eigenstates is not fictitious and ``real'', so long as it results in deducible phenomenon that can be falsified experimentally. I will show, in following sections, that the entanglement of the eigenstates does indeed result in experimentally verifiable effects; that of magnetic resonance transition rates and decoherence. This results from the Hamiltonian that we posit for the system, in which the environment acts locally on the electron and nuclear spin Hilbert spaces. 

Indeed, the magnetic resonance community agrees with the reality of such an entanglement, albeit by a different name of \emph{mixing}. The eigenstates shown in Eq.\eqref{nuc-elec entangled eigenstate}  are said to be ``mixed'' if they contain a ``mixture'' of the high-field eigenstates $|m_S\>\otimes |m_I\>$. In other words, the state is ``mixed'' if both $a_m$ and $b_m$ are non-zero, as it were. However, this is tantamount to saying that the eigenstates have a Schmidt-rank of two: that they are entangled. There is, therefore, a one-to-one correspondence between entanglement and  ``mixing'', and speaking of one necessarily implies the other.

\section{Nuclear-electronic magnetic resonance}
\subsection{Selection rules}\label{selection rules and spectra}
 In magnetic resonance experiments, a time-dependent  Hamiltonian -- due to the interaction of the spin with an oscillating  magnetic field, also referred to as the driving field--  is switched on  so as to allow for the establishment of superpositions between the eigenstates of the Breit-Rabi Hamiltonian in Eq.\eqref{nuclear-electronic spin system hamiltonian}. In the high-field regime, where the stationary states of the nuclear-electronic spin system are decoupled product states, superpositions between the eigenstates $\{\varphi_i,\varphi_j\}$ are possible by NMR if the transition is allowed by  the dynamical selection rule for $I_x$: if  $|\<\varphi_j|I_x\varphi_i\>|>0$ or, equivalently,    $\Delta m_I =\pm 1$ and $\Delta m_S=0$.  For ESR the transition must be allowed by the dynamical selection rule for $S_x$:  $|\<\varphi_j|S_x\varphi_i\>|>0$  or, equivalently, $\Delta m_S =\pm 1$ and $ \Delta m_I=0$. In both cases we can talk about changes in each individual spin  quantum number  because the electron and nuclear systems are in a pure product state, where each subsystem is itself a pure state with a deterministic outcome for a spin measurement; in other words, they are \emph{good quantum numbers}.

In the low-field regime, however,  with the exception of $\phi^\pm_{\pm( I+1/2)}$,  the eigenstates are entangled.  In general the dynamical selection rules cannot be categorised simply as ESR or NMR. Indeed, to avoid contradictions the magnetic resonance phenomenon  must  be called by a different name, where ESR and NMR are only valid descriptions  asymptotically  in the high-field limit. As the relevant selection rule allows transitions between eigenstates that obey $|\<\varphi_j|(S_x+I_x)\varphi_i\>|>0$, whereby $\Delta m=\pm1$, it is appropriate to call this nuclear-electronic magnetic resonance (NEMR). To see why this is the case, consider the reduced density operators for an eigenstate of the nuclear-electronic spin system in the low-field regime, $\mathrm{tr}_\mathfrak{n}[\Pi(\phi^\pm_m)]$ and $\mathrm{tr}_\mathfrak{e}[\Pi(\phi^\pm_m)]$,  provided by Eq.\eqref{reduced states nuc-elec}. These are, in general, convex combinations of pure states, and we  cannot say how the individual spin quantum numbers  change due to a transition between different eigenstates $\phi^\pm_m$, as there will be many contradictory answers depending on which component of the mixed states we consider. The amount by which the total spin magnetisation $m$ changes, however, is always well defined.  

  These statements notwithstanding, for all practical purposes we may qualitatively express the NEMR transition rates  in the low-field regime by ESR. This is because these transition rates are proportional to $|\<\varphi_j|S_x\varphi_i\>+\delta_\gamma\<\varphi_j|I_x\varphi_i\>|^2$ which can be approximated as $|\<\varphi_j|S_x\varphi_i\>|^2$ owing to the fact that $\delta_\gamma \ll1$. The nuclear spin effect on the Rabi frequency in pulsed spectroscopy is similarly negligible.

\subsection{NEMR Transitions}

We now look at all the magnetic resonance transitions permissible in nuclear-electronic spin systems, together with their relative c.w. transition  rates and frequencies.  The discussion shall be general so as to be able to describe the observable phenomenon, both in the high and low-field regimes, in a unified fashion. Recall the driving field Hamiltonian for a single spin species, given in Eq.\eqref{magnetic resonance driving field}. By extension, the spin operators $J_{x/y}$, describing a single spin species,  will be replaced by $F_{x/y}=S_{x/y}+\delta_\gamma I_{x/y}$ to give the  driving field 
\begin{equation}
H_I^\pm(t)=\omega_1f(t)\left(\cos[\omega t]F_x \pm \sin[\omega t]F_y\right))
\end{equation}
which interacts with both the electron and nuclear spins \emph{locally}. Here,  $\omega_1 = \gamma_e B_1$ is the strength of the driving field and  $f(t)$ is a function describing a  pulse of duration $\tau$.    Hence, following the arguments laid out in Sec.\ref{c.w. spectroscopy}, the transition rate between two eigenstates $\{\varphi_i,\varphi_j\}$ may be calculated as  $\mathcal{I}=2\pi|\omega_1\<\varphi_j|F_x \varphi_i\>|^2$. \footnote{It should be noted that we shall only consider   transitions allowed up to first-order perturbation theory, which obey the dynamical selection rule for $F_x$, and will ignore those allowed by second-order perturbation theory, namely those that obey the dynamical selection rule for $F_x^2$. These second-order transitions are reported by \citep{morishita-2009} using c.w. spectroscopy, but have very weak transition rates.} Furthermore, in a given magnetic field regime, the transitions    may  be classified as either
\begin{enumerate}[(a)]
\item ESR-allowed

when $|\<\varphi_j|S_x \varphi_i\>| \sim 1$.

\item ESR-forbidden but NMR-allowed

when $|\<\varphi_j|S_x \varphi_i\>| \ll 1$ but $|\<\varphi_j|I_x \varphi_i\>| \sim 1$.  
\item Dipole forbidden

when $|\<\varphi_j|S_x \varphi_i\>| \ll1$ and $ |\<\varphi_j|I_x \varphi_i\>|  \ll 1$.
 \end{enumerate}
There are four types of dipole allowed transitions, classified as either 
 $\phi^\pm_m \leftrightarrow \phi^\pm_{m-1}$ or  $\phi^\pm_m \leftrightarrow \phi^\mp_{m-1}$. Here, I shall expound upon the properties of each. In all the ensuing expressions for the transition rates, the common  factor of $\pi \omega_1^2/2$ will be omitted, and
we denote
$
C_{m_I}^{I-}=\sqrt{I(I+1)-m_I(m_I-1)}
$
.

\begin{enumerate}[(i)]
\item $\phi^\pm_m\leftrightarrow \phi^\pm_{m-1}$

There are  $2I$ of each of these  transitions. The transitions $\phi^+_m\leftrightarrow \phi^+_{m-1}$ have the associated transition rate   
\begin{align} 
\mathcal{I}^{+}_{m \leftrightarrow m-1}&\propto\left|a_mb_{m-1} +\delta_\gamma\left(C_{m-\frac{1}{2}}^{I-}a_ma_{m-1}+ C_{m+\frac{1}{2}}^{I-}b_mb_{m-1}\right) \right|^2,\nonumber \\
&\simeq \cos^2\left(\frac{\theta_{m}}{2}\right)\sin^2\left(\frac{\theta_{m-1}}{2}\right) \ \text{in the low-field regime},
\label{Iforbp} 
\end{align} 
 and the transitions  $\phi^-_m\leftrightarrow \phi^-_{m-1}$ have the  rate\begin{align} 
\mathcal{I}^{-}_{m \leftrightarrow m-1 } &\propto\left|-a_{m-1}b_{m} +\delta_\gamma\left(C_{m+\frac{1}{2}}^{I-}a_ma_{m-1}+ C_{m-\frac{1}{2}}^{I-}b_mb_{m-1}\right) \right|^2,\nonumber \\
&\simeq \cos^2\left(\frac{\theta_{m-1}}{2}\right)\sin^2\left(\frac{\theta_{m}}{2}\right) \ \text{in the low-field regime}. \label{Iforbm} 
\end{align}
Both these transitions are ESR-forbidden but NMR-allowed in the high-field regime. In the low-field regime, however, they become ESR-allowed.

\item
  $\phi^+_m\leftrightarrow \phi^-_{m-1}$ 
  
There are $2I+1$ of such transitions.  These  have the  rate \begin{align} 
\mathcal{I}^{+ \leftrightarrow -}_{m \leftrightarrow m-1} &\propto\left|a_{m}a_{m-1} +\delta_\gamma\left(-C_{m-\frac{1}{2}}^{I-}a_mb_{m-1}+ C_{m+\frac{1}{2}}^{I-}b_ma_{m-1}\right) \right|^2,\nonumber \\
&\simeq \cos^2\left(\frac{\theta_{m}}{2}\right)\cos^2\left(\frac{\theta_{m-1}}{2}\right) \ \text{at all magnetic fields}, \label{Idip} 
\end{align}
and are ESR-allowed at all field values.
\item $\phi^-_m\leftrightarrow \phi^+_{m-1}$

There are $2I-1$ of such transitions. Hence, unlike the other transition types, these can only be observed for systems with $I \geqslant 1$. These  have the rate\begin{align}
\mathcal{I}^{- \leftrightarrow +}_{m \leftrightarrow m-1 } &\propto\left|-b_{m}b_{m-1} +\delta_\gamma\left(C_{m+\frac{1}{2}}^{I-}a_mb_{m-1}- C_{m-\frac{1}{2}}^{I-}b_ma_{m-1}\right) \right|^2,\nonumber \\
&\simeq \sin^2\left(\frac{\theta_{m}}{2}\right)\sin^2\left(\frac{\theta_{m-1}}{2}\right) \ \text{at all magnetic fields}. \label{Iforbidden}
\end{align}Such transitions are ESR-allowed in the low-field regime, but dipole forbidden at high fields. Furthermore, unlike the other three transition types,
the uncoupled eigenstates $\phi^\pm_{\pm(I + \frac{1}{2})}$ are never involved here. \end{enumerate}

Here we see the first of the  experimentally verifiable effects of entanglement in the eigenstates of the free Hamiltonian $H_0$ that was alluded to in Sec.\ref{nuc-elec entanglement}: we predict that transitions that are either dipole forbidden or only NMR-allowed at high fields, where the eigenstates are separable, have transition rates in the low-field regime that are comparable with those considered as ESR.

Each of the dipole-allowed transitions have associated with them a magnetic field dependent frequency $\Omega$, determined by the difference in energies of the associated eigenstates, and can most generally be given as
\begin{align}
\Omega^{\pm,m \leftrightarrow \pm,m-1}=\frac{A_{\mathrm{iso}}}{2}\left|R_m-R_{m-1}   \mp2\tilde \omega_0 \delta_\gamma\right|, \nonumber \\
  \Omega^{\pm,m \leftrightarrow\mp,m-1}=\frac{A_{\mathrm{iso}}}{2}\left|R_m+R_{m-1}   \mp2\tilde \omega_0 \delta_\gamma\right| .\label{transition frequencies}
\end{align}
  
\subsection{Frequency stationary points}
The transition frequencies \eqref{transition frequencies} for any pair of eigensates 
\begin{equation}
\{\varphi_i,\varphi_j:\<\varphi_i|H_0\varphi_i\>>\<\varphi_j|H_0\varphi_j\>, |\<\varphi_i|F_x\varphi_j\>|>0\}
\end{equation}
 can also be  represented as
\begin{equation}
\Omega :=\<\varphi_i|H_0\varphi_i\>-\<\varphi_j|H_0\varphi_j\>.
\end{equation}
As the Hamiltonian is a function of $B_0$, we may differentiate it with respect to this parameter to determine
\begin{equation}
\frac{dH_0(B_0)}{dB_0}:=\lim_{\Delta B_0 \to 0}\frac{H_0(B_0+\Delta B_0)-H_0(B_0)}{\Delta B_0}=\gamma_eS_z-\gamma_n I_z.
\end{equation}
Consequently, as 
\begin{equation}
\frac{d\<\varphi_i|H_0\varphi_i\>}{dB_0}=\<\varphi_i|\frac{dH_0}{dB_0}\varphi_i\>,
\end{equation} the derivative of frequency with respect to the magnetic field can be easily shown to be
\begin{equation}
\frac{d\Omega}{dB_0}= \<\varphi_i|(\gamma_eS_z-\gamma_n I_z)\varphi_i\>-\<\varphi_j|(\gamma_eS_z-\gamma_n I_z)\varphi_j\>.
\end{equation}
The \emph{frequency stationary points} (FSPs) of the spectra are defined as the magnetic field values where this derivative vanishes. Namely, when \begin{equation}
\<\varphi_i|(\gamma_eS_z-\gamma_n I_z)\varphi_i\>=\<\varphi_j|(\gamma_eS_z-\gamma_n I_z)\varphi_j\>.
\end{equation}
Provided a nuclear spin satisfying $I > 1$, nuclear-electronic spin systems will have FSPs in the low-field regime, which are frequency minima for transitions of type $\phi^\pm_m\leftrightarrow \phi^\mp_{m-1}$, and frequency maxima for transitions of type  $\phi^\pm_m\leftrightarrow \phi^\pm_{m-1}$, if  $-I+3/2 \leqslant m \leqslant 0$. In what follows, I shall provide proofs for this claim, and show the conditions under which the FSPs are obtained.

\begin{description}
\item[Transitions of type $\phi^\pm_m\leftrightarrow \phi^\mp_{m-1}$] 
\

 Firstly, we note that the transitions  $ \phi^\pm_{\pm(I + 1/2)} \leftrightarrow\phi^\mp_{\pm I \mp 1/2}$ have the frequency 
\begin{equation}
\Omega=\frac{1}{2}\left( \mp2  \omega_0 \delta_\gamma +A_{\mathrm{iso}}\left[\mathcal{W}_{m={\pm(I+1/2)}}+R_{m={\pm I\mp1/2}}\right]\right). 
\end{equation}
Here, the gradient of $\Omega$ with respect to $B_0$ is given as 
\begin{equation}
\frac{d\Omega}{dB_0}=\frac{\gamma_e(1+ \delta_\gamma)}{2}\cos(\theta_{m=\pm I\mp1/2})+\frac{\gamma_e}{2}(1+\delta_\gamma \mp2\delta_\gamma)
\end{equation}
and vanishes only if 
\begin{equation}
\cos(\theta_{m=\pm I\mp1/2})=-\frac{1+\delta_\gamma \mp2\delta_\gamma}{1+\delta_\gamma}\simeq-1
\end{equation}
which cannot be satisfied since $\theta_{m=\pm I\mp1/2}<\pi \ \forall \ B_0$.
 For the rest of the transitions, where $-I+3/2\leqslant m \leqslant I-1/2$, the frequency is given by \begin{equation}
 \Omega=\frac{1}{2}\left(\mp2\omega_0\delta_\gamma+A_{\mathrm{iso}}[R_m+R_{m-1}] \right). 
 \end{equation} 
 In this case, the gradient of the frequency with respect to $B_0$ is calculated as 
 \begin{equation}
\frac{d\Omega}{dB_0}=\mp\gamma_e\delta_\gamma+\frac{\gamma_e(1+ \delta_\gamma)}{2}(\cos [\theta_m ]+\cos [\theta_{m-1}])\label{frequency gradient for minima condition}
\end{equation}
and vanishes when\begin{equation}
\cos(\theta_m)+\cos(\theta_{m-1})=\pm\frac{2\delta_\gamma}{1+\delta_\gamma}.\label{freq minima condition}
\end{equation}
Determining the  values of $B_0$ that satisfy the exact  Eq.\eqref{freq minima condition} requires solving a  high-order polynomial, the analytic solution to which will not be very instructive. However, taking into account that $\delta_\gamma$ is small, we may determine $B_0$ approximately by taking the limit of $\delta_\gamma \to 0$ to obtain
\begin{equation}
\lim_{\delta_\gamma \to 0}B_{0}^{\mathrm{FSP}}= -\frac{A_{\mathrm{iso}}}{\gamma_e}\frac{(m-1)\oo_m   + m\oo_{m-1}  }{\oo_{m-1} + \oo_m}.
\end{equation} 
 This gives a positive, and therefore valid solution, only if $ -I+3/2\leqslant m \leqslant 0$. To show that this is a minimum we simply calculate
\begin{equation}
\lim_{\delta_\gamma\to 0}\frac{d^2\Omega}{dB_0^2}\bigg|_{B_0^{\mathrm{FSP}}}=\frac{R_m+R_{m-1}}{R_mR_{m-1}}\left(1 - \cos^2[\theta_m] \right) 
\end{equation}
which is positive.

\item[Transitions of type $\phi^\pm_m\leftrightarrow \phi^\pm_{m-1}$] 
\

  Firstly, we note that the transitions  $ \phi^\pm_{\pm(I + 1/2)}\leftrightarrow\phi^\pm_{\pm I \mp 1/2} $ have a frequency 
\begin{equation}
\Omega=\frac{1}{2}\left( \mp2  \omega_0 \delta_\gamma +A_{\mathrm{iso}}\left[\pm\mathcal{W}_{m={\pm(I+1/2)}}\mp R_{m=\pm I \mp 1/2 }\right]\right)
\end{equation}
with the gradient
\begin{equation}
\frac{d\Omega}{dB_0}=\mp\frac{\gamma_e(1+ \delta_\gamma)}{2}\cos(\theta_{\pm I\mp1/2})\pm\frac{\gamma_e}{2}(1-\delta_\gamma ).
\end{equation}
Here, $d\Omega/dB_0=0$ if 
\begin{equation}
\cos(\theta_{m=\pm I \mp 1/2})=\frac{1-\delta_\gamma}{1+\delta_\gamma}
\end{equation}
which, although cannot be satisfied in the low-field regime, is still actualised  at finite magnetic fields and satisfies the criterion for FSPs.
 The magnetic field value which satisfies this condition is calculated as
\begin{equation}
B_0^{\mathrm{FSP}}=A_{\mathrm{iso}}\frac{-2\delta_\gamma(1+\delta_\gamma)m+\oo_m\sqrt{\delta_\gamma(\delta_\gamma^2-1)^2}}{2\delta_\gamma(1+\delta_\gamma^2)\gamma_e}.
\end{equation} 
To determine whether the transition is a minimum or maximum, we  evaluate
 \begin{equation}
\frac{d^2\Omega}{dB_0^2}\bigg|_{B_0^{\mathrm{FSP}}}=\mp\frac{\frac{\gamma_e^2}{2}(1+\delta_\gamma)^2}{R_{m=\pm I \mp 1/2}}\left(1-\left[\frac{1-\delta_\gamma}{1+\delta_\gamma}\right]^2\right) 
\end{equation}
 which implies that the FSP for transition $ \phi^+_{I + 1/2} \leftrightarrow\phi^+_{I - 1/2} $ is a frequency maximum, whilst that for the transition $ \phi^-_{-I - 1/2} \leftrightarrow\phi^-_{- I + 1/2} $ is a frequency minimum.

 For the remainder of the transitions, where $-I+3/2\leqslant m \leqslant I-1/2$, the frequency is given by 
\begin{equation}
\Omega=\frac{1}{2}\left(\mp2\omega_0\delta_\gamma+A_{\mathrm{iso}}(R_m-R_{m-1}) \right).
 \end{equation} 
 In this case, the gradient of the frequency with respect to $B_0$ is calculated as 
\begin{equation}
\frac{d\Omega}{dB_0}=\mp\gamma_e\delta_\gamma+\frac{\gamma_e(1+\delta_\gamma)}{2}(\cos [\theta_m ]-\cos [\theta_{m-1}])\label{frequency gradient for maxima condition}
\end{equation}
which vanishes when the condition 
  \begin{equation}
\cos \theta_m -\cos \theta_{m-1}=\pm\frac{2\delta_\gamma}{1+\delta_\gamma}\label{freq maxima condition}
\end{equation}  is met.  The values of $B_0$ that satisfy Eq.\eqref{freq maxima condition} are, in the limit $\delta_\gamma \to 0$, given by
\begin{equation}
\lim_{\delta_\gamma \to 0}B_0^{\mathrm{FSP}}=\frac{A_{\mathrm{iso}}}{\gamma_e}\frac{(m-1)\oo_m   - m\oo_{m-1}  }{\oo_{m-1} - \oo_m} 
\end{equation} 
which give positive, and therefore valid, solutions  only   if $-I+3/2\leqslant m \leqslant 0$.
 To show that this is a maximum we simply calculate
\begin{equation}
\lim_{\delta_\gamma\to 0}\frac{d^2\Omega}{dB_0^2}\bigg|_{\mathrm{FSP}}=\frac{R_{m-1}-R_m}{R_mR_{m-1}} (1-\cos^2[\theta_m]) 
\end{equation}
which is negative.

\end{description}

\subsection{Differences in transition frequency}\label{transition frequencies section}

  Another  quantity to consider is the \emph{difference} between transition frequencies, $\Delta \Omega,$ at a given magnetic field value. This is important when considering issues of control, where   the presence of degeneracies in transition frequency must be taken into account. Also, even in the absence of degeneracies,  to determine the strength of  a driving field needed for the rotating wave approximation to be valid, we need to know the smallest relevant $\Delta \Omega$ in the system. 

First, let us consider the frequency difference between transitions of the same type. For  arbitrary $m$ and $m'$, these are evaluated as   
  \begin{align}
|\Omega^{\pm,m \leftrightarrow \pm,m-1} - \Omega^{\pm,m' \leftrightarrow \pm,m'-1}|&=\frac{A_{\mathrm{iso}}}{2}\left|(R_{m}-R_{m-1})-(R_{m'}-R_{m'-1})\right| \label{same time freqdiff small}  \end{align}
  for transitions of type $\phi^\pm_m\leftrightarrow \phi^\pm_{m-1}$ and
  \begin{align}
  |\Omega^{\pm,m \leftrightarrow \mp,m-1} - \Omega^{\pm,m' \leftrightarrow \mp,m'-1}|&=\frac{A_{\mathrm{iso}}}{2}\left|(R_{m}+R_{m-1})-(R_{m'}+R_{m'-1})\right| \label{same time freqdiff large}\end{align}
for transitions of type $\phi^\pm_m\leftrightarrow \phi^\mp_{m-1}$. With the exception of one of the transitions being $\phi^\pm_{-I+1/2}\leftrightarrow\phi^-_{-I-1/2}$, $\Delta \Omega$ for both of these cases vanishes when $B_0=0$. Additionally, $\Delta \Omega$ given by Eq.\eqref{same time freqdiff small}  is always smaller than that given by Eq.\eqref{same time freqdiff large} at any  field value greater than zero. Furthermore, $\Delta \Omega$ for both cases  becomes stationary when 
\begin{equation}
\cos(\theta_m)\pm \cos(\theta_{m-1})=\cos(\theta_{m'})\pm \cos(\theta_{m'-1})
\end{equation} which can occur both in the  low-field and high-field regimes. In the high-field regime, this shows that the transition frequency differences stabilise to a given value. To determine this value, we note that 
\begin{equation}
\lim_{(\gamma_e+\gamma_n)B_0/A_{\mathrm{iso}}\to \infty}(R_m\pm R_{m-1})=m\pm (m-1).
\end{equation}
 Hence, in the high-field regime, $\Delta \Omega$ for transitions of type $\phi^\pm_m\leftrightarrow \phi^\mp_{m-1}$ stabilise to $A_{\mathrm{iso}}|m-m'|$, whereas   $\Delta \Omega$ for transitions of type $\phi^\pm_m\leftrightarrow \phi^\pm_{m-1}$   become smaller with $B_0$ and vanish altogether as $(\gamma_e+\gamma_n)B_0/A_{\mathrm{iso}}\to \infty $. The latter, however, does maximise to an appreciable value of order $A_{\mathrm{iso}}$ in the low-field regime.

Now let us consider frequency differences given two different transition types. The first way we do this is the following where $\Delta \Omega$,  given arbitrary $m$ and $m'$,  can be given as
\begin{equation}
\left|\Omega^{\pm,m \leftrightarrow \pm,m-1} - \Omega^{\pm,m' \leftrightarrow \mp,m'-1}\right|=\frac{A_{\mathrm{iso}}}{2}\left|R_m-(R_{m-1}  + R_{m'} + R_{m'-1})    \right|\label{parity +1 frequency difference large}
  \end{equation}
and
\begin{equation}
\left|\Omega^{\pm,m \leftrightarrow \pm,m-1} - \Omega^{\mp,m' \leftrightarrow \pm,m'-1}\right|=\frac{A_{\mathrm{iso}}}{2}\left|R_m-(R_{m-1}   + R_{m'}+R_{m'-1}  ) \mp4\tilde \omega_0 \delta_\gamma\right|\label{parity -1 frequency difference large} .
  \end{equation}
 Such frequencies are never degenerate when $m \ne m'$. In the high-field limit and for any $m$ and $m'$ these both grow  linearly with $B_0$, with the rates  
\begin{align}
&\lim_{(\gamma_e+\gamma_n)B_0/A_{\mathrm{iso}}\to \infty}\frac{d\Delta \Omega}{dB_0}=\gamma_e(1+\delta_\gamma) \ \ \ \ \ \ \text{and} &\lim_{(\gamma_e+\gamma_n)B_0/A_{\mathrm{iso}}\to \infty}\frac{d\Delta \Omega}{dB_0}=\gamma_e(1+(1\pm2)\delta_\gamma)
\end{align}
 respectively. There is only one case, where $m=m'$, that results in transition frequency degeneracy.  The transitions $\phi^+_{-I+1/2}\leftrightarrow \phi^-_{-I-1/2}$ and $\phi^-_{-I+1/2} \leftrightarrow \phi^-_{-I-1/2}$ will have a  frequency difference given by
\begin{equation}
\Delta \Omega=\frac{A_{\mathrm{iso}}}{2}\left|-2\mathcal{W}_{-(I+1/2)}     +4\tilde \omega_0 \delta_\gamma\right|   
\end{equation}
which vanishes when $\tilde\omega_0=(I+1/2)/(1-\delta_\gamma)$. This is the  type I cancelation resonance of $H_{m=-(I+1/2)}$ in the limit $\delta_\gamma \to 0$.

     The second way to consider $\Delta \Omega$ for two different transition types is given by \begin{align}
|\Omega^{+,m \leftrightarrow +,m-1} - \Omega^{-,m' \leftrightarrow-,m'-1}|&=\frac{A_\mathrm{iso}}{2}\left|(R_m-R_{m-1})-(R_{m'}-R_{m'-1})-4\tilde\omega_0\delta_\gamma\right|, \nonumber \\
|\Omega^{+,m \leftrightarrow-,m-1} - \Omega^{-,m'\leftrightarrow+,m'-1}|&=\frac{A_\mathrm{iso}}{2}\left|(R_m+R_{m-1})-(R_{m'}+R_{m'-1})-4\tilde\omega_0\delta_\gamma\right|.\label{parity -1 frequency difference small} \end{align}
For $m=m'$, these  are both equal to $2\omega_0\delta_\gamma\equiv 2B_0\gamma_n$ which is  independent of $A_{\mathrm{iso}}$ and only varies with  $B_0$. Due to the small value of $\delta_\gamma$, and excluding the special case of two transitions becoming degenerate in frequency,   this is the smallest amount by which any two transitions may differ in the low-field regime and hence it imposes a lower bound therein. At values of the magnetic field where $\omega_0 \sim A_{\mathrm{iso}}$ and  $2\omega_0\delta_\gamma > \omega^{\mathrm{th}}$, then $|\Omega-\Omega'| > \omega^{\mathrm{th}}$ for all pairs of transition frequencies $\{\Omega,\Omega'\}$. In the high-field regime, however, as $\Delta \Omega$ for  transitions $\phi^\pm_m\leftrightarrow \phi^\pm_{m-1}$ becomes very small, this argument will not hold.

\section{Coherent control with pulsed magnetic resonance}\label{nuc-elec coherent control}

This section aims to expand upon what was established in  Sec.\ref{magnetic resonance control}  to account for controllability of nuclear-electronic spin systems with magnetic resonance. The dynamics between the eigenstates of $H_0$  are governed by a  right-handed ($+$) or left-handed ($-$) circularly polarised magnetic field that couples to both the electron and nuclear spins according to their respective gyromagnetic ratios. The interaction Hamiltonian, then, can be expressed as
\begin{equation}
H_I^\pm(t)=\omega_1f(t)\left(\cos[\omega t]F_x \pm \sin[\omega t]F_y\right))
\end{equation}
where $\omega_1 = \gamma_e B_1$, $f(t)$ is a function describing a  pulse of duration $\tau$, and $F_{x/y}=S_{x/y}+\delta_\gamma I_{x/y}$.   We may write these spin operators in the eigenbasis of $H_0$, $\{\varphi_i\}$,  as $S_{x/y}=\sum_\Omega S_{x/y}(\Omega)$ and $I_{x/y}=\sum_\Omega I_{x/y}(\Omega)$, where 
\begin{align}
S_{x/y}(\Omega) &=\sum_{i,j}\delta(\Omega -\omega_{ij})\<\varphi_j|S_{x/y} \varphi_i\>|\varphi_j\>\<\varphi_i|,\nonumber \\
I_{x/y}(\Omega) &=\sum_{i,j}\delta(\Omega -\omega_{ij})\<\varphi_j|I_{x/y} \varphi_i\>|\varphi_j\>\<\varphi_i|.
\end{align}
Here $\omega_{ij}=E_i-E_j$ where $E_i$ is the energy of eigenvector $\varphi_i$.  This notation is useful because, as was encountered previously in Sec.\ref{master equation derivation}, the interaction picture operators $\tilde S_{x/y}(\Omega,t)$ are simply $e^{-\imag \Omega t}S_{x/y}(\Omega)$, and similarly for $\tilde I_{x/y}(\Omega,t)$. Because the transition frequencies are almost always non-degenerate \footnote{ with the  exception of $\phi^+_{-I+1/2} \leftrightarrow \phi^-_{-I-1/2}$ and $\phi^-_{-I+1/2} \leftrightarrow \phi^-_{-I-1/2}$ close to the type I cancelation resonance of $H_{m=-(I+1/2)}$} the summation for $S_{x/y}(\Omega)$ and $I_{x/y}(\Omega)$ will almost always contain only one term. Consequently, summing the terms with frequency $\Omega$ and $-\Omega$, where we have redefined $\Omega$ as the absolute value $|\Omega|$ which is positive, gives \begin{align}
\sum_{\omega \in \{\Omega,-\Omega\}} S_x(\omega)+\delta_\gamma I_x(\omega)&=    [\eta(\Omega)+\delta_\gamma \xi(\Omega)]  \sigma_x(\Omega), \nonumber \\
\sum_{\omega \in \{\Omega,-\Omega\}}S_y(\omega) +\delta_\gamma I_y(\omega)&=\mathrm{sign}(\Omega) [\eta(\Omega)+\delta_\gamma \xi(\Omega)]\sigma_y(\Omega),
\end{align}
where, for a given pair  $\{\varphi_i,\varphi_j\}$ that satisfies $E_i-E_j=\Omega >0$, we have
\begin{align}
&\sigma_x(\Omega)=|\varphi_i\>\<\varphi_j|+|\varphi_j\>\<\varphi_i|, & \sigma_y(\Omega)=-\imag|\varphi_i\>\<\varphi_j|+\imag|\varphi_j\>\<\varphi_i|.
\end{align}
In addition
 \begin{align}
\eta(\Omega)& = \langle\varphi_j| S_x \varphi_i \rangle , \nonumber \\ \xi(\Omega)& = \langle\varphi_j| I_x \varphi_i \rangle, 
\end{align} 
and 
\begin{equation}
\mathrm{sign}(\Omega)=\langle\varphi_i|( S_z +  I_z )\varphi_i\rangle -\langle\varphi_j |( S_z +  I_z)\varphi_j \rangle \in\{1,-1\}.
\end{equation}

It  should be noted that equations \eqref{Iforbp}-\eqref{Iforbidden},  determining the relative c.w. transition rates, are proportional to $(\eta[\Omega]+\delta_\gamma \xi[\Omega])^2$.
Also, at magnetic fields lower than $\sim A_{\mathrm{iso}}/2\gamma_n$, $\mathrm{Sign}_y(\Omega)=+1$ for transitions  $\phi^+_m\leftrightarrow \phi^\pm_{m-1}$  while $\mathrm{Sign}_y(\Omega)=-1$ for transitions  $\phi^-_m\leftrightarrow \phi^\pm_{m-1}$;  here the former will utilise a RH driving field, whereas the latter will employ a LH driving field. However, as mentioned earlier, because the energy ordering of the states $\{\phi^+_m\}$ reverses at fields larger than $\sim  A_{\mathrm{iso}}/2\gamma_n$, here the transitions $\phi^+_m\leftrightarrow \phi^+_{m-1}$ will also utilise a LH driving field.

 We wish to  make the two-level approximation, where  a single two-dimensional subspace can be considered in isolation.  This  subspace is $\{\varphi_0,\varphi_1\}$  which satisfies $E_0- E_1=\Omega_0 >0$. Such selectivity can be achieved by tuning the frequency of the driving field to be in resonance with $\Omega_0$, provided that we may make the rotating wave approximation, covered in Sec.\ref{rotating wave approximation section}.
We can think of the issue of selectivity in terms of the bandwidth of the pulse in the frequency domain. The length of a pulse is limited by the strength $\omega_1$; it is the ``area'' under the pulse that determines the amount by which the state is evolved, with stronger pulses requiring shorter pulse lengths and vice versa. The Fourier transform of a square pulse of duration $\tau$, frequency $\omega$, and strength $\omega_1$ is given by $\omega_1\mathrm{sinc}[\omega\frac{\tau}{2}]$. The shorter $\tau$ is, due to  $\omega_1$ being stronger, the larger the bandwidth of frequencies the sinc function will act upon.  

Let us  consider the interaction Hamiltonian $H_I(t)$ in the rotating frame of $ H_0$ where we have tuned the frequency of the driving field to be $\Omega_0$, and the polarity of the driving field is appropriately set to $\mathrm{sign}(\Omega_0)$. Hence, we have 
\begin{align}
\tilde H_I(t)&=\omega_1f(t)\sum_\Omega e^{-\imag t \Omega}   \left( \cos[\Omega_0 t]F_{x}(\Omega) +\mathrm{sign}(\Omega_0) \sin[\Omega_0 t]F_{y}(\Omega)\right)      , \nonumber \\
&=\omega_1f(t)\sum_{\Omega>0} [\eta(\Omega)+\delta_\gamma \xi(\Omega)] \nonumber \\ & \ \ \ \ \ \ \ \ \ \ \ \ \  \ \ \ \times\left(\cos[(\Omega_0-\mathds{P}_\Omega\Omega)t] \sigma_{x}(\Omega)  +\mathds{P}_\Omega  \sin[(\Omega_0-\mathds{P}_\Omega\Omega)t] \sigma_{y}(\Omega)\right)        \label{nuclear-electronic rotating frame Hamiltonian}
\end{align}
where we have assigned the \emph{parity} as  $\mathds{P}_\Omega:=\mathrm{sign}_y (\Omega_0)\mathrm{sign}_y (\Omega) \in \{\pm1\}$.  Assuming for a square pulse and a constant driving field strength $\omega_1$, the Rabi frequency  within each subspace is determined by 
\begin{equation}
\eta(\Omega)+\delta_\gamma \xi(\Omega)\label{Rabi frequency nuc-elec}.
\end{equation}  In the low-field regime the Rabi frequencies for all dipole-allowed transitions are of the same order of magnitude, being dominated by the   $\eta(\Omega)$ term. 

The rotating wave approximation, when valid, will allow us to omit all terms other than that where $\Omega_0-\mathds{P}_\Omega \Omega=0$: where $\mathds{P}_\Omega=+1$ and $\Omega=\Omega_0$. Here, the interaction Hamiltonian reduces to 
\begin{align}
\tilde H_I(t)&=\begin{cases} \omega_1[\eta(\Omega)+\delta_\gamma \xi(\Omega)]f(t)\sigma_x(\Omega_0) & \mathrm{when} \ \omega t\mapsto\Omega_0 t, \\
\omega_1[\eta(\Omega)+\delta_\gamma \xi(\Omega)]f(t)\sigma_y(\Omega_0) & \mathrm{when} \ \omega t \mapsto \Omega_0 t + \frac{\pi}{2},  \\
\end{cases}        
 \end{align}
thereby providing the generators for $\mathfrak{su}(2)$ within our selected subspace.
As discussed in Sec.\ref{rotating wave approximation section}, assuming for a square pulse,  the RWA is valid up to an arbitrarily small error $\epsilon$  when
\begin{equation}
\max_\Omega\left(\frac{\omega_1[\eta(\Omega)+\delta_\gamma \xi(\Omega)]}{| \Omega_0-\mathds{P}_\Omega\Omega|}\right) \ll 1. \label{nuc-elec coherent controll RWA condition}
\end{equation}
  A heuristic way of ensuring this is by setting $\omega_1/\min_\Omega(\Delta \Omega)\ll 1$. Because of our choice of a circularly polarised driving field, coupled with the fact that we are choosing to ignore the unique scenario where transition frequency degeneracy occurs,  and that   $\min_\Omega(\Delta \Omega)$ is realised when $\mathds{P}_\Omega =+1$, we need only consider $\min_\Omega(\Delta \Omega)$ from within the same transition type as the chosen one of frequency $\Omega_0$. \footnote{This, of course, only applies in the case of $I \geqslant 1$ where there are indeed more than one transition of a certain type. The case of $I=1/2$ is different, and as shall become apparent, results in the lack of an ability to achieve accurate speed-up in the low-field regime.} These values of $\min_\Omega(\Delta \Omega)$ are given by Equations \eqref{same time freqdiff small}-\eqref{same time freqdiff large} when $m'=m-1$. For all the transition types,  $\Delta \Omega$ increases with $B_0$ and reaches its maximal  value of order $\sim A_{\mathrm{iso}}$ in the low-field regime. In the case of transitions  $\phi^\pm_m\leftrightarrow \phi^\mp_{m-1}$, $\min_\Omega(\Delta \Omega)$ will have a value of $ A_{\mathrm{iso}}$ as $(\gamma_e+\gamma_n)B_0/A_{\mathrm{iso}}\to \infty$, so the advantage of the low-field regime rests on the fact that the transition $\phi^-_m\leftrightarrow \phi^+_{m-1}$ is dipole-forbidden at high fields. More strikingly, $\min_\Omega(\Delta \Omega)$ for transitions of type $\phi^\pm_m\leftrightarrow \phi^\pm_{m-1}$ in the high-field regime becomes increasingly smaller with  stronger fields, vanishing altogether as $(\gamma_e+\gamma_n)B_0/A_{\mathrm{iso}}\to \infty$. Therefore,  even though these transitions are driven by NMR, with consequently much slower Rabi frequencies than ESR given a constant $\omega_1$,     \emph{accurate} control at sufficiently large magnetic fields will require \emph{even slower} Rabi frequencies.

As a final remark, let us note that we require the ability to  perform any $U \in \su(2[2I+1])$ in order to attain controllability within the state space of the nuclear-electronic spin system $\mathcal{S}(\co^{2(2I+1)})$.  This is indeed possible in all magnetic field regimes, and here is a simple proof of principle argument:  take for example a desired unitary map of the form  $U:\phi^+_m\mapsto \alpha\phi^+_m+\beta\phi^-_{m'}$ where $\alpha, \beta \in \mathds{C}$ and $m-m'\leqslant2I$. All that is required is an initial set of  pulses that prepare the state $\alpha\phi^+_m+\beta\phi^-_{m-1}$, followed by a string of  $\pi$ pulses that take $\phi^-_{m-1}\mapsto\phi^-_{m-2}\mapsto \dots \phi^-_{m'}$.

\subsection{ Linearly polarised fields and selectivity}

As we have already covered in  Sec.\ref{rotating wave approximation section}, in most experimental situations one cannot establish a circularly polarised field. Instead a linearly polarised magnetic field is used, which is composed in equal parts of a RH and a LH field. In such a case, then, Eq.\eqref{nuclear-electronic rotating frame Hamiltonian} will be altered so as to include terms with both parity components $\{\pm1\}$ to give
\begin{align}
\tilde H_I(t) =&\omega_1f(t)\sum_{\Omega>0, \mathds{P}_\Omega\in\{\pm1\}} [\eta(\Omega)+\delta_\gamma \xi(\Omega)]  \nonumber \\
& \ \ \ \ \ \ \ \ \ \ \ \ \ \ \ \times \left(\cos[(\Omega_0-\mathds{P}_\Omega\Omega)t] \sigma_{x}(\Omega)  +\mathds{P}_\Omega  \sin[(\Omega_0-\mathds{P}_\Omega\Omega)t] \sigma_{y}(\Omega)\right).        \label{nuclear-electronic rotating frame Hamiltonian linearly polarised}
\end{align}
As a result  the requirements for the rotating wave approximation will become more stringent because, excluding the special instance of transition frequency degeneracy and the transitions of type $\phi^\pm_m\leftrightarrow \phi^\pm_{m-1}$ in the high-field limit,   $\min_\Omega(\Delta \Omega)$ will be determined by Eq.\eqref{parity -1 frequency difference small}, which is  $2B_0\gamma_n$. In the low-field regime this is much smaller than $\min_\Omega(\Delta \Omega)$ utilising a circularly polarised field, and will require much slower Rabi frequencies for accurate control.  

\section{Summary}
In this chapter we have investigated  nuclear-electronic spin systems, with an electron spin of one-half coupled to an arbitrary nuclear spin via an isotropic hyperfine interaction, and their closed system dynamics due to magnetic resonance. The eigenstates of these systems, save for two, are generally entangled with respect to the nuclear and electronic spin subspaces, becoming separable  asymptotically at large magnetic fields. The larger the hyperfine coupling between the two spins, the greater the   magnetic fields at which an appreciable entanglement is present become. This entanglement forces us to revisit the standard magnetic resonance selection rules,  ESR and NMR, and conceive of a more generalised selection rule for such coupled systems, which we call nuclear-electronic magnetic resonance (NEMR). The consequence of this is that transitions between eigenstates that, at sufficiently large magnetic fields are to a good approximation described as NMR, have, in the low-field regime, transition rates of the same order of magnitude as that which is characteristic of ESR. 

Even more interesting phenomena occur when the nuclear spin $I$ is greater than  one. First is the possibility of low-field regime transitions between eigenstates that, at high magnetic fields, are forbidden by both the NMR and ESR selection rules.  Secondly, there exist transitions that have frequency stationary points (FSPs), defined as finite magnetic fields where the gradient of the transition frequency with respect to the magnetic field vanishes. These occur when the expectation values of the operator $\gamma_eS_z-\gamma_n I_z$ on the two involved eigenstates equalise.

We showed that, provided the ability of tuning the frequency of the driving field to the NEMR resonances, and given access to both right-handed and left-handed circularly polarised driving fields, we can achieve full controllability of the system's Hilbert space by magnetic resonance. The two concomitant factors affecting  the speed of \emph{accurate} control are the NEMR transition rates, and the gap between the desired transition frequency and all other transition frequencies. Again, provided a nuclear spin that is greater than or equal to one, in general both of these factors are  more favourable in the low-field regime, making it possible to gain a speed-up of accurate control.

\spacing{1}                                  
 \bibliographystyle{plainnat}
 \bibliography{references}

\spacing{1}
\chapter{Open system dynamics}\label{open systems chapter}

\section{Introduction}

In the previous chapter we considered nuclear-electronic spin systems as closed, wherein their dynamics is governed by unitary evolution alone. Of course in reality such systems will be embedded in a physical environment where they will interact with external degrees of freedom which lie outside the ability of the experimentalist to control. Such open quantum system dynamics will, generally, lead to decoherence. The nature of the environment can, for the most part,  be dichotomised as being either bosons or spins. The former is applicable to, amongst others,     spins in the solid-state coupled to vibrational modes, or phonons, of the lattice. The dynamics here,  usually Markovian, can be described by  the spin-boson model which is covered extensively in the literature such as \citep{two-state-dissipative,dissipative,solidstateadvances} to name a few. The latter case is usually found in spins in the solid-state, where undesirable spin species in the lattice interact with the spin system of interest. The theory of the spin bath is covered exemplarily in \citep{spin-bath}. In many cases, the spin bath is described by the central-spin model \citep{Witzel-spectral-diffusion,Yang-spectral-diffusion} where the open system of interest is a central spin coupled, independently, to many surrounding spins which may or may not be interacting with one another. 

In this chapter I will consider open system dynamics for nuclear-electronic spin systems.  I aim to see how, all else being equal, the entanglement of the nuclear and electronic spins found in the eigenstates of the system Hamiltonian in the low-field regime affect the decoherence. Consequently, when studying the effects of Markovian dynamics  I will not consider the spin-boson model, responsible for the thermal noise in solid-state spin impurities, but rather a  phenomenological Markovian model where the system experiences effective fluctuations in the external magnetic field $B_0$. This is because the decay rates in the spin boson model depend on the spectral density function which is itself contingent upon the transition frequency, and hence the value of the external field, thus making the prospect of comparison between the low-field regime and high-field regime tenuous, if not impossible. This work is largely based on the publication from my colleagues and me in \citep{Mohammady-2012}.
After this  I shall conceive the   nuclear-electronic spin system as a central spin under the influence of a   spin bath, where  the cause of decoherence is the entanglement generated between the  two. This work is the analytic companion to the numerical analysis presented in \citep{Balian-diffusion}. It will be apparent that, although the two models have a fundamentally different microscopic basis, the qualitative properties of the two are in good accord.

In both studies, we always consider the  system  as initially being prepared in a superposition established by a single resonance of NEMR, namely the transitions $\phi^\pm_m\leftrightarrow \phi^\pm_{m-1}$ and $\phi^\pm_m\leftrightarrow \phi^\mp_{m-1}$ , so that the experimentally accessible effects of decoherence by means of conventional magnetic resonance spectroscopy are described in terms of a two-dimensional subspace of the system.    I will show that if the interaction of the system and its environment is sufficiently weak so as to result in pure decoherence  with respect to the eigenbasis of the system Hamiltonian, in a perturbative limit, the coherence time of the system can increase by orders of magnitude at specific values of the external magnetic field $B_0$. These are the so-called  \emph{optimal working points} (OWPs), established when
\begin{align}
\<\phi^\pm_m|S_z\phi^\pm_m\>=\<\phi^\pm_{m-1}|S_z\phi^\pm_{m-1}\> \ \ \ \text{and} \ \ \ \<\phi^\pm_m|S_z\phi^\pm_m\>=\<\phi^\mp_{m-1}|S_z\phi^\mp_{m-1}\>
\end{align}
 which occur for superpositions between subspaces $m$ and $m-1$ if and only if $-I+3/2\leqslant m\leqslant0$. These are possible only in systems with $I>1$.  The OWPs  are identified with the FSPs   in the limit $\delta_\gamma \to 0$. These two-level subspaces,  at the OWPs, constitute decoherence free subspaces (DFSs) \citep{DFS}.  Owing to the magnetic field dependence of establishing these DFSs, which is a parameter of the system Hamiltonian, we call this phenomenon \emph{parametric decoupling}.

\section{The basic model}

\begin{figure}[!htb]
\centering
\includegraphics[width=5in]{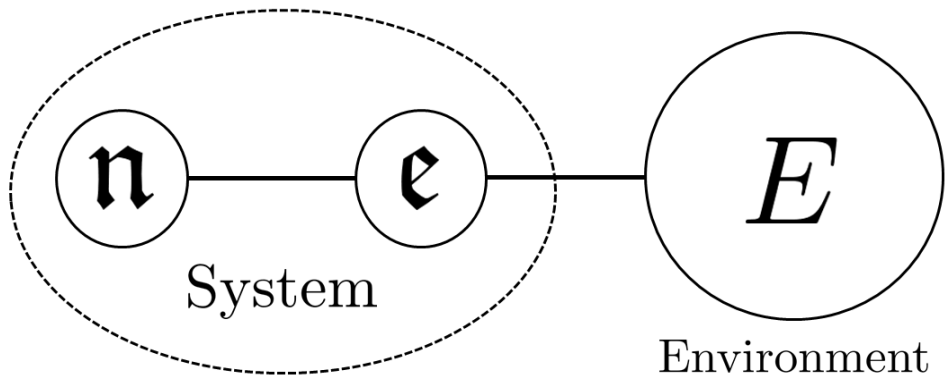} \caption{ The coupled nucleus and electron spins constitute the nuclear-electronic spin system, and the environment interacts with the system via the electron spin.}
\end{figure}
As the gyromagnetic ratio of the nuclear  spin is between three and four orders of magnitude smaller than that of the electron, and since the nuclear spin wavefunction is much more localised than that of the electron, the open system dynamics will be dominated by the interaction of the electron with its environment. We are therefore justified in constructing a model wherein only the electron spin interacts with its environment, such that the total Hamiltonian is
given by\begin{equation}
H=H_0^{\mathfrak{n+e}}+H_I^{\mathfrak{e}+E} + H^{^E}.
\end{equation}

As mentioned in Sec.\ref{decoherence in open systems section}, a sufficient condition for decoherence to be pure with respect to the eigenbasis of the system Hamiltonian is for the interaction Hamiltonian to commute with $H_0^{\mathfrak{n+e}}$.

We may also be interested in considering the nuclear spin independently, and talk about decoherence in this subspace. Due to the interaction between the nuclear and electronic spins, even though we posit that  only the electron spin interacts with the environment, the reduced system dynamics need not necessarily  act locally on the electron spin; the nuclear spin will generally be affected as well.  We assume that at some initial time the nuclear-electronic spin system and  the environment are separable, and then evolve the  composite system due to the total Hamiltonian. The reduced state of the nucleus after the joint evolution is
\begin{equation}
\rho^{\mathfrak{n}}(\tau)=\mathrm{tr}_{E+\mathfrak{e}}[U_\tau\rho^{\mathfrak{n+e}} \otimes \rho^E U_\tau^\dagger].
\end{equation} 
where
\begin{equation}
U_\tau=e^{-\imag\tau(H_0^{\mathfrak{n+e}}+H_I^{\mathfrak{e}+E}+H^E)}.
\end{equation}
This is, in general, different to the initial reduced state of the nucleus. A series of conditions which, when all are met,  form a sufficient criterion for  $\rho^\mathfrak{n}(\tau)=\mathrm{tr}_{\mathfrak{e}}[\rho^{\mathfrak{n+e}}]$, up to a local unitary transformation,  are 
\begin{enumerate}[(i)]
\item$[H_0^{\mathfrak{n+e}},H_I^{\mathfrak{e}+E}]_-=\mathds{O}$.
\item The eigenstates of $H_0^{\mathfrak{n+e}}$ are product states. 
\item $\rho^{\mathfrak{n+e}}= (1-\lambda)\rho_1^\mathfrak{n}\otimes\Pi_{+1/2}^\mathfrak{e} + \lambda\rho_2^\mathfrak{n}\otimes\Pi_{-1/2}^\mathfrak{e}$ where $\lambda \in [0,1]$.
\end{enumerate}
Condition (i), which implies that the total nuclear-electronic spin system undergoes pure decoherence, ensures that the joint  unitary operator  may be written as \begin{equation}
U_\tau=U_\tau^{\mathfrak{n+e}}U_\tau^{\mathfrak{e}+E}=e^{-\imag\tau H_0^{\mathfrak{n+e}}}e^{-\imag\tau (H_I^{\mathfrak{e}+E}+H^E)}
\end{equation}where the order of application is unimportant. Consequently, if  $U_\tau^{\mathfrak{n+e}}$ does not generate entanglement between the electron and nuclear spins, as is ensured by  conditions (ii) and (iii), then the action of  $U_\tau^{\mathfrak{e}+E}$ will have no effect on the nuclear spin. 

The eigenstates of $H_0^{\mathfrak{n+e}}$ are product states in the asymptotic high-field limit, where the operator that commutes with the system Hamiltonian must be of the form $H_I^{\mathfrak{e}+E}=S_z\otimes\mb$, where $\mb$ is some operator acting on the environment. If we prepare, in the high-field limit, our state to initially  be in the   superposition $\psi=\alpha\phi^\pm_m+\beta\phi^\pm_{m-1}$  we do indeed have a product state of the form
\begin{align}
\lim_{(\gamma_e+\gamma_n)B_0/A_{\mathrm{iso}}\to \infty}\psi&=\left|\pm1/2\right\>\otimes\left(\alpha\left|m\mp1/2\right\>+\beta\left|m-1 \mp1/2\right\>\right)
 \end{align} 
where the state of the electron is an eigenstate of $S_z$. Clearly, the action of $U_\tau^{\mathfrak{n+e}}$ on the system will only result in the superposition in the nuclear subspace picking up a phase, leaving the  system in a separable state. Consequently, the action of $U_\tau^{\mathfrak{e}+E}$ will not result in decoherence in the nuclear spin subspace.
 
The considerations above are tantamount to saying that, in the high-field limit, if the decoherence mechanism results in pure dephasing with respect to the electron spins, then the NMR superpositions will not decohere. If, on the other hand, the electron spins were to undergo depolarising noise, the nuclear spin subspace would still be affected.   

\section{Phenomenological model for Markovian dynamics}\label{phenomenological Markovian model}

\subsection{The master equation}
In this section, I will develop a phenomenological Markovian master equation for nuclear-electronic spin systems. The techniques used here have already been expressed previously in Sec.\ref{master equation derivation} and, as a result, I will not go in too much depth about the individual steps made. 

We propose the stationary state of the environment to be $\varrho$ and the interaction Hamiltonian to have a Schmidt-rank of one . Two cases shall be considered; where the interaction Hamiltonian couples $S_z$ with the bath, and where the interaction Hamiltonian couples $S_x$ with the bath. \footnote{Indeed, we may equivalently consider the case where the operator is any linear combination of $S_x$ and $S_y$ as both will have the same effect.} 
\begin{align}
&H_I^n=V_n S_n\otimes\mb   & n \in \{x,z\}.
\end{align}
The resulting non-unitary dynamics will be referred to as $Z$ noise and $X$ noise henceforth. Here, $V_x$ and $V_z$ are the strengths of the interaction, defined as $2\|H_I^n\|$ so that $\|S_n \otimes \mb\|=\|S_n\|=1/2$, and $\mb$ is a self-adjoint operator acting on the bath. If $V_{n} \tau_{\mathrm{env}} \ll 1$, where $ \tau_{\mathrm{env}}$ is the correlation time of the bath, we are justified in making the weak-coupling approximation, up to a certain accuracy $\epsilon$, and hence  coarse-grain the dynamics by the time scale $\Delta t$. The smaller $V_{n}$ is, the longer $ \tau_{\mathrm{env}}$ and by association $\Delta t$ can be for the weak-coupling approximation to remain valid. By following the procedures delineated in Sec.\ref{master equation derivation}, we arrive at the differential equation
\begin{align}
\frac{d}{dt}\rho(t)&= \imag\left[\rho(t),H_0\right]_- \nonumber \\ &+V_{n}^2\sum_{\Omega,\Omega'}G(\Omega)J(\Omega,\Omega')\left(S_n(\Omega) \rho(t)S_n^\dagger(\Omega')-S_n^\dagger(\Omega')S_n(\Omega)\rho(t)\right)+\mathbb{H.C} \end{align}
where \begin{align}
G(\Omega)&=\int_{0}^{\Delta t}d\tau e^{\imag \Omega \tau}\<\tilde\mb^\dagger(\tau)\tilde \mb(0)\>, \nonumber \\
J(\Omega,\Omega')&=e^{\imag(\Omega'-\Omega)\frac{\Delta t}{2}}\frac{\sin[(\Omega'-\Omega)\Delta t/2]}{(\Omega'-\Omega)\Delta t/2}.
\end{align}

We are justified in making the secular approximation $J(\Omega,\Omega')=\delta(\Omega-\Omega')$ if the condition $  \Delta t\gg \ 1/|\Omega-\Omega'| \ \forall \ \Omega \ne \Omega'$ is satisfied,  which can be restated as the requirement $V_{n}/|\Omega-\Omega'|\ll 1   \ \forall \ \Omega \ne \Omega'$. In the low-field regime, the smallest value by which any two frequencies can differ  is $2B_0\gamma_n$; only in the high-field regime do we find a smaller frequency difference, which vanishes as $B_0 \to \infty$.  Consequently, in the extreme limits of   $(\gamma_e+\gamma_n)B_0/A_{\mathrm{iso}}\to 0$ and $(\gamma_e+\gamma_n)B_0/A_{\mathrm{iso}}\to \infty $,   $V_n$ must be made infinitely small. The secular approximation is therefore most valid in the low-field regime conditional on the relation  $V_n/2B_0\gamma_n \ll 1$ being satisfied.

  Thus far, our derivation has not involved any phenomenology. It is in the evaluation of $G(\Omega)$ where this occurs. We require that
the correlation function be a Gaussian
\begin{equation}
 \<\tilde\mb^\dagger(\tau)\tilde \mb(0)\>=\frac{1}{2\sqrt{\pi \chi_n}}e^{-\frac{\tau^2}{4 \chi_n}}
\end{equation} 

\noindent which, in the limiting case of $\chi_n\to 0$,  tends to the Dirac delta function $\delta(\tau)$. Because of our coarse-graining approximations where $\Delta t \gg \Delta \tau_{\mathrm{env}}$, the correlation function vanishes sufficiently fast so that we may take the upper limit of the integrand of $G(\Omega)$ to infinity. The weaker $V_{n}$ is, the larger $\chi_n$ can be for this approximation to remain valid. To evaluate this function, we make the change in notation  

\begin{equation}
G(\Omega)=\frac{1}{2}\Gamma_n(\Omega) + \imag\Lambda_n(\Omega)
\end{equation}
where
\begin{align}
\Gamma_n(\Omega) &=   \int_{-\infty}^{\infty}\frac{d\tau}{2\sqrt{\pi\chi_n}}e^{-\frac{\tau^2}{4\chi_n}}e^{\imag \Omega\tau}=e^{-\chi_n\Omega^2}, \nonumber \\
\Lambda_n(\Omega) &=\frac{1}{2\imag}\int_0^{\infty} \frac{d\tau}{2\sqrt{\pi \chi_n}}e^{-\frac{\tau^2}{4\chi_n}}\left( e^{\imag \Omega\tau} - e^{-\imag \Omega\tau} \right).
\end{align}
Thus, we finally have our Markovian master equation in Lindblad form 
\begin{align}
\mathscr{L}[\rho(t)] &= \imag\left[\rho(t),H_0 +  H_{LS} \right]   \nonumber \\ &+V_n^2\sum_{\Omega} e^{-\chi_n \Omega^2}  
\left(   S^\dagger_n(\Omega)\rho(t)  S_n(\Omega)-\frac{1}{2}\left[ \rho(t),  S^\dagger_n(\Omega) S_n(\Omega)\right]_+  \right)\label{master equation nuclear-electronic}
 \end{align}
where $\mathscr{L}$ is the Liouville super-operator that generates the dynamical semigroup $e^{\mathscr{L}}$, and
\begin{equation}
H_{LS}=V_n^2\sum_\Omega\Lambda_n(\Omega) S_n^\dagger(\Omega) S_n(\Omega)
\end{equation}

\noindent is the Lamb shift and changes the energy levels of the system. This is a negligible effect here and  we shall ignore it henceforth.

\subsubsection{On adiabaticity} 
The above treatment is almost identical with that in \citep{Mohammady-2012}, except that I have not explicitly used the notion of a fluctuating magnetic field, but instead a quantum mechanical ``bath'' with Gaussian correlation functions that are independent of the external field $B_0$. Nevertheless, the action of the bath on the system results in effective fluctuating magnetic fields. As such, we may still make the identification $\chi_n\equiv |1/\dot B_n(t)|$, whereby we may treat the exponent of the correlation function as an indicator for \emph{adiabaticity}. The condition for adiabatic evolution can be quantitatively expressed as 
\begin{align}
 \left |\<\varphi_j|S_n \varphi_i\> \frac{\dot B_n(t) }{(\Omega^{i\leftrightarrow j})^2} \right|\ll 1.
\end{align}
where, given a sufficiently slow magnetic field fluctuation, transitions between  states $\{\varphi_i,\varphi_j\}$ are suppressed. Hence, it follows that the Markovian dynamics is
\begin{enumerate}[(i)]
\item \emph{Diabatic}

If $\chi_n\Omega^2 \to 0 \ \forall \ \Omega$, then $\Gamma(\Omega)=e^{-\chi_n\Omega^2}\to 1 \ \forall \ \Omega$

\item \emph{Adiabatic}

If $\chi_n\Omega^2 \to \infty \ \forall \ \Omega>0$, then $\Gamma(\Omega)=e^{-\chi_n\Omega^2}\to 0 \ \forall \ \Omega > 0 $

\end{enumerate}

 Under the adiabatic assumption, we may drop all terms of the master equation where $\Omega >0$. This will result in pure decoherence, with the preferred basis being that in which $H_0$ is diagonal. Only $Z$ noise can be treated adiabatically, as $X$ noise does not have any   $S_x(\Omega=0)$ terms.

How large $\chi_n\Omega^2 $ needs to be for the adiabatic approximation to hold depends on the degree of precision $\epsilon$ we require, and the time-scale we wish to consider. If $\chi_n\Omega^2 <\infty$, then at infinite time the resultant state $\rho(t=\infty)$ will not be given by the adiabatic master equation where only terms with $\Omega=0$ are considered. For finite times $\tau$, however, the accuracy with which the adiabatic master equation predicts the state evolution can be made arbitrarily high given arbitrarily large values of $\chi_n\Omega^2 $.
Furthermore, the adiabatic condition is compatible with the criteria for Markovian dynamics only in the case where $V_n\chi_n\ll1$. Hence, $\chi_n \to \infty$ implies $V_n \to 0$, meaning that  the dynamics will tend to be entirely unitary. It is therefore more a matter of $\Omega$, rather than $\chi_n$, being large that results in adiabatic, pure decoherence. 

\subsubsection{Magnetic resonance studies of the Markovian dynamics}

Now that we have determined our Markovian master equation, we may use it to analyse the dynamics of our nuclear-electronic spin systems under the influence of a fluctuating magnetic field. The tools at our disposal are the magnetic resonance experiments discussed in Sec.\ref{pulsed dynamics measurement scheme}. We may establish, with a series of pulses at a single resonant frequency, any coherence within a dynamical selection rule allowed two-dimensional subspace. These superpositions correspond to the transitions discussed in  Eqs.\eqref{Iforbp}-\eqref{Iforbidden}.    The two observable phenomenon of interest are the dephasing and depolarisation rates. Accordingly, we may  discard the free Hamiltonian term of the master equation and  consider the dynamics in the interaction picture. \footnote{Owing to the fact that the magnetic resonance measurements available to us are the weak ensemble measurements, we  strictly cannot ignore ensemble effects on the measured dephasing. However, by assuming either perfectly homogeneous magnetic fields, or at least a perfectly executable Hahn echo, we can safely ignore such contributions. } The dephasing and depolarisation rates would, respectively, be  determined by the time behaviour of $\sqrt{\mathrm{tr}[\sigma^{01}_x \tilde\rho(\tau)]^2+\mathrm{tr}[\sigma^{01}_y \tilde\rho(\tau)]^2}$ and $\mathrm{tr}[\sigma^{01}_z \tilde\rho(\tau)]$. Here $\{\sigma^{01}_x,\sigma^{01}_x,\sigma^{01}_z\}$ are in the basis of the initial superposition established, with $\varphi_0$ being the excited state and $\varphi_1$ the ground state.

\subsection{Analysis of $Z$ noise}

We wish to evaluate Eq.\ref{master equation nuclear-electronic} in the case of $n=z$, which we call $Z$ noise.  Before we begin let us note what $S_z$, in the basis that $H_0$ is diagonal, looks like. This is
\begin{equation}
S_z^m=\begin{cases}\frac{1}{2}\left[\cos(\theta_m)\sigma_z^m-\sin(\theta_m)\sigma_x^m\right] & \mathrm{when} \ -I+1/2\leqslant m\leqslant I -1/2, \\
\pm\frac{1}{2}\Pi(\phi^\pm_m) & \mathrm{when} \ m = \pm (I + 1/2), \\
\end{cases}
\end{equation} where the Pauli operators $\{\sigma_i^m\}$ are written in the basis  $\{\phi^+_m,\phi^-_m\}$.   $Z$ noise can be seen to confine interaction picture dynamics within $m$ subspaces. Hence, for an initial pure state $\Pi(\psi)$ where $\psi$ is a superposition of two $H_0$ eigenvectors, one from subspace $m$ and another from subspace $m-1$, we need only consider the truncated \footnote{Of course, when a subspace $m=\pm(I+1/2)$ is involved, we need only consider $\lo_s(\co^3)$, but considering this within the larger space $\lo_s(\co^4)$ will not prove problematic.} operator $S_z^m \oplus S_z^{m-1} \in \lo_s(\co^4)$.  We may therefore characterise the dynamics within the state space $\s(\co^4)$, such that every density operator may be written as
\begin{equation}
\rho(t)=\frac{1}{4}\left(\sum_{i,j=0}^3 n_{ij}(t) \sigma_i\otimes\sigma_j \right) \text{,}\ n_{00}(t)=1
\end{equation}
where the basis for $\rho(t)$ is $\{\phi^+_m,\phi^-_m,\phi^+_{m-1},\phi^-_{m-1}\}$. The truncated master equation Eq.\eqref{master equation nuclear-electronic}, in the interaction picture, can therefore be expressed as
\begin{align}
\mathscr{L}[\tilde \rho(t)]=&\frac{V_z^2}{4}\sum_{j,k\in\{m,m-1\}} \cos(\theta_j)\cos(\theta_k) \left(  \sigma_z^{j}\tilde\rho(t)\sigma_z^k- \frac{1}{2}\left[\delta_{jk}\Pi(j),\tilde\rho(t)\right]_+  \right) \nonumber \\&+\frac{V_z^2}{4}\sum_{j\in\{m,m-1\}} e^{-\chi_z\Omega_j^2}\sin^2(\theta_j)\left(\sigma_+^j\tilde\rho(t)\sigma_-^j+\sigma_-^j\tilde \rho(t)\sigma_+^j- \frac{1}{2}\left[\Pi(j),\tilde\rho(t)\right]_+ \right)  \label{master equation Z noise}
 \end{align}
where $\Pi(m)=\Pi(\phi^+_m)+\Pi(\phi^-_m)$ is a projector onto the $m$ subspace, $\sigma_{\pm}^m\equiv|\phi^\pm_m\>\<\phi^\mp_m|$ is an exchange operator, and $\Omega_m=E^+_m-E^-_m$. In the case that one of the subspaces is $m=\pm(I+1/2)$,  as shown above, the electron spin operator  is not given by a Pauli matrix, but a  rank-1 projector. Notwithstanding, we may still consider such a case in our general treatment by envisioning these  one-dimensional $m$ spaces as being a subspace of a fictitious two-dimensional space.
     
Because the density operator is characterised completely by the vector $\bold{\underline{n}}(t)$, which has sixteen elements, we may determine the dynamics by solving the sixteen simultaneous differential equations encapsulated by 
\begin{equation}
\frac{d \bold{\underline{n}}(t)}{dt}=\mathscr{L} \bold{\underline{n}}(t)\label{liouville super-operator differential equation z noise}
\end{equation}
where $\mathscr{L}$ is the Liouville super-operator in matrix form.  Furthermore, due to the fact that $\mathscr{L}$ is Markovian, and hence generates  a dynamical semigroup,
the solution to this is 
\begin{equation}
\bold{\underline{n}}(t)=e^{\mathscr{L}} \bold{\underline{n}}(t_0).\label{liouville super-operator differential equation z noise 1}
\end{equation}
 We may therefore write $\bold{\underline{n}}(t)$ in the eigenbasis of $\mathscr{L}$, with eigenvectors $\{\bold{\underline{n}}_l\}$ and eigenvalues $\{\lambda_l\}$, to evaluate the long-time dynamics of the system. In this basis, $\bold{\underline{n}}(t)$ is given as
\begin{equation}
\bold{\underline{n}}(t)=\sum_{l=0}^{15} c_l \bold{\underline{n}}_le^{t\lambda_l}
\end{equation}
where $\{c_l\} \in \mathds{C}$ are determined by the initial conditions. The imaginary component of $\lambda_l$ will simply lead to unitary dynamics, whereas the real components are responsible for decay. \footnote{In fact, the real component of $\lambda_l$ must always be  negative.} As $t \to \infty$, $\bold{\underline{n}}(t)$ will consist only of terms  $\{\bold{\underline{n}}_l:\mathrm{Re}(\lambda_l)=0\}$. The stationary states are thus any superposition of such eigenvectors. 

We shall be considering  general initial superposition states corresponding to the transition types discussed in Eqs.\eqref{Iforbp}-\eqref{Iforbidden}. In the case of equal superpositions, the corresponding  non-zero elements of $\bold{\underline{n}}(t_0)$ are given by

\begin{enumerate}[(i)]
\item $ \psi(t_0)=\frac{1}{\sqrt{2}}\left(\phi^+_m+\phi^-_{m-1}\right)$: 

\ \ \ \ \ \ \ \ \ $n_{00}(t_0)=1,n_{11}(t_0)=1,n_{22}(t_0)=-1,n_{33}(t_0)=1.$
\item $\psi(t_0)=\frac{1}{\sqrt{2}}\left(\phi^-_m+\phi^+_{m-1}\right)$: 

\ \ \ \ \ \ \ \ \ $n_{00}(t_0)=1,n_{11}(t_0)=1,n_{22}(t_0)=1,n_{33}(t_0)=-1.$
\item $ \psi(t_0)=\frac{1}{\sqrt{2}}\left(\phi^+_m+\phi^+_{m-1}\right)$: 

\ \ \ \ \ \ \ \ \ $n_{00}(t_0)=1,n_{03}(t_0)=1,n_{10}(t_0)=1,n_{13}(t_0)=1.$
\item $ \psi(t_0)=\frac{1}{\sqrt{2}}\left(\phi^-_m+\phi^-_{m-1}\right)$: 

\ \ \ \ \ \ \ \ \ $n_{00}(t_0)=1,n_{03}(t_0)=-1,n_{10}(t_0)=1,n_{13}(t_0)=-1.$
\end{enumerate}

\subsubsection{Adiabatic $Z$ noise}
The noise process is adiabatic in the limit $ \chi_z\Omega^2 \to \infty$, where  $e^{-\chi_z\Omega^2}\to 0 \ \forall \ \Omega>0$.   Hence,  the $\sigma_{\pm}^m$ terms in Eq.\eqref{master equation Z noise} are omitted, and  the resulting master equation in the interaction picture takes the form
\begin{align}
\mathscr{L}[\tilde \rho(t)]=&\frac{V_z^2}{4}\sum_{j,k\in\{m,m-1\}} \cos(\theta_j)\cos(\theta_k) \left(  \sigma_z^{j}\tilde\rho(t)\sigma_z^k- \frac{1}{2}\left[\delta_{jk}\Pi(j),\tilde\rho(t)\right]_+\right) \label{master equation adiabatic Z noise}
 \end{align}

which can also be written in the equivalent form 
\begin{align}
\mathscr{L}[\tilde \rho(t)]=V_z^2&\sum_{\phi_j,\phi_k\in\{\phi^\pm_m,\phi^\pm_{m-1}\}} \<\phi_j|S_z\phi_j\>\<\phi_k|S_z\phi_k\> \nonumber \\& \ \ \ \ \ \ \ \ \ \ \  \times\left( \Pi(\phi_j)\tilde\rho(t)\Pi(\phi_k)- \frac{1}{2}\left[\Pi(\phi_k)\Pi(\phi_j),\tilde\rho(t)\right]_+\right). \label{master equation adiabatic Z noise second form}
 \end{align}
 
The ensuing Markovian dynamics will only result in exponential dephasing, which we calculate as $\sqrt{\mathrm{tr}[\sigma^{eg}_x \tilde\rho(\tau)]^2+\mathrm{tr}[\sigma^{eg}_y \tilde\rho(\tau)]^2}=e^{-\tau/T_2}$. $1/T_2$ is the dephasing rate, where $T_2$ is the dephasing time. 
 The dephasing rate is given by
\begin{equation}
\frac{1}{T_2}=\frac{V_z^2}{8}\left(\cos[\theta_m]+\cos[\theta_{m-1}] \right)^2 \equiv\frac{V_z^2}{2}\left(\<\phi^\pm_m|S_z\phi^\pm_m\>-\<\phi^\mp_{m-1}|S_z\phi^\mp_{m-1}\> \right)^2\label{ESR adiabatic dephasing}
\end{equation}
for the initial states $\psi=\alpha\phi^\pm_m+\beta\phi^\mp_{m-1}$, and 
\begin{equation}
\frac{1}{T_2}=\frac{V_z^2}{8}\left(\cos[\theta_m]-\cos[\theta_{m-1}] \right)^2 \equiv\frac{V_z^2}{2}\left(\<\phi^\pm_m|S_z\phi^\pm_m\>-\<\phi^\pm_{m-1}|S_z\phi^\pm_{m-1}\> \right)^2\label{NMR adiabatic dephasing}
\end{equation}
for the initial states  $\psi=\alpha\phi^\pm_m+\beta\phi^\pm_{m-1}$. In both cases, the initial superpositions in the high-field limit are, respectively, given by $\psi^\mathfrak{e} \otimes| m_I\>$ and $|m_S\>\otimes\psi^\mathfrak{n}$. Because, in the high-field limit, adiabatic $Z$ noise results in a pure dephasing channel acting  on the electronic spin subspace the dephasing rate of Eq.\eqref{ESR adiabatic dephasing} maximises to $V_z^2/2$, whereas that of Eq.\eqref{NMR adiabatic dephasing} vanishes entirely. 

However, in subspaces $\{m,m-1\}$ where   $-I+3/2\leqslant m\leqslant 0$, the dephasing rates in both Eq.\eqref{ESR adiabatic dephasing} and Eq.\eqref{NMR adiabatic dephasing} vanish when $\<\phi^\pm_m|S_z\phi^\pm_m\>=\<\phi^\mp_{m-1}|S_z\phi^\mp_{m-1}\>$ and respectively when $\<\phi^\pm_m|S_z\phi^\pm_m\>=\<\phi^\pm_{m-1}|S_z\phi^\pm_{m-1}\>$, even in the low-field regime. We label the magnetic fields at which these conditions are satisfied in the low-field regime as optimal working points (OWPs).

Furthermore, the steady state solution for adiabatic $Z$ noise within the $\{m,m-1\}$ state space $\s(\co^4)$ is given by
\begin{equation} 
\bold{\underline{n}}(\infty)=\mathds{1}\otimes\mathds{1}+c_1\mathds{1}\otimes \sigma_z+c_2\sigma_z\otimes\mathds{1}+c_3\sigma_z\otimes\sigma_z. 
\end{equation}

\subsubsection{Diabatic $Z$ noise}
The noise process is diabatic in the limit $ \chi_z\Omega^2 \to 0$, where  $e^{-\chi_z\Omega^2}\to 1 \ \forall \ \Omega$.  Hence,  the resulting master equation in the interaction picture  is
\begin{align}
\mathscr{L}[\tilde \rho(t)]=&\frac{V_z^2}{4}\sum_{j,k\in\{m,m-1\}} \cos(\theta_j)\cos(\theta_k) \left(  \sigma_z^{j}\tilde\rho(t)\sigma_z^k- \frac{1}{2}\left[\delta_{jk}\Pi(j),\tilde\rho(t)\right]_+  \right) \nonumber \\&+\frac{V_z^2}{4}\sum_{j\in\{m,m-1\}} \sin^2(\theta_j)\left(\sigma_+^j\tilde\rho(t)\sigma_-^j+\sigma_-^j\tilde \rho(t)\sigma_+^j- \frac{1}{2}\left[\Pi(j),\tilde\rho(t)\right]_+ \right).  \label{master equation diabatic Z noise}
 \end{align}
 Since the terms with $\Omega >0$ are included, there will generally be both dephasing and depolarisation.  There is, however, a nuance to be considered as regards to the measurement of the depolarisation rate. The depolarisation process affects individual $m$ subspaces, where $\mathrm{tr}[\sigma_z^m \tilde\rho(t)]=e^{-\tau/T_1}$, with the depolarisation rate being given by 
\begin{equation}
\frac{1}{T_1}  =\frac{V_z^2}{2}\sin(\theta_m)^2\equiv2V_z^2\<\phi^-_m|S_z\phi^+_m\>^2
\end{equation}
which maximises at the  type I cancelation resonance, and vanishes in the high-field limit. However, we do not observe this directly with magnetic resonance. Recall that the measurement procedure for detecting depolarisation is $\mathrm{tr}[\sigma_z^{01}\tilde \rho(t)]$, which probes two different subspaces, each of which is depolarising at an exponential rate. Consequently, given  $T_1^{0}$ as the depolarisation time of the excited state and $T_1^1$ as that of the ground state, and $P^0$ as the initial population of the excited state and $P^1$ as that of the ground state, the measured depolarisation would be
\begin{equation}
\mathrm{tr}[\sigma_z^{01}\tilde\rho(t)] =\frac{1}{2}P^0\left(1 +e^{-t/T_1^0}\right) - \frac{1}{2}P^1\left(1 +e^{-t/T_1^1}\right).  \label{effectivedepolarisation}
\end{equation}
Consequently, we cannot, in general, ascribe an exponential depolarisation time for the dynamical selection rule allowed subspaces.

As regards to the dephasing measurement, we may repeat the approach used for the adiabatic case. The dephasing rate is given by
\begin{equation}
\frac{1}{T_2}=\frac{V_z^2}{4}\left(1+\cos[\theta_m]\cos[\theta_{m-1}] \right) \equiv V_z^2\left(\frac{1}{4}-\<\phi^\pm_m|S_z\phi^\pm_m\>\<\phi^\mp_{m-1}|S_z\phi^\mp_{m-1}\> \right)\label{ESR diabatic dephasing}
\end{equation}
for the initial states $\psi=\alpha\phi^\pm_m+\beta\phi^\mp_{m-1}$, and 
\begin{equation}
\frac{1}{T_2}=\frac{V_z^2}{4}\left(1-\cos[\theta_m]\cos[\theta_{m-1}] \right) \equiv V_z^2\left(\frac{1}{4}-\<\phi^\pm_m|S_z\phi^\pm_m\>\<\phi^\pm_{m-1}|S_z\phi^\pm_{m-1}\> \right)\label{NMR diabatic dephasing}
\end{equation}
for the initial states  $\psi=\alpha\phi^\pm_m+\beta\phi^\pm_{m-1}$. As with the case of adiabatic $Z$ noise, in the high-field limit the dephasing rate for Eq.\eqref{ESR diabatic dephasing} maximises to $V_z^2/2$, whereas that for Eq.\eqref{NMR diabatic dephasing} vanishes. However, due to the presence of depolarisation in the low-field limit, there are no longer any OWPs for diabatic $Z$ noise, and the dephasing never vanishes in the low-field regime. Indeed, the dephasing rate of both Eq.\eqref{ESR diabatic dephasing} and Eq.\eqref{NMR diabatic dephasing} reach  $\sim V_z^2/4$ near the optimal working point of $\cos(\theta_m)=-\cos(\theta_{m-1})$, which is half of the maximum value that is realised for Eq.\eqref{ESR diabatic dephasing} in the high-field limit, but the maximum value that Eq.\eqref{NMR diabatic dephasing} attains at any field.

Additionally, we note that the stationary states, given an initial superposition that does not involve subspaces $m=\pm(I+1/2)$, are
\begin{equation}
\bold{\underline{n}}(\infty)=\mathds{1}\otimes\mathds{1}+c_1\sigma_z\otimes\mathds{1} \end{equation}
whereas those involving such subspaces are
\begin{equation}
\bold{\underline{n}}(\infty)=\mathds{1}\otimes\mathds{1}+c_1(\sigma_z\otimes\sigma_z-\one\otimes\sigma_z)+c_2\sigma_z\otimes\one. \end{equation}

\subsection{A comment on $X$ noise}
The treatment of $X$ noise, where the interaction Hamiltonian of the open system couples $S_x$  to the bath, is not so simple to treat analytically in general. First of all, there are no $S_x(\Omega=0)$ terms and, hence, we can only consider the diabatic case. More importantly $S_x$ couples the entire Hilbert space. As a result we are unable to truncate $S_x$ into a subspace of a more manageable size.   As  $I$ increases, thereby enlarging the dimension of the system Hilbert space, the number of simultaneous Bloch equations needed to be considered would quickly grow untractable.  

However, by simple observation we can infer certain properties of $X$ noise. Firstly, in the low-field regime    the stationary state for $X$ noise is the maximally mixed state $\frac{1}{d} \mathds{1}$. This is because both $\<\phi^\pm_{m-1}|S_x  \phi^\pm_m\>$ and $\<\phi^\mp_{m-1}|S_x  \phi^\pm_m\>$ are non-zero in the low-field regime given any $m$.  In the high-field regime, on the other hand, where the electron and nuclear spins in the eigenstates of $H_0$ are separable,  $X$ noise is  restricted to the $\{m,m-1\}$ subspace.  The Lindblad master equation, in the interaction picture,  is
\begin{equation}
\lim_{(\gamma_e+\gamma_n)B_0/A_{\mathrm{iso}}\to \infty}\mathscr{L}[\tilde \rho(t)]=\frac{V_x^2}{4}\left(\sum_{\Omega}   
\left[  \sigma_+(\Omega)\tilde\rho(t)  \sigma_-(\Omega)+\sigma_-(\Omega)\tilde\rho(t)  \sigma_+(\Omega)  \right] - \tilde \rho(t)\right)
\end{equation}
where $\sigma_\pm(\Omega)=|\pm1/2\>\<\mp1/2|\otimes\Pi_{m_I}$ such that 
\begin{equation}
\mathrm{tr}[H_0(\Pi_{{\pm1/2}}\otimes\Pi_{{m_I}}-\Pi_{{\mp1/2}}\otimes\Pi_{{m_I}})]=\Omega.
\end{equation} Note that the Lindblad operators contain projectors on the nuclear spin subspace, and not the identity operator; $X$ noise does not act locally on the electron spin even in the high-field limit. This shouldn't be surprising as $[H_0,S_x]_- \ne \mathds{O}$.

 Given an initial ESR superposition 
\begin{equation}
\lim_{(\gamma_e+\gamma_n)B_0/A_{\mathrm{iso}}\to \infty}\alpha\phi^+_m+\beta\phi^-_{m-1}=\left(\alpha\left|+1/2\right\>+\beta\left|-1/2\right\>\right)\otimes\left|m-1/2\right\>\equiv\psi^{\mathrm{ESR}}\otimes\left|m-1/2\right\>
\end{equation}
the depolarising  channel $\e$ will result in the statistical mixture 
\begin{align}
\e: \Pi(\psi)^{\mathrm{ESR}}\otimes\Pi_{{m-1/2}}&\mapsto\left(\frac{1}{2}\Pi_{{+1/2}}+\frac{1}{2}\Pi_{{-1/2}}\right)\otimes\Pi_{{m-1/2}},\nonumber \\
&= \frac{1}{2}\Pi(\phi^+_m)+\frac{1}{2}\Pi(\phi^-_{m-1}).
\end{align}
Here, the exponential depolarisation rate is
\begin{equation}
\frac{1}{T_1}=\frac{V_x^2}{2}
\end{equation}
and the dephasing rate is half this
\begin{equation}
\frac{1}{T_2}=\frac{V_x^2}{4}.
\end{equation}

Let us now turn to the case of an NMR superposition 
\begin{align}
\lim_{(\gamma_e+\gamma_n)B_0/A_{\mathrm{iso}}\to \infty}\alpha\phi^\pm_m+\beta\phi^\pm_{m-1}&=  \left|\pm1/2\right\> \otimes \left(\alpha\left|m\mp1/2\right\>+\beta\left|m-1 \mp1/2\right\>\right)\nonumber \\ &\equiv\left|\pm1/2\right\>\otimes\psi^{\mathrm{NMR}}.
\end{align}
In this case, the  depolarising channel $\e$ will lead to a statistical mixture
 \begin{align}
\e:\Pi_{{\pm1/2}}\otimes \Pi(\psi^{\mathrm{NMR}})&\mapsto\frac{1}{2}\Pi_{{\pm1/2}}\otimes\left(\Pi_{{m\mp1/2}} + \Pi_{{m-1\pm1/2}}\right)\nonumber \\ & \ \ \ \ \ \ +\frac{1}{2}\Pi_{{\mp1/2}}\otimes\left(\Pi_{{m\pm1/2}}+\Pi_{{m-1\pm1/2}}\right),\nonumber \\
&=\frac{1}{4}\left(\Pi(\phi^+_m)+\Pi(\phi^-_m)+\Pi(\phi^+_{m-1})+\Pi(\phi^-_{m-1}) \right).
\end{align}

\section{Non-Markovian dynamics due to a spin bath}
\subsection{Pure decoherence due to weak spin-bath coupling}
Let the nuclear-electronic spin system be system $A$  governed by the Hamiltonian $H^A \in \lo_s(\h^A)$ given by Eq.\eqref{nuclear-electronic spin system hamiltonian}, and let the  bath of $N$ spin objects be system $B$ governed by the Hamiltonian $H_B \in \lo_s(\h^B)$. Finally, let  the electron spin of the nuclear-electronic spin system interact with  the spin bath with an interaction Hamiltonian $H_I \in \lo_s(\h^A\otimes \h^B)$.  Hence,  the total Hamiltonian describing the system-bath evolution is 
\begin{align}
 H &=  H^A +  H_I +  H^B \nonumber \\
 H_I &=  \sum_{i \in {\{x,y,z\}}}S_i\otimes  \mb_{i}  
\end{align} 
where $\mb_{i}$ is  some self-adjoint operator acting on the whole bath conditional on   the operator $S_i$ acting on the electron of the nuclear-electronic spin system. As stated previously, the decoherence will be pure with respect to the eigenbasis of $H^A$ only if $[H^A,H_I]_-=\mathds{O}$. Here we show how the decoherence can be pure in a perturbative sense if this commutation relation is not satisfied.

 By expressing the operators in the eigenbasis of the free Hamiltonians $H^A$ and $H^B$ as 
\begin{align}
S_i&=\sum_\Omega S_i(\Omega) & \mb_{i}=\sum_\omega \mb_{i}(\omega)
\end{align} 
 we may write  the interaction Hamiltonian in the interaction picture as  \begin{align}
\tilde H_I(t)&= \sum_{i \in \{x,y,z\}} \sum_{\omega,\Omega}e^{-\imag (\Omega + \omega) t}S_i(\Omega)\otimes\mb_{i}(\omega)
\end{align}
and a RWA interaction Hamiltonian, including only the terms diagonal with respect to the eigenbasis of $H^A$
 \begin{align}
\tilde H_I(t)^{\mathrm{RWA}}&=  \sum_{\omega,m}e^{-\imag \omega t}\left(\<\phi^+_m|S_z\phi^+_m\>\Pi(\phi^+_m)+\<\phi^-_m|S_z\phi^-_m\>\Pi(\phi^-_m)\right)\otimes\mb_{z}(\omega).
\end{align}

 The unitary operator obtained by the Dyson series for $\tilde H_I(t)$ is 
\begin{align}
\tilde U_{t,t_0}=\mathds{1}+\frac{1}{l!}\sum_{l=1}^\infty \left(\sum_{i \in \{x,y,z\}}\sum_{\omega,\Omega}\frac{(e^{-\imag (\Omega + \omega) t}-e^{-\imag (\Omega + \omega) t_0})}{\Omega + \omega}S_i(\Omega)\otimes\mb_{i}(\omega)\right)^l
\end{align} 
which is convergent by $\tilde U_{t,t_0}^{\mathrm{RWA}}$ in the sense that $d(\tilde U_{t,t_0}^{\mathrm{RWA}}\psi, \tilde U_{t,t_0}\psi) < \epsilon \ \forall \ \psi \in \h^A\otimes \h^B$ if
\begin{align}
\left|\sum_{i \in \{x,y,z\}}\sum_{\omega,\Omega}\frac{(e^{-\imag (\Omega + \omega) t}-e^{-\imag (\Omega + \omega) t_0})}{\Omega + \omega}\<\psi^\perp |S_i(\Omega)\psi\>\<\varphi|\mb_{i}(\omega)\phi\>\right|  \ll 1  \  \ 
\end{align}

for all  pairs of orthogonal $H^A$ eigenstates $\{\psi,\psi^\perp\} \in \h^A$ and all pairs of $H^B$ eigenstates $\{\phi,\varphi\} \in \h^B$.  \footnote{In other words, $\phi$ and $\varphi$ may be the same eigenstate.} This condition is satisfied if  $\|H^A\| \gg \|H_I\|$.

The unitary operator $\tilde U_{t,t_0}^{\mathrm{RWA}}$  can be shown to be of the form \begin{align}
\tilde U_\tau^{\mathrm{RWA}}&=\sum_m\Pi(\phi^+_m)\otimes U_\tau^{\phi^+_m}+\Pi(\phi^-_m)\otimes U_\tau^{\phi^-_m}, \nonumber \\
U_\tau^{\phi^\pm_m}&=e^{-\imag\tau\left(H_B+\<\phi^\pm_m|S_z\phi^\pm_m\>\mb_{z}   \right)}.\label{RWA spin-bath entangling unitary}
\end{align} 
We may identify  $\{U_\tau^{\phi^\pm_m}\}$ as unitary operators on the bath, \emph{conditional} on the system being in the state $\phi^\pm_m$. Because all terms of the Hamiltonian permitting transitions between the $H^A$ eigenstates were removed by the RWA, the quantum channel on the system Hilbert space that results from $\tilde U_\tau^{\mathrm{RWA}}$ leads to  only pure decoherence.

\subsection{Parametric decoupling}
As was the case for the Markovian model of decoherence,  we are interested in establishing a superposition between two eigenstates of $H^A$ with magnetic resonance. These correspond to the transitions of types $\phi^\pm_m \leftrightarrow \phi^\pm_{m-1}$ and $\phi^\pm_m \leftrightarrow \phi^\mp_{m-1}$. We may generally consider such a superposition as
\begin{equation}
\psi=\alpha\varphi_0+\beta\varphi_1 . 
\end{equation}
 Additionally, let us assume that the system and environment are initially in the product state $\rho=\Pi(\psi)\otimes\varrho$ which we may express  as
\begin{equation}
\rho=\sum_iP(i)|\phi_i\>\Pi(\psi) \<\phi_i|.
\end{equation}
 Furthermore, the terms of the unitary operator in Eq.\eqref{RWA spin-bath entangling unitary} that we need to consider here are 
\begin{equation}
(\Pi(\varphi_0)+\Pi(\varphi_1))\tilde U_\tau ^{\mathrm{RWA}}(\Pi(\varphi_0)+\Pi(\varphi_1))=\Pi(\varphi_0)\otimes U_\tau^{\varphi_0}+\Pi(\varphi_1)\otimes U_\tau^{\varphi_1}\label{spin bath unitary operator}
\end{equation}
as, due to the orthogonality of the projectors $\Pi(\phi^\pm_m)$, the other terms are not supported by this subspace.    This (generally) entangling unitary operation, together with measurement of the environment  as per Stinespring's dilation theorem discussed in Sec.\ref{Stinespring's dilation theorem}, will lead to the  dephasing channel
\begin{align}
\e_\tau[\Pi(\psi)]&=\sum_{i,j}P(i)\<\phi_j|\tilde U_\tau|\phi_i\>\Pi(\psi) \<\phi_i|\tilde U_\tau^\dagger |\phi_j\>.\nonumber \\
&=\sum_{i}P(i)\left(\<U_\tau^{\varphi_0}\phi_i| U_\tau^{\varphi_1}\phi_i\>\Pi(\varphi_1)\Pi(\psi) \Pi(\varphi_0) + \mathbb{H.C} \right) \nonumber \\ &  \ \ \   +\Pi(\varphi_0)\Pi(\psi) \Pi(\varphi_0) + \Pi(\varphi_1)\Pi(\psi) \Pi(\varphi_1).\label{spin bath quantum channel}
\end{align}
 The dephasing, such that it can be measured in magnetic resonance experiments, is calculated at arbitrary evolution time $\tau$ as
\begin{align}
\sqrt{\mathrm{tr}[\sigma_x^{01}\e_\tau[\Pi(\psi)]]^2+\mathrm{tr}[\sigma_y^{01}\e_\tau[\Pi(\psi)]]^2}&\equiv2|\<\varphi_0|\e_\tau[\Pi(\psi)]\varphi_1\>|,\nonumber \\&=2\sum_{i}P(i)\left| \<U_\tau^{\varphi_1} \phi_i|U_\tau^{\varphi_0}\phi_i\>\<\varphi_0|\psi\>\<\psi|\varphi_1\>\right|, \nonumber \\
&=2|\alpha\beta|\sum_{i}P(i)\left|\<U_\tau^{\varphi_1} \phi_i| U_\tau^{\varphi_0}\phi_i\>\right|.\label{spin-bath decoherence time equation}
\end{align}
 The exact determination of this quantity for any arbitrary evolution time $\tau$ will require long computation times, which grow exponentially with the size of the bath.  There are numerical approximation techniques available, however, such as the cluster correlation expansion \citep{Witzel-spectral-diffusion,Yang-spectral-diffusion} that, provided the convergence criteria are satisfied,  enable this value to be estimated at much shorter computation times.

By reflecting upon the nature of the quantum channel in Eq.\eqref{spin bath quantum channel}, two methods of removing the decoherence present themselves; either ensure no measurements are carried out on the environment such that the composite system will evolve unitarily, and hence reversibly, or decouple the evolution of the system from that of the environment. The former is impossible to achieve, as the spin bath is designated the term ``environment'' precisely because processes taking place therein are beyond the scope of the experimentalist to control; the ``measurements'' on the environment are not carried out by any person, but are simply  physical processes where the bath spins themselves interact with other degrees of freedom. The only avenue left open is thus decoupling of the evolution. 

In the special case where the interaction Hamiltonian commutes with the free evolution Hamiltonian of the bath, $[H_I,H_B]_-=\mathds{O}$, we may use the Hahn echo to carry out dynamical decoupling as was discussed earlier in Sec.\ref{Hahn echo sec}. This is because the commutation of the two Hamiltonians will allow us to identify $U_\tau^{\varphi_0}U_\tau^{\varphi_1}=U_\tau^{\varphi_1}U_\tau^{\varphi_0}$.  More generally, however, this commutation relation will not be satisfied, and the Hahn echo cannot achieve dynamical decoupling. There is, however, a method of decoupling the dynamics, not by dynamical intervention, but rather by tuning of the parameter of the external magnetic field $B_0$. Hence, we shall label this appropriately as \emph{parametric decoupling}.  

Consider again the unitary operator Eq.\eqref{spin bath unitary operator} from our example above. This is an operator with a Schmidt-rank of two, and hence is inseparable, if and only if the conditional unitary operators are not equal: $U_\tau^{\varphi_0}\ne U_\tau^{\varphi_1}$. However, these unitaries are respectively functions of   $\<\varphi_0|S_z\varphi_0\>$ and $\<\varphi_1|S_z\varphi_1\>$ which  are equal  at the  optimal working point. At such a magnetic field value, therefore, the two conditional unitaries will be identical, and Eq.\eqref{spin bath unitary operator} will simplify to
\begin{align}
(\Pi(\varphi_0)+\Pi(\varphi_1))\tilde U_\tau^{\mathrm{RWA}}({B_0^{\mathrm{OWP}}})(\Pi(\varphi_0)+\Pi(\varphi_1))
&=(\Pi_\tau^{\varphi_0}+\Pi_\tau^{\varphi_1})\otimes U
\end{align} 
where $U=U_\tau^{\varphi_0}=U_\tau^{\varphi_1}$. This is decoupled with respect to the system-bath partition, and cannot establish entanglement between the subspaces in question. 

It should be noted, however, that decoherence has not been removed entirely at the OWPs. Our arguments have rested upon the structure of the unitary operator $\tilde U_\tau^{\mathrm{RWA}}$ which is generated by the approximative interaction Hamiltonian that is diagonal with respect to the system Hamiltonian. At the optimal working points, the effect of these non-diagonal terms in the Hamiltonian will become dominant and limit the coherence time.

\section{Summary}
In this chapter we studied the open system dynamics of nuclear-electronic spin systems, where we posited that only the electron spin interacts with the environment.  Both a phenomenological Markovian model and a microscopic model of a non-Markovian spin bath were investigated. In both cases, if the interaction Hamiltonian involves only the $S_z$ operator acting on the electron spin, and is weak enough in comparison with the energy splitting of the system Hamiltonian, what results is  pure decoherence with respect to the system Hamiltonian eigenbasis in a perturbative limit. Furthermore, two-dimensional subspaces spanned by $\{\phi^\pm_m,\phi^\mp_{m-1}\}$ or $\{\phi^\pm_m,\phi^\pm_{m-1}\}$ such that $-I+3/2\leqslant m \leqslant 0$ provide decoherence free subspaces at specific magnetic fields where, respectively, the conditions $\<\phi^\pm_m|S_z\phi^\pm_m\>=\<\phi^\mp_{m-1}|S_z\phi^\mp_{m-1}\>$ and $\<\phi^\pm_m|S_z\phi^\pm_m\>=\<\phi^\pm_{m-1}|S_z\phi^\pm_{m-1}\>$ are satisfied. We call these magnetic field values the optimal working points (OWPs) which  coincide with the FSPs in the limit of $\delta_\gamma \to 0$. The reason pure decoherence is suppressed at the OWPs is that at such values of the magnetic field, which is a parameter of the system Hamiltonian, the joint unitary channel acting on the system and its environment decouples with respect to the system-environment partition, and hence does not establish any entanglement between them. We therefore call this phenomenon parametric decoupling, in contrast with dynamical decoupling which uses dynamical intervention to decouple a system's evolution from that of its environment.  

However, this decoupling of the interaction applies to the approximate unitary evolution operator acting on the system and its environment; the un-perturbative unitary operator does not decouple, and at the OWPs the coherence time may be increased by orders of magnitude, but is not removed entirely.

\spacing{1}                                  
 \bibliographystyle{plainnat}
 \bibliography{references}

\spacing{1}
\chapter{Bismuth doped silicon and phosphorus doped silicon: a comparative study}
\section{Introduction}
Now that we have completed our general study of closed and open system dynamics of nuclear-electronic spin systems, we may begin to apply our findings to two concrete examples: phosphorus doped silicon (Si:P) and bismuth doped silicon (Si:Bi). Although the former has been subject to extensive study, and hence is of little interest here, we present it alongside the more novel system of Si:Bi as a means of comparison. To this end, we require numerical values for the parameters defining the system Hamiltonian $H_0$, which are provided in Table \ref{Si:Bi constants}.
\begin{table}[!htb]
\centering
\begin{tabular}{|c||c||c|}\hline
&Si:P & Si:Bi \\\hline
$S$& 1/2 & 1/2 \\\hline
$I$&1/2 & 9/2 \\\hline
$\gamma_e$&27.974 GHz/T & 27.997 GHz/T \\\hline
$\gamma_n$&17.251 MHz/T & 6.963 MHz/T \\\hline
$\delta_\gamma$ &$6.167\times 10^{-4}$ & $2.487\times 10^{-4}$ \\\hline
$A_{\mathrm{iso}}$ &117.5 MHz & 1.4754 GHz \\\hline
\end{tabular}
\caption{Numerical constants for Si:P and Si:Bi }\label{Si:Bi constants}
\end{table}

The electron gyromagnetic ratio is given by $\gamma_e=\beta_e g_e$, where $\beta_e=13.9962$ GHz/T is the Bohr magneton, and $g_e$ is the electron g-factor which depends on the substance. For Si:P and Si:Bi, the electron g-factor is provided by   \citep{Si:P-gfactor} and \citep{FeherEndor} respectively. Similarly, the nuclear gyromagnetic ratio is given as $\gamma_n=\beta_n g_n$ where $\beta_n=7.6$ MHz/T is the nuclear magneton and $g_n$ is the nuclear g-factor the value of which, for both bismuth and phosphorus, was taken from NMR tables provided by \url{www.webelements.com}. 

 In what follows I shall take the general equations presented in the previous two sections and provide numerical solutions that compare the magnetic resonance properties of Si:Bi and Si:P.  
\section{Energy spectrum and entanglement}

\begin{figure}[!htb]
\centering
\subfloat[ Si:P.]{\label{SiPenergyentropy}\includegraphics[width=0.5\textwidth]{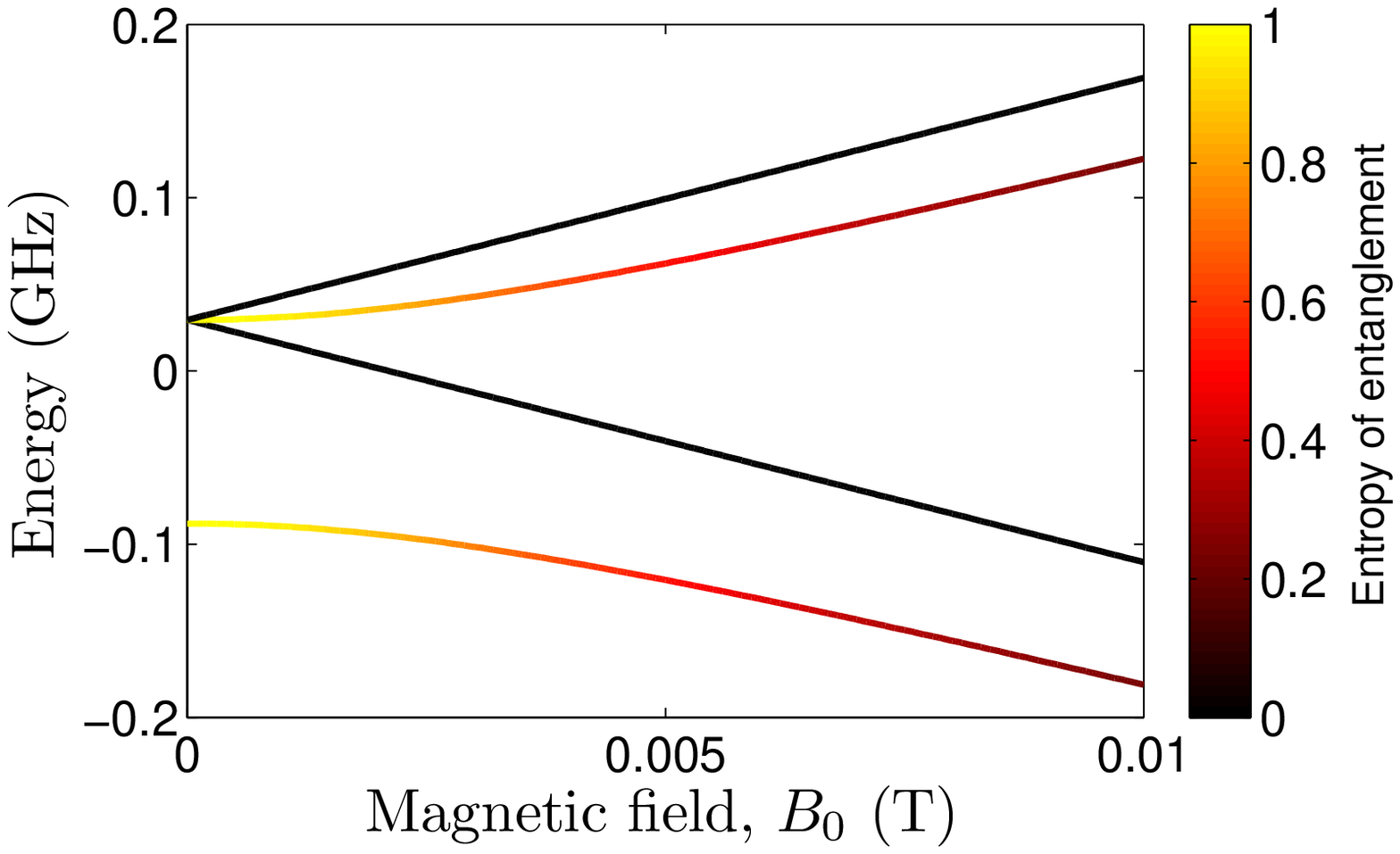}}  \subfloat[Si:Bi. ]{\label{SiBienergyentropy}\includegraphics[width=0.5\textwidth]{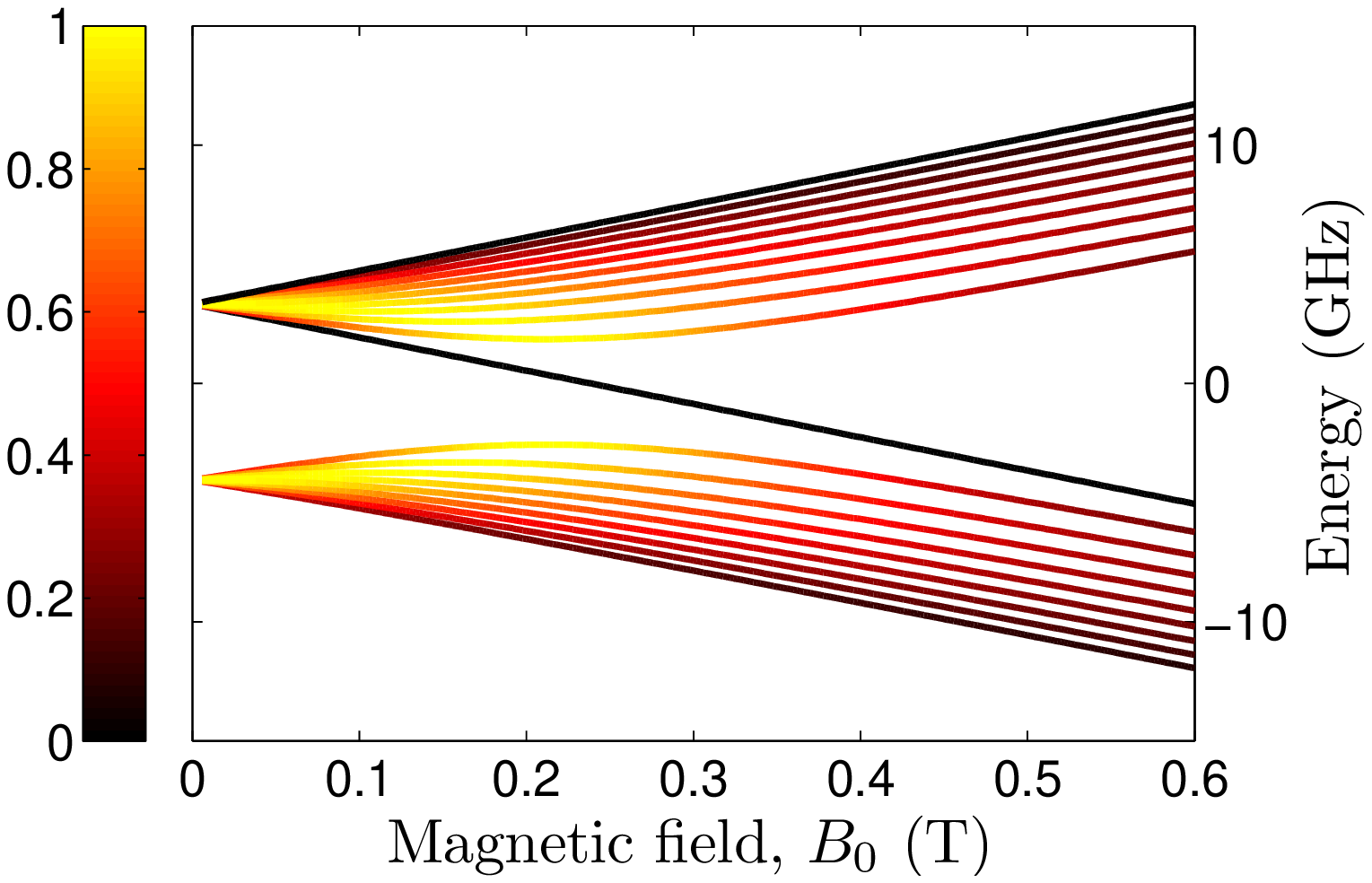}}
\caption{The energy spectrum of Si:P and Si:Bi together with the entropy of entanglement between the electron and nuclear spin of the   eigenstates. This is represented by the colour scale. The entanglement between the electron and nuclear spins can be seen to maximise near the avoided crossings, where the type I cancelation resonance takes place.}
\end{figure}
Figures \ref{SiPenergyentropy} and \ref{SiBienergyentropy} depict, for Si:P and Si:Bi respectively, the eigenvalues of the Breit-Rabi Hamiltonian, provided by equations \eqref{H2 energies} and \eqref{H1 energies}, together with the entanglement between the electron and nuclear spin   of each eigenstate, calculated by the entropy of entanglement given by Eq.\eqref{nuc-elec entropy of entanglement}. Let us now compare the key features of the two systems.
\begin{enumerate}[(i)]
\item Si:P has four eigenstates, and is thus described by the state space $\s(\co^4)$, whereas Si:Bi owing to its larger nuclear spin has twenty eigenstates, exhibiting a state space $\s(\co^{20})$.
\item At  magnetic fields that are larger than zero, the eigenstates separate into two clusters with respect  to their relative energies; the bottom cluster are the states $\{\phi^-_m\}$ and the top cluster are the states $\{\phi^+_m\}$.

\item The larger energies of Si:Bi compared with Si:P are due to the order of magnitude larger isotropic hyperfine interaction strength. This, together with the value of the nuclear spin $I$, also determines what magnetic field ranges constitute the low-field regime; for Si:P this is approximately $0-0.01$ T, whereas for Si:Bi it is  $0-0.6$ T. 

\item The low-field regime is  where the entanglement of the eigenstates, other than states $\phi^\pm
_{ \pm( I+1/2)}$, is appreciably large.  As $\delta_\gamma$ is $\sim 10^{-4}$ in both cases the  energy levels become stationary, with respect to the magnetic field, very close to the type I cancelation resonances. These are the points at which the eigenstates of subspaces $-I+1/2\leqslant m \leqslant0$ are the maximally entangled Bell states $\Psi^\pm$. Si:P has only one subspace, $m=0$, which exhibits this maximisation of the entanglement at zero field. Si:Bi on the other hand has five subspaces, $m=\{0,-1,-2,-3,-4\}$, which have a type I cancelation resonance at the increasingly larger field values of $\{0 \ \mathrm{T},0.05\ \mathrm{T},0.11\ \mathrm{T},0.16\ \mathrm{T},0.21\ \mathrm{T}\}$. 
\end{enumerate}

\begin{figure}[!htb]
\centering
\subfloat[Si:P. ]{\label{SiPthermalconcurrence}\includegraphics[width=0.5\textwidth]{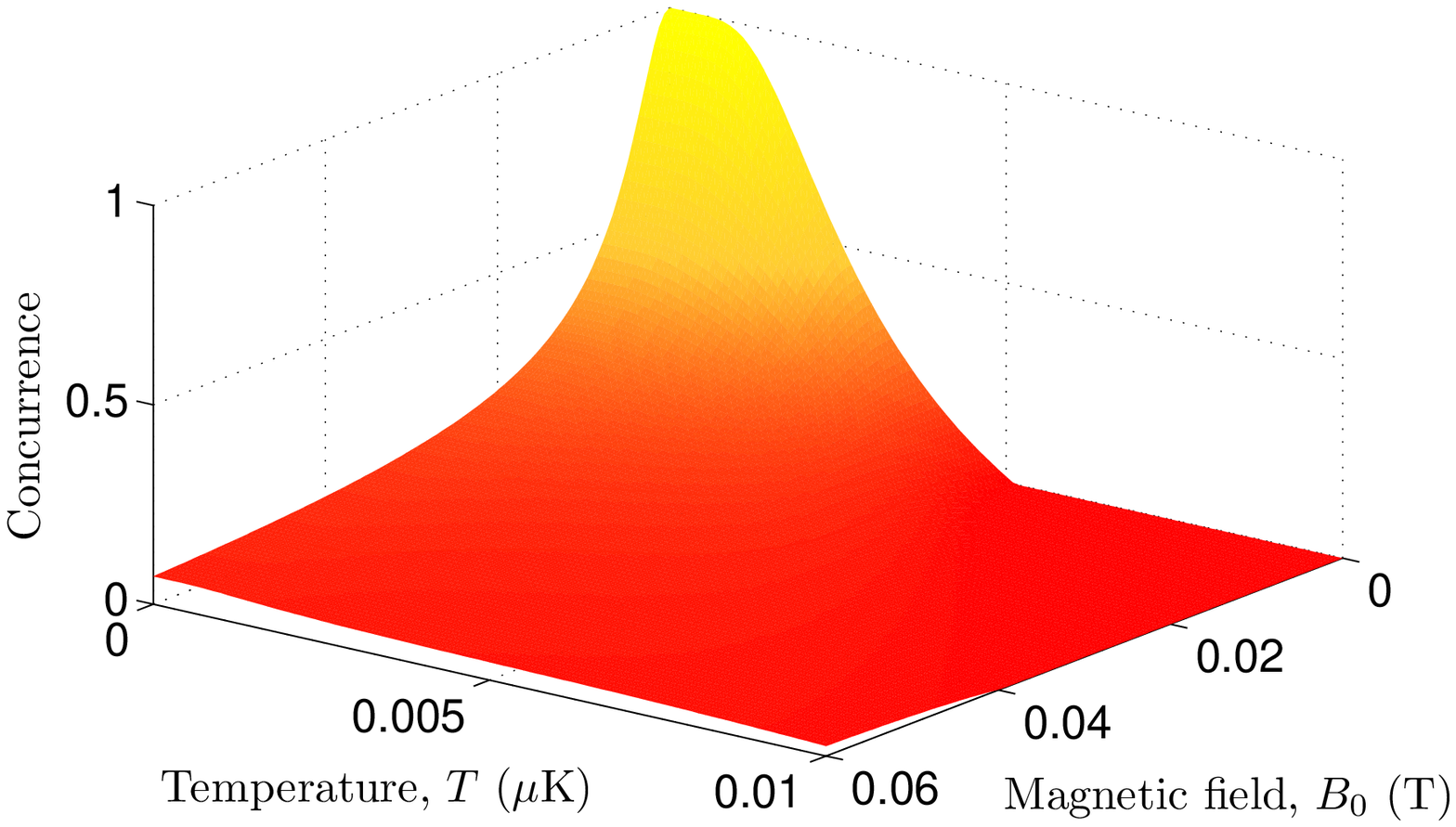}}
 \subfloat[Si:Bi.]{\label{SiBithermalnegativity}\includegraphics[width=0.5\textwidth]{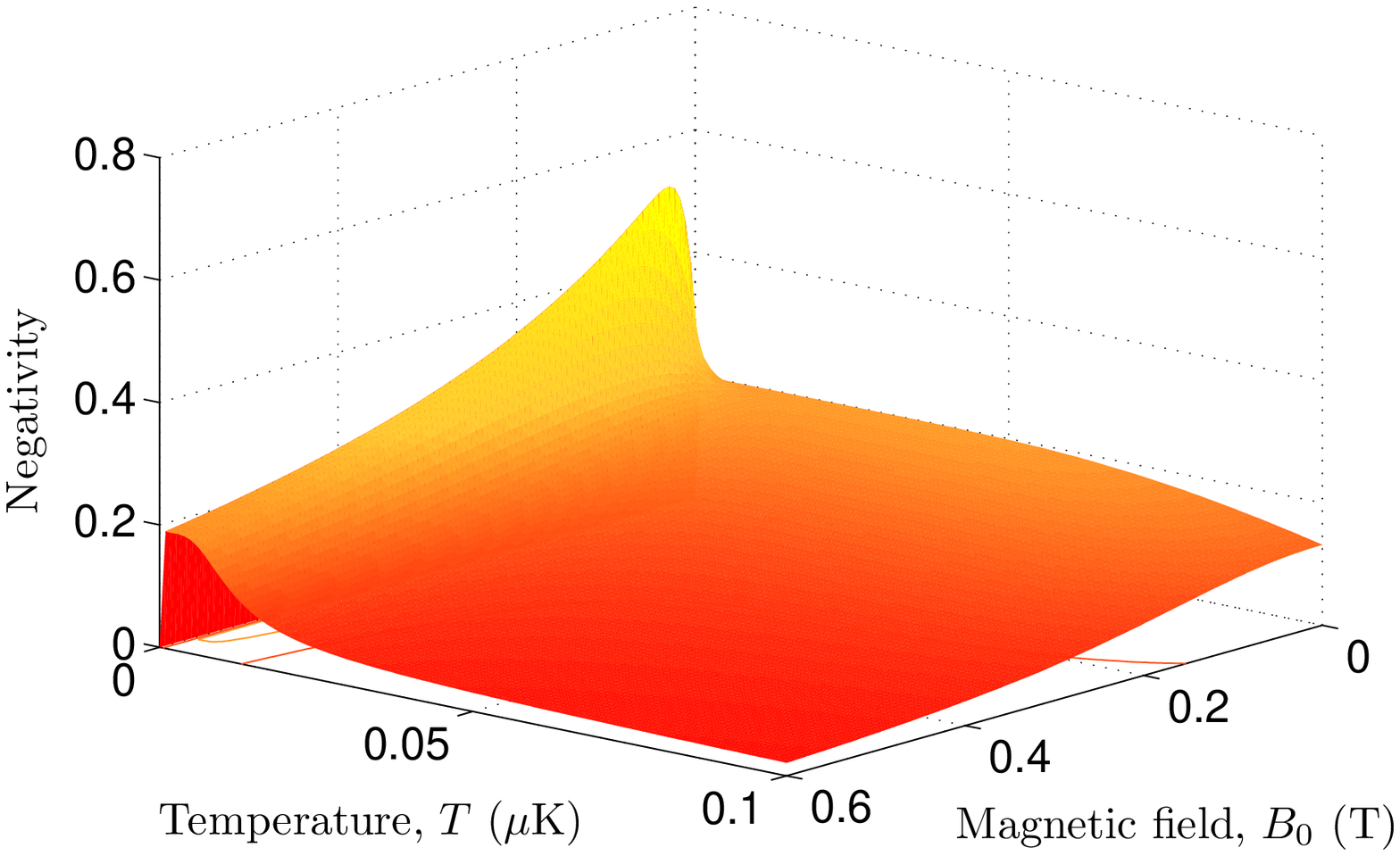}}
\caption{Entanglement between the electron and nuclear spin of the thermal ensemble state $\rho_{\mathrm{th}}$ for Si:P and Si:Bi as a function of temperature and the external magnetic field. }
\end{figure}

In most experimental situations the initial preparation that nature provides \footnote{This can be attributed to open system dynamics that interact the system with a thermal bath, such that $\rho_{\mathrm{th}}$ is the stationary state of the evolution after the system and the bath have reached thermal equilibrium. } is the thermal state $\rho_{\mathrm{th}}$, defined earlier in Eq.\eqref{thermal state}.
  Such states can, given a sufficiently small magnetic field and temperature, exhibit entanglement. This is shown in Figures \ref{SiPthermalconcurrence} for Si:P and \ref{SiBithermalnegativity} for Si:Bi. The measure of entanglement used for Si:P is the concurrence, as it can be considered as  two  coupled qubits.   The composite state space of Si:Bi, on the other hand,   is given by $\co^2\otimes \co^{10}$, for which the concurrence is inappropriate. Consequently the negativity was used instead. As the ground state of Si:P is maximally entangled at zero field, the concurrence of the thermal state is also maximal at zero field and zero temperature. Staying at zero field, an increase of temperature decreases the entanglement due to the reduction in purity of the ensemble state. An increase in the magnetic field will also result in a loss of entanglement. Si:Bi, on the other hand, is more complex. Here, the ground state  $\phi^-_4$ has a negligible entanglement at zero field, and thus a small increase in temperature increases the probability of sampling from the higher energy, more entangled eigenstates. This in turn  results in an increase of entanglement of the thermal state which, at higher temperatures still, vanishes just as with Si:P.

\section{Spectroscopic properties}

\subsection{Continuous wave spectroscopy}

\begin{figure}[!htb]
\centering
\subfloat[c.w. spectra of Si:P.  ]{\label{SiPspectrum}\includegraphics[width=0.5\textwidth]{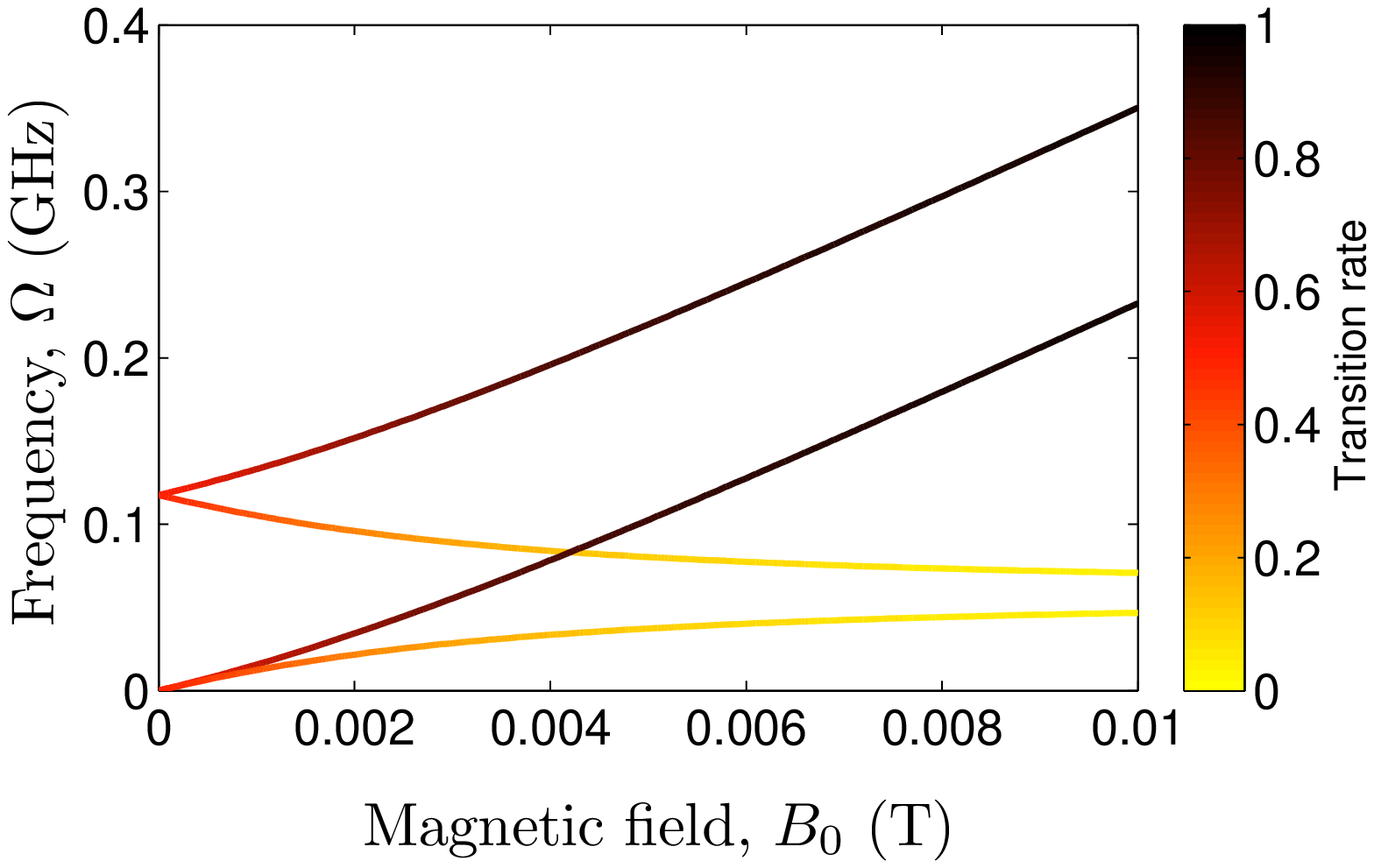}}
      \subfloat[c.w. spectra of Si:Bi]{\label{SiBispectrum}\includegraphics[width=0.5\textwidth]{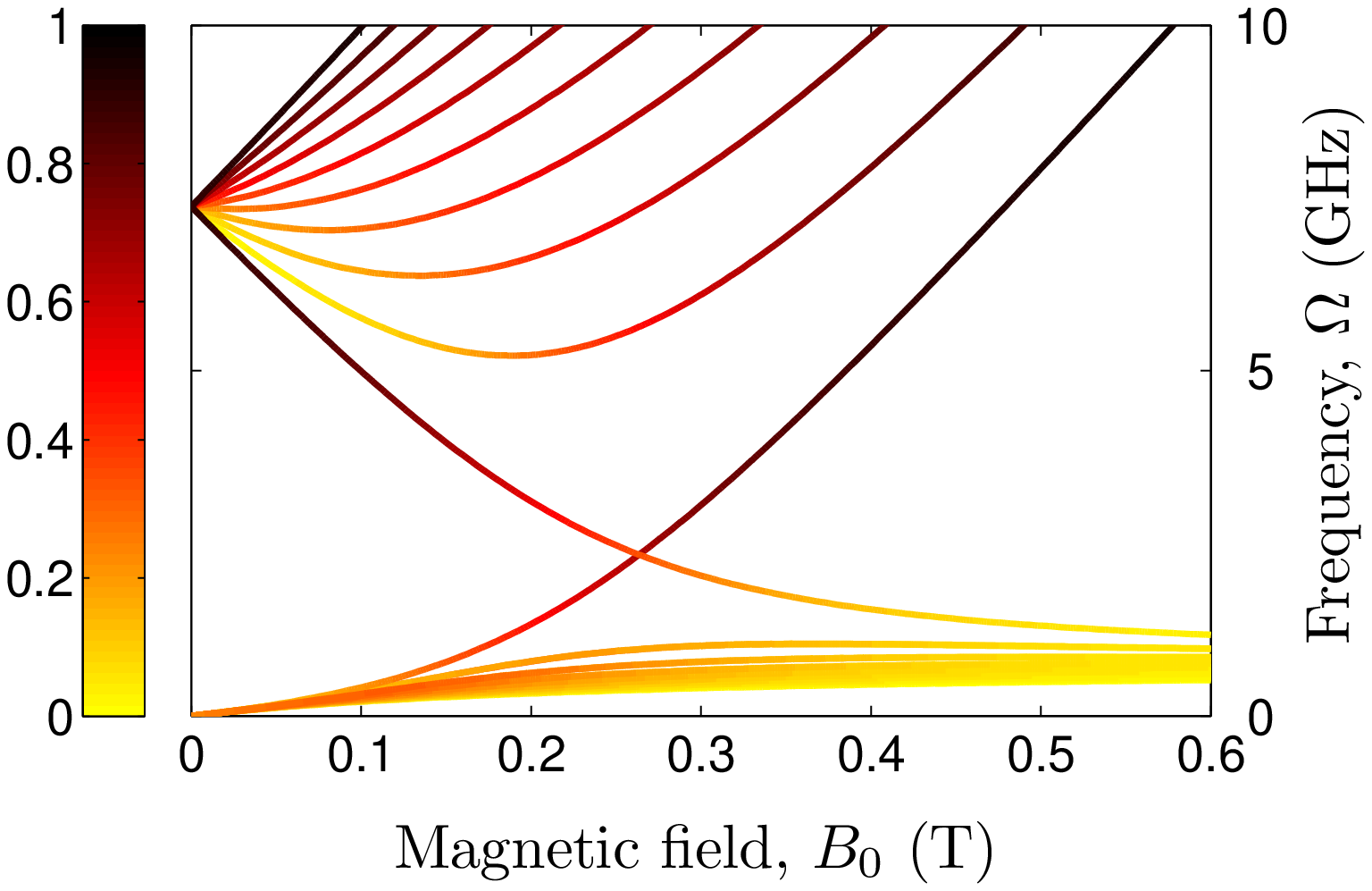}}
\caption{The c.w. spectra of Si:P and Si:Bi. The colour bar shows the relative rates of each transition calculated using the equations \eqref{Iforbp}-\eqref{Iforbidden} normalised so that ESR transition rates are unity. The group of transitions around 60 MHz at 0.01 T for Si:P and 1 GHz at 0.6 T for Si:Bi are, at high fields, described approximately by NMR selection rules. Hence, their transition rates relative to those of the ESR lines become negligibly small at fields larger than $\sim0.01$ T and $\sim0.6$ T respectively.  }
\end{figure}

We may use the relative transition rates of all dipole allowed transitions, calculated in Eqs.\eqref{Iforbp}-\eqref{Iforbidden},  coupled with their frequencies, shown by Eq.\eqref{transition frequencies}, to determine the c.w. spectroscopic properties of both Si:P and Si:Bi. These are shown, respectively, in Figures \ref{SiPspectrum} and \ref{SiBispectrum}. As before, let us consider the key features of the two spectra.

\begin{enumerate}[(i)]
\item  Si:P only has four first order dipole-allowed transitions. These are the ESR-allowed transitions $\{\phi^+_1\leftrightarrow \phi^-_0 ,\phi^+_0\leftrightarrow \phi^-_{-1}\}$ and the transitions that are ESR-allowed at low fields but ESR-forbidden and NMR-allowed at high fields, $\{\phi^+_1 \leftrightarrow \phi^+_0,\phi^-_0 \leftrightarrow \phi^-_{-1}\}$. Of course, as demonstrated experimentally by \citep{morishita-2009}, the double-transitions $\phi^+_0\leftrightarrow\phi^-_0$ and $\phi^+_1\leftrightarrow\phi^-_{-1}$ are also  allowed in second order perturbation theory, albeit with much weaker rates. Our discussion here, however, concerns only first order transitions so these shall be ignored.  Note that the transitions of type $\phi^-_m\leftrightarrow \phi^+_{m-1}$ which are dipole-forbidden at high field are not present here. On the other hand, Si:Bi has a rich spectra  of a total of  thirty six transitions, eight of which are those that are dipole-forbidden at high field.

\item
Si:P, with $I=1/2$, does not contain any FSPs in the low-field regime. \footnote{The Si:P transitions $\phi^+_1\leftrightarrow\phi^+_0$ and $\phi^-_0\leftrightarrow \phi^-_{-1}$ do have FSPs at the field value $B_0\simeq 0.08 $ T. This, however, is in the high-field regime for Si:P.} This is because there is only one subspace where $-I+1/2\leqslant m\leqslant0$ is satisfied, and low-field regime FSPs, as with the OWPs, require at least two such subspaces. Si:Bi on the other hand has, corresponding to $m \in \{0,-1,-2,-3\}$, four frequency minima  at the increasingly larger magnetic fields of \newline$\{0.03 \ \mathrm{T}, 0.08 \ \mathrm{T}, 0.13 \ \mathrm{T},0.19 \ \mathrm{T}\}$, and also the four frequency maxima  at the increasingly smaller field values $\{2.61 \ \mathrm{T}, 0.87 \ \mathrm{T}, 0.52 \ \mathrm{T},0.37 \ \mathrm{T}\}$. \footnote{The frequency maximum that occurs at 2.61 T, however, also constitutes the high-field regime for Si:Bi. } 

\item

Due to the larger hyperfine interaction strength and  nuclear spin of Si:Bi, the transition from    low-field to high-field regime occurs at larger magnetic field values in   Si:Bi than for  Si:P. Consequently the transitions  of type $\phi^\pm_m\leftrightarrow \phi^\pm_{m-1}$ in Si:Bi continue to be ESR-allowed at magnetic fields where the same transitions are  approximately only NMR-allowed in Si:P. 

\end{enumerate}

\subsection{Pulsed spectroscopy}\label{pulsed spectroscopy bismuth}
For a given driving field strength $\omega_1$ the Rabi frequency, determined by the  nutation experiment in pulsed spectroscopy (see Sec.\ref{nutation measurement scheme}), varies according to Eq.\eqref{Rabi frequency nuc-elec}, the transition rates of c.w. spectroscopy being the square of which. Therefore in the high-field limit the Rabi frequencies of NMR transitions $\phi^\pm_m\leftrightarrow \phi^\pm_{m-1}$ are approximately three orders of magnitude slower than those of the ESR  transitions $\phi^+_m\leftrightarrow \phi^-_{m-1}$. In the low-field regime, however, due to the entanglement in the eigenstates,  the  transitions $\phi^\pm_m\leftrightarrow \phi^\pm_{m-1}$ become ESR-allowed and consequently gain a three-orders of magnitude speed-up. While this may seem to suggest that \emph{accurate} control of these subspaces is also sped up by three orders of magnitude in the low-field regime, we must not neglect the issue of selectivity. If the transition frequency gaps $\Delta(\Omega)$, which limit the Rabi frequency due to the relationship in Eq.\eqref{nuc-elec coherent controll RWA condition}, get smaller in the low-field regime, then in order to achieve the same level of accuracy as in the high-field regime \emph{slower} pulses must be used. As discussed in  Sections \ref{transition frequencies section} and \ref{nuc-elec coherent control}, however, the situation is in fact usually the opposite of this; the differences in transition frequency for the high-field NMR transitions, given a nuclear spin $I \geqslant 1$,  are actually maximised in the low-field regime.

 It may be illustrative to consider some concrete examples from Si:P and Si:Bi. Here, we wish to demonstrate the relationship between the magnetic field regime, in the sense that it affects the gap between transition frequencies, and accuracy of control. As such, we must ensure that the  Rabi frequency remains constant throughout. Seeing as the Rabi frequency is given by $\omega_1 [\eta(\Omega)+\delta_\gamma \xi(\Omega)]$, then, it follows that we must vary $\omega_1$.\footnote{It should be noted, however, that this arbitrary control of $\omega_1$ is unfeasible in practice, and we only do this to allow for a systematic comparison of Rabi frequencies, field regime, and accuracy of control.} Furthermore, the discussion here only considers circularly polarised driving fields, such that the smallest \emph{relevant} frequency differences are, in the case of Si:Bi, given by Equations \eqref{same time freqdiff small}-\eqref{same time freqdiff large}. Si:P, on the other hand, only has one transition of  type $\phi^+_m\leftrightarrow \phi^+_{m-1}$ and similarly with $\phi^-_m\leftrightarrow \phi^-_{m-1}$. As such,  the relevant smallest frequency differences are determined, depending on the case, by other combinations discussed in Section \ref{transition frequencies section}.  

\subsubsection{Si:Bi}
\begin{figure}[!htb]
\centering
\subfloat[ Si:Bi $\Delta \Omega$ for transition $\phi^+_0\leftrightarrow\phi^+_{-1}$ ]{\label{SiBiNMRfreqdiff}\includegraphics[width=0.51\textwidth]{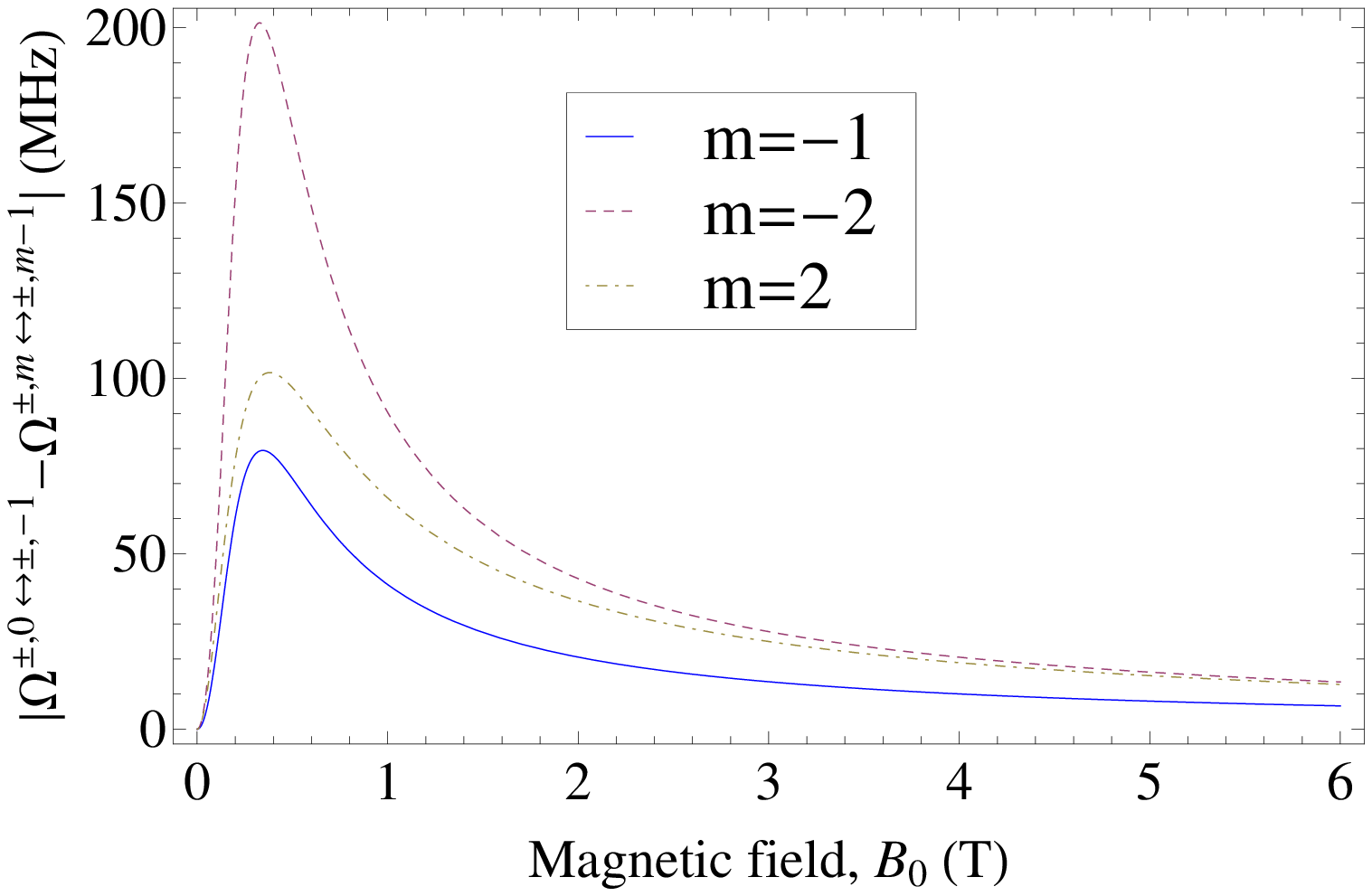}}
 \subfloat[Si:Bi $\Delta \Omega$ for transition $\phi^+_0\leftrightarrow\phi^-_{-1}$]{\label{SiBiESRfreqdiff}\includegraphics[width=0.51\textwidth]{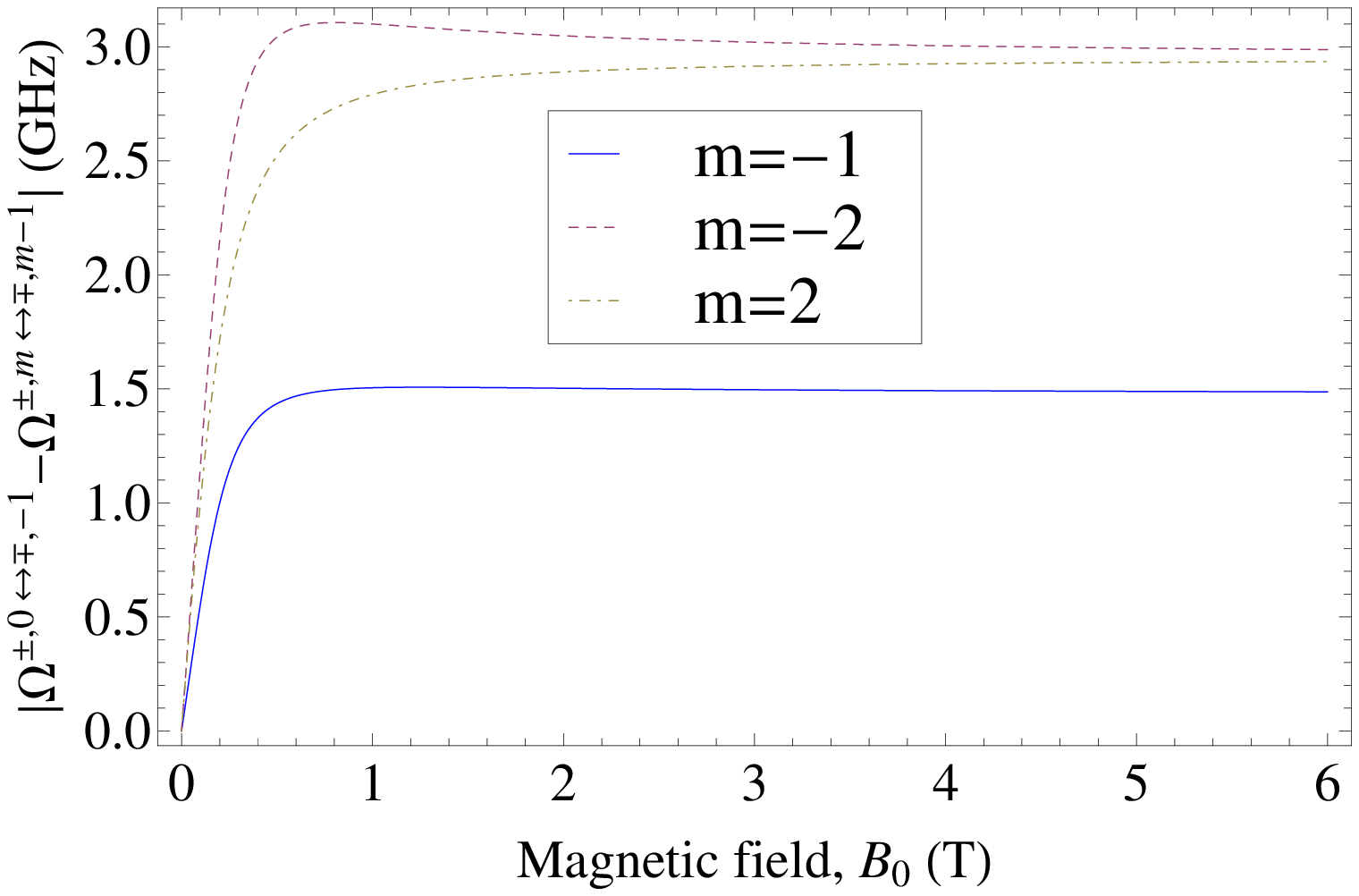}} \\ \subfloat[Si:Bi  control accuracy for transitions $\phi^+_0\leftrightarrow\phi^\pm_{-1}$]{\label{SiBiselectiverotationmzero}\includegraphics[width=0.52\textwidth]{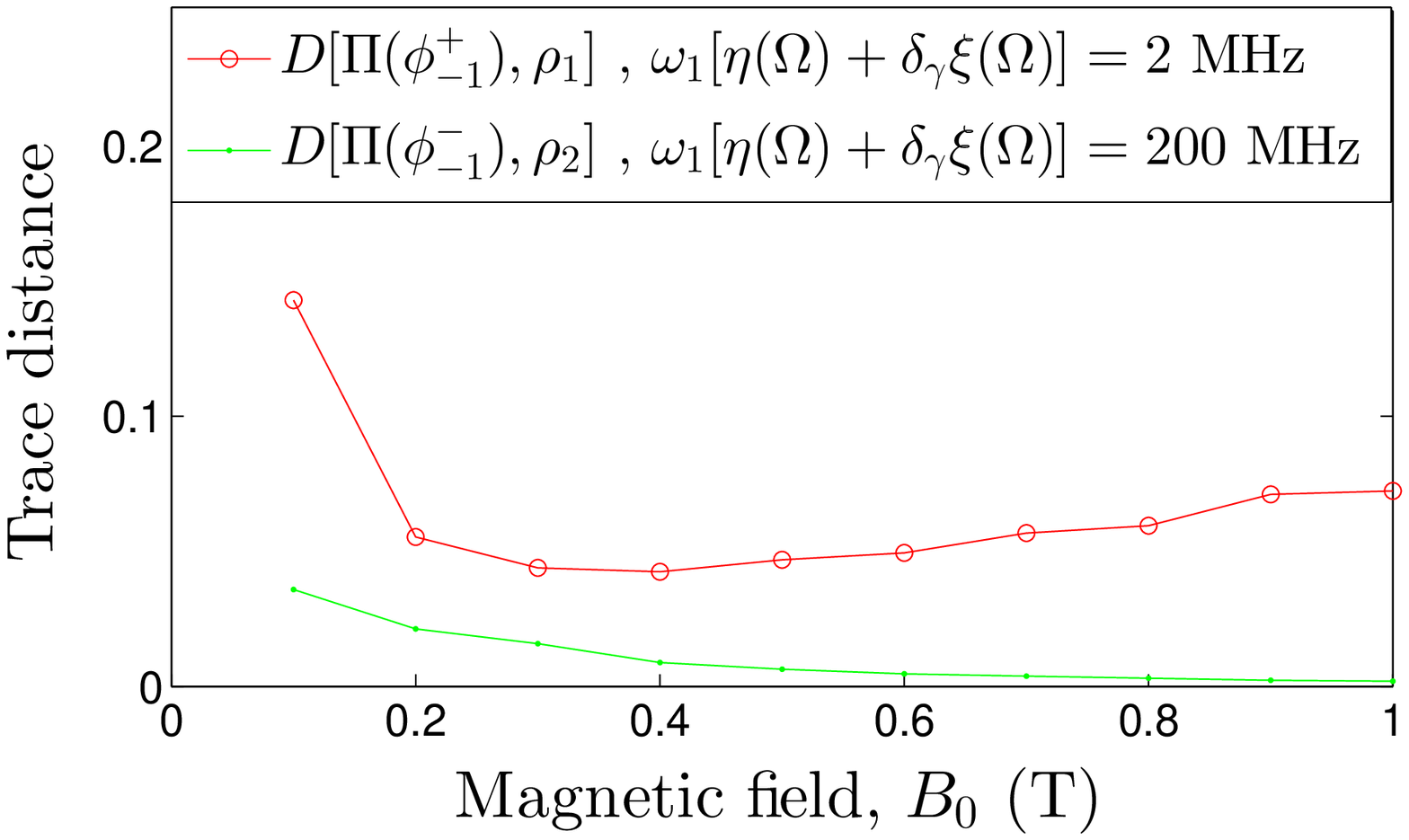}}
 \ \  \subfloat[Si:Bi  control accuracy for transitions $\phi^-_{-5}\leftrightarrow\phi^\pm_{-4}$ with $\omega_1(\eta(\Omega)+\delta_\gamma\xi(\Omega))=200$ MHz.]{\label{SiBiselectiverotationSband}\includegraphics[width=0.52\textwidth]{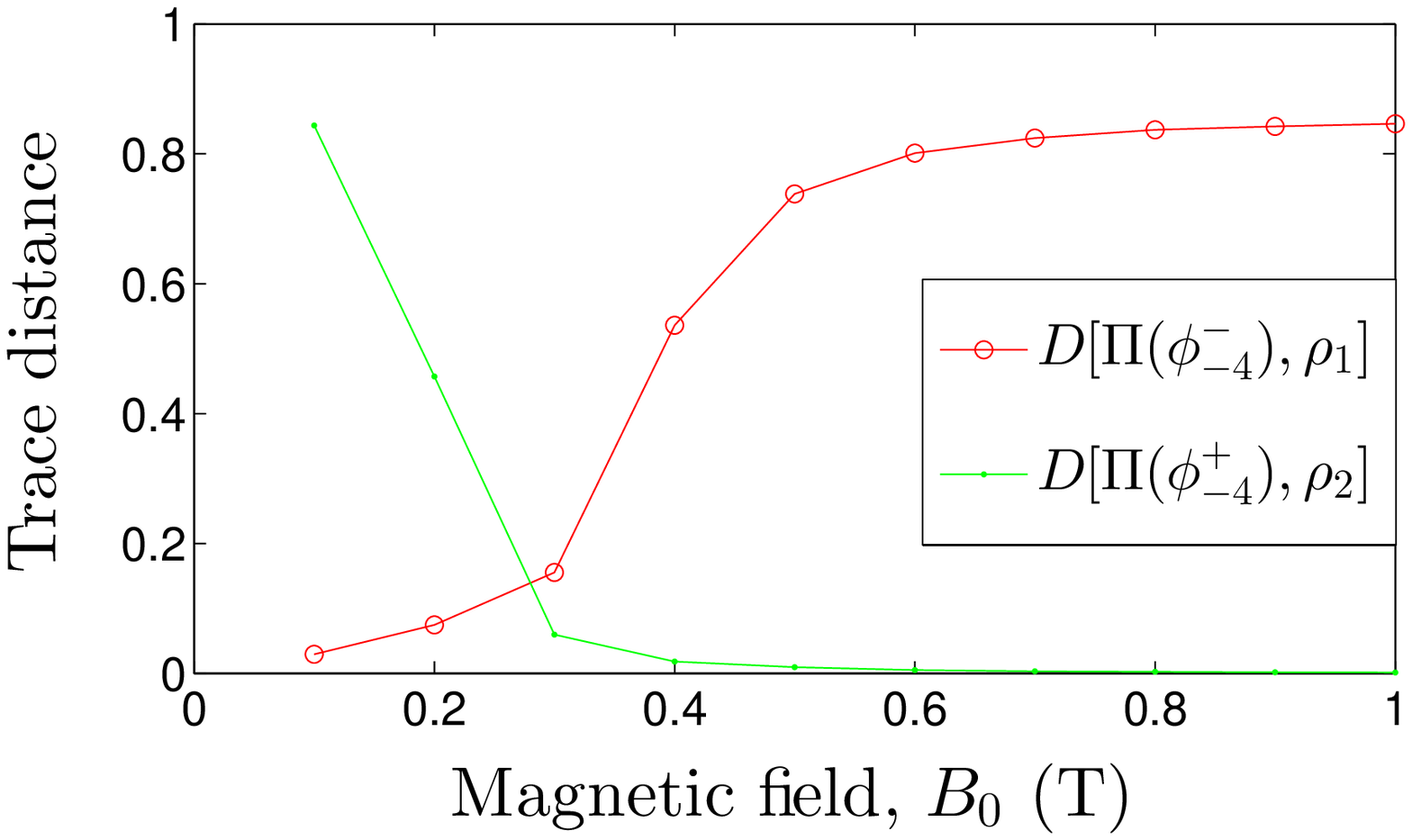}}
\caption{Accuracy of control in Si:Bi}
\end{figure} 

For Si:Bi, Figures \ref{SiBiNMRfreqdiff} and \ref{SiBiESRfreqdiff} show the  smallest frequency difference $\min(\Delta \Omega)$  where the desired transitions are, respectively, the high-field NMR transition  $\phi^+_0\leftrightarrow\phi^+_{-1}$ and the ESR transition $\phi^+_0\leftrightarrow\phi^-_{-1}$. The accuracy of control for these transitions, given a constant Rabi frequency,  is demonstrated by Fig.\ref{SiBiselectiverotationmzero}. Here, the system is initially set to $\Pi^+_0$ and the  Liouville-von Neumann equation is numerically integrated, using the Runge-Kutta-Fehlberg method, for a period of  $\tau = \pi/(2 \omega_1 [\eta(\Omega)+\delta_\gamma \xi(\Omega)])$ which prepares the state $\rho_1$(or $\rho_2$) by tuning the frequency of the driving field to be in resonance with $\Omega^{+,0\leftrightarrow+,-1}$(or $\Omega^{+,0\leftrightarrow-,-1}$). In different magnetic field regimes, the driving field is adjusted so that the Rabi frequency is always $ \omega_1 [\eta(\Omega)+\delta_\gamma \xi(\Omega)]=2$ MHz (or 200 MHz).   After completion, the trace distance is calculated between the solution $\rho_1$(or $\rho_2$) and the desired state $\Pi(\phi^+_{-1})$(or $\Pi(\phi^-_{-1}))$. Evidently, $\mathrm{D}[\rho_1,\Pi(\phi^+_{-1})]$ is minimised at $\sim 0.4$ T, where the relevant $\min(\Delta \Omega)$ shown in   Fig.\ref{SiBiNMRfreqdiff} maximises. At higher and lower magnetic fields, accurate control for this transition requires slower pulses. In contrast, as  $\min(\Delta \Omega)$ in Fig.\ref{SiBiESRfreqdiff} plateaus at its maximal value after $\sim 1$ T, $\mathrm{D}[\rho_2,\Pi(\phi^-_{-1})]$ continues to decrease up to this field value, and does not change considerably beyond this. Also notice that, while the Rabi frequency used for the transition $\phi^+_0\leftrightarrow \phi^-_{-1}$ is a hundred times greater than that used for  $\phi^+_0\leftrightarrow \phi^+_{-1}$, the former is always achieved with greater accuracies. This shows that, even though both transitions have approximately the same transition rate in the low-field regime, for any arbitrary  accuracy the high-field NMR transition must be more than a hundred times slower; the speed-up in the low-field regime is not as great as we would like it to be.   

 This is not very impressive, but I had chosen the worst-case scenario in Si:Bi. Fig.\ref{SiBiselectiverotationSband}, on the other hand, compares the accuracy and rate of control of the ESR transition $\phi^+_{-4}\leftrightarrow \phi^-_{-5}$ with the high-field NMR transition $\phi^-_{-4} \leftrightarrow \phi^-_{-5}$. Note that  $\phi^-_{-5}$ is the maximal $m$ state $\phi^-_{-I-1/2}$ of Si:Bi. As before, the system is initially set to $\Pi(\phi^-_{-5})$ and the von-Neumann equation with the appropriate driving field is integrated for the required period so as to constitute a $\pi$ pulse and thereby prepare states $\rho_1$(or $\rho_2$). In this case, the Rabi frequency is set to be 200 MHz for both instances. The trace distance is then calculated between the prepared state and the desired state $\Pi(\phi^-_{-4})$(or $\Pi(\phi^+_{-4})$). As can be seen, the two transitions can be achieved with the same degree of accuracy and rate at $\sim 0.3$ T. In such a situation we do have genuine speed-up for the high-field NMR transition.

\subsubsection{Si:P}
\begin{figure}[!htb]
\centering
\subfloat[ Si:P $\Delta \Omega$ for transition $\phi^+_1\leftrightarrow\phi^+_0$]{\label{SiPNMRfreqdiff}\includegraphics[width=0.52\textwidth]{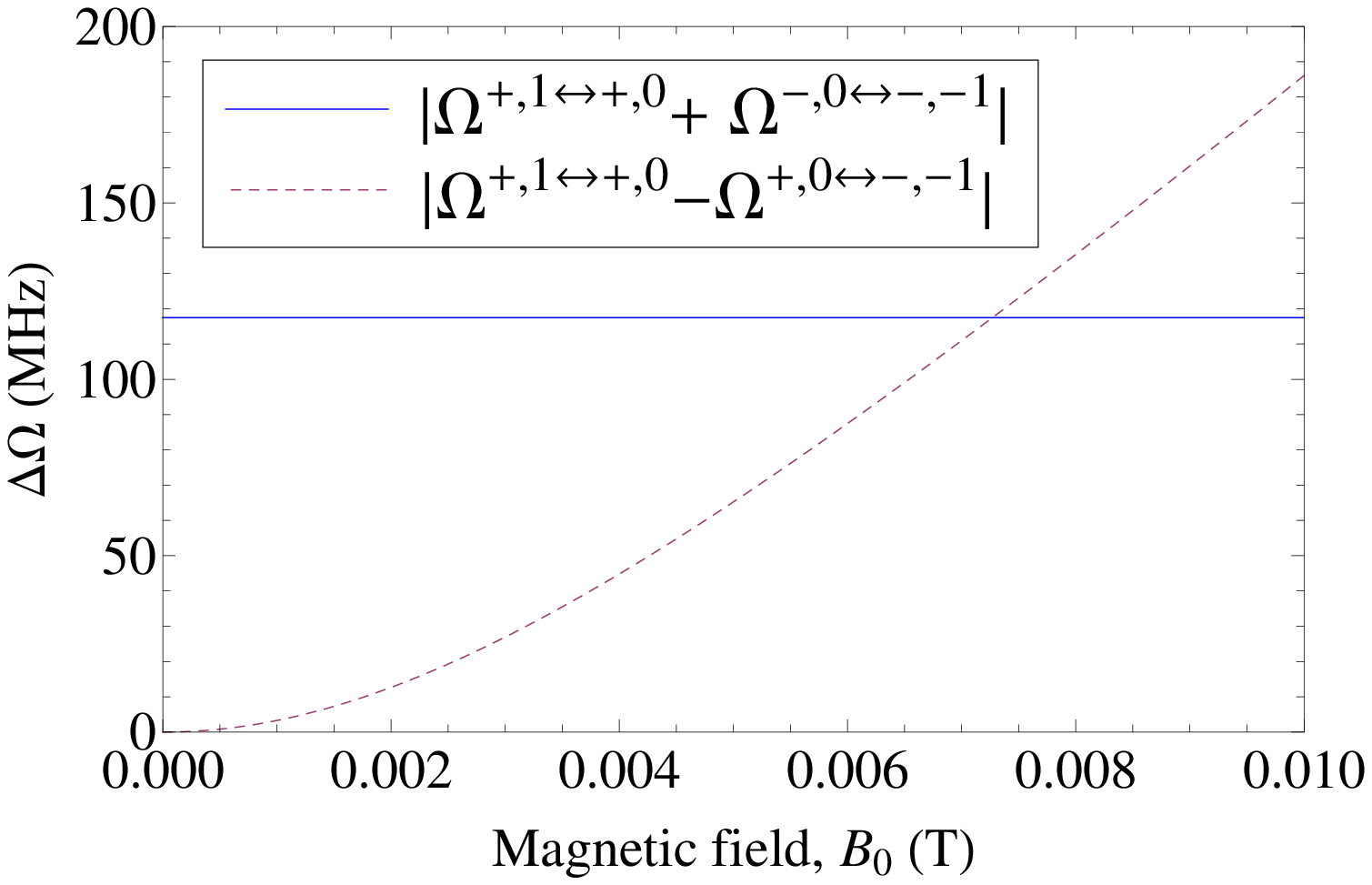}}
 \subfloat[ Si:P $\Delta \Omega$ for transition $\phi^+_1\leftrightarrow\phi^-_0$]{\label{SiPESRfreqdiff}\includegraphics[width=0.52\textwidth]{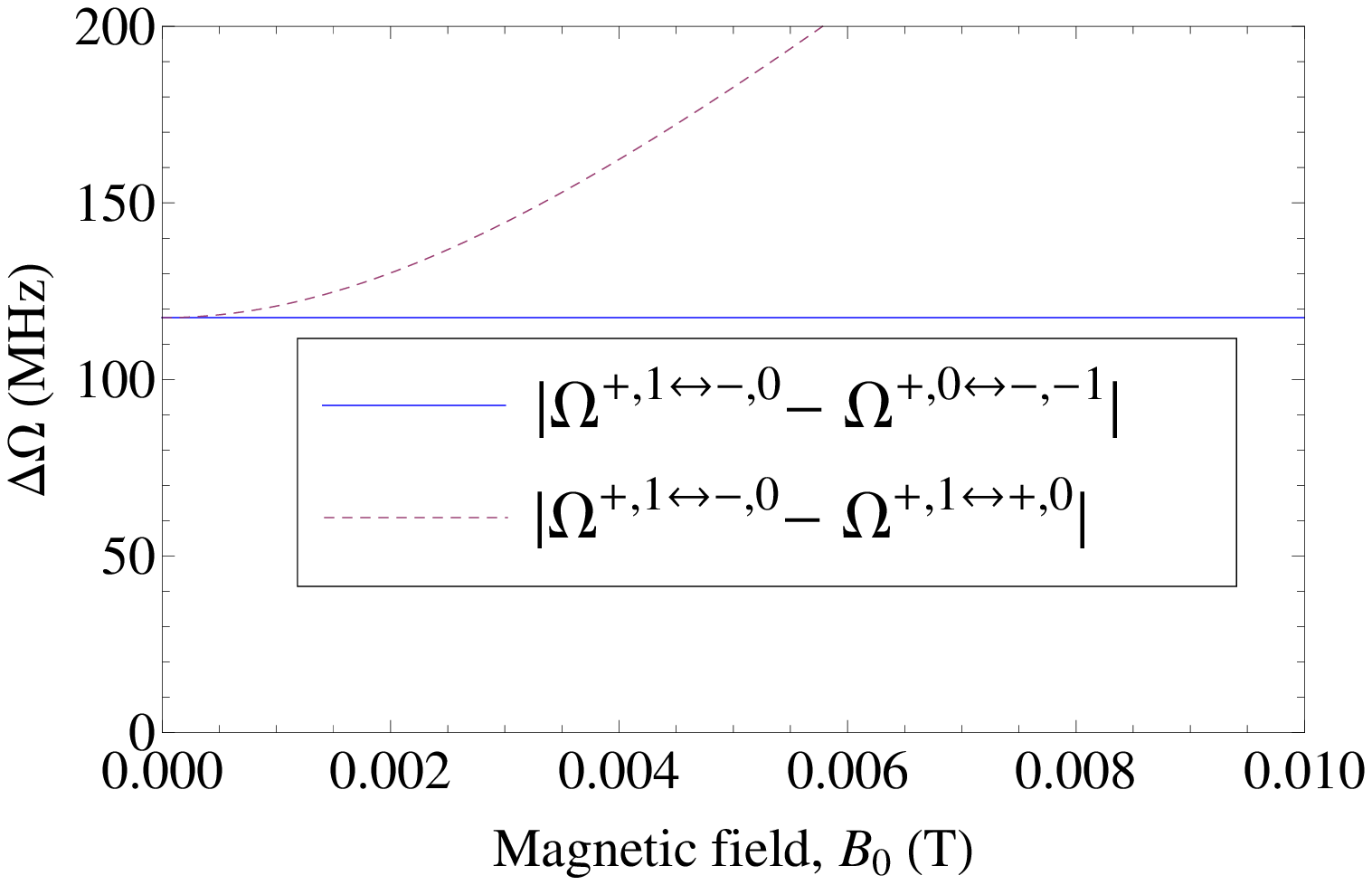}}\\ \subfloat[ Si:P  control accuracy for transitions $\phi^+_1\leftrightarrow\phi^\pm_0$]{\label{SiPselectiverotation}\includegraphics[width=0.6\textwidth]{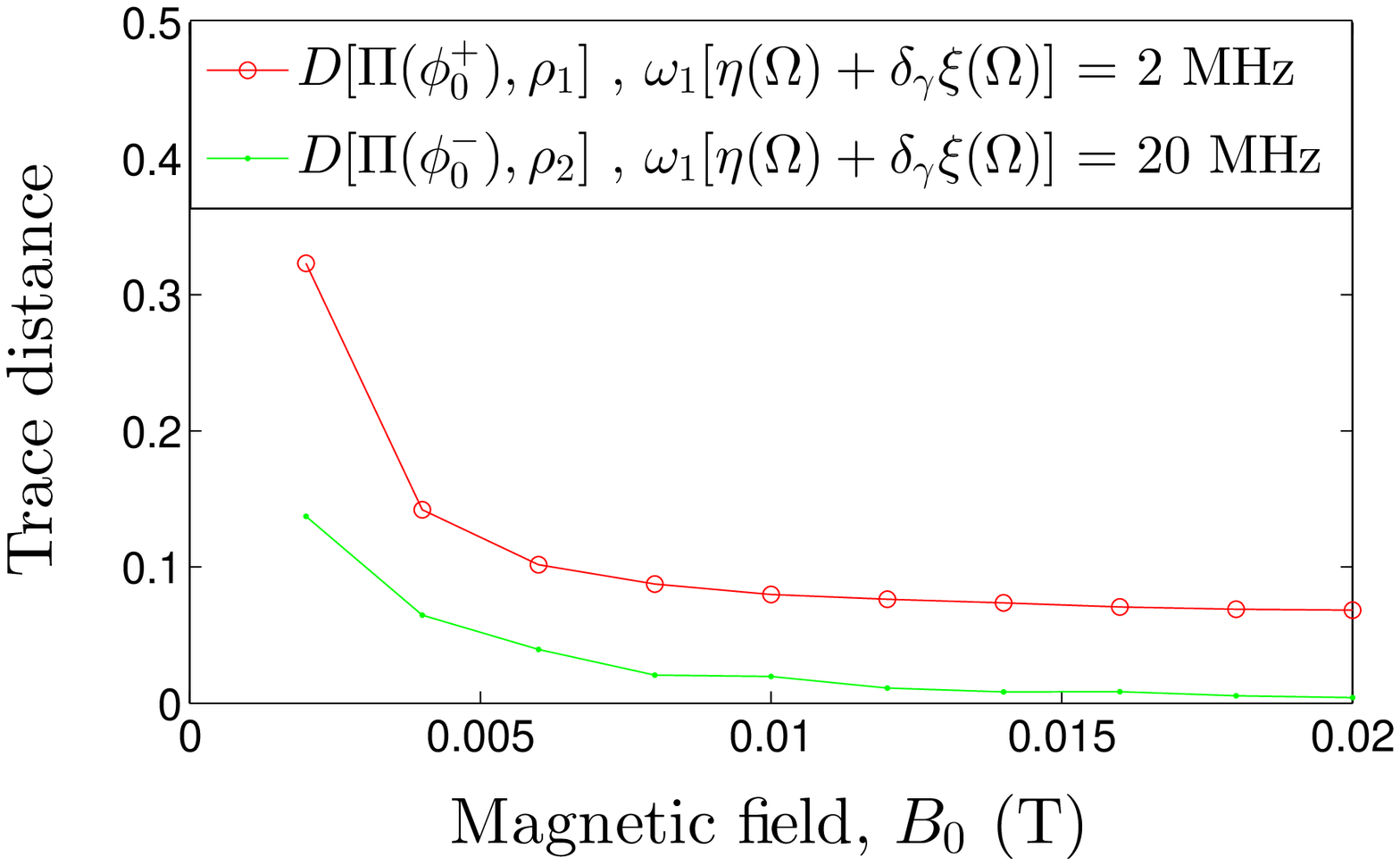}}
 
 \caption{Accuracy of control in Si:P}
\end{figure}

For Si:P, Figures \ref{SiPNMRfreqdiff} and \ref{SiPESRfreqdiff} show the  relevant $\min(\Delta \Omega)$ for the ESR transition $\phi^+_1\leftrightarrow\phi^-_0$ and the high-field NMR transition $\phi^+_1 \leftrightarrow \phi^+_0$, whilst Fig.\ref{SiPselectiverotation} compares the accuracy of control between the two. Here, the system is initially prepared in state $\Pi(\phi^+_1)$, and with a constant Rabi frequency of 2 MHz (or 20 MHz), the system is evolved so as to constitute a $\pi$ pulse so as to prepare  $\rho_1$(or $\rho_2$). The trace distance is then calculated between the prepared state and the desired state $\Pi(\phi^+_0)$(or $\Pi(\phi^-_0)$). In both cases accuracy of control improves with the magnetic field, where one of the relevant $\Delta\Omega$ in the system grows larger from its vanishingly small value at zero field. However, the other $\Delta\Omega$ is  constant at all fields, and poses a lower bound for speed of accurate control.

\subsubsection{Si:Bi in comparison with Si:P}
Two differences between Si:P and Si:Bi  now become apparent. First of all, due to the larger hyperfine coupling in Si:Bi, resulting in larger ESR transition frequency differences,  accurate control within ESR subspaces can be achieved at much faster speeds in Si:Bi than can be done in Si:P, at all magnetic fields. More interestingly, however, is the potential for speed-up  of transitions $\phi^\pm_m\leftrightarrow \phi^\pm_{m-1}$, which are classified as NMR in the high-field limit,  in the low-field regime. For Si:P, moving to the low-field regime results in a  diminution of $\Delta \Omega$.  As such, even though the relevant  transition rates increase three-fold in the low-field regime,  in order to maintain some level of accuracy one must compensate by decreasing $\omega_1$ so as to achieve an overall slower pulse. The situation in Si:Bi is different; moving to the low-field regime results in an increase in the relevant $\Delta \Omega$ for the high-field NMR transitions, such that the increase in transition rates may be utilised to achieve a genuine speed-up of accurate control, albeit not always by three orders of magnitude.

\section{Decoherence properties}
According to the theoretical model of decoherence we discussed in Chapter \ref{open systems chapter} Si:P, owing to its nuclear spin of $I=1/2$, does not have any OWPs and, consequently, is of little interest.  We will therefore focus only on Si:Bi in this regard. Additionally as the qualitative properties of our Markovian model of decoherence and the more microscopically well-motivated, yet more arduous, model of spin-bath decoherence are very similar, we shall only consider the prior. 

\begin{figure}[!htb]
\centering
\includegraphics[width=0.8\textwidth]{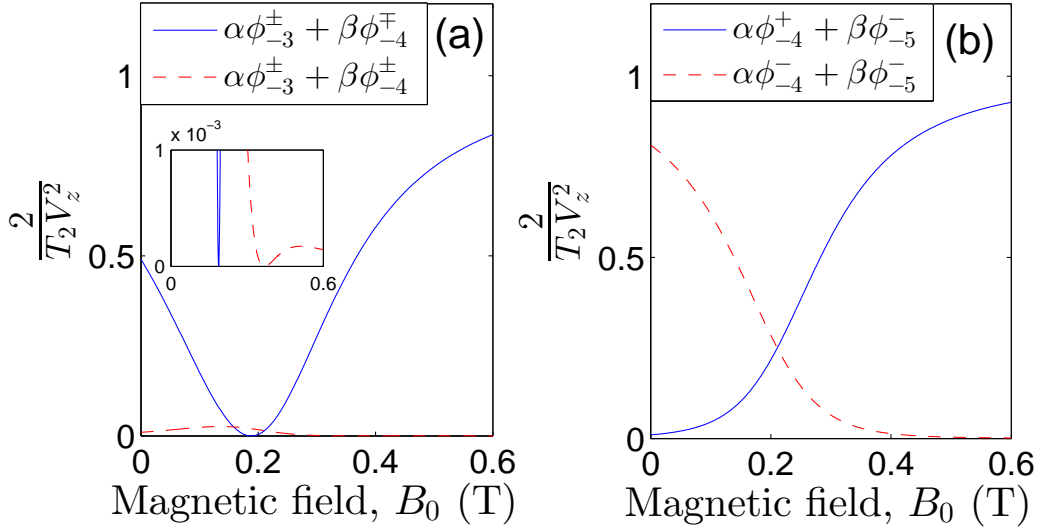}
\caption{Dephasing rates of Si:Bi given adiabatic $Z$ noise}\label{SiBiadiabaticZnoise}
\end{figure}

Provided the presence of adiabatic $Z$ noise,  the OWPs of Si:Bi which are very close to the low-field regime FSPs will offer decoherence free subspaces. This is shown in Fig.\ref{SiBiadiabaticZnoise}(a). Notice that the maximum dephasing rate of  $\alpha\phi^\pm_{-3}+\beta\phi^\pm_{-4}$ is much smaller than that of  $\alpha\phi^\pm_{-3}+\beta\phi^\mp_{-4}$. Figure \ref{SiBiadiabaticZnoise}(b) shows the dephasing rates for the superpositions involving the maximal $m$ state $\phi^-_{-5}$, given adiabatic $Z$ noise. As is to be expected, the dephasing rate for the  superposition $\alpha\phi^-_{-4}+\beta\phi^-_{-5}$ vanishes asymptotically in the high-field limit, whereas that of the superposition $\alpha\phi^+_{-4}+\beta\phi^-_{-5}$ maximises asymptotically in the high-field limit. Such transitions do not have any OWPs and, as such, the dephasing rate never vanishes in the low-field regime. 

\begin{figure}[!htb]
\centering
\includegraphics[width=0.8\textwidth]{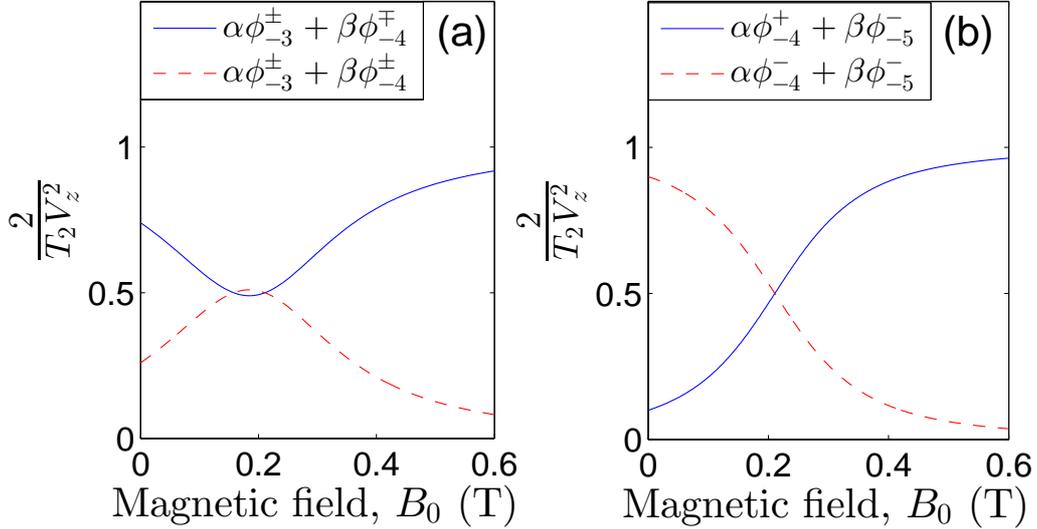}
\caption{Dephasing rates of Si:Bi given diabatic $Z$ noise}\label{SiBidiabaticZnoise}
\end{figure}

Figures \ref{SiBidiabaticZnoise}(a) and \ref{SiBidiabaticZnoise}(b) compare the dephasing rates for the same transitions as before, except  with diabatic $Z$ noise. The differences seen in Fig.\ref{SiBidiabaticZnoise}(b) are marginal, but it is Fig.\ref{SiBidiabaticZnoise}(a) which has the most striking features. Here, the  superposition $\alpha\phi^\pm_{-3}+\beta\phi^\mp_{-4}$ minimises its dephasing rate, and the  superposition $\alpha\phi^\pm_{-3}+\beta\phi^\pm_{-4}$ maximises its dephasing rate, at $B_0=0.1846$ T. This is close, but not identical to, the OWP $\cos(\theta_{-3})=-\cos(\theta_{-4})$ at $B_0=0.1882$ T. The reason that the dephasing rates here are so much larger in the low-field regime than was the case for adiabatic $Z$ noise is that, at such field values, the depolarisation of each $m$ subspace is maximal. This in turn leads to dephasing. 

\begin{figure}[!htb]
\centering
 \subfloat[ Diabatic $Z$ noise]{\label{Znoisespectrum}\includegraphics[width=0.5\textwidth]{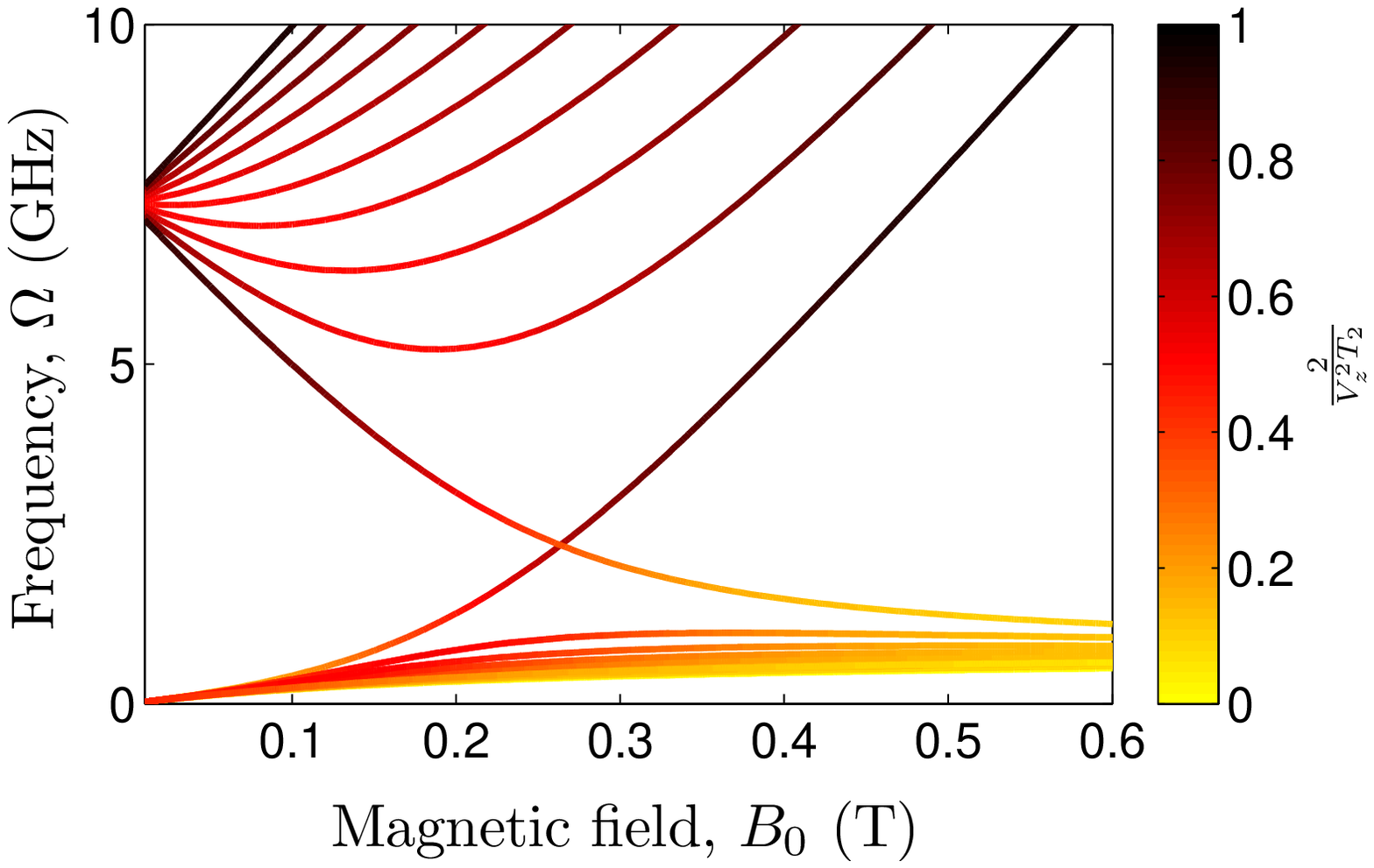}}  \subfloat[Diabatic $X$ noise]{\label{Xnoisespectrum}\includegraphics[width=0.5\textwidth]{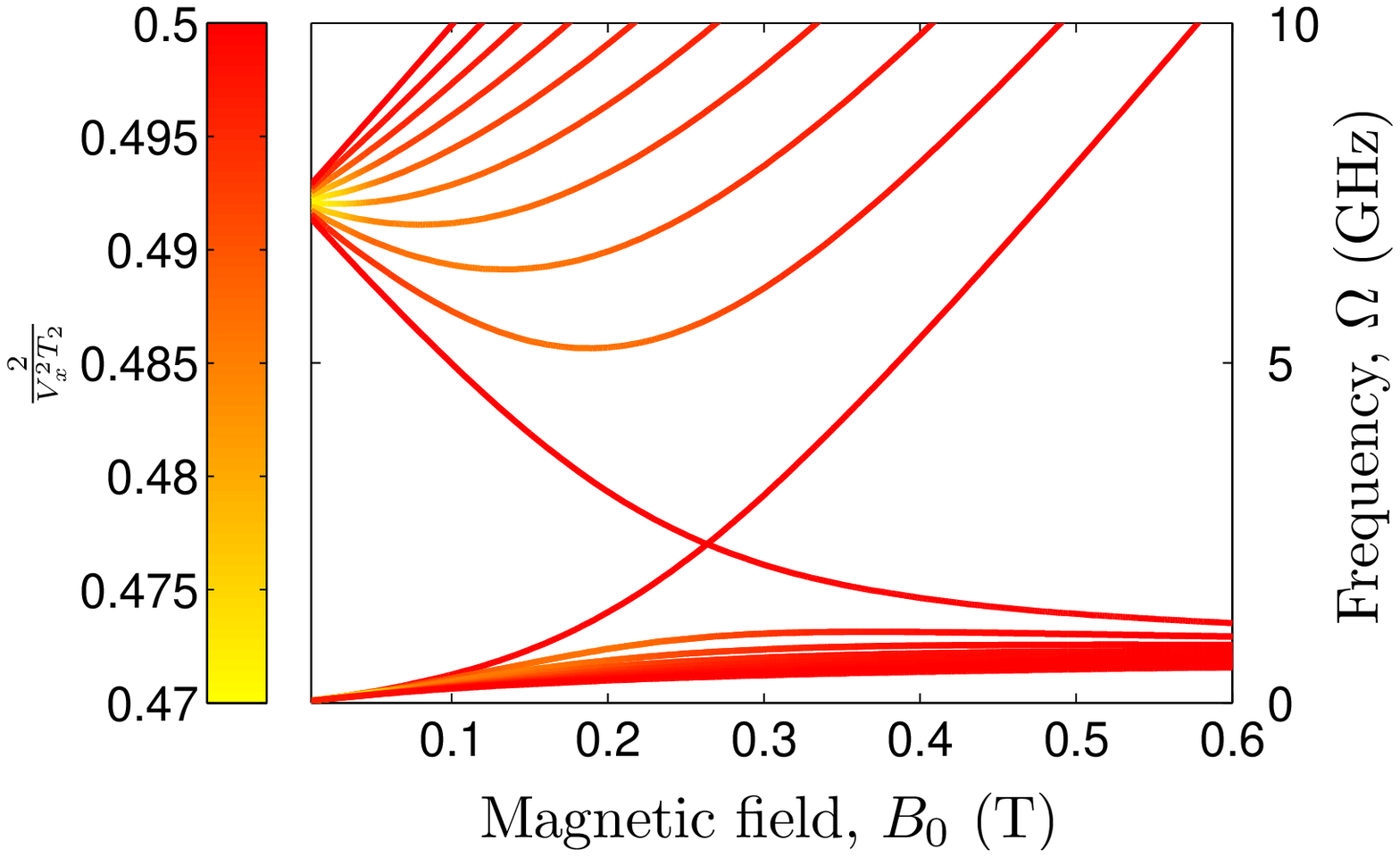}}
\caption{Si:Bi dephasing rates in units of $V_{x/z}^2/2$ for diabatic Z and X noise.}
\end{figure}

Figures \ref{Znoisespectrum} and \ref{Xnoisespectrum} compare the numerically determined dephasing rates of diabatic $Z$ noise and $X$ noise, for all superpositions. Although the dephasing rates given $Z$ noise vary largely depending on magnetic field regime and initial superposition, $X$ noise yields dephasing rates of $\sim V_x^2/4$ at all magnetic fields and for all superpositions.   

\begin{figure}[!htb]
\centering
\subfloat[Density matrix at initial time]{\label{densitymatrixt0}\includegraphics[width=0.5\textwidth]{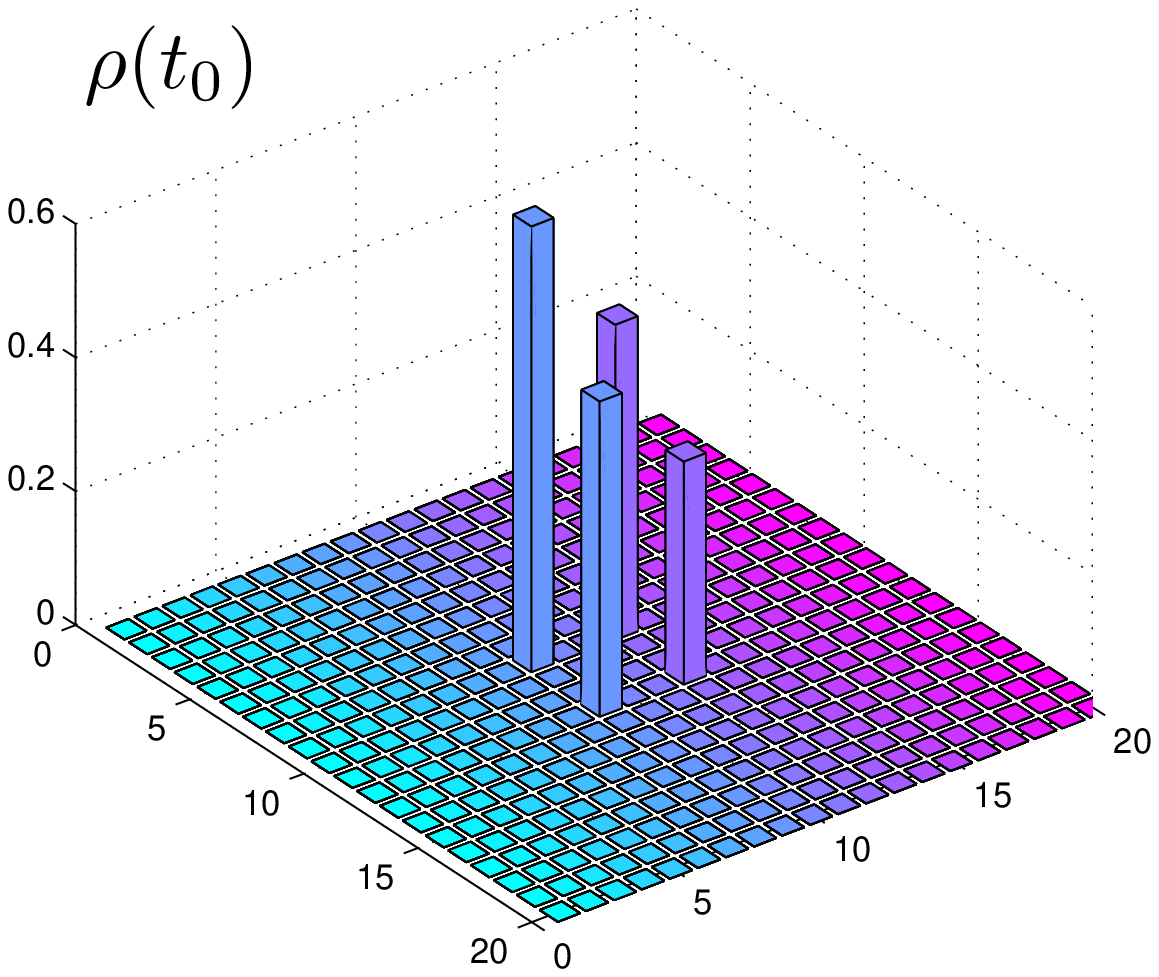}}
\\ \subfloat[Density matrix after the application of diabatic $Z$ noise]{\label{densitymatrixZnoise}\includegraphics[width=0.7\textwidth]{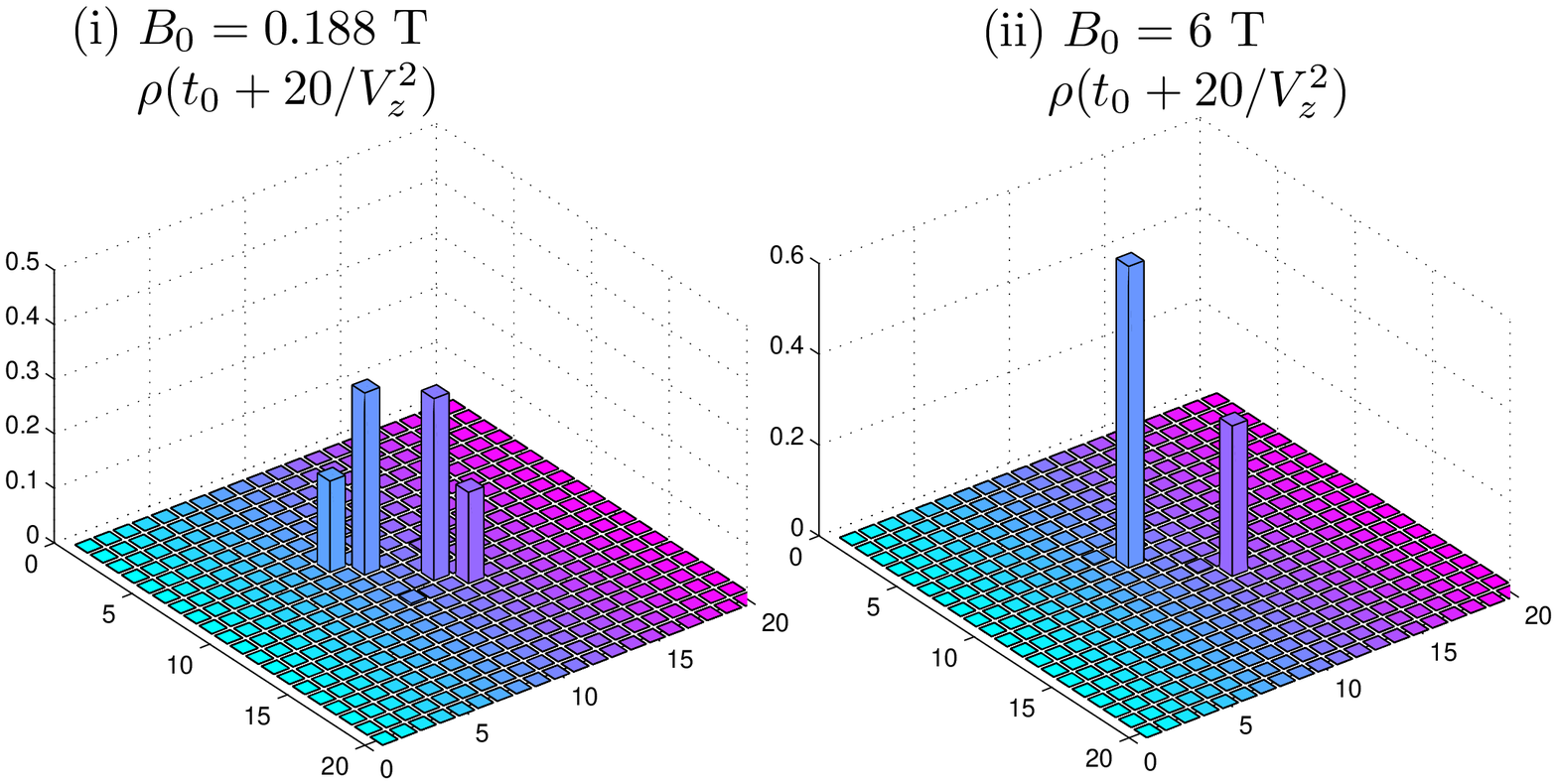}} \\ \subfloat[Density matrix after the application of $X$ noise]{\label{densitymatrixXnoise}\includegraphics[width=0.7\textwidth]{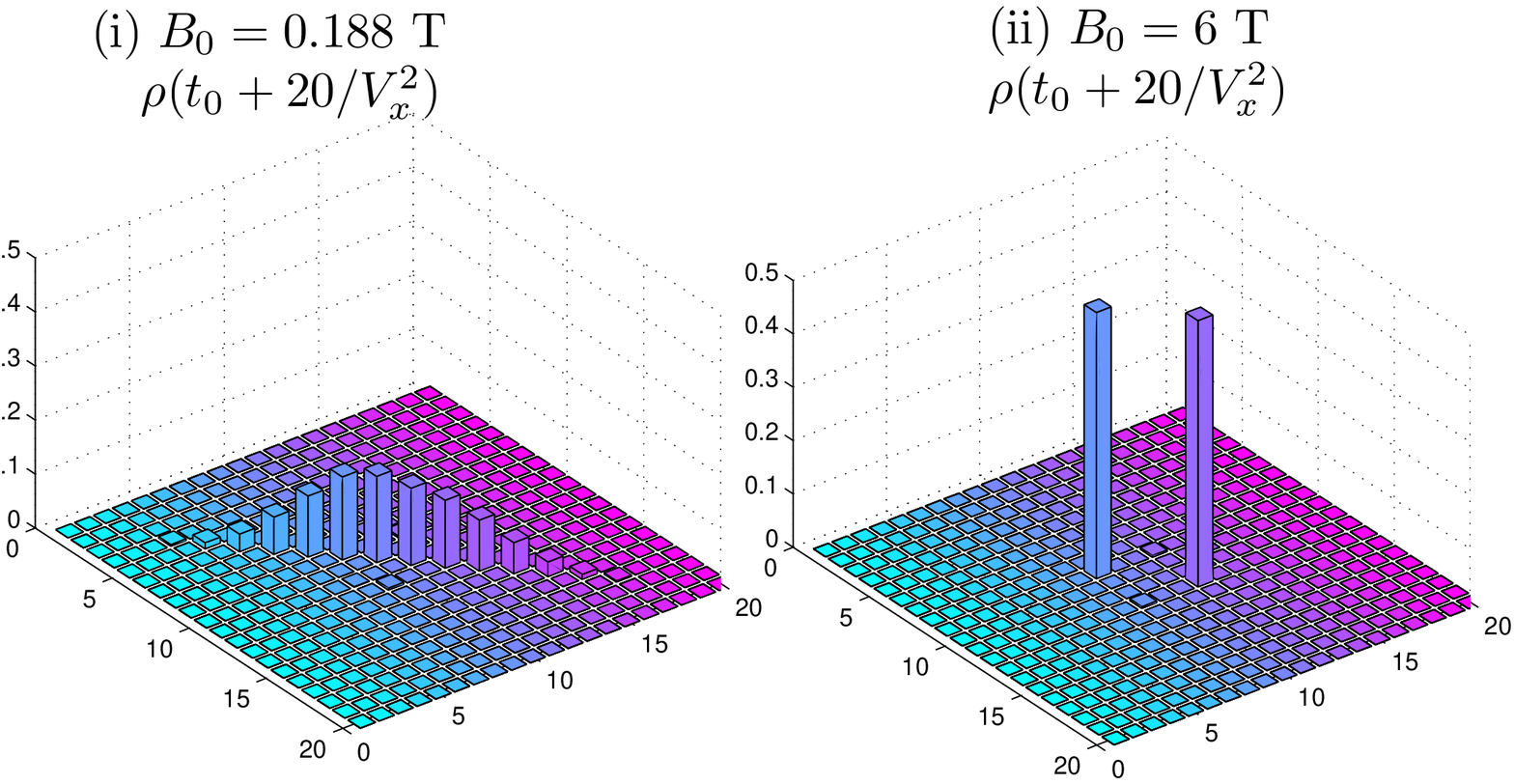}}
\caption{Density matrix tomography before and after the application of diabatic $Z$ and $X$ noise in the low-field and high-field regimes of Si:Bi.}
\end{figure}

 It is, however, in the density matrix tomography that the distinction between diabatic $X$ and $Z$ noise becomes most apparent. Figure \ref{densitymatrixt0} shows the population of the matrix elements of the system density operator $\rho$ at initial time. This corresponds to the superposition 
\begin{equation}
\psi=\frac{1}{\sqrt{3}}\phi^+_{-3}+\frac{2}{\sqrt{3}}\phi^-_{-4}\equiv \frac{1}{\sqrt{3}}\varphi_{12}+\frac{2}{\sqrt{3}}\varphi_9
\end{equation}
where by $\varphi_i$ we refer to the $i^{\mathrm{th}}$ eigenstate, with $\varphi_1\equiv \phi^-_4$ being the ground state and $\varphi_{20}$ the maximally excited state. \footnote{Of course, at $B_0\gtrsim A_{\mathrm{iso}}/2\gamma_n\simeq 110$ T, the labeling of states $\varphi_{11}$ to $\varphi_{20}$ will reverse, with what was initially labeled $\varphi_{11}$ now being $\varphi_{20}$ and so on.  } The following figures \ref{densitymatrixZnoise} and \ref{densitymatrixXnoise} show the matrix elements of $\rho$ after the system has evolved for a time $\tau = 20/V_{z/x}^2$ undergoing diabatic $Z$ noise or $X$ noise respectively. Figure \ref{densitymatrixZnoise}(i)  shows the case at the optimal working point $B_0=0.188$ T. Because $S_z$ does not commute with $H_0$ at this field regime, and we have not made the adiabatic approximation, the system does not undergo pure dephasing, but rather dissipative decoherence. Indeed, each $m$ subspace undergoes a depolarising channel. Hence, half the population of state $\phi^+_{-3}\equiv \varphi_{12}$ goes to state $\phi^-_{-3}\equiv \varphi_8$, and similarly half the population of state $\phi^-_{-4}\equiv \varphi_9$ goes to state $\phi^+_{-4}\equiv \varphi_{11}$. Figure \ref{densitymatrixZnoise}(ii) shows the density matrix elements if the state undergoes diabatic $Z$ noise at $B_0 = 6$ T, which is in the high-field regime. Here, $S_z$ does commute with the system Hamiltonian, and hence the system only undergoes pure dephasing.  it is therefore only the off-diagonal elements of $\rho$ which have disappeared.

Figure \ref{densitymatrixXnoise}(i) shows the case for $X$ noise at the optimal working point. Here, the exact tracking of the dynamics is a complicated affair, but we can see that many of the diagonal elements of $\rho$ are now populated, as the system is continuing towards becoming maximally mixed. Figure \ref{densitymatrixXnoise}(ii), on the other hand, shows the matrix elements of $\rho$ at $B_0 = 6$ T where, due to the separability of the electron and nuclear spins, the system undergoes a depolarising channel in the subspace spanned by the initial superposition. 

\section{Summary}
In this chapter we applied the general analytic results pertaining to nuclear-electronic spin systems to the specific cases of phosphorus-doped silicon (Si:P) and bismuth-doped silicon (Si:Bi). Owing to the larger nuclear spin and hyperfine constant, Si:Bi has many interesting properties that are not present in Si:P. These can be enumerated thusly:
\begin{enumerate}[(i)]
 
\item
Si:Bi is in the ``low-field regime'' at a large range of  experimentally accessible magnetic fields, $0-0.6$ T, whereas this regime for Si:P is attained at much smaller magnetic fields.
\item Magnetic resonance control of all two-dimensional subspaces allowed by the NEMR selection rules are achievable at faster speeds in Si:Bi than is the case for Si:P, owing to the larger hyperfine constant. Furthermore, the larger nuclear spin in Si:Bi means that, for this system, control of subspaces spanned by $\{\phi^\pm_m, \phi^\pm_{m-1}\}$ can be achieved at faster speeds in the low-field regime than is possible in the high-field regime. This is not so for Si:P.
\item
Si:Bi has several two-dimensional subspaces that, at specific values of the magnetic field, are robust against pure decoherence due to effective ``magnetic field fluctuations'' that affect only the electron spin. These ``optimal working points'' do not exist in Si:P.   
\end{enumerate}
The simultaneous presence of optimal-working points and potential speed-up of magnetic resonance control in the low-field regime presents Si:Bi as an attractive source of qubits for quantum computation.

\spacing{1}                                  
 \bibliographystyle{plainnat}
 \bibliography{references}

\spacing{1}
\chapter{Experimental investigations of Si:Bi at S-band}
\section{Introduction}
\begin{figure}[!htb]
\centering
\includegraphics[width=4in]{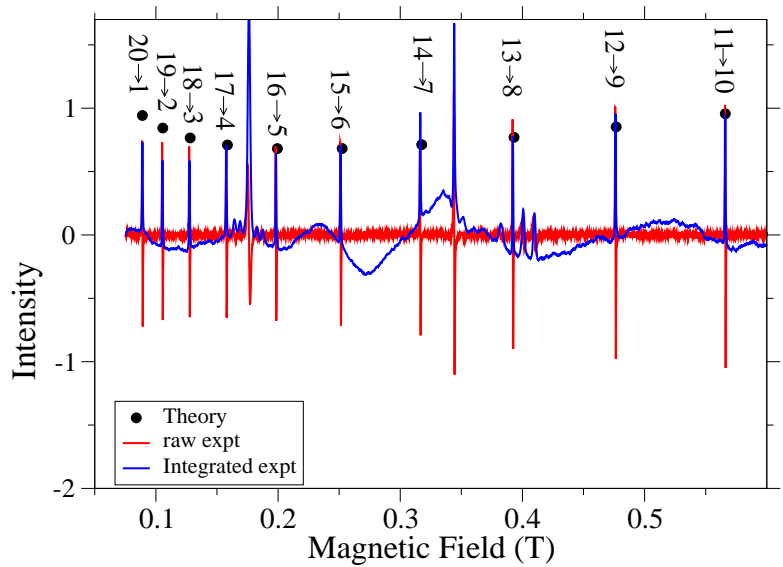} \caption{  C.w. experimental data for Si:Bi at 9.7 GHz, compared with theoretical predictions. The large sharp resonance at 0.35 T is due to silicon dangling bonds, while the remainder are caused by defects in the sapphire ring  used as a dielectric microwave resonator.    } \label{X-band Si:Bi}
\end{figure}

The  predictions made by our  model  are in good agreement with experimental data available for Si:P and Si:Bi at X-band ($9.7$ GHz) ESR. Figure \ref{X-band Si:Bi}, for example, shows that  the positions  of the Si:Bi transitions in c.w. spectroscopy matches well with our theoretical predictions. The relative strengths, however, agree to a lesser degree. The observed transitions here  correspond, in the most part, to the high-field ESR transitions $\phi^+_m\leftrightarrow\phi^-_{m-1}$. However we have claimed that the ESR and NMR selection rules, for nuclear-electronic spin systems, are asymptotic limits of the more general NEMR selection rules. Consequently our model predicts that, at X-band, we should also observe some  of the transitions that are dipole-forbidden at high field, $\phi^-_m\leftrightarrow \phi^+_{m-1}$, albeit with much smaller transition rates. For example, the X-band transition $\phi^-_1\equiv\varphi_4 \leftrightarrow \phi^+_0\equiv \varphi_{15}$ occurs at $B_0=200.54$ mT, which is only $0.14$ mT smaller than the field value of the X-band transition $\phi^+_1\equiv\varphi_{16} \leftrightarrow \phi^-_0\equiv \varphi_5$, with a transition rate that is roughly a factor of $\sim1/1000$ with respect to the latter. Unfortunately, the data used for the above figure exhibits broadening of approximately $0.42$ mT, which is  larger than the separation between the two transitions.  As such, we are unable to observe such transitions here, no matter how much we amplify the strength of the microwave field. However, recent studies by \citep{ForbiddenlineMorton} conducted on Si:Bi with isotopically purified silicon, where the lack of spin-bath decoherence mechanisms lower the broadening to a sufficient degree, have experimentally demonstrated the presence of  transitions such as these.   

My colleagues and I, in a study conducted in ETH Zurich \citep{Morley-hybrid}, have also tried to verify the theoretical predictions made by our model using magnetic resonance spectroscopy, but at S-band (4.044 GHz). Here, we predict to see the transition $\phi^+_{-4}\equiv \varphi_{11} \leftrightarrow \phi^-_{-5}\equiv \varphi_{10}$ at $B_0=345.02$ mT and the transition $\phi^-_{-4}\equiv \varphi_9 \leftrightarrow \phi^-_{-5}\equiv \varphi_{10}$ at $B_0=145.63$ mT, with the ratio of the c.w. transition rates, calculated by equations  \eqref{Iforbp}-\eqref{Iforbidden}, given as 
\begin{equation}
\frac{\mathcal{I}_{-4\leftrightarrow -5}^{+\leftrightarrow -}}{\mathcal{I}_{-4\leftrightarrow -5}^-}\bigg|_{\Omega=4.044 \ \mathrm{GHz}}\simeq\left( \frac{\cos(\frac{\theta_{-4}}{2})\bigg|_{B_0=345.02 \ \mathrm{mT}}}{\sin(\frac{\theta_{-4}}{2})\bigg|_{B_0=145.63 \ \mathrm{mT}}}\right)^2  =1.2\label{S-band cw rate ratio}
\end{equation}
and the ratio of the Rabi frequencies, determined by Eq.\eqref{Rabi frequency nuc-elec}, given by 
\begin{equation}
\frac{\eta(\Omega)+\delta_\gamma \xi(\Omega)\bigg|_{B_0=345.02 \ \mathrm{mT}}}{\eta(\Omega)+\delta_\gamma \xi(\Omega)\bigg|_{B_0=145.63 \ \mathrm{mT}}}\simeq \frac{\cos(\frac{\theta_{-4}}{2})\bigg|_{B_0=345.02 \ \mathrm{mT}}}{\sin(\frac{\theta_{-4}}{2})\bigg|_{B_0=145.63 \ \mathrm{mT}}} =1.1.\label{S-band pulsed rate ratio}
\end{equation}
At X-band, the transition $\varphi_{10}\leftrightarrow \varphi_{11}$ is observed at $B_0=566.86$ mT, at which it is characterised well by the ESR selection rule. The transition $\varphi_{10}\leftrightarrow\varphi_9$, on the other hand, is characterised well by the NMR selection rule at this field value. However, as is evident by Eq.\eqref{S-band cw rate ratio}, both these transitions have, at S-band, a similar transition rate. This is where the ESR and NMR selection rules no longer apply, and the true characteristics of the NEMR selection rules make themselves manifest.

The S-band spectrometer used in this study was home-made by the ESR research group at ETH-Zurich \citep{S-band-Zurich}, whose function is described, in abstract terms, in Sec.\ref{ESR equipment}. This spectrometer has a frequency range of $\sim 2 - 4$ GHz, and has four microwave channels with a power output of $\sim$ 1 kW. The Si:Bi sample used was a single float-zone crystal of natural silicon, bulk-doped in the melt with bismuth atoms at a concentration of $3 \times 10^{15}$ Bi cm$^{-3}$. 
\section{C.w. spectroscopy}

\begin{figure}[!htb]
\centering
\includegraphics[width=0.6\textwidth]{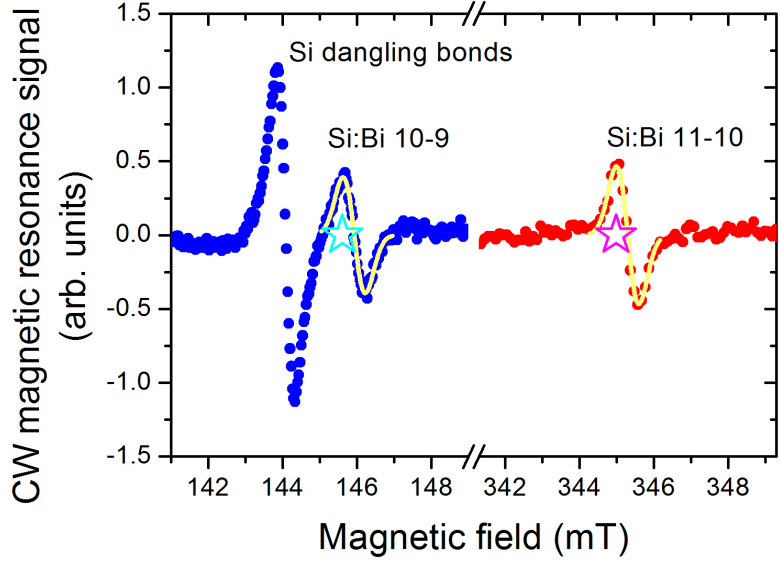}
\caption{Si:Bi c.w. spectra at 4.044 GHz. The blue dots correspond with the low-field transition $\varphi_{10}\leftrightarrow \varphi_9$, whereas the red dots are associated with the high-field transition $\varphi_{11} \leftrightarrow \varphi_{10}$. The yellow line is a differentiated Gaussian used to fit the experimentally acquired data, while the stars indicate the theoretically calculated positions of the transitions. }\label{S-band cw}
\end{figure}
In order to ascertain the spectrum of Si:Bi at 4.044 GHz, we first conducted a c.w.  experiment.
Initially the power to the strong magnet of the spectrometer was engaged, subsequent to  the water cooling being activated.  In tuning mode,  the resonance frequency of the cavity was set to 4.044 GHz. The Si:Bi sample, held in place within a quartz tube  by a teflon rod and vacuum grease, was then placed inside the cryostat, and its presence inside the cavity was confirmed when the resonance frequency of the cavity was  offset. The sample was then cooled to the desired temperature by pumping liquid helium, and controlled to an accuracy of $\pm 0.05$ K.  Subsequent to this,  the microwave frequency was tuned so as to become resonant with the cavity and sample. Next, the iris size was changed so as to ensure that  the cavity was critically coupled; that all the incident radiation entered the cavity.  After engaging the reference arm, the power and phase thereof were tuned so as to maximise the diode current's zero-value at around 200 $\mu A$ to ensure operation in the linear regime, where a linear increase in the incident microwave power results in a linear increase in detected current. The modulation frequency was then set to 0.1 MHz, whilst the time constant and repetition times were set to 30 ms and 100 ms respectively.

The experiment was conducted at 42 K because lower temperatures  result in $T_1$ times  so long that, even with low microwave power, the spectra becomes saturated. At  42 K, however, some of the bismuth donors will donate their electrons into the conduction band of silicon which, due to their mobility, significantly reduce the spin-lattice relaxation time and lower the saturation.    At temperatures significantly greater than 42 K, and as low as 60 K, the number of bismuth electrons that enter the conduction band grows so large that  the sample becomes conducting and resonance is not established.

Figure \ref{S-band cw} shows the c.w. spectrum of Si:Bi at 4.044 GHz. The broad ``Drude'' resonance, owing to the large number of conduction electrons which absorb the electric field and not the magnetic field of the microwave, have been subtracted. The stars indicate the theoretical position of the transitions, given the constants provided in Table \ref{Si:Bi constants}. These are in good agreement with the measured transitions. The large peak at $\sim 144 $ mT is, as with the X-band data shown in Fig.\ref{X-band Si:Bi}, attributed to the silicon dangling bonds \citep{Pb-centers}.  The Si:Bi resonance signals are fitted by differentiated Gaussian functions which, after integration, yield Gaussian functions whose area is proportional to the transition rate. Dividing the area of the $\varphi_{10}\leftrightarrow\varphi_{11}$ Gaussian  by that of the $\varphi_{10}\leftrightarrow \varphi_9$ Gaussian gives a transition rate ratio of $\sim$ 1.2, which agrees very well with our prediction.

\section{Pulsed spectroscopy}
Upon ascertaining the spectrum of Si:Bi at 4.044 GHz, pulsed spectroscopic techniques were used to investigate the dynamics. After  the magnetic field was set so as to be in resonance with our desired transition, the pulsed mode of the spectrometer was activated. Because the measurement process here is different to that employed in c.w. spectroscopy, and saturation is not a problem, we were able to operate at temperatures as low as 8 K. 
    
\subsection{Nutation experiment}

\begin{figure}[!htb]
\centering
\includegraphics[width=0.6\textwidth]{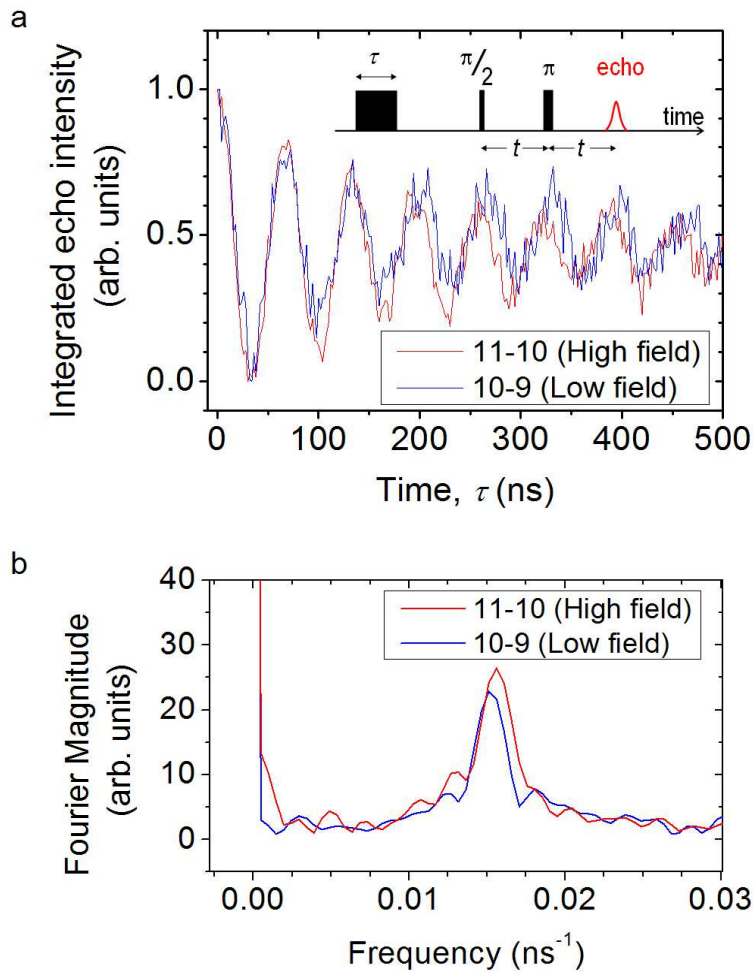}
\caption{Si:Bi Rabi oscillations at 4.044 GHz. Here,  given a constant microwave power, both transitions have approximately the same Rabi frequency.  }\label{S-band pulsed}
\end{figure}

To ascertain the relative Rabi frequencies for the two transitions the nutation experiment, described in Sec.\ref{nutation measurement scheme}, was employed. The top of figure \ref{S-band pulsed} shows the measured Rabi oscillation for each of the two transitions. The microwave power used to drive the nutation was unknown, but constant for both transitions. The lower half of figure \ref{S-band pulsed} shows the Fourier transform of the Rabi oscillations. The ratio of the Rabi frequencies is $\sim$ 1.1 which, again, agrees with our predictions. Both of the transitions, given the microwave power used, have a Rabi frequency of $\sim 0.015$ ns$^{-1}$, meaning that a $\pi$ pulse is achieved in $\sim 30$ ns.

\subsection{Relaxation times}

\begin{figure}[!htb]
\centering
\includegraphics[width=0.6\textwidth]{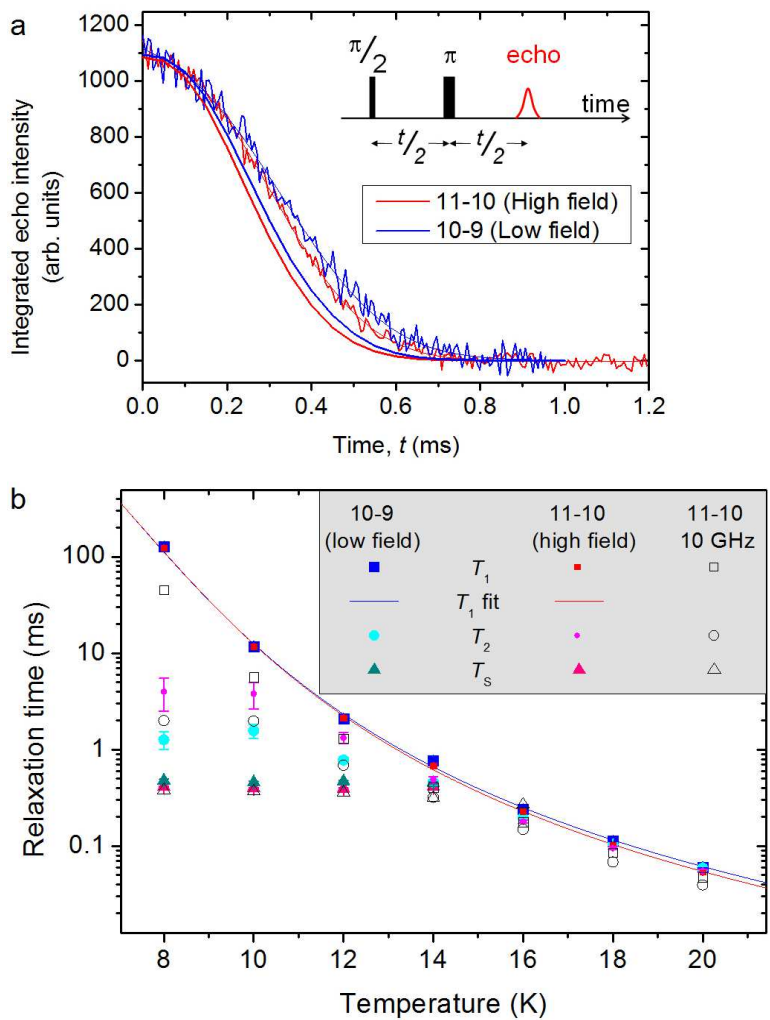}
\caption{Relaxation time scales of the two Si:Bi transitions at 4.044 GHz, as a function of temperature. a) shows the measured dephasing at 8 K compared with spin-bath simulations using the cluster correlation method performed by my colleagues in \citep{Morley-hybrid}. b) shows the experimentally extracted $T_1$, $T_2$ and $T_S$ times as a function of temperature for both of these transitions, as well as for the  high-field transition $\phi_{10}\leftrightarrow\phi_{11} $ at 9.7 GHz.}\label{S-band pulsed relaxation}
\end{figure}

The relaxation times of the system are determined using the dephasing measurement scheme described in Sec.\ref{dephasing  measurement scheme} and the amplitude damping measurement scheme described in Sec.\ref{Amplitude damping measurement scheme}. The amplitude damping time-scale, also called the spin-lattice relaxation time, is characterised by  $T_1$ using the fitting function
$e^{-t/T_1},$
 whereas the dephasing time-scale is characterised by the two parameters $T_2$ and $T_S$ via the fitting function $e^{-t/T_2-(t/T_S)^n}$. Figure \ref{S-band pulsed relaxation}(b) shows these values, measured for both the  transitions at 4.044 GHz, at various temperatures. Additionally, the relaxation times for transition $\phi_{10}\leftrightarrow\phi_{11}$ are shown at 9.7 GHz for comparison.

    At large temperatures the dominant source of decay is  the spin-lattice relaxation, due to the interaction between the system and the phonons of the silicon lattice. The details of this process are beyond the scope of this thesis, but roughly speaking the spin-lattice relaxation rate is dependent on 
\begin{enumerate}[(i)]
\item  The interaction strength of the system with the phonons, being stronger for electrons than for nuclei.   
\item  The temperature,  due to the greater thermal vibrations of the lattice at higher temperatures. 
\item The   transition frequency of the system involved in the relaxation process, which determines the relaxation rate due to the spectral density function $\omega e^{-\omega/\omega_c}$ where $\omega_c$ is a cut-off frequency.
\end{enumerate}
  The relevant time-scale here is  $T_1$ which is roughly the same for both of the 4.044 GHz transitions at all temperatures. If we  posit that the dominant transitions of the system involved in the relaxation process are within the two-dimensional subspace of the initial pseudo-pure state,   this observation can be attributed to the facts that $|\<\varphi_{10}|S_x \varphi_9\>| \simeq |\<\varphi_{10}|S_x \varphi_{11}\>|$ and that both of these transitions operate at the same frequencies. This explanation is further supported by the fact that the  transition $\varphi_{10}\leftrightarrow \varphi_{11}$  has shorter $T_1$ times at 9.7 GHz than it does at 4.044 GHz, which is correlated with the  larger transition  frequency and the fact that $|\<\varphi_{10}|S_x \varphi_{11}\>|$ is larger at 9.7 GHz than it is at 4.044 GHz.
   
Because the dephasing is limited by $T_1$, at temperatures above $\sim 18$ K the dephasing will follow an exponential fit where $T_S=0$ and $T_2\approx T_1$. At lower temperatures, the $T_1$ ceases to be the limiting factor on the dephasing, and the $T_2$ and $T_S$ times begin to plateau, not changing significantly below $10$ K. We may infer that the limiting decoherence mechanism at low temperatures is one whereby the spins have a temperature-independent interaction with the environment. Furthermore, as the $T_1$ time grows as large as 100 ms at 10 K, whereas $T_2$ and $T_S$ for both transitions are around 1 ms, we may infer that this mechanism leads, at least approximately, to pure decoherence. Because of the non-zero value of $T_S$, however, which leads to non-exponential dephasing, we know that it is not a Markovian process and, strictly speaking, our treatment in Sec.\ref{phenomenological Markovian model} does not apply here. 

The limiting coherence time at such ``cryogenic'' temperatures is the $T_S$ which,   in natural silicon, is identified with spectral diffusion due to the spin bath of $^{29}$Si isotopes. This hypothesis is strengthened by Fig.\ref{S-band pulsed relaxation}(a), where the dephasing measurement scheme is simulated given a spin-bath interacting only with the electron spin of the bismuth donor. This is simulated, effectively, by solving Eq.\eqref{spin-bath decoherence time equation} using the numerical approximation techniques of the cluster expansion.  The reason for the Gaussian-like shape of this decay can be qualitatively attributed to the non-Markovian unitary evolution of the  system and its spin-bath, just as the Rabi oscillation in a two-level system follows a sinusoidal shape.  

  Figure \ref{S-band pulsed relaxation} shows that, at 4.044 GHz, the $T_S$ timescale for the transition $\varphi_{10}\leftrightarrow\varphi_9$ is slightly longer than that of $\varphi_{10}\leftrightarrow\varphi_{11}$, which is corroborated by the simulation of the spin-bath model by the cluster expansion. A naive application of our Markovian model of adiabatic $Z$ noise, using equations \eqref{ESR adiabatic dephasing} and \eqref{NMR adiabatic dephasing}, gives the ratio of dephasing times for the two transitions as
\begin{equation}
\frac{T_S^{10 \leftrightarrow 9}\bigg|_{B_0=145.63 \ \mathrm{mT}}}{T_S^{10 \leftrightarrow 11}\bigg|_{B_0=345.02 \ \mathrm{mT}}}=\frac{\cos^4(\frac{\theta_{-4}}{2})\bigg|_{B_0=345.02 \ \mathrm{mT}}}{\sin^4(\frac{\theta_{-4}}{2})\bigg|_{B_0=145.63 \ \mathrm{mT}}}=1.4
\end{equation}    

which is also in good agreement with the experimentally determined $T_S$ coefficients. This is unsurprising as in both models the effective cause of decoherence is due to the elements of the  $S_z$ operator that are diagonal with respect to the Hamiltonian's eigenbasis. The only difference between the two models is that one is Markovian and the other is not, which only results in a difference in the \emph{shape} of the dephasing, with one being exponential and the other Gaussian-like.

\section{Summary}
In this chapter we surveyed the experimental studies of Si:Bi at S-band (4.044 GHz) conducted in ETH Zurich \citep{Morley-hybrid}. We observed that the  high-field ESR transition $\varphi_{10}\leftrightarrow \varphi_{11}$ and high-field NMR transition $\varphi_{10}\leftrightarrow \varphi_9$ have similar transition rates at S-band, with a ratio that agrees well with our theoretical predictions. This is attributed to the entanglement between the nucleus and electron spins present in the Hamiltonian eigenstates in the low-field regime. 

Furthermore, the decoherence properties of these two transitions were studied at low temperatures, and it was found to agree well with a non-Markovian spin-bath model where the electron spin interacts with the spin bath, leading to approximately pure decoherence.

\spacing{1}                                  
 \bibliographystyle{plainnat}
 \bibliography{references} 

\part{Nuclear-electronic spin systems for quantum information}
\spacing{1}
\chapter{Towards a scalable  silicon quantum computer: bismuth vis a vis phosphorus   }
\section{Introduction}
In the preceding four chapters we studied nuclear electronic spin systems, focusing on the dynamical properties that can be probed by currently available magnetic resonance technology, and provided a simple analytic model that was corroborated experimentally for the novel system of Si:Bi. In this concluding chapter I shall take a step back and, with the knowledge gained from the aforementioned analysis, reflect upon the prospects of using the silicon-donor instantiation of nuclear-electronic spin systems for the purpose of quantum information processing. We shall see that the advantages of Si:Bi over Si:P  depends strongly upon several concomitant factors that we have not had time to consider in any detail. The resolution of such queries are, as a consequence,  left as open questions.     
\section{Meeting DiVincenzo's criteria}

It is useful to construct our discussion within the paradigm of  DiVincenzo's criteria, discussed in Sec.\ref{DiVincenzo's criteria}, and see how Si:P and Si:Bi fare in each. Although these only apply within the circuit model of quantum computation, they will nevertheless present a qualitative measure of the advantages and disadvantages between the two systems. 
\subsection{Access to a scalable Hilbert space}

By definition, nuclear-electronic spin systems exist in a Hilbert space, so the first half of this condition is automatically satisfied; a single donor of Si:P is exhibited by the Hilbert space $\co^4$ whereas Si:Bi has the larger space $\co^{20}$. 
 In order to have a \emph{scalable} Hilbert space, on the other hand, we need to be able to have many systems,  between which we can establish an interaction. Such an interaction may be effected by the overlap of the respective wavefunctions, or by dipole interactions. Because the nuclear spins are much more localised than the electrons, and that they also have a much weaker gyromagnetic ratio, direct interaction between them is much more difficult to realise than is the case for electrons.    We may, therefore,   set up an exchange Hamiltonian, often also referred to as a Heisenberg or $JJ$-interaction,
between the electrons of adjacent donors\begin{equation}
H^{JJ}=\mathcal{J}\sum_{i \in \{x,y,z\}}S_i\otimes S_i\equiv \mathcal{J}S_z\otimes S_z+\frac{\mathcal{J}}{2}\left(S_-\otimes S_++S_+\otimes S_- \right)\label{two-qubit kane interaction Hamiltonian}
\end{equation}
   where the strength $\mathcal{J}$ is a controllable parameter. Another proposal is that  made by \citep{Stoneham-2003}, purporting the use of control atoms between the systems of interest which, using optical means, can be excited to their Rydberg states. These excited states have a much larger wavefunction, and  would overlap with the electronic wavefunctions of the system of interest, and constitute an intermediated interaction between the two.  The interaction Hamiltonian for the system is thus
\begin{equation}
H^{JJ-JJ}=\Pi_e^C\left(\sum_{i \in \{x,y,z\}}\mathcal{J}_1S_i^A\otimes S_i^C + \mathcal{J}_2S_i^B\otimes S_i^C\right)\label{two-qubit stoneham interaction Hamiltonian}
\end{equation}
where $A$ and $B$ are the systems of interest, and $C$ is the control system whose excited Rydberg state is denoted as $\Pi_e^C$.

A less technologically demanding proposal would be  to establish a permanent  interaction between adjacent donors. This can be achieved by, for example, placing the donors close enough such that their electronic wavefunctions are constantly overlapping, leading to an always-on exchange interaction that permits  quantum computation schemes such as that proposed by \citep{Sougato-Benjamin-2003}.  
\subsection{Initialisation}

We require the ability to initialise all of our qubits to some fiducial state prior to initialising any algorithm.  In many physical systems the most readily available method of initialisation is by cooling, or polarisation, of the ensemble to the ground state. For nuclear electronic spin systems this is the state  $\phi^-_{I-1/2}$. Yet owing to the small nuclear spin gyromagnetic ratio of systems such as Si:P and Si:Bi, the amount of polarisation, achievable under the available temperatures and desirable magnetic fields, is not satisfactory. It has been demonstrated, however, that the system may be \emph{hyperpolarised} \footnote{A composite
 system is hyperpolarised to a state $\psi$ if the percentage of the population in this state is larger than would be given simple thermal polarisation. If the system $\bigotimes_{n=1}^N \rho^n$ is fully hyperpolarised, then it is brought to the state $\psi^{\otimes N}$.} by optical means to the state $\phi^-_{-I-1/2} $. For  Si:P this hyperpolarisation has been demonstrated to be as large as $\sim 68$ \% \citep{mccamey-hyperpolarisation}, achieved  at a temperature of 1.37 K and a magnetic field of 8.5 T.  Si:Bi, on the other hand, has been hyperpolarised to   $\sim 90$ \% \citep{sekiguchi-2010} at 1.5 K and 6 T. Of course, these numbers may be brought higher still with further refinement of the techniques used, as is evident by the recent work of \citep{Steger-Si-180s} who, using isotopically purified silicon with less than 50 ppm $^{29}$Si isotopes, allowing for donor densities of less than $10^{12} \ \mathrm{cm}^{-3}$, managed to hyperpolarise Si:P to $\sim  90 \ \%$.

\subsection{Universal set of quantum gates} 

\begin{figure}[!htb]
\centering
\includegraphics[width=3in]{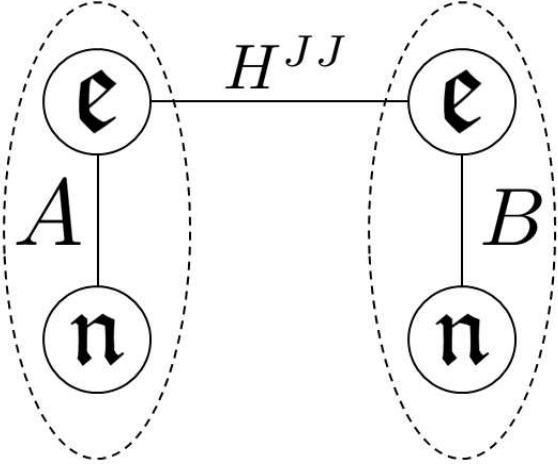} \caption{  The CZ gate between electrons can be achieved using the Heisenberg interaction Hamiltonian.  } \label{dimer}
\end{figure}  
With magnetic resonance we are capable of performing any unitary operator from the group $\su(2)$ within any two-dimensional subspace, comprised of  eigenstates where a transition between the two is allowed by the dynamical selection rule for $F_x=S_x+\delta_\gamma I_x$. Consequently any single-qubit gate is achievable with magnetic resonance and, additionally, a natural designation for the logical basis would be the eigenstates of the Hamiltonian itself. This claim comes with the caveat that we are able to either focus our driving magnetic field so as to only affect the desired donor, or establish different resonant frequencies for all the systems, perhaps by a gradient in the static magnetic field, so that we achieve single-donor selectivity even with a non-resolving driving field. The single-qubit gates from the standard set, for example, can be constructed as
\begin{align}
P_{\pi/8}&= e^{\imag \frac{\pi}{8}}e^{-\imag \frac{3\pi}{2}\bar \sigma_x}e^{-\imag \frac{7\pi}{4}\bar \sigma_y}e^{-\imag \frac{\pi}{2}\bar \sigma_x}, \nonumber \\
H&=e^{\imag \frac{\pi}{2}}e^{-\imag 2\pi\bar \sigma_x}e^{-\imag \frac{3\pi}{2}\bar \sigma_y}e^{-\imag \pi\bar \sigma_x},
\end{align}
where the basis on which they act are the two eigenstates $\{\varphi_0,\varphi_1\}$ that obey $|\<\varphi_0| F_x\varphi_1\>| >0$. Of course it is possible to use more than a two-level subspace provided by the system in question. The entire Hilbert space may, in principle, be manipulated by magnetic resonance pulses that are tuned, in sequence, to any of the permissible transitions; we may perform any unitary operation on a single system from the group $\su(d)$. As such Si:P may provide two qubits whereas Si:Bi can provide up to four qubits, leaving four energy levels to spare. Even if we were inclined to do this, true two-qubit gates between adjacent donors are still required for scalability.  Assuming for a controllable interaction Hamiltonian of the form in Eq.\eqref{two-qubit kane interaction Hamiltonian}, as shown by \citep{quantum-dots,XYtwoqubit}, it is possible to establish the root-swap gate
\begin{equation}
\sqrt{\mathrm{SWAP}}=e^{\imag \frac{\pi}{8}}e^{-\imag\frac{\pi}{2\mathcal{J}} H^{JJ}}
\end{equation}
between two adjacent donor electrons by activating the coupling  for a period $\tau = \pi/2\mathcal{J}$. This  can in turn be used to construct the CZ gate, for the electrons, with the sequence
\begin{equation}
\mathrm{CZ}=e^{-\imag \frac{\pi}{4}}\left(e^{-\imag \frac{\pi}{2}\bar \sigma_z }\otimes e^{-\imag \frac{3\pi}{2}\bar \sigma_z } \right)\sqrt{\mathrm{SWAP}}\left(e^{-\imag \frac{\pi}{2}\bar \sigma_z }\otimes\mathds{1}\right)\sqrt{\mathrm{SWAP}}
\end{equation}  
where the single qubit gates can, as previously stated, be realised by means of magnetic resonance. 

It should be noted that this scheme requires $\mathcal{J}$ to be strong in comparison with the relevant energy differences of the states involved. However, if the energy differences far outweigh the coupling strength, then the Hamiltonian of Eq.\eqref{two-qubit kane interaction Hamiltonian} simplifies to an Ising interaction
\begin{equation}
H^{JJ} \approx H^{ZZ}= \mathcal{J} S_z\otimes S_z
\end{equation}
which can be used to establish a CZ gate with the sequence
\begin{equation}
\mathrm{CZ}=e^{-\imag \frac{\pi}{4}}\left(e^{-\imag \frac{3\pi}{2}\bar \sigma_z }\otimes e^{-\imag \frac{3\pi}{2}\bar \sigma_z } \right)e^{-\imag \frac{\pi}{\mathcal{J}}S_z\otimes S_z}.
\end{equation}
 Note that this implementation of the CZ gate  requires only one use of the Ising interaction between the electrons, whereas that using the full exchange Hamiltonian requires  two root-swap gates \citep{Makhlin}.  

A non-trivial question pertaining to quantum control now becomes apparent.   It is clear that we are capable, using such  exchange interactions, to achieve controllability in the composite \emph{electron} system. It is not self evident, however, that we are also capable of doing this, at all magnetic fields, in the composite  \emph{nuclear-electronic} spin system. To elaborate on this we refer to Fig.\ref{dimer} where the nuclear-electronic spin system on the left is labeled system $A$ and that on the right system $B$, with an $H^{JJ}$  interaction being established between the respective electronic subsystems of each.
Because of our ability to perform any unitary from $\su(d)$ on a single donor with magnetic resonance, the issue of controllability simplifies to the ability of preparing a maximally entangled state between two donors. In the high-field limit, then, the two donors may  initially be prepared in the product state \begin{equation}
|+\>^{A,\mathfrak{e}}\otimes|m_I\>^{A,\mathfrak{n}}\otimes|+\>^{B,\mathfrak{e}}\otimes|m_I'\>^{B,\mathfrak{n}}
\end{equation}
  where $|\pm\>=\frac{1}{\sqrt{2}}(|+1/2\>\pm|-1/2\>)$, such that after the application of the CZ gate  the composite system is transformed to 
\begin{align}
&\frac{1}{\sqrt{2}}\left(\left|-1/2\right\>^{A,\mathfrak{e}}\otimes |+\>^{B,\mathfrak{e}}+\left|+1/2\right\>^{A,\mathfrak{e}}\otimes|-\>^{B,\mathfrak{e}}\right)\otimes|m_I\>^{A,\mathfrak{n}}\otimes |m_I'\>^{B,\mathfrak{n}}\nonumber \\ &=\Phi^+_{\mathfrak{e}}\otimes|m_I\>^{A,\mathfrak{n}}\otimes |m_I'\>^{B,\mathfrak{n}}
\end{align}
which is maximally entangled with respect to the electronic Hilbert spaces. \footnote{  The nuclear spins here act as ancillary systems, and we may transfer the entanglement to these spins using local NMR pulses.} Clearly, by relabeling  in the adiabatic basis, this state can also be written as 
\begin{equation}
\Phi^+=\frac{1}{\sqrt{2}}\left({\phi^-_m}^A\otimes\varphi_+^B+{\phi^+_m}^A\otimes\varphi_-^B \right)
\end{equation}
where $\varphi_\pm^B=\frac{1}{\sqrt{2}}(\phi^+_{m'}\pm \phi^-_{m'-1})$, so that we have established a CZ gate in the adiabatic basis also. Consequently we know that, at high field,  the Heisenberg interaction suffices for controllability within at least a four-dimensional subspace of the full composite system. The question remains as to the possibility of controllability within the full Hilbert space of the composite system, and/or the feasibility of doing so in the low-field regime where the entanglement between the nuclear and electronic spins of a single donor comes into play.    A concrete resolution of this question requires careful analysis of the Lie algebra provided by the Heisenberg interaction and magnetic resonance pulses. The answer to this question will greatly affect any possible advantage Si:Bi may  have over Si:P; if we can only establish controllability within a four-dimensional subspace of Si:Bi, even at high fields, then its larger Hilbert space may not be of much use, and if it is impossible to obtain controllability in the low-field regime then the OWPs will  not offer  an advantage either.

\subsection{Coherence times compared with gate times}
For fault-tolerant quantum computation, the coherence times of the system must be sufficiently long in comparison with the gate times. 
 At sufficiently low temperatures  the limiting cause of decoherence, which is  approximately pure,  is attributed to  the interaction of the central system of interest's electron spin  with the  ``bath'' of surrounding spin-half $^{29}$Si nuclei and other donor electron spins.  As demonstrated by \citep{Tyryshkin-2011} the limiting coherence time of Si:P with less than 50 ppm $^{29}$Si isotopes, in the high-field limit and at approximately 5 Kelvin,  was the $T_2$ time, taking the  value of $\sim1$ s. This is very long indeed, and as shown by \citep{Steger-Si-180s} who also used a lower concentration of donors, the nuclear spin $T_2$ time was brought to the even larger value of $\sim 180$ s. Due to the similarity of the decoherence mechanisms in both Si:P and Si:Bi, such  impressively long $T_2$ times can be expected for isotopically enriched Si:Bi as well. Furthermore, due to the smaller active region in Si:Bi, resulting from the greater localisation of the electron spin wavefunction, we may expect even longer coherence times where the contribution of the residue $^{29}$Si isotopes are reduced further still. In addition the quantum gates by magnetic resonance can be achieved at much faster rates in Si:Bi than is possible for Si:P, owing to the larger gap between the transition frequencies present in the former system as a result of the larger hyperfine coupling strength. Therefore, Si:Bi is the more advantageous system with respect to the gate time to coherence time ratio.

An issue that has been discussed previously is the advantage of working in the low-field regime. In the high-field limit, quantum gates in the ESR subspaces spanned by $\{\phi^+_m,\phi^-_{m-1}\}$ are much faster than  those in the NMR subspaces spanned by $\{\phi^\pm_m,\phi^\pm_{m-1}\}$. This is due to both the greater interaction strength between the driving field and the electron spin, and the fact that ESR frequencies are much more widely separated than is the case for the NMR frequencies. By moving to the low-field regime, the speed of the quantum gates associated with ESR transitions in the high-field limit decrease by a factor of one-half, but for Si:Bi it is possible to gain a greater speed-up for gates in the high-field limit NMR subspaces. This, coupled with the presence of OWPs in Si:Bi, would suggest that working in the low-field regime is preferable overall. However, when we have approximately removed the spin-bath sources of decoherence,  thermal noise becomes the dominant source of decoherence, and we must ask how its effects vary for different subspaces and different magnetic field regimes.  Fig.\ref{S-band pulsed relaxation} demonstrates that, for Si:Bi at 4.044 GHz, the $T_1$ time of the high-field NMR subspace $\{\phi^-_{-4},\phi^-_{-5}\}$  is as short as that of the ESR subspace $\{\phi^+_{-4},\phi^-_{-5}\}$. This can be attributed to the similar rates of both these transitions, as well as their similar transition frequencies.  Therefore, if we are to limit ourself to just a two-dimensional subspace of Si:Bi, it will always be preferable to work in the high-field ESR subspaces $\{\phi^+_m,\phi^-_{m-1}\}$ and not the high-field NMR subspaces $\{\phi^\pm_m,\phi^\pm_{m-1}\}$, because while both will have similar $T_2$ times at the OWPs, and are expected to also have  $T_1$ times of the same order of magnitude, the former will permit much faster gate times due to the larger gap in its transition frequencies.

\subsection{Measurement}
\begin{figure}[!htb]
\centering
\includegraphics[width=5in]{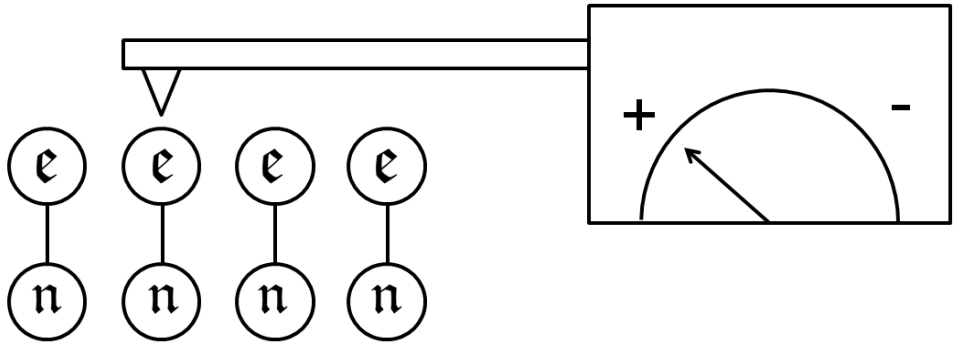} \caption{  An apparatus that performs strong, projective measurements on the electronic subspace of a single nuclear-electronic spin system.  } \label{electron PVM}
\end{figure}

The issue of measurement is central to the experimental realisation of QIP in silicon. In the work done in this thesis all ``measurements'' were in fact weak ensemble measurements which only reveal the expectation values of sharp observables. What we require, for quantum computation, is the ability to perform strong measurements, described in Sec.\ref{observables}, where the different outcomes of a measurement can be distinguished from one another.   Various techniques for strongly measuring single donor systems in silicon are concurrently being developed by several research groups. An issue of great importance pertaining to our work is that, because of the difficulty in establishing controlled interactions between the nuclear spin and a measuring apparatus, especially in a way that leaves the electron spin unaltered, the research to date has focused on either measuring with respect to the Hamiltonian eigenbasis, or just the electron spin $S_z$ operator. Some of these approaches are electrical spin to charge conversion \citep{kane1998,single-spin-transistor,resonant-tunneling,single-electron-transistor}, electrically detected magnetic resonance (EDMR) \citep{non-demolition-measurement-donor,Morley-2008-1}, magnetic resonance force microscopy \citep{MRFM-single-spin} and optical read-out \citep{optical-spin-readout,optical-spin-to-charge-readout}.  

One  example is the scheme proposed by  \citep{single-shot-electron-measurement,Jarryd-2012-single-atom}, which in an abstract, ideal setting , can be described by   a von Neumann-L\"uders measurement, described in Sec. \ref{Measurement model}, which uses a  unitary interaction $U$ between the system and the apparatus' charge degree of freedom, initially set to the state $\varphi$, such that \begin{align}
U:\begin{cases}\phi^-_m\otimes\varphi &\mapsto \phi^-_m\otimes\varphi_- \ \forall \ m,  \\
\phi^+_m\otimes\varphi &\mapsto \phi^+_m\otimes\varphi_+ \ \forall  \ m.  \\
\end{cases} 
\end{align} 
This will allow us to store our quantum information within a two-dimensional subspace $\{\phi^+_m,\phi^-_{m-1}\}$ and measure projectively with respect to this basis. However, we may also use   magnetic resonance to store our quantum information in the $\{\phi^-_m\}$ subspace, and with a $\pi$ pulse of the correct frequency convert the joint object-apparatus state 
\begin{equation}
\psi\otimes\varphi=\sum_m\alpha_m \phi^-_m\otimes\varphi \label{low-field regime measurement}
\end{equation}
to
\begin{equation}
\left(\alpha_n \phi^+_{n+1} + \sum_{m\ne n}\alpha_m \phi^-_m\right)\otimes\varphi
\end{equation}   
which is mapped by the measurement unitary $U$ to 
\begin{equation}
\alpha_n \phi^+_{n+1}\otimes\varphi_+ + \sum_{m\ne n}\alpha_m \phi^-_m\otimes\varphi_-.
\end{equation}
The reduced density operator of the apparatus is then given by
\begin{equation}
|\alpha_n|^2\Pi(\varphi_+)+ \sum_{m_I \ne n}|\alpha_{m}|^2 \Pi(\varphi_-)
\end{equation}
 and measurement of the apparatus with respect to the orthonormal basis $\{\varphi_+,\varphi_-\}$ describes the sharp observable with the two effects $\{\Pi(\phi^-_n), \mathds{1}-\Pi(\phi^-_n)\}$, and the procedure can be repeated (given the same initial states of object and probe) for all the resonance frequencies to achieve the sharp observable  with the $2I+1$ effects $\{\Pi(\phi^-_m)\}$. To ensure that this measurement is repeatable however, we must perform a second $\pi$ pulse to send the post-measurement state $\Pi(\phi^+_{n+1})$ to the state $\Pi(\phi^-_n)$. In the limit of $(\gamma_e+\gamma_n)B_0/A_{\mathrm{iso}}\to \infty$, this measurement describes a projective measurement on the nuclear spin.

\section{Summary}
In this chapter we considered the prospects of using the systems Si:Bi and Si:P as platforms for quantum computation, comparing their merits with respect to  DiVincenzo's criteria. The larger transition frequency gaps and the presence of decoherence free subspaces in the form of optimal working points indicate Si:Bi as being the more advantageous system. Furthermore, current experimental techniques permit us to perform strong projective measurements on Si:Bi within either a two-dimensional subspace $\{\phi^\pm_m,\phi^\mp_{m-1}\}$, or a ten-dimensional subspace  $\{\phi^-_m\}$, at all magnetic fields, compared with  only a two-dimensional subspace of Si:P. This poses the possibility of utilising a larger Hilbert space in a single site of Si:Bi.

Restricting ourselves to just a two-dimensional subspace $\{\phi^\pm_m,\phi^\mp_{m-1}\}$, it is preferable to operate in the  low-field limit. This is because the gap in transition frequencies do not change by much, and the transition rates only decrease by one-half, but the presence of OWPs increase the spin-bath induced coherence times by a much higher amount.  

The main issue that has yet to be solved is that of scalability;  can we perform entangling gates between two sites of Si:Bi such that we are able to achieve full controllability in the composite ``qubit'' subspace, and do so at all magnetic fields? If we are unable to do so in the low-field regime, then the advantages of the optimal working points in Si:Bi will no longer exist for scalable quantum computation. This shall be left as an open question. 

\spacing{1} 
  \bibliographystyle{plainnat}                                 
  \bibliography{references}

\spacing{1}
\chapter{Conclusions}

\section{Comprehensive summary}
In this thesis it was shown that  materials such as group V dopants in silicon --  Si:P and Si:Bi for example -- at low temperatures can be modeled effectively as nuclear-electronic spin systems. Within this paradigm,   simple analytic expressions regarding the magnetic resonance properties of such systems were developed which have been shown to match well with experiment. One of the more interesting consequences of this analysis is that it shows that conceptualising  magnetic resonance in nuclear-electronic spin systems as being either NMR or ESR is only valid in the asymptotic high-field limit. The more general notion of nuclear-electronic magnetic resonance (NEMR) was thus developed, of which NMR and ESR are asymptotic limits. 

The possibility of achieving full quantum control, with magnetic resonance pulses alone, of nuclear-electronic spin systems was also demonstrated. In high fields, pulses within NMR subspaces take three orders of magnitude longer to perform than those in ESR subspaces, owing to the weaker gyromagnetic ratio of the nuclear spin. Accuracy of control, limited by the gap between the desired transition frequency and all unwanted frequencies, can be mitigated to some extent by switching between right-handed and left-handed circularly polarised driving fields.   Furthermore, it was shown that, in Si:P, accurate control of all subspaces is achieved with optimal speed in the high-field limit. In Si:Bi on the other hand, ignoring the effects of temperature-dependent relaxation processes, a speed up of gates in the high-field NMR subspaces can be realised by operating in the low-field limit.  This is because, at such regimes, the limiting transition frequency gap for these subspaces are maximised.        

In addition, it was demonstrated that nuclear-electronic spin systems with nuclear spins greater than one, such as Si:Bi but excluding Si:P, have certain magnetic field values called optimal working points,  which are closely associated with the field values where the derivative of the transition frequency with respect to the magnetic field vanishes, in the low-field regime. Each of these optimal working points in turn has associated with it a  two-dimensional  decoherence free subspace, wherein pure decoherence mechanisms due to the interaction of the environment with  the electron spin may be  suppressed.  These processes include, but need not be limited to, the interaction of a bath of spin one-half $^{29}$Si nuclei with the donor electron spin. 

It was also shown that, using an energy-dependent measuring scheme suggested by recent work, we may perform strong projective measurements on up to a ten-dimensional subspace of Si:Bi, compared with only a two-dimensional subspace of Si:P, at all magnetic fields. 

In summary, then, the larger hyperfine interaction strength and the nuclear spin of Si:Bi, compared with those of Si:P, suggest it as an advantageous system for QIP. In the low-field regime, excluding thermal noise contribution to decoherence, the speed of our quantum gates are much faster than the coherence time of the system.

\section{Open questions}
There are several avenues left open for future research  ranging from engineering concerns such as building of magnetic resonance equipment that can operate at different  frequencies, and  instruments that can   measure and control a single spin object, to more fundamental questions relating to the quantum mechanical phenomenon of nuclear-electronic spin systems beyond magnetic resonance. 

To assess the true benefits of working in the  subspaces spanned by $\{\phi^-_m\}$ in the low-field regime, for example, both experimental and theoretical work must be done with aims at understanding the behaviour of thermalisation across the different magnetic field regimes. In other words, will the benefits of gate speed-up and increase of $T_2$ times at the OWPs  outweigh the concomitant decrease in $T_1$ times?  

 Another important question that needs to be answered  is the  possibility of generating entanglement between adjacent  donors with respect to the coupled nuclear-electronic, or adiabatic, eigenbasis. We saw in the previous chapter that, by coupling the electron spins of two adjacent sites with a Heisenberg interaction, we may generate entangling gates -- such as the CNOT gate -- that act on the electrons. It does not follow, however, that this will allow for establishing a CNOT that acts on  the adiabatic basis in the low-field regime, where the eigenstates of the Hamiltonian of a single system show entanglement between the electron and nucleus. Nor does this suggest the possibility of generating entanglement between arbitrary subspaces of the two systems; utilising the full ten-dimensional subspace of Si:Bi, that can be measured projectively, in a scalable quantum computer would require a method of entangling all ten of these degrees of freedom with the counterparts of an adjacent system.

Interesting phenomenon may also be investigated by considering nuclear-electronic spin systems with a Hamiltonian that couples to electric fields; this would be useful for hybrid quantum computation architectures which involve superconducting circuits that would be perturbed by the presence of a magnetic field.


\appendix
\spacing{1}                                                      

\chapter{Angular momentum}\label{angular momentum appendix}
From classical mechanics, and the theorem named after Emmy Noether, we know that any continuous and differentiable coordinate transformation, that leaves the equations of motion invariant, i.e.  is a \emph{symmetry} of the system, has associated with it a conserved quantity. Conversely, as shown by use of the Poisson brackets, such a conserved quantity is the \emph{generator} of the aforementioned continuous and differentiable coordinate transformation. Angular momentum is the quantity that is conserved by rotational symmetries; it generates infinitesimal rotations in a system. Although such a concept has an intuitive explanation regarding the rotation of bodies in classical mechanics,  the analogue to angular momentum in quantum mechanics only has meaning in terms of the generators themselves, which are self-adjoint operators on a Hilbert space. In this section, then, I shall briefly cover the basics of angular momentum in quantum mechanics to the extent needed for this thesis; for the intrinsic spin of a system. Further details can be found in \citep{Sakurai}.

The operators $\{ J_x,  J_y,  J_z\}$ form the set of  mutually incompatible observables for the    spin of a particle. These obey the commutation relations
\begin{equation}
\left[ J_x, J_y  \right]_- = \imag\epsilon_{xyz} J_z \ 
\end{equation}
 and generate the unitary transformations on the Hilbert space, that can be thought of as rotations, by the Taylor series
\begin{equation}
\sum_{n=0}^\infty=\frac{(-\imag \theta J_i)^n}{n!}\equiv e^{-\imag J_i \theta}
\end{equation}
where we note that $J_i^0=\mathds{1}$. 

We may define the operator $\vec{\mathbf{J}}^2:=J_x^2+J_y^2+J_z^2$ which commutes with each of $\{J_x,J_y,J_z\}$; this operator is diagonal in the same basis as each of the angular momentum operators. By convention, we consider the system in the basis in which $J_z$ and $\vec{\mathbf{J}}^2$ are both diagonal, labeled $|m_J\>$  such that 
\begin{align}
&J_z|m_J\>=m_J|m_J\> &\vec{\mathbf{J}}^2|m_J\>=J(J+1)|m_J\>.
\end{align} To determine the  possible values $m_J$ can take given $J$, which itself can take only integer or half-integer values, we introduce the ladder operators
\begin{equation}
 J_\pm :=  J_x \pm \imag J_y \ 
\end{equation}
which have the properties
\begin{align}
[J_+,J_-]_-&=2J_z, \nonumber \\  [J_z,J_\pm]_-&=\pm J_\pm, \nonumber \\ [\vec{\mathbf{J}}^2,J_\pm]_-&= \mathds{O}. 
\end{align}
Because of the last of these commutation relations, the ladder operators do not cause a change in $J$; we may therefore consider this as the total angular momentum that is conserved.
Due to this angular momentum conservation, given a state with total angular momentum $J$, there are states with a maximum and minimum value of $m_J$ ; $|m_J^{\max}\>$ and $|m_J^{\min}\>$, such that
\begin{align}
&J_+|m_J^{\max}\> = \phi_\mathrm{null} &
J_-|m_J^{\max}\> = \phi_\mathrm{null} \ .
\end{align} 
It therefore follows that there are a finite number of values $m_J$ can take, with the range
\begin{equation}
m_J= -J,-J+1,...,J-1,J \ .
\end{equation}

That is, $m_J$ values are separated by unit intervals, and range between $\pm|J|$. Therefore, the Hilbert space of a system of spin $J$ has dimension $2J + 1$. 

With some algebra utilising the commutation relations, the rules governing the application of the  ladder operators  on the $J_z$ eigenbasis can  be expressed thusly:
\begin{align}
& J_\pm|{m_J}\> = C_\pm|m_J\pm1\> \nonumber \\
C_\pm^J := &\sqrt{J(J+1)-m_J(m_J\pm1)} \ .
\end{align}

\chapter{Bipartite entanglement measures}\label{entanglement measures}
A bipartite system is inseparable, or entangled, if it cannot be written as a convex combination of product states
\begin{equation}
\rho=\sum_iP(i) \rho_i^A\otimes\rho_i^B.
\end{equation} 
To quantify entanglement, we  work in the paradigm of local operations and classical communication (LOCC). A bipartite state $\rho$ is said to be more entangled than the state $\varrho$ if, using the class of operations allowed by LOCC, we can transform $\rho$ to $\varrho$, but not the converse. Entanglement, then, is a quantity that cannot be increased by LOCC. It follows that a good entanglement measure must depict such an ordering; it must be an entanglement monotone.  Two good review articles on entanglement are \citep{plenio-2007-7,Horodecki2007}. In what follows I shall give a brief description of the most commonly used entanglement measures which are computationally easy to perform.
\section{Entropy of entanglement}\label{von Neumann entropy}

\subsection{Von Neumann entropy and information}

 Analogous to classical information, the information in a  quantum system can be quantified by its von Neumann entropy \citep{quantum-coding}. The von Neumann entropy of a state $\rho \in \s(\co^d)$ is given by
\begin{equation}  
S(\rho) := -\mathrm{tr}\left[\rho \ \mathrm{log}_d ( \rho )\right]\equiv -\sum_{i=1}^dP(i)\ \mathrm{log}_d[P(i)]
\end{equation}
where the probabilities $\{P(i)\}$ are the  eigenvalues in the spectral decomposition  of $\rho$. The maximum value of the von Neumann entropy is $\mathrm{log}_d(d)=1$, realised for a maximally mixed state.  To see how the von Neumann entropy gives quantum information an operational meaning, consider a preparation device that    produces a  product state $\varrho \in \s(\co^{d^N})$ composed of $N$ pure  states stochastically chosen from the  orthonormal basis $\{\varphi_i\}_{i=1}^d$.   The preparation device can thus be described by the statistical ensemble $\rho=\sum_iP(i)\Pi(\varphi_i)$. The composite system is then sent to a receiver with access to a  measuring device that can unambiguously distinguish each of the orthogonal pure states, and who also has a priori knowledge of the statistical ensemble.     In the thermodynamic limit of $N \to \infty$, the law of large numbers may be applied to say that in any $\varrho$ that the preparation device produces, each pure state $\Pi(\varphi_i)$ appears $NP(i)$ times; these are the so-called typical sequences, each of which can be used to encode a message. There exists a unitary operator $U$ that can be used to compress the information content in all of the typical sequences $\varrho$ to a subspace  
\begin{equation}
\s\left(\co ^{d^{NS(\rho)}}\right)
\end{equation} such that the rest of the state space is redundant. The larger the von Neumann entropy of the statistical ensemble $\rho$, the greater the information content is and, consequently,  the less  $U$ can compress the information.  For example, if $\rho=\frac{1}{d}\mathds{1}$ then $S(\rho)=1$ and no compression is allowed. In the other extreme case, if $\rho=\Pi(\varphi_i)$ then $S(\rho)=0$ and no data need be transmitted at all, as the a priori knowledge of the receiver suffices for him to know what the sent message will be!   

\subsection{Entropy of entanglement}

A unique measure of entanglement for pure bipartite states is given by the von Neumann entropy of the reduced density operators of said state, called the entropy of entanglement. For a pure state $\Pi(\psi)\in \s(\h^A\otimes\h^B)$, this is given as
\begin{equation}
E[\Pi(\psi)] := S[\rho^{B}] \equiv S[\rho^{A}]
\end{equation}
which is normalised to one for a maximally entangled state.
The reason that the entropy of entanglement is symmetric for both subsystems is that it depends solely on the Schmidt coefficients of the pure bipartite state in question.       The  vector  $\psi \in \h^A\otimes \h^B$ associated with the  pure state $\Pi(\psi)$ can be written in its Schmidt form
\begin{equation}
\psi = \sum_{i=1}^k \sqrt{P(i)}\phi_i\otimes\varphi_i \ \ 
\end{equation}

\noindent where $k \leqslant \min(d_A,d_B)$ is the Schmidt-rank of $\psi$, and $\{\phi_i\}_{i=1}^{d_A}$ and $\{\varphi_i\}_{i=1}^{d_B}$ are an orthonormal basis of $\h^A$ and $\h^B$ respectively in which the reduced density operators $\rho^A$ and $\rho^B$ are diagonal. The Schmidt coefficients are   $\{\sqrt{P(i)}\}$  where $\{P(i)\}$ are the eigenvalues of both the reduced density operators. As such, the entropy of entanglement can simply be calculated as 
\begin{equation}
E[\Pi(\psi)]=-\sum_{i=1}^kP(i)\mathrm{log}_N[P(i)] \ \ \text{where} \ \ N = \min(d_A,d_B).
\end{equation}
The maximally entangled states represented in the Schmidt form are, equivalent up to local unitary transformations, given by  
\begin{equation}
\Phi^+:=\frac{1}{\sqrt{\min(d_A,d_B)}}\sum_{i=1}^{\min(d_A,d_B)}\phi_i^A\otimes\varphi_i^B.
\end{equation}
 
The entropy of entanglement gives entanglement an operational meaning in terms of non-local information. A pure product state will have pure reduced density operators, which in turn have zero entropy. Here, there is no non-local information. On the other hand a maximally entangled state will in turn have  maximally mixed reduced density operators and, as such, a maximal amount of non-local information contained in the correlations between the subsystems. This motivates the definition of entanglement as the \emph{ebit}, where a maximally entangled state of two qubits is one ebit. With access to the LOCC operation class, Alice may then \emph{teleport} a qubit of quantum information  to Bob  if each has a part of a maximally entangled state \citep{Teleportation}; in other words, if the two shared one ebit of quantum information. 

   Another method of operationally defining entanglement is by how many maximally entangled states we may \emph{distill}. Given $N$ copies of a pure bipartite state $\Pi(\psi)$, the number $M$ of maximally entangled states that can be obtained by LOCC, in the thermodynamic limit of $N\to \infty$, is given by
\begin{equation}
\lim_{N\to \infty}\frac{M}{N}=E[\Pi(\psi)].
\end{equation}

\section{Entanglement of formation and concurrence}\label{concurrence}
The von Neumann entropy satisfies the conditions for an entanglement measure  for pure states. But what about mixed states, or more specifically, a convex combination of entangled pure states? These are not separable and hence also entangled, although they can never be maximally entangled as these are always pure. However,  we cannot use the von Neumann entropy to determine \emph{how} entangled they are.  We need to look for a more robust measure of entanglement. The \emph{entanglement of formation} of a bipartite state is defined as the minimum average entanglement -- as given by the von Neumann entropy -- of an ensemble of pure states that would produce $\rho$ 
\begin{equation}
E_F[\rho]=\min_{\{\Pi(\psi_i)\}}\sum_iP(i)E[\Pi(\psi_i)].
\end{equation}
Clearly, as there are an infinite  such pure state decompositions, performing this minimisation  is no easy task! However, an identity exists called the \emph{concurrence} of the density operator, $\mathcal{C}[\rho]$ \citep{woottersconc},  which is defined  for any  $\rho \in\s(\mathds{C}^2\otimes \mathds{C}^2)$  that simplifies the problem.
The concurrence is given by
\begin{align}
{\cal C}[\rho] &= \max\{0,\lambda_1 -\lambda_2-\lambda_3-\lambda_4 \} \nonumber \\
\lambda_i&=\mathbb{R}\left[\sqrt{\mathrm{eig}_i[\rho\breve\rho]}\right], \lambda_1>\lambda_2>\lambda_3>\lambda_4
\nonumber \\ \breve\rho&:=(\sigma_y\otimes\sigma_y) {\rho^*}(\sigma_y\otimes\ \sigma_y)
\end{align}
where $\mathrm{eig}_i[T]$ is the $i^{\mathrm{th}}$ eigenvalue of the operator $T$, and $\mathbb{R}[\cdot]:\mathds{C}\to\mathds{R}$ gives the real component of any complex number $c$. Here, $\rho^*$ is the complex conjugate of $\rho$ \footnote{A density operator is self-adjoint. But an adjoint of a matrix is the complex conjugate of its transpose. Therefore $\rho\ne \rho^*$ unless all the matrix elements are real.}. The concurrence is an entanglement monotone and provides real values in the range of $[0,1]$. The entanglement of formation  is calculated from the concurrence of a state as follows:
\begin{align}
E_F[\rho]&= \mathcal{F}[\mathcal{C}(\rho)], \nonumber \\
\mathcal{F}[\mathcal{C}]&=h\left[\frac{1+\sqrt{1-\mathcal{C}^2}}{2} \right], \nonumber \\
h[x] &=-x\mathrm{log}_2x-(1-x)\mathrm{log}_2(1-x).
\end{align} 
clearly, for a pure state the above formula will just give the entanglement of the state as is determined by the entropy of entanglement, and a concurrence of 1 is achieved for the maximally entangled pure states.

\section{The positive partial transposition  criteria and negativity }\label{negativity}
Although the concurrence is a good measure for entanglement of an arbitrary state $\rho \in \mathds{C}^2\otimes\mathds{C}^2$, we would like to determine whether or not a state in a higher dimensional state space is entangled, and also preferably develop an entanglement monotone. As the state space $\s(\co^d)$ is convex and the trace operation continuous,  we may use a self-adjoint operator $W$ to  identify a hyperplane of density operators that bisect the state space.  This hyperplane is defined by states $\varrho$ such that 
\begin{equation}
\mathrm{tr}[W\varrho] = 0 \ .
\end{equation}
We may choose $W$ such that all separable states are on one side, and all states on the other are entangled \footnote{Of course, some entangled states could be on the same side of the hyperplane as the separable states, and as such that particular entanglement witness $W$ will not detect the entanglement of those states.}. Such a witness must satisfy the following:
\begin{enumerate}[(i)]
\item $W$ has at least one negative eigenvalue such that $W < \mathds{O}$
\item If $\rho$ is separable,  $\mathrm{tr}[W\rho]\geqslant 0$
 \end{enumerate}
  
Hence,  a state $\rho$ is \emph{witnessed} to be entangled if and only if $\mathrm{tr}[W\rho] < 0$.
We may use the Choi-Jamio\l kwoski isomorphism, relating a quantum operation $\e^B:\h^B \to \h^{B'}$, where the dimension of $\h^B$ and $\h^{B'}$ need not be the same, to the operator $\varrho(\e)^{A+B}$ on $\h^{A+B}$ according to the relation 
\begin{align}
\varrho(\e)^{A+B}=\mathds{1}^A\otimes \e^B\left[\Pi(\Phi^+)\right] \end{align}
to determine the nature of $W$. As  $W$ is negative, the operator related to it is not completely positive. For a positive but not completely positive map $P^B$ we have
\begin{align}
\mathrm{tr}[W\rho]&=\mathrm{tr}[\mathds{1}^A\otimes P^B(\Pi(\Phi^+))\rho] \nonumber \\
&= \mathrm{tr}\left[\sum_m\mathds{(1}^A\otimes K_m^{B})\Pi(\Phi^+)(\mathds{1}^A \otimes K_m^{B\dagger})\rho\right] \nonumber \\
&= \mathrm{tr}\left[\sum_m\mathds{(1}^A\otimes K_m^{B\dagger})\rho(\mathds{1}^A \otimes K_m^B)\Pi(\Phi^+)\right] \nonumber \\
&=  \mathrm{tr}[\mathds{1}^A\otimes P^{B\dagger}(\rho)\Pi(\Phi^+)] \nonumber \\ &=\langle \Phi^+|(\mathds{1}^A\otimes P^{B\dagger}[\rho] )\Phi^+\rangle  <0 \ \text{ only if} \ \rho \ \text{is entangled} .
\end{align} 
Here, $P^{B\dagger}$ is the dual map of $P^B$. 
Unfortunately, the positive but not completely positive maps are not well characterised for all dimensions, and determining the appropriate entanglement witness $W$ is not easy. However, for a bipartite system $\co^d$ with $d\leqslant 6$, i.e. for $\mathds{C}^2\otimes\mathds{C}^2$ or $\mathds{C}^2\otimes\mathds{C}^3$, a necessary and sufficient condition for entanglement is given by the positive partial transpose condition (PPT).  This follows from the fact that a positive but not completely positive map $P:\s(\mathds{C}^d)\to\s(\mathds{C}^{d'})$ where $d=2$ and $d'=2$ or $3$  can be written as $P=\e_1+\e_2\circ T$, with the completely positive quantum operations $\e_1$ and $\e_2$, and where $T$ is the transposition operator. \footnote{The transpose of $\rho$ gives $T[\rho]=\rho^T$ such that $\<\phi_i|\rho^T\phi_j\>=\<\phi_j|\rho\phi_i\>$ for all vectors in the orthonormal basis $\{\phi_i\}$.} As such, the non-positive entanglement witness is solely determined by the non-complete positivity of the     partial transposition.  This map is defined as 
\begin{align}
T^B[\cdot]:T^B[\rho^{A+B}]&\mapsto(\rho^{A+B})^{T_B} \nonumber \\
\langle \phi_i^A\otimes\varphi_i^B|(\rho^{A+B})^{T_B}\phi_j^A\otimes\varphi_j^B\rangle &:= \langle \phi_i^A\otimes\varphi_j^B|\rho^{A+B}\phi_j^A\otimes\varphi_i^B\rangle
\end{align}
for the orthonormal basis $\{\phi_i\}$ in $\h^A$ and $\{\varphi_i\}$ in $\h^B$. As can be determined easily, given a separable state, the partial transposition thereof gives a positive operator and hence a valid quantum state.\begin{equation}
T^B[\rho^A\otimes\rho^B]:=(\rho^A\otimes[\rho^B]^T)\ge \mathds{O}.
\end{equation}

This means that the separability of $\rho$ is sufficient for $\rho^{T_B}$ to be positive.  The converse claim, that the separability of $\rho$ is necessary for $\rho^{T_B}$ to be positive, is limited to the dimension of the composite system. According to the PPT criterion, the separability of  $\rho$ is necessary and sufficient for $\rho^{T_B}$ to be positive, given a composite system with $d \leqslant 6$.  For higher dimensions however, although the presence of negative eigenvalues for $\rho^{T_B}$ determines that $\rho$ is entangled, the positivity of this operator does not guarantee separability. As with the von Neumann entropy and concurrence, the PPT criterion determines entanglement or separability in a symmetric manner  for both $\rho^{T_B}$ as well as $\rho^{T_A}$. 
  
The \emph{negativity} \citep{vidal-2002-65} of a bipartite density operator gives an entanglement monotone based on the PPT criterion. As the partial transposition does not affect the trace of the density operator $\rho$, we can label the positive and negative eigenvalues of the matrix $\rho^{T_A}$ (or $\rho^{T_B}$) as $\lambda_i$ and $\mu_j$ respectively, such that they always add to one. The negativity can therefore be defined as the amount by which the absolute sum of the eigenvalues of $\rho^{T_A}$ differ from one.  

\begin{align}
{\cal{N}}[\rho^{A+B}]&:= \Vert (\rho^{A+B})^{T_A}\Vert_{\mathrm{tr}}-1 \nonumber \\ &=\sum_i \lambda_i +| \mu_i| -\lambda_i+|\mu_i| \nonumber \\
&=\sum_i2|\mu_i|.
\end{align}

  The maximum negativity of a system with Hilbert space of dimension $d_A\times d_B$ is $\min(d_A,d_B)-1$, and so we use the normalised version
\begin{equation}
{\cal{N}}[\rho^{A+B}]:=\frac{ \Vert (\rho^{A+B})^{T_A}\Vert_{\mathrm{tr}}-1}{\min(d_A,d_B)-1}.
\end{equation}
 The negativity and concurrence are related by the relation  \citep{concnegcomparison}
\begin{equation}
\sqrt{(1-{\cal C)}^2+{\cal C}^2}-(1-{\cal C})\leq {\cal N} \leq \cal C.
\end{equation}

\chapter{Von Neumann-L\"uders measurements with a Gaussian probe state}\label{von Neumann measurement Gaussian pointer}

When trying to measure a sharp observable on a finite dimensional Hilbert space, we need not use a finite dimensional probe Hilbert space. Indeed, one interesting example of the von Neumann-L\"uders measurement model uses a probe that exists in the infinite dimensional Hilbert space $\ell^2(\mathds{R})$. The probe can be acted on by  a coordinate observable $Q$, with eigenvalues $q$, and a conjugate momentum observable $P=-\imag \partial_q$, with eigenvalues $p$, which obey the canonical commutation relation $[Q,P]_-=\imag\mathds{1}$.    If the interaction Hamiltonian  contains one of  these observables acting on the probe, the change in the value of the conjugate observable reveals the eigenvalue of the observable we wished to measure.
  
Consider a sharp observable, given by the self-adjoint operator $O$, that we wish to measure on a pure state with an associated vector  $\varphi$. We add a probe state $\psi$ which, in the $Q$-representation, has the  Gaussian wavefunction 
\begin{align}
\psi(q) &=  \frac{1}{\sqrt{\Delta_Q \sqrt{2\pi}}}e^{-\frac{q^2}{4 \Delta_Q^2}}, \end{align}
and, in the $P$-representation, the Gaussian wavefunction
\begin{align}
\psi(p)&= \frac{1}{\sqrt{\Delta_P \sqrt{2\pi}}}e^{-\frac{ p^2}{4\Delta_P^2}}.
\end{align}
Utilising the relations 
\begin{align}
&\mathds{1}:=\int_{\mathds{R}}dq |q\>\<q|=\int_{\mathds{R}}dp |p\>\<p| &  \<p|q\>=\frac{1}{\sqrt{2\pi}}e^{-\imag pq},
\end{align}  we may map between  the $Q$-representation and  $P$-representation wavefunctions as
 \begin{align}
 \psi(p)&=\<p|\mathds{1} \psi\>=\int_\mathds{R}dq\<p|q\>\<q|\psi\>=\frac{1}{\sqrt{2\pi}}\int_\mathds{R}dq\psi(q)e^{-\imag pq} \nonumber \\
 \psi(q)&=\<q|\mathds{1} \psi\>=\int_\mathds{R}dp\<q|p\>\<p|\psi\>=\frac{1}{\sqrt{2\pi}}\int_\mathds{R}dp\psi(p)e^{\imag pq}
 \end{align}
 which are Fourier transforms.
It is simple to show that given a $Q$ standard deviation $\Delta_Q$, the $P$ standard deviation is $\Delta_P=1/(2\Delta_Q)$. 

The measurement Hamiltonian is of the form 
\begin{equation}
H=gf(t) O\otimes   P
\end{equation}
where $f(t)$ is  non vanishing only during the time of the measurement and is normalised such that $\int dt f(t) =1$. An example of such a function is a Dirac delta function.  $g$ is the strength of this interaction. We assume that either the system  Hamiltonian commutes with $O$  or else the measurement process is fast enough that the free evolution of the system may be neglected. 

Let $O$ have the eigenstates $\{\phi_{j}\}$ with the corresponding eigenvalues $\{a_j\}$ such that we may write the initial object state as   $\varphi=\sum_j \alpha_j\phi_{j}$.      The state transformation after the interaction is given as
\begin{align}
e^{-\imag g O\otimes P}\varphi\otimes\psi(p)&= \sum_j\alpha_j\phi_{j}\otimes  e^{-(\Delta_Q^2 p^2+\imag ga_jp)}   \end{align}
where we have omitted the normalisation constants of $\psi(p)$ for clarity. Applying a Fourier transformation on the probe wavefunction  gives the state transformation  in the  $Q$ representation as
\begin{align}
\sum_j \alpha_j\phi_{j}\otimes e^{-\frac{(q-ga_j)^2}{4 \Delta_Q^2}} =\sum_j \alpha_j\phi_{j}\otimes  \psi(q-ga_j) . \end{align} 
 The composite state has some entanglement, with the $Q$ value of the probe state correlated with the eigenvalue of $O$. For a measurement to be ideal in a coarse grained picture, such that measurement of the probe state unambiguously determines the eigenstate of the observable $O$, the overlap between the different $\psi(q-ga_j)$  must be brought arbitrarily low so they can be considered as \emph{effectively orthogonal} \footnote{Strictly speaking, as all states are Gaussian their tails will continue to infinity and hence they will never be exactly orthogonal. But we may consider them as affectively orthogonal in a coarse grained picture.}. 
 Therefore, we require $g$ to be strong compared to    $\Delta_Q$    such that  $\Delta_Q \ll g \Delta_O$, where $\Delta_O$ is the standard deviation in the eigenvalues of $O$.

The reduced state of the probe, after the measurement interaction, is then given by the statistical mixture 
\begin{equation}
\rho=\sum_j |\alpha_i|^2 \Pi(\psi(q-ga_j))  
\end{equation}
where $\Pi(\psi(q-ga_j))$ is the pure projector density operator of the probe. The probability distribution gained by the measurement of $Q$ on the probe will consist of widely separated Gaussian functions centered on the values $ga_j$.  Consequently, a single measurement will reveal which eigenstate the system has been measured to be in,  but generally\footnote{A single measurement can be used to accurately ascertain the measured eigenvalue in the limit of $\psi(q)$ being a delta function.} there will be an error in the evaluation of the eigenvalue due to the uncertainty $\Delta_Q$ for the Gaussian probe state. For either  $N$ runs of the experiment or a single run on an ensemble of $N$ identical systems (each with their own probe state) this error  scales as $1/\sqrt{N}$ due to the central limit theorem.     Because of the strong value of $g$, this ideal measurement of a sharp observable is also referred to as a \emph{strong measurement}. 

In the case of weak $g$ where $\Delta_Q \gg g\Delta_ O$, however, there will be a large overlap between the probe vectors. The resultant probability distribution for $Q$ can then be approximated by the first order expansion of the Taylor series to provide a single broad Gaussian centered on the expectation value of $O$ given by $\<O\>=\sum_{j=1}^d|\alpha_j|^2a_j$.
\begin{align}
P(q|\rho,Q)&=\frac{1}{\Delta_Q\sqrt{2\pi}}\sum_j|\alpha_j|^2e^{-\frac{(q-ga_j)^2}{2\Delta_Q^2}}\approx \frac{1}{\Delta_Q\sqrt{2\pi}} \sum_j|\alpha_j|^2\left(1-\frac{(q-ga_j)^2}{2\Delta_Q^2} \right) \nonumber \\ &= \frac{1}{\Delta_Q\sqrt{2\pi}} \left(1-\frac{q^2-2qg\<O\>+g^2\<O\>^2+g^2\Delta_ O^2}{2\Delta_Q^2} \right)\nonumber \\&\approx \frac{1}{\Delta_Q\sqrt{2\pi}} e^{-\frac{\left (q-g \<O\>\right)^2}{2\Delta_Q^2}}.
\end{align}

A single measurement of $Q$ in this case will give practically no information about the system;  with a large ensemble of $N$ identically prepared systems  we may only ascertain  the expectation value of $O$ with arbitrary accuracy, but we can not determine the eigenstate of $O$ the system has collapsed to.  
 This is usually referred to as a \emph{weak measurement}, introduced by \citep{weak-measurements-Aharanov-1988,weak-measurements-Aharanov-1990}.

 \spacing{1} 
  \bibliographystyle{plainnat}                                 
  \bibliography{references}

\end{document}